\documentclass[11pt,fleqn]{article}
\usepackage{amsfonts}
\usepackage{amsmath}
\usepackage{amssymb}
\usepackage{a4wide}
\usepackage{float}
\usepackage[multiple]{footmisc}
\usepackage{graphicx}
\usepackage{authblk}

\usepackage{lscape}
\usepackage{mathpazo}
\usepackage[hidelinks]{hyperref}
\usepackage{chngcntr}
\usepackage[round]{natbib}
\usepackage{rotating}
\numberwithin{equation}{section}
\usepackage{color}
\usepackage{setspace}
\usepackage{dcolumn}
\usepackage{booktabs}
\usepackage{diagbox}
\usepackage{subcaption}
\usepackage[ruled,vlined]{algorithm2e}
\allowdisplaybreaks

\usepackage{enumitem}
\usepackage{siunitx}[=v2] 
\newcolumntype{d}[1]{D{.}{.}{#1}}

\def\*#1{\mathbf{#1}}
\def\+#1{\boldsymbol{#1}}
\newcommand{\norm}[1]{\left\lVert#1\right\rVert}

\newtheorem{theorem}{Theorem}
\newtheorem{lemma}{Lemma}

\usepackage[margin=0.75in,paper=letterpaper]{geometry}
\begin{document}
\setlength{\baselineskip}{0.5cm}
\title{\bf Robust Tests for Factor-Augmented Regressions with an Application to the Novel EA-MD-QD Dataset   }
	{
  \author[1]{Alessandro Morico}
		\author[2]{Ovidijus Stauskas\thanks{Corresponding author: \hyperref[ovidijus.stauskas@bi.no]{ovidijus.stauskas@bi.no} (Nydalsveien 37, Oslo). We are grateful to the participants of seminars at BI Norwegian Business School, Norges Bank and CREST. In particular, Hilde Bjørnland, Emil Stoltenberg, Jonas Moss and Christian Gouriéroux and Knut Are Aastveit.}}
		\affil[1]{University of Bologna}
  \affil[2]{BI Norwegian Business School}
		\maketitle
	}
\maketitle

\begin{abstract}
\onehalfspacing
    We present four novel tests of equal predictive accuracy and encompassing à la \cite{pitarakis2023direct,pitarakis2025novel} for factor-augmented regressions, where factors are estimated using cross-section averages (CAs) of grouped series. Our inferential theory is asymptotically normal and robust to an overspecification of the number of factors. Our tests are empirically relevant as they accommodate for different degrees of predictor persistence and remain invariant to the location of structural breaks in the loadings. Monte Carlo simulations indicate that our tests exhibit excellent local power properties. Finally, we apply our tests to the novel EA-MD-QD dataset by \cite{EAdataset} - which covers the Euro Area as a whole and its primary member countries - and show that factors estimated by CAs offer substantial predictive power. 
\end{abstract}

\textbf{JEL Classification:} C12; C13; C33.

\textbf{Keywords:} Forecasting; factor-augmented regression model; common factor model; common correlated effects.
\vspace{1cm}
\onehalfspacing
\section{Introduction}
Nowadays, forecasting by using so-called diffusion indexes has become increasingly popular thanks to the growing availability of large datasets. In fact, this parsimonious approach allows forecasters to extract the predictive content of many potential predictors into a reduced number of indexes, also referred to as latent factors, when the data follows an approximate factor structure proposed by \cite{chamberlain1983funds}. Because the factors are unobserved, they must be estimated in a first step before augmenting the smaller forecasting model (possibly of autoregressive nature). The dominant method to estimate the factors is Principal Components (PC) (see e.g. \citealp{bai2002determining}). The simple yet powerful motivation behind this setup is that the estimated factors have been shown to enhance the accuracy of economic forecasts empirically (see \citealp{stock1999forecasting}; \citealp{forni2003financial}; \citealp{Ludvigson_Ng_2011} or \citealp{ciccarelli2010global} from an extensive literature). Although the rich content of predictive information embedded in the factors represents an opportunity to make more accurate forecasts, it is essential to formally evaluate their predictive power. Fundamentally, the approach entails (pseudo) out-of-sample forecast evaluation between the factor-augmented model and its nested baseline. \\
\indent In this article, we build on recent advances in the forecasting literature: both methodological and data advancements. We develop novel tests of forecast accuracy and encompassing for nested regressions with latent factors. Our contribution is to extend the innovative and extremely useful inferential theory for pseudo out-of-sample comparisons of \cite{pitarakis2023direct,pitarakis2025novel} to the case of generated regressors, where asymptotic normality of the proposed statistics holds under many empirical regularities: 1) uncertainty about the number of factors, 2) presence of structural breaks, or 3) persistent series. We achieve this by revisiting the methodology by \cite{karabiyik2021forecasting} and \cite{stauskas2022tests} who were the first to abandon the usual PC and instead adopted cross-section averages (CAs) of grouped series - as first proposed by \cite{pesaran2006estimation} - in a forecast comparison setup.\footnote{Note that \cite{pesaran2006estimation} proposed the Common Correlated Effects (CCE) estimator for the parameters of the panel linear regression model, where CAs constitute only its first step to proxy factors. In our time series setting, we only rely on the CAs.} In contrast to extracting the factors as eigenvectors corresponding to the largest eigenvalues of the covariance matrix of the data, CAs average the noise and settle on the so-called low-rank component of the data, which are the factors identified up to a linear transformation. \\
\indent The notion of CAs comes naturally in practice, as it is becoming progressively standard to organize the predictors into large blocks of variables, such as consumption, money aggregates, prices or exchange rates. In fact, \cite{eickmeier2008successful} consider 52 studies that use factor-augmented regression models to forecast inflation and/or output, and the datasets used in these studies all have the same structure of blocks of variables. Two notable examples of such structure are the FRED-MD (monthly) database by \cite{mccracken2016fred} and its quarterly counterpart FRED-QD (see \citealp{mccracken2020fred}). \cite{EAdataset} recently presented their first European counterpart - namely the EA-MD-QD dataset - which covers the Euro Area as a whole and the main member countries with 100+ series organized in 11 blocks. As such datasets offer a large number of estimated factors and cover a long time horizon, a robust procedure to evaluate their predictive content becomes key. \\
\indent The CAs procedure in relation to the properties of the statistics of \cite{pitarakis2023direct,pitarakis2025novel} brings advantages relative to PC. First, as averages are taken at each point in time, factors can be integrated with their order being unknown, while such information is integral to the optimization procedure in PC (see \citealp{bai2004estimating}); second, CAs can be combined with structural breaks in factor loadings which can occur in the out-of-sample period, because a single break in loadings is algebraically equivalent to doubling the number of factors.; third, if the number of CAs is greater than the number of factors, controlling for overspecification in the parameter estimation step is simpler and does not require restrictions on the dimensions of the dataset or restrictive normality assumptions on model errors (see \citealp{moon2015linear}). Importantly, we do not require that a single block loads on a single factor only. Indeed, each block can load on all the factors in the system, or many blocks may not load on any factors at all, as long as the average loading matrix has a rank equal to the true number of factors. This relates to and incorporates the notion of local factors, which affect only a subset of series, in the PC setting (see \citealp{freyaldenhoven2022factor}). Finally, missing values are not a problem for CAs as they can still be easily computed in unbalanced panels (see \citealp{chudik2015large}). On the other hand, such case is difficult in PC as it should be coupled with an alternating maximization algorithm (see \citealp{bai2009panel}, or \citealp{su2025estimation}).  \\
\indent The studies of \cite{stauskas2022tests} and \cite{gonccalves2017tests} (in the PC setting) explore the statistics of \cite{clark2001tests} and \cite{mccracken2007asymptotics}, which solve the degeneracy problem faced by the seminal tests of \cite{west1996asymptotic} in the nested environment (see also a survey in \citealp{diebold2015comparing}). However, the solution comes at the expense of highly non-standard limiting distributions based on functionals of stochastic integrals of Brownian motion and the relative growth rate of in-sample versus out-of-sample observations. Although asymptotically valid critical values can be obtained by simulation (see \citealp{clark2012reality}; or \citealp{hansen2015equivalence}), their practical implementation remains challenging. \\
\indent The encompassing and forecast accuracy tests in \cite{pitarakis2023direct,pitarakis2025novel} have two key features: 1) they put an end to the joint problem of variance degeneracy and nonstandardness of asymptotics in nested models, and asymptotically follow a standard normal distribution, and 2) the observed predictors can be non-stationary. Therefore, we naturally fill in the gap in the literature by allowing CAs-generated regressors. If combined with the framework of \cite{pitarakis2023direct,pitarakis2025novel}, the mechanics of CAs solve many issues that arise in \cite{stauskas2022tests} and \cite{gonccalves2017tests}. In stark contrast to all existing tests of equal accuracy for factor-augmented regressions, our user-friendly theoretical framework can flexibly accommodate for: 1) different degrees of persistence in the regressors, 2) structural breaks in the factor loadings in the full sample and, most importantly, 3) overestimation of the number of factors. To the best of our knowledge, this is the first result in the CAs literature where an overestimation of the number of factors does not have any asymptotic effect, irrespective of the expansion rates of $N$ and $T$. In spirit, this result resembles \cite{moon2015linear} in the PC context. Furthermore, extensive Monte Carlo experiments on size and local power exhibit excellent finite sample properties of our tests under a plethora of realistic economic scenarios, thus making our methodology relevant and applicable in many empirical applications. Ultimately, this study is the first to examine the predictive capacity of factors in the EA-MD-QD dataset by \cite{EAdataset} in a nested setup. We show that they offer powerful predictive insight in the context of macroeconomic forecasting especially in the presence of large shocks such as the Covid pandemic. \\
\indent We also highlight that this work goes in parallel to, and draws some interim results from, the PC-based theory of \cite{margaritella2024new} for the same statistics. The latter study is the first in the PC setting, and it utilizes the fact that the method can be applied to any dataset with large dimensions. The current study, on the other hand, leverages the block structure that is natural in large macroeconomic datasets. This suggests a simpler estimator with empirically relevant properties, which are not typically possessed by PC.   \\
\indent The rest of the paper is organised as follows. Section 2 introduces our forecasting setup, the CCE approach and the tests under the observed predictors together with a list of assumptions. Section 3 dives into the CCE-based statistics and presents the main results. In Section 4, we conduct an extensive Monte Carlo study. The notation adopted in the paper is as follows. Scalars, vectors and matrices are denoted by $a \in \mathbb{R}$, $\*a\in \mathbb{R}^p$ and $\* A\in \mathbb{R}^{c \times p}$, respectively. For any matrix $\* A$, the Frobenius norm is defined as $\left\|\*A\right\| = \sqrt{\textrm{tr}(\* A'\* A)}$ where $\mathrm{tr}(.)$ is the trace operator. Moreover, $\mathrm{rk}(\*A)$ and $\*A^+$ represent rank and Moore-Penrose inverse, respectively, while $\lambda_{\mathrm{min}/\mathrm{max}}(\*A)$ is the minimum/maximum eigenvalue of $\*A$. Next, $\lfloor x \rfloor$ represents the integer part of $x$, $\sup$ denotes the supremum, $k_0$ represents the number of in-sample observations, while $T - k_0 = n$ is the number of the out-of-sample observations. With respect to the notion of convergence, $\to$ and $\to_p$ denote the limit and convergence in probability, respectively.

\section{Econometric Setup}\label{sec_2}
Consider a vector of predictors $\*z_t=(\*w_t',\*f_t')'\in \mathbb{R}^{q+r}$ such that we have the following forecasting models:
\begin{align}
&y_{t+1}=\+\theta'\*w_t+u_{1,t+1}, \label{model1}\\
    &y_{t+1}=\+\theta'\*w_t+\+\alpha'\*f_t+u_{2,t+1}=\+\delta'\*z_t+u_{2,t+1},  \label{model2}
\end{align}
where $\*w_t$ represents a small set of \textit{known} predictors. While we may choose the observables $\*w_t$ based on economic theory or our experience as forecasters, it is important to remember that nowadays we face an abundance of potential predictors that may or may not improve our forecasts. In order to exploit the information embedded in large dimensional datasets, one could extract the potential predictive content of the many regressors into just a few series of latent factors $\*f_t$. This situation is reflected in (\ref{model2}). Because $\*f_t$ is typically unobserved, we define the \textit{infeasible} least squares (LS) estimator of $\+\delta$ for each $t=1,\ldots,T$ as 
\begin{align*}
\widetilde{\+\delta}_t=\left(\sum_{s=1}^{t-1}\*z_s\*z_s' \right)^{-1}\sum_{s=1}^{t-1}\*z_s y_{s+1},
\end{align*}
where the out-of-sample forecast error is defined as $\widetilde{u}_{2,t+1}=y_{t+1}-\widetilde{\+\delta}_t'\*z_t$ ($\widetilde{\+\theta}_t$ and $\widetilde{u}_{1,t+1}$ are defined analogously).
Throughout, $\*z_t$ can be stationary or  non-stationary. We are not interested in specific time series properties of $\*z_t$, but we want to vary its degree of persistence and its effect on the forecast comparison problem. Large databases, such as FRED-MD or EA-QD-MD, offer ``recipes'' for how to standardise the long series (see e.g. Appendix A in \citealp{EAdataset}). Some persistence may remain and this can be flexibly modeled by moderate integration in the spirit of \cite{magdalinos2009limit}, which we formalize in our assumptions. \\
\indent To extract $\*f_t$, we assume that our panel of $M$ predictors can be divided into a finite number $m$ of blocks exogenously. The economic rationale underlying this approach lies in the idea that variables with similar economic features are expected to be driven fundamentally by a series of (possibly unobserved) common shocks (see e.g. \citealp{hallin2011dynamic}; \citealp{moench2013dynamic}; \citealp{ando2016panel}) and, as a result, are to be grouped into one block. The FRED-MD database of the Federal Reserve (see \cite{Ludvigson_Ng_2011} and \cite{mccracken2016fred} for more details) is among the primary examples, where the series are grouped in 8 blocks. The new EA-MD-QD dataset by \cite{EAdataset} contains 11 blocks: (1) National Accounts, (2) Labor Market Indicators, (3) Credit aggregates, (4) Labor Costs, (5) Exchange Rates, (6) Interest Rates (7) Industrial Production and Turnover, (8) Prices, (9) Confidence Indicators, (10) Monetary Aggregates and (11) Others. It is then a reasonable to assume that every series in each block is, due to their very characteristics, sensitive to the same changes in the economic environment. \\
 \indent To reflect such data configuration formally, let $I_j = \{1_j, \dots, N_j\}$ be the set of indexes of the series contained in each block for $j = 1, \dots, m$. We denote the predictor panel data variable as $\* x_{i,t} = [x_{i_1,t}, \dots, x_{i_m,t}]^\prime \in \mathbb{R}^m$, where $x_{i_j,t}$ can be seen as the predictor $i$ of the block $j$. That is, we can allow for blocks of different sizes. For simplicity of notation and without loss of generality, let us assume that every block has the same number of predictors $N = M/m$. The data generating process of $\*x_{i,t}$ is assumed to be
\begin{align} \label{eq: factor model}
\* x_{i,t} = \boldsymbol{\Lambda}_i^\prime \* f_t + \* e_{i,t}, \quad t = 1, \dots, T,
\end{align}
where $\* f_t \in \mathbb{R}^r$ is a vector of common factors, $\boldsymbol{\Lambda}_i \in \mathbb{R}^{r \times m}$ is the matrix of factor loadings, and $\* e_{i,t}\in \mathbb{R}^{m}$ are the idiosyncratic components. In this work, we estimate the latent factors using CAs as suggested in \cite{pesaran2006estimation}: 
\begin{align}\label{CAs}
\widehat{\*f}_t = \overline{\* x}_t=\overline{\+\Lambda}'\*f_t+\overline{\*e}_t,
\end{align}
where $\overline{\* A} = N^{-1} \sum_{i=1}^N \* A_i$ denotes an average for any matrix $\* A_i$. Note that for every fixed $t=1,\ldots, T$, we have $\left\|\overline{\*e}_t\right\|=O_p(N^{-1/2})$ under many empirically relevant assumptions (see e.g. \citealp{pesaran2011large}), and so the  factors are identified up to a linear transformation, which is $\overline{\+\Lambda}$. Therefore, given that $\mathrm{rk}(\overline{\+\Lambda})=r$, we obtain 
\begin{align}\label{MP_solved}
    \*f_t= (\overline{\+\Lambda}')^+\widehat{\*f}_t+O_p(N^{-1/2}),
\end{align}
which justifies using averages to approximate the factors up to a linear transformation. Clearly, if $m=r$, then $\overline{\+\Lambda}^+=\overline{\+\Lambda}^{-1}$, which is equivalent to knowing the number of factors $r$.  Therefore, the overall set of \textit{feasible} predictors is $\widehat{\*z}_t=(\*w_t',\widehat{\*f}_t')'\in \mathbb{R}^{q+m}$, meaning that the feasible estimator is
\begin{align}\label{feas_estm}
\widehat{\+\delta}_t=\left(\sum_{s=1}^{t-1}\widehat{\*z}_s\widehat{\*z}_s' \right)^{-1}\sum_{s=1}^{t-1}\widehat{\*z}_s y_{s+1},
\end{align}
where $\widehat{u}_{2,t+1}=y_{t+1}-\widehat{\+\delta}_t'\widehat{\*z}_t$. Notice how (\ref{MP_solved}) does not restrict one factor to one block, but it is allowed. Similarly, each block can load on all $r$ factors. In fact, we can have $\overline{\+\Lambda}=\left[\overline{\+\Lambda}_r,\+0_{r\times (m-r)} \right]$, so that $m-r$ blocks are driven purely by idiosyncratics. As long as $\mathrm{rk}(\overline{\+\Lambda}_r)=r$, (\ref{MP_solved}) holds and we have $r$ ``local'' factors that underlie only a subset of blocks. Therefore, CAs seamlessly accommodate the notion of local factors from the PC setting (see \citealp{freyaldenhoven2022factor}).   \\
\indent Note that $\left\|\overline{\*e}_t\right\|=O_p(N^{-1/2})$ implies that $\frac{1}{T}\sum_{t=1}^T\widehat{\*f}_t\widehat{\*f}_t'$ is asymptotically singular $\overline{\+\Lambda}'\left(\frac{1}{T}\sum_{t=1}^T\*f_t\*f_t'\right)\overline{\+\Lambda}$ is, unless $m=r$. Consider the case when $m>r$, which is equivalent to overspecifying the number of factors. Indeed, based on the discussion above, EA-MD-QD would provide us with $m=11$ estimated factors ($m=8$ in FRED-MD), while the average number detected in the literature is smaller.\footnote{For instance, in the case of FRED-MD, \cite{stauskas2022tests} estimate $r=2$ by using the criterion of \cite{ahn2013eigenvalue}. \cite{margaritella2024new} estimate 7 on average by using less conservative criteria from \cite{bai2002determining}.} In order to proceed with the analysis of (\ref{feas_estm}), we need to re-define our target object since we estimate $\*f^0_t=[\*f_t', \*0_{1\times(m-r)}]'$. Hence, similarly to \cite{stauskas2022tests}, we introduce a so-called \textit{rotation} matrix $\overline{\*H}\in \mathbb{R}^{m\times m}$, such that
\begin{align}
\overline{\*H}'\widehat{\*f}_t=\overline{\*H}'\overline{\+\Lambda}'\*f_t+\overline{\*H}'\overline{\*e}_t=\*f_t^0+\overline{\*H}'\overline{\*e}_t. 
\end{align}
Next, we let $\*D_N=\mathrm{diag}(\*I_r, \sqrt{N}\*I_{m-r})$, such that 
\begin{align}\label{f0}
\widehat{\*f}_t^0=\*D_N\overline{\*H}'\widehat{\*f}_t=\*D_N\overline{\*H}'\overline{\+\Lambda}'\*f_t+\*D_N\overline{\*H}'\overline{\*e}_t=\*f^0_t+\overline{\*e}^0_t.
\end{align}
The \textit{transformed} error is $\overline{\mathbf{e}}^0_t=[\overline{\mathbf{e}}^{0\prime}_{r,t}, \overline{\mathbf{e}}^{0\prime}_{-r,t}]'$, where $\left\|\overline{\mathbf{e}}^{0}_{r,t} \right\|=O_p(N^{-1/2})$, but $\left\|\overline{\mathbf{e}}^{0}_{-r,t} \right\|=O_p(1)$. This means that $\left(\frac{1}{T}\sum_{t=1}^T\widehat{\*f}_t^{0}\widehat{\*f}_t^{0\prime}\right)^{-1}$ exists asymptotically. Note that if $m=r$, then $\overline{\*H}=\overline{\+\Lambda}^{-1}$ and $\*D_N=\*I_r$, and so (\ref{f0}) reduces to $\*f_t+\overline{\+\Lambda}^{-1\prime}\overline{\*e}_t$. We leave the definition of $\overline{\*H}$ in the Supplement. \\
\indent We stress that $\overline{\*H}$ serves in (\ref{f0}) as a purely analytical device, which exists as long as $\overline{\+\Lambda}$ does. It provides means to track the contribution of the redundant $m-r$ averages on $\widehat{\+\delta}_t$ and, in turn, $\widehat{u}_{2,t+1}$, which is the key component in the tests. Importantly, we do not need to know or apply $\overline{\*H}$ in practice. Note that by applying the Frisch-Waugh-Lovell Theorem to the second coordinate of (\ref{feas_estm}) (in stacked notation, where we suppress dependence on $t$), we obtain the following: 
\begin{align}
   \widehat{\+\theta}_t=(\*W'\*M_{\widehat{\*F}}\*W)^{-1}\*W'\*M_{\widehat{\*F}}\*y=(\*W'\*M_{\widehat{\*F}^0}\*W)^{-1}\*W'\*M_{\widehat{\*F}^0}\*y,
\end{align}
where $\*M_\*A=\*I-\*A(\*A'\*A)^{-1}\*A'=\*I-\*A\*B(\*B'\*A'\*A\*B)^{-1}\*B'\*A'=\*M_{\*A\*B}$ is a projection matrix onto the orthogonal complement of $\*A$ or $\*A\*B$ for a full rank $\*B$ ($=\overline{\*H}\*D_N$). It is essential that projecting onto $\widehat{\*f}_t$ is equivalent to projecting onto $\widehat{\*f}_t^0$ since $\overline{\*H}\*D_N$ is a full-rank matrix. \\
\indent In this step, it is natural to compare the current study with \cite{margaritella2024new} in the PC setting. Although PC and CAs have rather distinct theoretical bases, CAs have technical advantages beyond those differences. Firstly, as seen from (\ref{f0}), the overestimation of $r$ is embedded in our analysis. We pursue this route instead of assuming the existence of a consistent estimator of $r$ (the usual step in the PC setting; see e.g. \citealp{bai2006confidence}) to carefully characterize the asymptotic behavior of $m-r$ redundant CAs. Moreover, we effectively work with the predictors of the mixed integration order, which can be handled by CAs (see \citealp{westerlund2018cce}). Although $\*z_t$ is modeled as a mildly integrated process, $\overline{\*e}_{-r,t}^0$ is still stationary. Consequently, redundant $m-r$ CAs can be seen as additional predictors, whose asymptotic contribution needs to be handled very carefully given the overall non-stationary environment. Note that a mixed order of predictors is not allowed in the original studies of \cite{pitarakis2023direct} and \cite{pitarakis2025novel}, but we show that this is not necessary in our problem. In the next section, we introduce the tests under $\*z_t$, and then give their representations under $\widehat{\*z}_t$ in Section 3. 
\subsection{Test Statistics in Nested Regressions (Observed Factors)}
The goal of our forecast accuracy tests is to assess which of the two models is more informative in predicting the dependent variable: if $\+\alpha=\*0_{r \times 1}$, then $\*f_t$ does not have any predictive power towards $y_{t+1}$ and it is sufficient to make forecasts with the model in (\ref{model1}). This also means that under the null of $\+\alpha=\*0_{r \times 1}$, the model in (\ref{model2}) nests the ''smaller'' model in (\ref{model1}) so that $u_{1,t+1}=u_{2,t+1}$. Such null implies that $\mathbb{E}(u_{1,t+1}(u_{1,t+1}-u_{2,t+1}))=0$. The latter equality can be tested in numerous ways, where one of the most popular tests is given by \cite{clark2001tests}, which relies on the \textit{estimated} forecast errors $\widetilde{u}_{j,t+1}$ for $j\in \{1,2\}$. They are based on the \textit{recursive} estimators $\widetilde{\+\theta}_t$ and $\widetilde{\+\delta}_t$. The test has a highly non-standard asymptotic distribution, which also depends on tuning parameters. \cite{pitarakis2023direct}, on the other hand, proposed a modification that results in an asymptotically normal distribution free of nuisance/tuning parameters. To compare the forecasting performance of two models in a nested environment, we propose four new tests of forecast nesting and accuracy by extending the work of \cite{pitarakis2023direct} and \cite{pitarakis2025novel} to the case of predictive regressions with generated regressors estimated by CAs. Firstly, let us consider the test for forecast nesting from \cite{pitarakis2023direct}. Assuming the factors are observed, the test statistic for forecast nesting from \cite{pitarakis2023direct} is given by
 \begin{align} \label{eq: PIT}
     \hspace{-1.0cm} s_{f,1}(m_0)=\frac{1}{\widetilde{\omega}_1}\left(\frac{1}{\sqrt{n}}\sum_{t=k_0}^{T-1}\widetilde{u}^2_{1,t+1}-\frac{1}{2}\left[\frac{n}{m_0}\frac{1}{\sqrt{n}}\sum_{t=k_0}^{k_0+m_0-1}\widetilde{u}_{1,t+1}\widetilde{u}_{2,t+1} + \frac{n}{n-m_0}\frac{1}{\sqrt{n}}\sum_{t=k_0+m_0}^{T-1}\widetilde{u}_{1,t+1}\widetilde{u}_{2,t+1} \right] \right)
 \end{align}
 where $m_0=\lfloor n \mu_0\rfloor = \lfloor (T-k_0) \mu_0\rfloor$ is a cut-off point to split the average for $\mu_0\in (0,1)$, $\mu_0\neq 1/2$, and $\widetilde{\omega}_1^2$ is the estimated variance of the limiting distribution. As shown in \cite{pitarakis2023direct}, values of $\mu_0$ closer to 1/2 yield better local power in finite sample, so we set it equal to 0.45 hereafter. Still, the true factors are not observed and must be estimated instead, so the subscript ''$f$'' indicates that the test is \textit{infeasible}. Secondly, \cite{pitarakis2025novel} proposed two additional tests of forecasting accuracy whose corresponding infeasible statistics are: 
\begin{align}\label{eq: ET1}
     &s_{f,2}(l^0_1, l^0_2)=\frac{1}{\widetilde{\omega}_2}\frac{n}{l^0_1}\left(\frac{1}{\sqrt{n}}\sum_{t=k_0}^{k_0+l^0_1-1}\widetilde{u}^2_{1,t+1}-\frac{l^0_1}{l^0_2}\frac{1}{\sqrt{n}}\sum_{t=k_0}^{k_0+l^0_2-1}\widetilde{u}^2_{2,t+1} \right)\\
 &s_{f,3}(\nu_0,\lambda^0_2)=\frac{1}{\widetilde{\omega}_3}\frac{1}{n(1-\nu_0)}\sum_{l_1=\lfloor n\nu_0 \rfloor+1}^n   s_{f,2}(l_1, \lfloor n\lambda^0_2 \rfloor )\label{eq: ET2}.
 \end{align}
 where $l^0_j=\lfloor n\lambda_j^0 \rfloor$ for $j=1,2$, with $\lambda _j^0\in (0,1)$ representing two portions of the out-of-sample period. That is, the  forecast MSE loss differentials of both models are computed over partially overlapping out-of-sample segments, i.e. $l^0_1>l_2^0$ or vice-versa, and $\widetilde{\omega}_j^2$ for $j=2,3$ are the corresponding variance estimators. Note that if the two segments were fully overlapping then their variance would be asymptotically degenerate as in \cite{clark2001tests}, so $l^0_1=l_2^0$ is ruled out to avoid incurring into this exact issue. Further, note from the formulation of (\ref{eq: ET2}) that, once $l_2^0$ (or $\lambda_2^0$) is fixed, new test statistics can be obtained by averaging (\ref{eq: ET1}) over some chosen feasible set of $l_1$ based on the tuning parameter $\nu_0\in (0,1)$. As shown by \cite{margaritella2024new}, among other averaging possibilities is also a fourth statistic of the form
\begin{align}\label{eq: ET3}
 s_{f,4}(\lambda^0_1,\nu_0)=\frac{1}{\widetilde{\omega}}_{4}\frac{1}{n(1-\nu_0)}\sum_{l_2=\lfloor n\nu_0 \rfloor+1}^n   s_{f,2}(\lfloor n\lambda^0_1 \rfloor, l_2),
 \end{align}
 which is obtained by fixing $l_1^0$ (or $\lambda_1^0$) and averaging (\ref{eq: ET1}) over some chosen feasible set of $l_2$. Notice that \cite{pitarakis2025novel} provide recommendations for $l_1^0$, $l_2^0$ and $\nu_0$. Finally, \cite{margaritella2024new} also show that $\widetilde{\omega}^2_j$ for $j \in \{1,2,3,4\}$ can be written as
  \begin{align} \label{conventional_estm}
     \widetilde{\omega}_j^2 \propto \widetilde{\phi}^2 =  \frac{1}{n} \sum_{t=1}^{n} \left(\widetilde{u}_{2,t+1}^2 - \frac{1}{n} \sum_{t=1}^{n}\widetilde{u}_{2,t+1}^2\right)^2
 \end{align}
 up to a known constant specific to each statistic. In all four cases, we test the null against the one-sided alternative since $s_{f,j}\to_d \mathcal{N}(0,1)+c$ for $c>0$ that depends on the model parameters. 

\subsection{Assumptions} \label{Assumptions}
Throughout our analysis, we employ the following set of assumptions. \\

\noindent \textbf{Assumption 1.} $\mathbb{E}(u_{t+1}|\mathcal{F}_t)=0$, $\mathbb{E}(u^2_{t+1}|\mathcal{F}_t)=\sigma^2$, and $\mathbb{E}(u^4_{t+1}) < \infty$ for all $t$, where $\mathcal{F}_t$ is the sigma-algebra generated by $\{\mathbf{z}_t,\mathbf{z}_{t-1},...,\*X_t, \*X_{t-1},\ldots, y_t, y_{t-1},\ldots\}$.
\bigskip 

\noindent \textbf{Assumption 2.} \begin{itemize}
    \item[(a)] $\{\*z_t \}$ is a mildly integrated process as defined in \cite{magdalinos2009limit}. In particular,
\begin{align*}
   \*z_t = \*R_{z,T}\*z_{t-1} + \*u_{z,t}, \quad \*R_{z,T} = \*I_{q+r} + \*CT^{-\tau}, \quad \tau \in (0,1), \quad \*C <0 \hspace{2mm} (diagonal),
\end{align*}
where $\*u_{z,t}$ is a zero-mean linear process. 
\item[(b)] If $\tau=0$, then $|\lambda_{\mathrm{min}}(\*C)|<2$ in order to ensure that $\*R_z=\*I_{q+r}+\*C$ is inside the unit circle, and $\*u_{z,t}$ is such that $\{\*z_t\}$ has absolute summable autocovariances.
 \item[(c)] For a given $\tau \in [0,1)$, some $\kappa \in (0,1)$ and $\boldsymbol{\Sigma}_{ZZ} = \left[\begin{array}{cc}\boldsymbol{\Sigma}_{ff} & \boldsymbol{\Sigma}_{wf}'\\
   \boldsymbol{ \Sigma}_{wf} & \boldsymbol{\Sigma}_{ww} \end{array}\right]$, we have as $T\to \infty$
 \begin{align*}
   \sup_{\kappa \in (0,1)}\left\|\frac{1}{T^{1+\tau}}\sum_{s=1}^{\lfloor \kappa T\rfloor}\*z_s\*z_s' - \kappa^{1+\tau} \+\Sigma_{ZZ} \right\|=o_p(1),  \quad\sup_{\kappa\in (0,1)}\left\|\frac{1}{T^{(1+\tau)/2}}\sum_{s=1}^{\lfloor\kappa T\rfloor}\*z_su_{s+1} \right\|=O_p(1), 
\end{align*}
\end{itemize}
\bigskip
\noindent \textbf{Assumption 3.}
\begin{itemize}
  \item[(a)] 
  \begin{itemize}
      \item[(i)]  If $\tau \in (0,1)$, then $\{\*e_{i,t}\}$ is uncorrelated over time with $\mathbb{E}(\*e_{i,t})=\*0_{m\times 1}$, $\mathbb{E}(\*e_{i,t}\*e_{i,t}')=\+\Sigma_{ee,i,t}$ positive definite and $\mathbb{E}(\|\*e_{i,t} \|^4)<\infty$. 
      \item[(ii)]   If $\tau =0$, then we let $\*e_{i,t}= \mathbf{C}_i(L)\boldsymbol{\epsilon}_{i,t} = \sum_{j=0}^{\infty}\mathbf{C}_{i,j}\boldsymbol{\epsilon}_{i,t-j}$, where $\boldsymbol{\epsilon}_{i,t}$ is independent across $t$ with $\mathbb{E}(\boldsymbol{\epsilon}_{i,t})=\mathbf{0}_{m\times 1}$, $\mathbb{E}(\boldsymbol{\epsilon}_{i,t}\boldsymbol{\epsilon}_{i,t}') =\boldsymbol{\Sigma}_{\epsilon\epsilon,i,t}$ positive definite,  $\mathbb{E}(\|\boldsymbol{\epsilon}_{i,t}\|^4)<\infty$, and $\sum_{j=0}^{\infty}j^{1/2}\| \mathbf{C}_{i,j} \|<\infty$. Also, $\+\xi_t=\mathrm{vec}\left(N\overline{\*e}_t\overline{\*e}_t'-\mathbb{E}\left(N\overline{\*e}_t\overline{\*e}_t' \right) \right)$ is strong mixing with coefficients of size $-bd(b-d)$ with $b>4$ and $b>d>2$, $\mathbb{E}(\|\+\xi_t\|^b)<\infty$, and $\lim_{T\to\infty}T^{-1}\sum_{t=1}^{T}\sum_{s=1}^{T}\mathbb{E}(\+\xi_t\+\xi_s')$ is positive definite.
  \end{itemize}

  \item[(b)] For $\tau \in [0,1)$ and $\eta \in (0,1)$, we have the following limiting behavior as $(N,T) \to \infty$:
\begin{align*}
\hspace{-1.4cm} \sup_{\eta \in (0,1)}\left\|\frac{1}{NT}\sum_{i=1}^N\sum_{j=1}^N\sum_{k=1}^{\lfloor \eta T\rfloor}\mathbb{E}(\*e_{i,k}\*e_{
j,k}') - \eta\+\Sigma_{ee}\right\|=o(1), \quad \sup_{\eta \in (0,1)}\left\|\frac{1}{NT^{(1+\tau)/2}}\sum_{i=1}^N\sum_{s=1}^{\lfloor\eta T \rfloor}\*z_s\*e'_{i,s} \right\|=O_p(1)
\end{align*}
for a positive definite matrix $\+\Sigma_{ee}$.
\end{itemize}

\bigskip

\noindent \textbf{Assumption 4.} $\mathbf{z}_t$ and $u_{t}$ are independent of $\mathbf{e}_{i,s}$ for all $t$, $s$ and $i$.

\bigskip

\noindent \textbf{Assumption 5.} $\boldsymbol{\Lambda}_i$ is a non-random matrix, such that $\overline{\boldsymbol{\Lambda}}\to \boldsymbol{\Lambda}$ as $N\to \infty$ and $\overline{\boldsymbol{\Lambda}}=[\overline{\boldsymbol{\Lambda}}_r, \overline{\boldsymbol{\Lambda}}_{-r}]$, where $\overline{\boldsymbol{\Lambda}}_{-r}\in \mathbb{R}^{r\times (m-r)}$ and $\overline{\boldsymbol{\Lambda}}_r\in\mathbb{R}^{r \times r}$ is full rank for all $N$, including $N\to \infty$, and $\|\boldsymbol{\Lambda}_i\|<\infty$. If $m=r$, then $\overline{\boldsymbol{\Lambda}}=\overline{\boldsymbol{\Lambda}}_r$.
\bigskip

\noindent \textbf{Assumption 6.} $\boldsymbol{\alpha} = T^{-1/4-\tau/2}\boldsymbol{\alpha}^0$, where $\|\boldsymbol{\alpha}^0\|<\infty$. \\

\noindent Assumption 1 is standard in the literature and imposes a martingale difference sequence (MDS). It implies one-step-ahead forecasts (see e.g. \citealp{campbell2006efficient}, \citealp{hjalmarsson2010predicting}, or \citealp{breitung2015instrumental}). While some tests in \cite{pitarakis2023direct, pitarakis2025novel} and allow for $h$-step-ahead forecasts, we focus on the smallest set of conditions to accommodate all of them. However, we  explore the theory behind $h$-steps, and explain the extension in Remark 1. The effect of general serial correlation in $u_{t}$ is explored in simulations, where the power remains for low levels of correlation. Assumption 2 (a) provides comparative statics, as $\tau \to 1$ increases the persistence of the process. \cite{magdalinos2009limit} allow us to manipulate $\tau$ and give theoretical guarantees, such as $\frac{1}{T^{1+\tau}}\sum_{t=1}^T\*f_t\*f_t'\to_p\+\Sigma_{\*F}$ (constant positive definite matrix). The convenient feature of this specification is that the discussion below (\ref{feas_estm}) and (\ref{f0}) still holds. Part (b) ensures that we return to the usual stationarity conditions under $\tau=0$. Part (c) regulates the behavior of a recursive LS estimator under mildly integrated predictors. \\
\indent  Assumption 3 (a)  reveals that we cannot maintain an \textit{arbitrary} correlation of idiosyncratics and persistent predictors simultaneously. In the literature of CAs and PC under $\tau=0$, we require that terms of the form of $\frac{1}{T}\sum_{t=1}^T\*z_t\*e_{i,t}'$ vanish for all $i$ in order to demonstrate that the factor estimation error is negligible. When $\tau >0$ and $\{\*e_{i,t}\}_{t=1}^T$ is serially correlated, the rate is too slow, and this problem will remain in the recursive estimation. We have that 
 $\left\|\frac{1}{T^{1+\tau}}\sum_{t=1}^T\*z_t\*e_{i,t}'\right\|=O_p(T^{-\tau/2})$ (see Lemma 3.1 in \citealp{magdalinos2009limit}), but if $\{\*e_{i,t}\}_{t=1}^T$ follows an MDS process or a moving average (MA) process of finite order, then $\left\|\frac{1}{T^{1+\tau}}\sum_{t=1}^T\*z_t\*e_{i,t}'\right\|=O_p(T^{-(1+\tau)/2})$ as needed (see our Auxiliary Results in the Supplementary material). Under Assumption 2 (b), the rate is suitable even if the idiosyncratics exhibit more general time-dependence. Overall, this can be seen as a price for conducting ``comparative statics'' by altering $\tau$, as this restriction is inherent in the process of \cite{magdalinos2009limit}.\footnote{In the Supplement, we provide additional Monte Carlo evidence which reveal that correlation in $\*e_{i,t}$ is not really harmful for the statistical power of the tests.}\\
 \indent Assumption 3 (a) part (ii) is similar to the one in \cite{stauskas2022tests}, and it regulates the dependence under stationarity so that $\{ \+\xi_s\}_{s=1}^t$ obeys the invariance principle and is needed to make sure that (the feasible) $\widehat{\+\delta_t}$ does not dominate the asymptotic theory. Assumption 3 (b) allows for weak cross-section dependence together with unconditional time and cross-section heteroskedasticity. For example, a spatial dependence structure employed in \cite{stauskas2022tests} is a special case. Assumption 4 is standard, but it can be relaxed at the expense of higher moment requirements, while Assumption 5 treats the factor loadings as fixed parameters. They can be made stochastic as in, for instance, \cite{pesaran2006estimation}. It also ensures that the averages are informative about the factors. Finally, Assumption 6 adapts the parameterization necessary for our statistics in moderately integrated environments, because our asymptotic results will be derived under the local alternative setting.

\section{Main Results: CA-Based Statistics }\label{sec_3}
Notice that the expressions for the feasible versions of the statistics in (\ref{eq: PIT})-(\ref{eq: ET3}) are obtained by simply replacing $\widetilde{u}_{2,t+1}$ with $\widehat{u}_{2,t+1}$.
One can reconcile feasible and infeasible statistics using the following asymptotic equivalence decompositions by using 
\begin{align}
    s_{\widehat{f},j}=s_{f,j}+s_{f,j}\left(\widetilde{
\omega}_j/\widehat{\omega}_j -1\right) + \xi_{N,T,j},
\end{align}
for $j=1,\ldots, 4$, where explicit representations of $\xi_{N,T,j}$ come
from \cite{margaritella2024new}: 
\begin{align}
    &\hspace{-1.0cm}\xi_{N,T,1}=\frac{1}{2\widehat{\omega}_{1}}\left(\frac{n}{m_0}\frac{1}{\sqrt{n}}\sum_{t=k_0}^{k_0+m_0-1}\widetilde{u}_{1,t+1}(\widetilde{u}_{2,t+1}- \widehat{u}_{2,t+1}) + \frac{n}{n-m_0}\frac{1}{\sqrt{n}}\sum_{t=k_0+m_0}^{T-1}\widetilde{u}_{1,t+1}(\widetilde{u}_{2,t+1}- \widehat{u}_{2,t+1}) \right),\label{eq: FAT_decomp}\\
    &\hspace{-1.0cm}\xi_{N,T,2}= \frac{1}{\widehat{\omega}_{2}} \frac{n}{l_2^0}\left(\frac{2}{\sqrt{n}}\sum_{t=k_0}^{k_0+l_2^0-1}\widetilde{u}_{2,t+1}(\widetilde{u}_{2,t+1}-\widehat{u}_{2,t+1}) -\frac{1}{\sqrt{n}}\sum_{t=k_0}^{k_0+l_2^0-1}(\widetilde{u}_{2,t+1}-\widehat{u}_{2,t+1})^2\right),\label{eq: FET1_decomp}\\
    &\hspace{-1.0cm}\xi_{N,T,3}=\frac{1}{\widehat{\omega}_{3}} \frac{n}{l^0_2}\left(\frac{n-\lfloor n\nu_0\rfloor}{n(1-\nu_0)}\right) \left(\frac{2}{\sqrt{n}}\sum_{t=k_0}^{k_0+l^0_2-1}\widetilde{u}_{2,t+1}(\widetilde{u}_{2,t+1}-\widehat{u}_{2,t+1})- \frac{1}{\sqrt{n}}\sum_{t=k_0}^{k_0+l^0_2-1}(\widetilde{u}_{2,t+1}-\widehat{u}_{2,t+1})^2\right),\label{eq: FET2_decomp}\\
    &\hspace{-1.0cm}\xi_{N,T,4}=\frac{1}{\widehat{\omega}_{4}} \left(\frac{1}{n(1-\nu_0)}\right)  \sum_{l_2=\lfloor n\nu_0 \rfloor}^n \frac{n}{l_2}\left(\frac{2}{\sqrt{n}}\sum_{t=k_0}^{k_0+l_2-1}\widetilde{u}_{2,t+1}(\widetilde{u}_{2,t+1}-\widehat{u}_{2,t+1})- \frac{1}{\sqrt{n}}\sum_{t=k_0}^{k_0+l_2-1}(\widetilde{u}_{2,t+1}-\widehat{u}_{2,t+1})^2\right)\label{eq: FET3_decomp}.
\end{align}
It is clear from the expansions of the feasible statistics that $\widetilde{u}_{2,t+1}-\widehat{u}_{2,t+1}$ is at the core of the upcoming analysis. It can be shown that this difference admits the following representation: 
\begin{align}\label{u-u_diff}
\widetilde{u}_{2,t+1}-\widehat{u}_{2,t+1}=(\mathbf{Q}_N^{-1}\widehat{\boldsymbol{\delta}}_{t} - \widetilde{\boldsymbol{\delta}}_t^{0})'\widehat{\mathbf{z}}_{t}^0 + (\widetilde{\boldsymbol{\alpha}}_t-\boldsymbol{\alpha})'\overline{\mathbf{e}}^{0}_{r,t} +\boldsymbol{\alpha}'\overline{\mathbf{e}}^{0}_{r,t}=I+II+III,
\end{align}
where each of the components brings a distinct contribution to the difference between  infeasible and feasible forecast. For instance, $III$ takes into account the usual factor estimation error, which stems from the $r$ factors, while $II$ additionally involves an error from estimating $\+\alpha$ under the \textit{observed} $\*f_t$. The component $I$ is the most subtle one. In particular, here $\widetilde{\+\delta}_t^0=[ \widetilde{\+\alpha}_t',\*0_{1\times (m-r)}, \widetilde{\+\theta}_t' \*]'\in \mathbb{R}^m$, which is $\widetilde{\+\delta}_t$ appended with extra $m-r$ zeros in place of the redundant averages. With $\widehat{\+\delta}_t\in \mathbb{R}^m$ being the LS estimator of the model parameters with $\widehat{\*z}_t$, we have $\*Q_N=\mathrm{diag}(\overline{\*H}\*D_N,\*I_q)$ transforming $\widehat{\*f}_t$ into $\widehat{\*f}^0_t$ inside of (\ref{feas_estm}). In effect, $I$ tracks the asymptotic effect of the factor overestimation error via the $m-r$ redundant averages. Lemma 1 below formalises its behavior uniformly over $t$. 
\begin{lemma} \label{lemma1}
Under Assumptions 1-6 as $(N,T) \to \infty$, 
\begin{equation*}
    \left\|\* D_T(\* Q_N^{-1}\boldsymbol{\widehat{\delta}}_t - \boldsymbol{\widetilde{\delta}}_t^0)\right\| = o_p(1)
\end{equation*}
uniformly in $t$, where $\* D_T=\mathrm{diag}(T^{\frac{1}{4}+\frac{\tau}{2}}\* I_r, T^{1/4}\* I_{m-r}, T^{\frac{1}{4}+\frac{\tau}{2}}\* I_q)$.
\end{lemma}

\noindent There are significant differences between our Lemma 1 and its counterpart in \cite{stauskas2022tests}. Firstly, we undertake our analysis under non-stationary predictors $\*z_t$. The subtle detail is that the feasible predictors are $\widehat{\*z}_t$ with the excess averages driven by stationary idiosyncratics. Therefore, we effectively have a mixed integration order of the regressors, which is reflected by the normalization matrix $\*D_T$. Strikingly, the appropriately normalised $\* Q_N^{-1}\boldsymbol{\widehat{\delta}}_t - \boldsymbol{\widetilde{\delta}}_t^0$ is asymptotically negligible even if $m>r$. The intuition behind these results is as follows. Let $\tau=0$, such that $\*D_T=T^{1/4}\*I_{q+m}$, to demonstrate that this difference does not originate solely from the persistence of the regressors. Then, we can show that 
\begin{align}\label{Lemma1_int}
 T^{1/4}(\* Q_N^{-1}\boldsymbol{\widehat{\delta}}_t - \boldsymbol{\widetilde{\delta}}_t^0)=[
    \*0_r',
    T^{-1/4}\*v_t',
    \*0_{q}'
]' + o_p(T^{-1/4})=O_p(T^{-1/4}),
\end{align}
where 
\begin{align}\label{v_term}
    \*v_t=\left(\frac{1}{T}\sum_{s=1}^{t-1}\overline{\*e}_{-r,s}^{0}\overline{\*e}_{-r,s}^{0\prime}\right)^{-1}\frac{1}{\sqrt{T}}\sum_{s=1}^{t-1}\overline{\*e}_{-r,s}^{0}u_{s+1}\in \mathbb{R}^{m-r}
\end{align}
 and $\left\|\*v_t \right\|= O_p(1)$ which is the same martingale difference process as in \cite{stauskas2022tests}. The representation in (\ref{Lemma1_int}) immediately implies that 
$\left\|\sqrt{T}(\* Q_N^{-1}\boldsymbol{\widehat{\delta}}_t - \boldsymbol{\widetilde{\delta}}_t^0)\right\|=O_p(1)$ and so we are back to \cite{stauskas2022tests}. The reason why we only need to scale up by $T^{1/4}$ and not $\sqrt{T}$ is because the former characterises the local power of the tests in \cite{pitarakis2023direct} and \cite{pitarakis2025novel}. 
In particular, the asymptotic distribution of his infeasible statistics is generated by $\{u_{t+1}^2-\sigma^2 \}$, whereas the components involving (functions of) $\{\*z_tu_{t+1} \}$ - which are the distribution generators in \cite{clark2001tests} - are negligible. In other words, ``the weight'' is placed on the former martingale difference instead of the latter. Naturally, the same happens with the components of the feasible statistics, and so the influence of $\*v_t$ becomes negligible, as well.
In general, Lemma 1 suggests that the redundant $m-r$ cross-section averages should not interfere with the asymptotic normality of the feasible statistics. In practice, it means that practitioners can use all the available blocks in FRED-MD or EA-MD-QD datasets and stay agnostic about the true number of factors, as long as $m\geq r$. This stands in sharp contrast to the CAs literature, where the redundant $m-r$ averages result in an asymptotic bias, unless $TN^{-1}\to 0$, which is a substantial restriction (see e.g. \citealp{karabiyik2017role}; or \citealp{de2024cross}). To the best of our knowledge, this is the first result in this literature when the redundant $m-r$ CAs stay completely harmless even if $TN^{-1}\to c>0$. \\
\indent In connection to Lemma 1, we also use the facts that uniformly in $t$,
\begin{align}
    &\left\| (\widetilde{\boldsymbol{\alpha}}_t-\boldsymbol{\alpha})'\overline{\mathbf{e}}^{0}_{r,t} \right\|\leq \left\|\widetilde{\boldsymbol{\alpha}}_t-\boldsymbol{\alpha} \right\|\left\|\overline{\mathbf{e}}^{0}_{r,t} \right\|=O_p(T^{-1/2-\tau/2}N^{-1/2}),\\
    &\left\| \boldsymbol{\alpha}'\overline{\mathbf{e}}^{0}_{r,t}\right\|\leq \left\|\+\alpha\right\|\left\| \overline{\mathbf{e}}^{0}_{r,t} \right\|=O_p(T^{-1/4-\tau/2}N^{-1/2})
\end{align}
to ultimately demonstrate that uniformly in $t$ and a constant $C>0$
\begin{align}\label{u-u_ineq}
    \left|\widetilde{u}_{2,t+1}-\widehat{u}_{2,t+1} \right| &\leq T^{\tau/2-1/2}C\left\|T^{1/4}\*D_T(\* Q_N^{-1}\boldsymbol{\widehat{\delta}}_t - \boldsymbol{\widetilde{\delta}}_t^0) \right\|\left\|T^{-\tau/2}\widehat{\*z}_t^0 \right\| + o_p(1)=o_p(1),
\end{align}
since $\left\|T^{-\tau/2}\widehat{\*z}_t^0 \right\|$ is uniformly bounded by the results in Lemma 3.1 of \cite{magdalinos2009limit}.\footnote{Expanding $\*z_t$, it is $\sup_{k_0\leq t \leq T-1}\mathbb{E}\left(\left\|\frac{1}{T^{\tau/2}}\sum_{s=1}^t\*R_{z,T}^{t-s}\*u_{z,s} \right\|^2\right)=O(1)$, but $\left\|T^{-\tau/2}\widehat{\*z}_t^0 \right\|\leq \left\| T^{-\tau/2} \*z_t\right\|+\underbrace{\left\|T^{-\tau/2} \overline{\*e}_t^0 \right\|}_{o_p(1)}$ as $\overline{\*e}_t^0$ is stationary.} Note that (\ref{u-u_ineq}) holds for all $\tau \in (0,1)$, and we can show that no further restrictions are needed when we sum its squares, as required by the expansions of our statistics. These arguments can be used in the expansions (\ref{eq: FAT_decomp}) - (\ref{eq: FET3_decomp}) which are formalised by Lemma 2.

 
 \begin{lemma} \label{e}
Suppose that the conditions in Lemma 1 hold. Then, as $(N,T) \to \infty$
\begin{enumerate}[label=(\alph*)]
    \item $\left|\xi_{N,T,j} \right|=o_p(1)$
    \item $|\widetilde{\phi}^2-\widehat{\phi}^2| = o_p(1)$.
\end{enumerate}
\end{lemma}

\noindent Notice that a direct consequence of the last part of Lemma 2 is that the variance estimators of the test statistics (\ref{eq: FAT_decomp})-(\ref{eq: FET3_decomp}) are all consistent. In fact, it follows from the workings of \cite{pitarakis2023direct}, \cite{pitarakis2025novel} and \cite{margaritella2024new} that $\left|\widetilde{\omega}_j^2 - \widehat{\omega}_j^2\right| = \theta_j\left|\widetilde{\phi}^2 - \widehat{\phi}^2 \right| = o_p(1)$
for each $j$, where $\theta_j$ is a function of tuning parameters specific to the particular statistic. As a result, it follows from the Continuous Mapping Theorem that the respective expressions $\left|\widetilde{\omega}_j/\widehat{\omega}_j-1 \right|$ are all negligible. By using Lemma 2, we can show that $s_{\widehat{f},1}$ - $s_{\widehat{f},4}$ admit the asymptotic representations reported below in Theorem 1:

\begin{theorem} \label{theorem1}
Suppose that the conditions in Lemma 1, Lemma 2 hold. Then, as $(N,T) \to \infty$, and using the results in Corollary 1, we have   $s_{\widehat{f},j}=s_{f,j}+o_p(1)$.
\end{theorem} 
Since the decompositions in expressions (\ref{eq: FAT_decomp}) - (\ref{eq: FET3_decomp}) are functions of the quantities in Lemma \ref{e}, the results in Theorem \ref{theorem1} follow readily. $\blacksquare$ \\
\indent Three comments are in order. Firstly,  the asymptotic theory of our four test statistics follows directly from \cite{pitarakis2023direct} and \cite{pitarakis2025novel} together with their major features, such as the ability to handle non-stationary predictors. The asymptotic representation in Theorem 1 holds without any restrictions on the relative expansion rate of $N$ and $T$ so long as they both diverge to infinity, thus making our testing procedure applicable to many large-dimensional macroeconomic and financial datasets. Secondly, note from (\ref{u-u_ineq}) that the rate of convergence in Lemma 2 is $O_p(T^{(\tau-1)/2})$. This indicates that the contribution of $m-r$ redundant averages may still decay slowly. Under $\mathbb{H}_1:\+\alpha^0\neq \*0_{r\times1}$, power can be reduced due to an exacarbated factor estimation error if $\tau\approx 1$. The fact that the error is still negligible is natural, as a mildly integrated process is weaker than (local-to-) unit root process ($\tau=1$).  Lastly, the value of our statistics in empirical settings can be boosted even more due to their robustness to structural breaks in the factor loadings. We formulate this in Proposition 1.\\

\noindent \textbf{Proposition 1.} \textit{Let the loadings break at some time points $D_1=\lfloor \phi_1 T \rfloor,\ldots, D_d=\lfloor \phi_d T \rfloor$ with $\phi_1<\phi_2<\ldots<\phi_d \in (0,1)$ for a finite $d$, such that
\begin{align*}
\boldsymbol{\Lambda}_{i,t}'\mathbf{f}_t=\mathbb{I}(t<D_1)\boldsymbol{\Lambda}_{1,i}'\mathbf{f}_t +\left(\sum_{j=1}^{d} \mathbb{I}(t\geq D_j)\+\Lambda_{j+1,i}'\right)\*f_t.
\end{align*}
Let $\*Q_i= [\boldsymbol{\Lambda}_{1,i}', \boldsymbol{\Lambda}_{2,i}',\ldots, \+\Lambda_{d+1,i}']'\in \mathbb{R}^{(d+1) r\times m}$. Under Assumptions 1-6 together with $\mathrm{rk}(\overline{\*Q})=(d+1)\times r$, Theorem 1 continues to hold.} 
\\

\noindent Proposition 1 further increases applicability of the tests, because it can be combined with the overspecification and nonstationarity of factors under Theorem 1. It states that under $d$ breaks in the factor loadings, the starting model can be reformulated into a model with $(d+1)\times r$ factors. Letting $d=1$ (a single break), we obtain a simple expression: $\boldsymbol{\Lambda}_{i,t}'\mathbf{f}_t = \mathbb{I}(t<D)\boldsymbol{\Lambda}_{1,i}'\mathbf{f}_t + \mathbb{I}(t\geq D)\boldsymbol{\Lambda}_{2,i}'\mathbf{f}_t$,
which was also examined in \cite{stauskas2022tests}, where we need $\mathrm{rk}(\overline{\*Q})=2r$. However, in their case, the condition of $D\in (1, k_0]$ is necessary. In other words, the effect of the structural break must be subsumed in the initial estimation of the parameters before going into the out-of-sample analysis. The reason is that the asymptotic distributions of statistics in \cite{clark2001tests} are highly non-standard. If $D\in (k_0, T-1]$, there is a break in the asymptotic variance of the statistics. While it is typically possible to simulate the non-standard distribution, the break in the variance makes the simulation infeasible as the break date is unknown. In our Proposition 2, the break location does not matter. Intuition follows from the discussion below (\ref{v_term}). Our tests have the property that normality is generated by $\{u^2_{t+1}-\sigma^2\}$ independently of the predictors. Therefore, the factors and their loadings (even with breaks) do not interfere with asymptotic normality. Clearly, the predictors matter under $\mathbb{H}_1:\+\alpha\neq \*0_{r\times 1}$. However, we demonstrate that the factor estimation error is asymptotically negligible (Lemma 2) and does not affect the power function. It remains negligible even under breaks because we re-write the initial model with $r$ factors as the model with an inflated number of factors. \\
\indent Our last result takes into account the uncertainty around the tuning parameters in statistics $2$-$4$ from \cite{pitarakis2025novel} and \cite{margaritella2024new}. Since different choices may lead to varying results in statistical power, Proposition 2 offers feasible power enhancements. \\

\noindent \textbf{Proposition 2.} \textit{Let $s_{\widehat{f},j}^{P}=s_{\widehat{f},j}+\widehat{\triangle}_j$ for $j=2,3,4$ be a power-enhanced statistic with the enhancement term $\widehat{\triangle}_j$. Then, under Assumptions 1-6 as $(N,T)\to \infty$
\begin{align*}
    s_{\widehat{f},j}^{P}= s_{f,j}^{P} + o_p(1) \quad for\hspace{2mm} j=2,3,4.
\end{align*}
}
\noindent Here, the power-enhancing term $\triangle_j$ is based on $\breve{u}_{2,t+1}^2= \widetilde{u}_{2,t+1}^2-(\widetilde{u}_{1,t+1}-\widetilde{u}_{2,t+1})^2$. Clearly, we only have the feasible $\widehat{\triangle}_j$, which obeys $\widehat{\triangle}_j=\triangle_j+o_p(1)$ according to Proposition 2. Due to the asymptotic equivalence, we will focus on the power-enhanced statistics in our Monte Carlo experiments and in the empirical application. 
\\

\noindent \textbf{Remark 1 ($h$ steps).} \textit{We can consider multi-step forecasts. We can update Assumption 1 to $\mathbb{E}(u_{t+h}|\mathcal{F}_t)=0$ ($h>1$) similarly to \cite{bai2006confidence}. This implies that $\{u_{t+h}\}$ follows an $MA(h-1)$ process, which is sufficient for $\frac{1}{\sqrt{T}}\sum_{s=1}^{t-h}(u_{s+h}^2-\sigma^2)$ to generate Brownian motion, similarly to \cite{pitarakis2023direct}. In our proofs, many terms that should vanish can be written as $\frac{1}{T}\sum_{t=1}^{T-h}g_tu_{t+h}$. Here, $g_t$ depends on $T^{(1+\tau)/2}(\widetilde{\+\delta}_t-\+\delta), T^{(1+\tau)/2}(\widetilde{\+\alpha}_t-\+\alpha)$ and similar recursive LS deviations. As $MA(h-1)$ induces a very limited serial correlation of forecast errors, we can verify the convergence rates of the latter terms (see Section 2.2 and 5.1 in the Supplementary material). Effectively, $g_t=g_t(\ldots,u_{t-1},u_t)$, and $\mathbb{E}(g_tu_{t+h})=0$ by the law of iterated expectations. It follows that $\mathbb{E}\left(\left|\frac{1}{T}\sum_{t=1}^{T-h}g_tu_{t+h}\right|^2 \right)=\frac{1}{T^2}\sum_{t=1}^{T-h}\sum_{s=1}^{T-h}\mathbb{E}(g_{t+h}g_{s+h})\mathbb{C}ov(u_{t+h},u_{t
s+h})=O(T^{-1})$ due to $MA(h-1)$ structure and a truncated sum. In Section 4.5, we examine our statistics in simulations for $h\in \{1,4,12,24 \}$ and find $s_{\widehat{f},1}$ and $s_{\widehat{f},4}$ to show the best power across the horizons. Clearly, (\ref{conventional_estm}) is replaced by HAC estimator as recommended in \cite{pitarakis2025novel} to take serial correlation into account when estimating variances of our statistics.  }  \\

\noindent \textbf{Remark 2 (Number of factors).} \textit{We can set $m=r$ by estimating the number of factors similarly to the PC literature (see \citealp{ahn2013eigenvalue}), which is relevant to avoid possible small sample distortions of size and/or power when $m>r$ (we will document this for $s_{\widehat{f},1}$ and $s_{\widehat{f},4}$). This can be achieved by using the Information Criterion (IC) by \cite{de2024cross} (see their Proposition 3), which consistently selects an optimal set of CAs. Although their result holds for $\tau=0$, \cite{ditzen2025selection} extended it to $\tau>0$. If $\tau$ is not too large (e.g $\tau \leq 0.5$), the IC can be trusted without any modifications. If $\tau \approx 1$, a consistent selection is possible when both $N,T$ are very large. However, if data of such dimensions are available, our Theorem 1 can be trusted to eliminate the effect of $m-r$ redundant CAs as $(N,T)\to \infty$.
}
\section{Monte Carlo Simulations}
In this section, we introduce the data-generating process (DGP) - which is very general and encompasses a plethora of empirically relevant scenarios - of our Monte Carlo simulations. Then present a large number of simulation results in Section \ref{MC_results}.  

\subsection{Data-Generating Process} \label{DGP}
For the purpose of Monte Carlo simulation, we propose the same DGPs of \cite{stauskas2022tests} and set, in the factor-augmented regression, $\* w_t = y_t$, $\theta = 0.5$, $\alpha = 0$ in the size experiment and $\alpha \in \{0.1, 0.2, 0.3, 0.4, 0.5, 0.6\}$ in the (graphical) power experiments. Various cross-sectional and time dimensions are investigated depending on the simulation considered, while the choice of $\pi_0$ is 0.25 so that $k_0 = \lfloor\pi_0 T \rfloor$ becomes a quarter of the total sample, and the forecast horizon is $h=1$. We choose the number of true factors to be $r=1$ and and the number of blocks to be $m = 3$ (unless otherwise stated). Consider the following general setup:
\begin{align}
&y_{t+1}=\theta y_{t}+\alpha f_t+u_{t+1}, \quad u_{t+1} \sim ID(0,1) \label{eq: pred_reg} \\
& \* x_{i,t} = \* \Lambda_{i,t}' f_t + \* e_{i,t}, \label{eq: x_panel}\\
& f_t = \delta f_{t-1} + \sqrt{1-\delta^2} v_t, \quad v_t \overset{i.i.d}{\sim} \mathcal{N}(0,1) \label{eq: f_dynamics} \\
& \* e_{i,t} = \rho_i \* e_{i,t-1} + \sqrt{1-\rho_i^2} \* w_{i,t}, \label{eq: idiosyncratics}\\
& \* w_{i,t} =  \boldsymbol{\epsilon}_{i,t}+ \sum_{k = i+1}^{K} \beta( \boldsymbol{\epsilon}_{i-k,t} + \boldsymbol{\epsilon}_{i+k,t}), \quad \boldsymbol{\epsilon}_{1,0} = \dots = \boldsymbol{\epsilon}_{N,0}=\boldsymbol{0}_{m \times 1}, \quad K = 5. \label{eq: networ_dep}
\end{align}

\noindent We examine a large number of modelling specifications of $\*\Lambda_{i,t}$, $\boldsymbol{\epsilon_{i,t}}$, $u_t$, $\delta$, $\rho_i$, $\beta$ and $\tau$. For conciseness, we summarise our proposed DGPs - labelled (1) to (9) - in Table \ref{tab: DGP_table}. 
To start with Equation (\ref{eq: pred_reg}), we explore different distributional assumptions for $u_t$, which is drawn from $\mathcal{N}(0, 1)$, $t(10)$  or $\mathcal{N}(0, (1-\phi) + \phi u_{t-1}^2)$ for $\phi = \frac{1}{2}$ across the various DGPs. These three modelling choices allow us to investigate the effect differences of a standard normal distribution, fat tails and ARCH effects of first order on our tests. Further, Equation (\ref{eq: x_panel}) describes the DGP of the factor model. Notice here that the loadings specification encompasses both the time-invariant, nonzero mean 

\begin{table}[H]
\hspace*{-0.55cm}
  \centering
    \begin{tabular}{ccccccc}
    DGP   & $\*\Lambda_{i,t}$ & $\boldsymbol{\epsilon}_{i,t}$   & $u_t$     & $\delta $ & $\rho_i$  & $\beta$ \\
    \midrule
    \midrule
          &                                                                    &       &       &       &       &  \\
    (1)   &  $\mathcal{N}(\boldsymbol{1}_{m \times 1},\boldsymbol{I}_{m})$     & $\mathcal{N}(\boldsymbol{0}_{m \times 1},\boldsymbol{\Sigma}_{\epsilon \epsilon,i})$     &  $\mathcal{N}(0, 1)$     &  0       &  0     &  0 \\
    (2)   &  $\mathcal{N}(\boldsymbol{1}_{m \times 1},\boldsymbol{I}_{m})$     & $\mathcal{N}(\boldsymbol{0}_{m \times 1},\boldsymbol{\Sigma}_{\epsilon \epsilon,i})$     &  $\mathcal{N}(0, 1)$     &  0.4     &  $\mathcal{U}(0.2, 0.6)$     & 0.2 \\
    (3)   &  $\mathcal{N}(\boldsymbol{1}_{m \times 1},\boldsymbol{I}_{m})$     & $\mathcal{N}(\boldsymbol{0}_{m \times 1},\boldsymbol{\Sigma}_{\epsilon \epsilon,i})$     &  $\mathcal{N}(0, 1)$     &  0.4     &  $\mathcal{U}(0.6, 1)$     & 0.4 \\ 
    (4)   &  $\mathcal{N}(\boldsymbol{1}_{m \times 1},\boldsymbol{I}_{m})$     & $t(10)$     & $\mathcal{N}(0, 1)$      & 0.4      &  $\mathcal{U}(0.2, 0.6)$     & 0.2 \\
    (5)   &  $\mathcal{N}(\boldsymbol{1}_{m \times 1},\boldsymbol{I}_{m})$     & $\mathcal{N}(\boldsymbol{0}_{m \times 1},\boldsymbol{\Sigma}_{\epsilon \epsilon,i,t})$   & $\mathcal{N}(0, 1)$      &  0.4     &  $\mathcal{U}(0.2, 0.6)$     & 0.2 \\ 
    (6)   &  $\mathcal{N}(\boldsymbol{1}_{m \times 1},\boldsymbol{I}_{m})$     & $\mathcal{N}(\boldsymbol{0}_{m \times 1},\boldsymbol{\Sigma}_{\epsilon \epsilon,i})$     & $\mathcal{N}(0, (1-\phi) + \phi u_{t-1}^2)$      &  0.4     &  $\mathcal{U}(0.2, 0.6)$     & 0.2 \\
    (7)   &  $\mathcal{N}(\boldsymbol{1}_{m \times 1},\boldsymbol{I}_{m})$     & $\mathcal{N}(\boldsymbol{0}_{m \times 1},\boldsymbol{\Sigma}_{\epsilon \epsilon,i})$     & $t(10)$      &  0.4     &  $\mathcal{U}(0.2, 0.6)$     & 0.2 \\
    (8)   &  $\mathcal{N}(\mathbb{1}_{t<T/2} + 2\cdot \mathbb{1}_{t\geq T/2},\boldsymbol{I}_{m})$     & $\mathcal{N}(\boldsymbol{0}_{m \times 1},\boldsymbol{\Sigma}_{\epsilon \epsilon,i})$      &  $\mathcal{N}(0, 1)$      & 0.4      &  $\mathcal{U}(0.2, 0.6)$     & 0.2 \\
    (9)   &  $\mathcal{N}(\boldsymbol{0}_{m \times 1},\boldsymbol{I}_{m})$     & $\mathcal{N}(\boldsymbol{0}_{m \times 1},\boldsymbol{\Sigma}_{\epsilon \epsilon,i})$     & $\mathcal{N}(0, 1)$      &  0.4     &  $\mathcal{U}(0.2, 0.6)$     & 0.2 \\
          &       &       &       &       &       &  \\
    \bottomrule
    \end{tabular}
  \caption{Monte Carlo experiments}
  \vspace{10pt}
  {\raggedright \footnotesize \textit{Notes}: In this setup, $\rho$ and $\beta$ measure the degree of serial and cross-sectional dependence in $f_t$ and $\* e_{i,t}$ respectively, $\boldsymbol{\Sigma}_{\epsilon \epsilon,i}$ and $\boldsymbol{\Sigma}_{\epsilon \epsilon,i,t}$ are diagonal matrices with elements on the main diagonal drawn from $\mathcal{U}(0.5, 1.5)$, $\mathbb{1}_{(.)}$ is the usual indicator function, and $t(x)$ refers to a Student-t distribution with $x$ degrees of freedom. \par}
  \label{tab: DGP_table}
\end{table}

\noindent case as well as the time-varying cases in DGP (8), which incorporates a structural shift in the loadings mean from 1 to 2 at time $T/2$ as in \cite{breakschendoladogonzalo}. Notice that the break is in the out-of-sample period in order to test the location invariance predicted by our theory. On the contrary, DGP (9) has time-invariant loadings with mean zero, which implies that $\overline{\+\Lambda}$ is asymptotically zero and CAs are inconsistent. Finally, we choose in Equation (\ref{eq: f_dynamics}) a common autoregressive specification\footnote{Notice that if $\delta = 0$ then the factors can be considered as a 'static' sequence of i.i.d. shocks. Hence, even if $\alpha \neq 0$, the probability of rejecting the null hypothesis when it is false (test power) should converge to the desired significant level (as for the test size) because the factors are expected to be uninformative.} for the time dynamics of the factors (see \citealp{bai2006confidence}, \citealp{gonccalves2017tests}, and \citealp{StockWatson2002}, for instance). In a similar manner, in Equation (\ref{eq: idiosyncratics}), we allow the panel idiosyncratic components $\* e_{i,t}$ to be weakly dependent in $i$ and $t$ (similarly to \citealp{bai2002determining}, \citealp{BanerjeeMarcellinoMasten2008}, \citealp{boivin2006more}, and \citealp{breakschendoladogonzalo}), meaning that they exhibit serial correlation via the coefficient $\rho_i$ as well as cross-sectional correlation of spatial type by means of the $\beta$ coefficient in Equation $(\ref{eq: networ_dep})$. Note that the shocks $\boldsymbol{\epsilon}_{i,t}$ in this equation are drawn from $\mathcal{N}(\boldsymbol{0}_{m \times 1},\boldsymbol{\Sigma}_{\epsilon \epsilon,i})$, $\mathcal{N}(\boldsymbol{0}_{m \times 1},\boldsymbol{\Sigma}_{\epsilon \epsilon,i,t})$ or $t(10)$, meaning that the first distribution acts as a benchmark, whereas the second and third distributions introduce time-varying volatility and fat tails, respectively. Here, both covariance matrices $\boldsymbol{\Sigma}_{\epsilon \epsilon,i}$ and $\boldsymbol{\Sigma}_{\epsilon \epsilon,i,t}$ are simply diagonal and their non-zero elements are generated from $\mathcal{U}(0.5, 1.5)$ so to accommodate for cross-sectional heteroskedasticity, as permitted under CCE. Lastly, remember that we are working under the theory of moderately integrated systems \`a la \cite{magdalinos2009limit}, whereby we assume that the parameters $\theta$ and $\delta$ can be written as $\theta = 1-\theta_0/T^\tau$ with $\norm{\theta_0} < \infty$ and $\delta = 1-\delta_0/T^\tau$ with $\norm{\delta_0} < \infty$. Hence, we consider $\tau \in \{0, 0.2, 0.4, 0.6, 0.8\}$ in the next section. For $j=2,3,4$, we consider power-enhanced statistics $s_{\widehat{f},j}^P$. 

\subsection{Results} \label{MC_results}

In this section, we report size and local power experiments run on 1,000 MC replications across different DGPs and parameter settings. The significance level is set to 5\% and the critical values are obtained from the usual standard normal table. In detail, we present a comparison of test size  and local power of the four statistics $s_{\widehat{f},1}$, $s_{\widehat{f},2}^{P}$, $s_{\widehat{f},3}^{P}$ and $s_{\widehat{f},4}^{P}$ in Table \ref{tab: Size} and Table \ref{tab: LocalPower}, respectively. In both analyses, we consider a setup under DGP (2) and two levels of persistence, i.e. $\tau = 0.2$ (mild) and $\tau = 0.6$ (moderate). DGP (2) assumes moderate dependence in the idiosyncratics and so it could be regarded as an unsophisticated but realistic representation of economic data under a factor model structure. Hence, it should serve as a benchmark for all other MC settings. Throughout the paper, we set the tuning parameters to $\mu_0 = 0.45$ for $s_{\widehat{f},1}$, $\nu_0 = 0.8$, $\lambda_1^0 = 1$ and $\lambda_2^0 = 0.65$ for $s_{\widehat{f},2}^{P}$ and $s_{\widehat{f},3}^{P}$, $\lambda_1^0 = 0.65$ and $\lambda_2^0 = 1$ for $s_{\widehat{f},4}^{P}$ as in \cite{margaritella2024new}.


\begin{table}[htbp]
  \centering
    \begin{tabular}{rrrrrrrrrrrrrrr}
    \multicolumn{7}{c}{$\tau$ = 0.2}                         &       & \multicolumn{7}{c}{$\tau$ = 0.6} \\
\cmidrule{1-7}\cmidrule{9-15}    \multicolumn{1}{c}{$N$} & \multicolumn{1}{c}{$T$} &       & \multicolumn{1}{c}{$s_{\widehat{f},1}$} & \multicolumn{1}{c}{$s_{\widehat{f},2}^{P}$} & \multicolumn{1}{c}{$s_{\widehat{f},3}^{P}$} & \multicolumn{1}{c}{$s_{\widehat{f},4}^{P}$} &       & \multicolumn{1}{c}{$N$} & \multicolumn{1}{c}{$T$} &       & \multicolumn{1}{c}{$s_{\widehat{f},1}$} & \multicolumn{1}{c}{$s_{\widehat{f},2}^{P}$} & \multicolumn{1}{c}{$s_{\widehat{f},3}^{P}$} & \multicolumn{1}{c}{$s_{\widehat{f},4}^{P}$} \\
\cmidrule{1-7}\cmidrule{9-15}          &       &       &       &       &       &       &       &       &       &       &       &       &       &  \\
    \multicolumn{1}{c}{10} & \multicolumn{1}{c}{50} &       & \multicolumn{1}{c}{25.4} & \multicolumn{1}{c}{12.2} & \multicolumn{1}{c}{13.1} & \multicolumn{1}{c}{20.6} &       & \multicolumn{1}{c}{10} & \multicolumn{1}{c}{50} &       & \multicolumn{1}{c}{25.2} & \multicolumn{1}{c}{10.8} & \multicolumn{1}{c}{12.5} & \multicolumn{1}{c}{20.8} \\
    \multicolumn{1}{c}{20} & \multicolumn{1}{c}{50} &       & \multicolumn{1}{c}{26.6} & \multicolumn{1}{c}{11.6} & \multicolumn{1}{c}{14.4} & \multicolumn{1}{c}{21.9} &       & \multicolumn{1}{c}{20} & \multicolumn{1}{c}{50} &       & \multicolumn{1}{c}{25.6} & \multicolumn{1}{c}{11.5} & \multicolumn{1}{c}{14.1} & \multicolumn{1}{c}{22.4} \\
    \multicolumn{1}{c}{50} & \multicolumn{1}{c}{50} &       & \multicolumn{1}{c}{23.9} & \multicolumn{1}{c}{10.8} & \multicolumn{1}{c}{12.6} & \multicolumn{1}{c}{19.9} &       & \multicolumn{1}{c}{50} & \multicolumn{1}{c}{50} &       & \multicolumn{1}{c}{24.9} & \multicolumn{1}{c}{11.0} & \multicolumn{1}{c}{12.6} & \multicolumn{1}{c}{20.2} \\
    \multicolumn{1}{c}{100} & \multicolumn{1}{c}{50} &       & \multicolumn{1}{c}{24.5} & \multicolumn{1}{c}{12.5} & \multicolumn{1}{c}{14.1} & \multicolumn{1}{c}{21.6} &       & \multicolumn{1}{c}{100} & \multicolumn{1}{c}{50} &       & \multicolumn{1}{c}{25.7} & \multicolumn{1}{c}{12.0} & \multicolumn{1}{c}{13.8} & \multicolumn{1}{c}{22.3} \\
    \multicolumn{1}{c}{200} & \multicolumn{1}{c}{50} &       & \multicolumn{1}{c}{23.5} & \multicolumn{1}{c}{11.1} & \multicolumn{1}{c}{13.3} & \multicolumn{1}{c}{19.9} &       & \multicolumn{1}{c}{200} & \multicolumn{1}{c}{50} &       & \multicolumn{1}{c}{25.0} & \multicolumn{1}{c}{12.5} & \multicolumn{1}{c}{14.2} & \multicolumn{1}{c}{21.4} \\
    \multicolumn{1}{c}{10} & \multicolumn{1}{c}{100} &       & \multicolumn{1}{c}{19.6} & \multicolumn{1}{c}{8.5} & \multicolumn{1}{c}{9.9} & \multicolumn{1}{c}{16.1} &       & \multicolumn{1}{c}{10} & \multicolumn{1}{c}{100} &       & \multicolumn{1}{c}{20.3} & \multicolumn{1}{c}{7.8} & \multicolumn{1}{c}{9.7} & \multicolumn{1}{c}{15.1} \\
    \multicolumn{1}{c}{20} & \multicolumn{1}{c}{100} &       & \multicolumn{1}{c}{20.2} & \multicolumn{1}{c}{7.7} & \multicolumn{1}{c}{8.7} & \multicolumn{1}{c}{14.8} &       & \multicolumn{1}{c}{20} & \multicolumn{1}{c}{100} &       & \multicolumn{1}{c}{20.2} & \multicolumn{1}{c}{7.9} & \multicolumn{1}{c}{8.1} & \multicolumn{1}{c}{14.6} \\
    \multicolumn{1}{c}{50} & \multicolumn{1}{c}{100} &       & \multicolumn{1}{c}{17.9} & \multicolumn{1}{c}{8.8} & \multicolumn{1}{c}{8.9} & \multicolumn{1}{c}{15.5} &       & \multicolumn{1}{c}{50} & \multicolumn{1}{c}{100} &       & \multicolumn{1}{c}{19.3} & \multicolumn{1}{c}{8.6} & \multicolumn{1}{c}{8.4} & \multicolumn{1}{c}{15.3} \\
    \multicolumn{1}{c}{100} & \multicolumn{1}{c}{100} &       & \multicolumn{1}{c}{17.8} & \multicolumn{1}{c}{8.6} & \multicolumn{1}{c}{8.2} & \multicolumn{1}{c}{15.6} &       & \multicolumn{1}{c}{100} & \multicolumn{1}{c}{100} &       & \multicolumn{1}{c}{18.3} & \multicolumn{1}{c}{8.4} & \multicolumn{1}{c}{9.5} & \multicolumn{1}{c}{16.0} \\
    \multicolumn{1}{c}{200} & \multicolumn{1}{c}{100} &       & \multicolumn{1}{c}{18.1} & \multicolumn{1}{c}{8.5} & \multicolumn{1}{c}{9.2} & \multicolumn{1}{c}{15.1} &       & \multicolumn{1}{c}{200} & \multicolumn{1}{c}{100} &       & \multicolumn{1}{c}{18.6} & \multicolumn{1}{c}{9.2} & \multicolumn{1}{c}{11.0} & \multicolumn{1}{c}{16.2} \\
    \multicolumn{1}{c}{10} & \multicolumn{1}{c}{200} &       & \multicolumn{1}{c}{13.7} & \multicolumn{1}{c}{7.1} & \multicolumn{1}{c}{7.8} & \multicolumn{1}{c}{12.6} &       & \multicolumn{1}{c}{10} & \multicolumn{1}{c}{200} &       & \multicolumn{1}{c}{13.5} & \multicolumn{1}{c}{7.4} & \multicolumn{1}{c}{7.4} & \multicolumn{1}{c}{12.1} \\
    \multicolumn{1}{c}{20} & \multicolumn{1}{c}{200} &       & \multicolumn{1}{c}{14.1} & \multicolumn{1}{c}{7.0} & \multicolumn{1}{c}{7.3} & \multicolumn{1}{c}{9.8} &       & \multicolumn{1}{c}{20} & \multicolumn{1}{c}{200} &       & \multicolumn{1}{c}{15.1} & \multicolumn{1}{c}{7.1} & \multicolumn{1}{c}{8.3} & \multicolumn{1}{c}{11.1} \\
    \multicolumn{1}{c}{50} & \multicolumn{1}{c}{200} &       & \multicolumn{1}{c}{13.8} & \multicolumn{1}{c}{7.8} & \multicolumn{1}{c}{7.0} & \multicolumn{1}{c}{13.0} &       & \multicolumn{1}{c}{50} & \multicolumn{1}{c}{200} &       & \multicolumn{1}{c}{15.7} & \multicolumn{1}{c}{8.0} & \multicolumn{1}{c}{8.8} & \multicolumn{1}{c}{12.6} \\
    \multicolumn{1}{c}{100} & \multicolumn{1}{c}{200} &       & \multicolumn{1}{c}{13.6} & \multicolumn{1}{c}{8.1} & \multicolumn{1}{c}{8.8} & \multicolumn{1}{c}{12.0} &       & \multicolumn{1}{c}{100} & \multicolumn{1}{c}{200} &       & \multicolumn{1}{c}{15.3} & \multicolumn{1}{c}{8.2} & \multicolumn{1}{c}{8.7} & \multicolumn{1}{c}{12.3} \\
    \multicolumn{1}{c}{200} & \multicolumn{1}{c}{200} &       & \multicolumn{1}{c}{15.4} & \multicolumn{1}{c}{7.5} & \multicolumn{1}{c}{7.6} & \multicolumn{1}{c}{12.4} &       & \multicolumn{1}{c}{200} & \multicolumn{1}{c}{200} &       & \multicolumn{1}{c}{16.0} & \multicolumn{1}{c}{7.8} & \multicolumn{1}{c}{8.3} & \multicolumn{1}{c}{13.0} \\
    \multicolumn{1}{c}{500} & \multicolumn{1}{c}{500} &       & \multicolumn{1}{c}{10.2} & \multicolumn{1}{c}{4.6} & \multicolumn{1}{c}{5.4} & \multicolumn{1}{c}{8.0} &       & \multicolumn{1}{c}{500} & \multicolumn{1}{c}{500} &       & \multicolumn{1}{c}{9.9} & \multicolumn{1}{c}{4.6} & \multicolumn{1}{c}{5.2} & \multicolumn{1}{c}{8.6} \\
          &       &       &       &       &       &       &       &       &       &       &       &       &       &  \\
\cmidrule{1-7}\cmidrule{9-15}    \end{tabular}%
\caption{Size (\%) under DGP(2) and persistence.}
  \label{tab: Size}%
\end{table}%
\subsection{Size}
Starting from test size, Table \ref{tab: Size} reports the results in an environment with low persistence on the left-hand side and high persistence on the right-hand side. Observe that the test size converges to the significance level as $T \to \infty$ but it is seems to be insensitive to different levels of $N$. Further, it is evident that size does not deteriorate as $\tau$ grows, as anticipated by the theory of \cite{pitarakis2023direct}. With respect to the performance of the individual statistics, it is clear that $s_{\widehat{f},2}^{P}$ and $s_{\widehat{f},3}^{P}$ deliver the best test size, while both $s_{\widehat{f},1}$ and $s_{\widehat{f},4}^{P}$ suffer from oversize for smaller values of $T$. However, additional results in the Supplementary material indicate that this oversize experienced by the latter pair is due to our overspecification of the number of blocks relative to the true number of factors, i.e. $m=3$ while $r=1$. By ensuring that $m=r$, it is therefore possible to materially reduce the issue of oversize while maintaining excellent local power. This can be achieved by using the IC by \cite{de2024cross} - which remains valid under non-stationary factors, as demonstrated by \cite{ditzen2025selection} - unless persistence is very high, in which case very large $N$ and $T$ combinations are required. In essence, we conclude that reliable test size is obtained with larger $T$ at all levels of persistence and for all statistics. 
\subsection{Power}

Next, we turn our attention to the local alternative, where we set $\alpha=0.5$. Given our theoretical results, we should point out that, loosely speaking, our tests should not discern between estimated and observed factors for sufficiently large block sizes, meaning that the theoretical framework of \cite{pitarakis2023direct,pitarakis2025novel} is virtually restored. This is due to the fact that factor uncertainty inherent in CAs disappears with $N \to \infty$ in each block. The results are reported in Table \ref{tab: LocalPower}. Similarly to test size, we see a distinct difference in performance between the four statistics under both persistence environments, with $s_{\widehat{f},1}$ and $s_{\widehat{f},4}^{P}$ exhibiting significantly better local power relative to $s_{\widehat{f},2}^{P}$ and $s_{\widehat{f},3}^{P}$. Such constrast, however, becomes materially unimportant with larger values of $\alpha$. Contrary to our simulations on test size, whereas, we notice that higher persistence in the predictors lowers the local power of all statistics. This is perfectly consistent with our theory in Section \ref{sec_3} where the convergence rates under the local alternative depend on $\tau \in [0,1)$: as the system of predictors becomes increasingly persistent, the tests devised by Pitarakis - and, by natural extension, our tests under CA-estimated factors - will be diminishingly able to detect enhanced forecasting power in the augmented regression relative to its nested baseline. Overall, we suggest using $s_{\widehat{f},2}^{P}$ and $s_{\widehat{f},3}^{P}$ for better test size and $s_{\widehat{f},1}$ and $s_{\widehat{f},4}^{P}$ for enhanced local power.  

\begin{table}[htbp]
  \centering
    \begin{tabular}{rrrrrrrrrrrrrrr}
    \multicolumn{7}{c}{$\tau$ = 0.2}                         &       & \multicolumn{7}{c}{$\tau$ = 0.6} \\
\cmidrule{1-7}\cmidrule{9-15}    \multicolumn{1}{c}{$N$} & \multicolumn{1}{c}{$T$} &       & \multicolumn{1}{c}{$s_{\widehat{f},1}$} & \multicolumn{1}{c}{$s_{\widehat{f},2}^{P}$} & \multicolumn{1}{c}{$s_{\widehat{f},3}^{P}$} & \multicolumn{1}{c}{$s_{\widehat{f},4}^{P}$} &       & \multicolumn{1}{c}{$N$} & \multicolumn{1}{c}{$T$} &       & \multicolumn{1}{c}{$s_{\widehat{f},1}$} & \multicolumn{1}{c}{$s_{\widehat{f},2}^{P}$} & \multicolumn{1}{c}{$s_{\widehat{f},3}^{P}$} & \multicolumn{1}{c}{$s_{\widehat{f},4}^{P}$} \\
\cmidrule{1-7}\cmidrule{9-15}          &       &       &       &       &       &       &       &       &       &       &       &       &       &  \\
    \multicolumn{1}{c}{10} & \multicolumn{1}{c}{50} &       & 75.3  & \multicolumn{1}{c}{42.7} & \multicolumn{1}{c}{44.3} & \multicolumn{1}{c}{69.1} &       & \multicolumn{1}{c}{10} & \multicolumn{1}{c}{50} &       & \multicolumn{1}{c}{54.6} & \multicolumn{1}{c}{24.3} & \multicolumn{1}{c}{26.1} & \multicolumn{1}{c}{45.1} \\
    \multicolumn{1}{c}{20} & \multicolumn{1}{c}{50} &       & 80.8  & \multicolumn{1}{c}{46.4} & \multicolumn{1}{c}{46.9} & \multicolumn{1}{c}{71.7} &       & \multicolumn{1}{c}{20} & \multicolumn{1}{c}{50} &       & \multicolumn{1}{c}{61.3} & \multicolumn{1}{c}{27.8} & \multicolumn{1}{c}{30.7} & \multicolumn{1}{c}{49.7} \\
    \multicolumn{1}{c}{50} & \multicolumn{1}{c}{50} &       & 79.6  & \multicolumn{1}{c}{49.3} & \multicolumn{1}{c}{51.6} & \multicolumn{1}{c}{74.7} &       & \multicolumn{1}{c}{50} & \multicolumn{1}{c}{50} &       & \multicolumn{1}{c}{59.6} & \multicolumn{1}{c}{31.3} & \multicolumn{1}{c}{32.2} & \multicolumn{1}{c}{53.2} \\
    \multicolumn{1}{c}{100} & \multicolumn{1}{c}{50} &       & 81.1  & \multicolumn{1}{c}{49.7} & \multicolumn{1}{c}{50.9} & \multicolumn{1}{c}{76.0} &       & \multicolumn{1}{c}{100} & \multicolumn{1}{c}{50} &       & \multicolumn{1}{c}{62.7} & \multicolumn{1}{c}{33.2} & \multicolumn{1}{c}{33.3} & \multicolumn{1}{c}{54.5} \\
    \multicolumn{1}{c}{200} & \multicolumn{1}{c}{50} &       & 81.8  & \multicolumn{1}{c}{52.4} & \multicolumn{1}{c}{53.1} & \multicolumn{1}{c}{77.1} &       & \multicolumn{1}{c}{200} & \multicolumn{1}{c}{50} &       & \multicolumn{1}{c}{63.3} & \multicolumn{1}{c}{34.3} & \multicolumn{1}{c}{34.6} & \multicolumn{1}{c}{56.1} \\
    \multicolumn{1}{c}{10} & \multicolumn{1}{c}{100} &       & 95.8  & \multicolumn{1}{c}{62.5} & \multicolumn{1}{c}{65.1} & \multicolumn{1}{c}{91.0} &       & \multicolumn{1}{c}{10} & \multicolumn{1}{c}{100} &       & \multicolumn{1}{c}{78.7} & \multicolumn{1}{c}{35.5} & \multicolumn{1}{c}{35.8} & \multicolumn{1}{c}{61.9} \\
    \multicolumn{1}{c}{20} & \multicolumn{1}{c}{100} &       & 96.6  & \multicolumn{1}{c}{72.1} & \multicolumn{1}{c}{72.9} & \multicolumn{1}{c}{93.4} &       & \multicolumn{1}{c}{20} & \multicolumn{1}{c}{100} &       & \multicolumn{1}{c}{81.7} & \multicolumn{1}{c}{42.5} & \multicolumn{1}{c}{43.2} & \multicolumn{1}{c}{70.8} \\
    \multicolumn{1}{c}{50} & \multicolumn{1}{c}{100} &       & 97.0  & \multicolumn{1}{c}{77.7} & \multicolumn{1}{c}{76.9} & \multicolumn{1}{c}{95.2} &       & \multicolumn{1}{c}{50} & \multicolumn{1}{c}{100} &       & \multicolumn{1}{c}{82.7} & \multicolumn{1}{c}{47.2} & \multicolumn{1}{c}{49.0} & \multicolumn{1}{c}{77.4} \\
    \multicolumn{1}{c}{100} & \multicolumn{1}{c}{100} &       & 96.9  & \multicolumn{1}{c}{78.3} & \multicolumn{1}{c}{77.6} & \multicolumn{1}{c}{96.0} &       & \multicolumn{1}{c}{100} & \multicolumn{1}{c}{100} &       & \multicolumn{1}{c}{85.7} & \multicolumn{1}{c}{50.9} & \multicolumn{1}{c}{51.0} & \multicolumn{1}{c}{79.3} \\
    \multicolumn{1}{c}{200} & \multicolumn{1}{c}{100} &       & 97.4  & \multicolumn{1}{c}{79.6} & \multicolumn{1}{c}{79.5} & \multicolumn{1}{c}{95.7} &       & \multicolumn{1}{c}{200} & \multicolumn{1}{c}{100} &       & \multicolumn{1}{c}{84.3} & \multicolumn{1}{c}{53.2} & \multicolumn{1}{c}{53.1} & \multicolumn{1}{c}{81.3} \\
    \multicolumn{1}{c}{10} & \multicolumn{1}{c}{200} &       & 99.7  & \multicolumn{1}{c}{82.6} & \multicolumn{1}{c}{83.0} & \multicolumn{1}{c}{98.5} &       & \multicolumn{1}{c}{10} & \multicolumn{1}{c}{200} &       & \multicolumn{1}{c}{91.1} & \multicolumn{1}{c}{51.8} & \multicolumn{1}{c}{54.2} & \multicolumn{1}{c}{80.6} \\
    \multicolumn{1}{c}{20} & \multicolumn{1}{c}{200} &       & 99.9  & \multicolumn{1}{c}{89.5} & \multicolumn{1}{c}{90.2} & \multicolumn{1}{c}{99.6} &       & \multicolumn{1}{c}{20} & \multicolumn{1}{c}{200} &       & \multicolumn{1}{c}{93.5} & \multicolumn{1}{c}{61.0} & \multicolumn{1}{c}{60.3} & \multicolumn{1}{c}{86.9} \\
    \multicolumn{1}{c}{50} & \multicolumn{1}{c}{200} &       & 99.9  & \multicolumn{1}{c}{91.6} & \multicolumn{1}{c}{92.2} & \multicolumn{1}{c}{99.4} &       & \multicolumn{1}{c}{50} & \multicolumn{1}{c}{200} &       & \multicolumn{1}{c}{94.3} & \multicolumn{1}{c}{68.1} & \multicolumn{1}{c}{67.1} & \multicolumn{1}{c}{89.8} \\
    \multicolumn{1}{c}{100} & \multicolumn{1}{c}{200} &       & 100.0 & \multicolumn{1}{c}{93.7} & \multicolumn{1}{c}{93.5} & \multicolumn{1}{c}{99.7} &       & \multicolumn{1}{c}{100} & \multicolumn{1}{c}{200} &       & \multicolumn{1}{c}{95.0} & \multicolumn{1}{c}{70.8} & \multicolumn{1}{c}{71.5} & \multicolumn{1}{c}{92.5} \\
    \multicolumn{1}{c}{200} & \multicolumn{1}{c}{200} &       & 100.0 & \multicolumn{1}{c}{94.5} & \multicolumn{1}{c}{94.9} & \multicolumn{1}{c}{99.6} &       & \multicolumn{1}{c}{200} & \multicolumn{1}{c}{200} &       & \multicolumn{1}{c}{94.5} & \multicolumn{1}{c}{72.5} & \multicolumn{1}{c}{72.5} & \multicolumn{1}{c}{93.2} \\
    \multicolumn{1}{c}{500} & \multicolumn{1}{c}{500} &       & 100.0 & \multicolumn{1}{c}{99.8} & \multicolumn{1}{c}{99.9} & \multicolumn{1}{c}{100.0} &       & \multicolumn{1}{c}{500} & \multicolumn{1}{c}{500} &       & \multicolumn{1}{c}{99.9} & \multicolumn{1}{c}{94.1} & \multicolumn{1}{c}{93.4} & \multicolumn{1}{c}{99.6} \\
          &       &       &       &       &       &       &       &       &       &       &       &       &       &  \\
\cmidrule{1-7}\cmidrule{9-15}    \end{tabular}
\caption{Local power (\%) under DGP(2) and persistence.}
  \label{tab: LocalPower}%
\end{table}%

\noindent To demonstrate the robustness of our tests, we provide in Figure \ref{fig: LocalPowerG1G2G3G4} a comprehensive collection of power curves across DGPs (1) to (9) for all statistics, with the usual values $\alpha \in \{0.1, 0.2, 0.3, 0.4, 0.5,\allowbreak  0.6\}$ and $\tau = 0.5$. All statistics exhibit satisfactory power across the various DGPs but seem to suffer moderately in terms of both size and local power under ARCH effects and regression heavy tail settings. The former can be improved if (\ref{conventional_estm}) is substituted by a HAC estimator since $\{u_{t+1}^2-\sigma^2 \}$ is serially correlated under conditional heteroskedasticity.  

\begin{figure}[H]
\begin{subfigure}[b]{0.50\linewidth} 
\includegraphics[width = 1.0\textwidth]{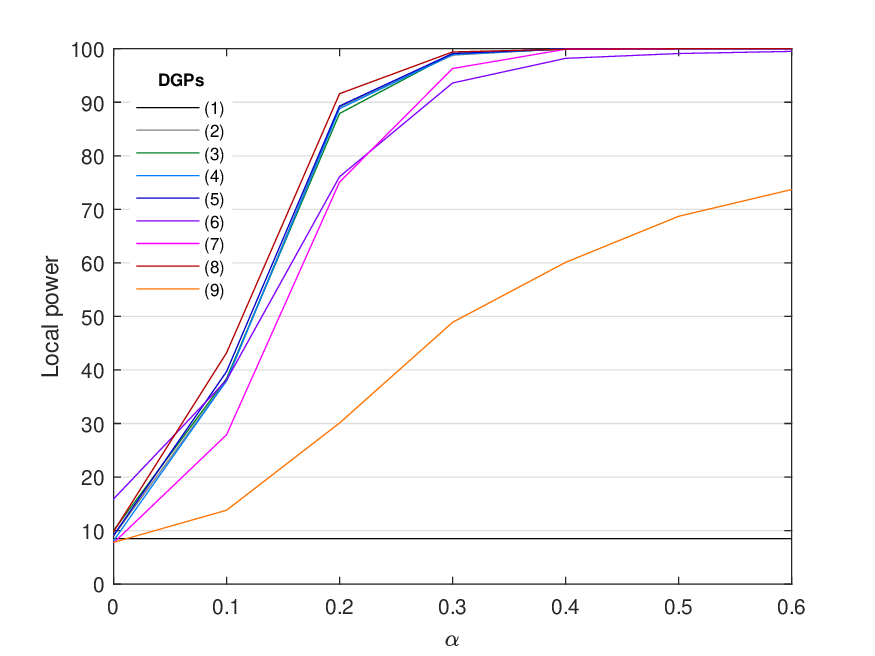}
\footnotesize \caption{$s_{\widehat{f},1}$ with $\mu_0 = 0.45$.}
\end{subfigure}
\begin{subfigure}[b]{0.50\linewidth}
\includegraphics[width = 1.0\textwidth]{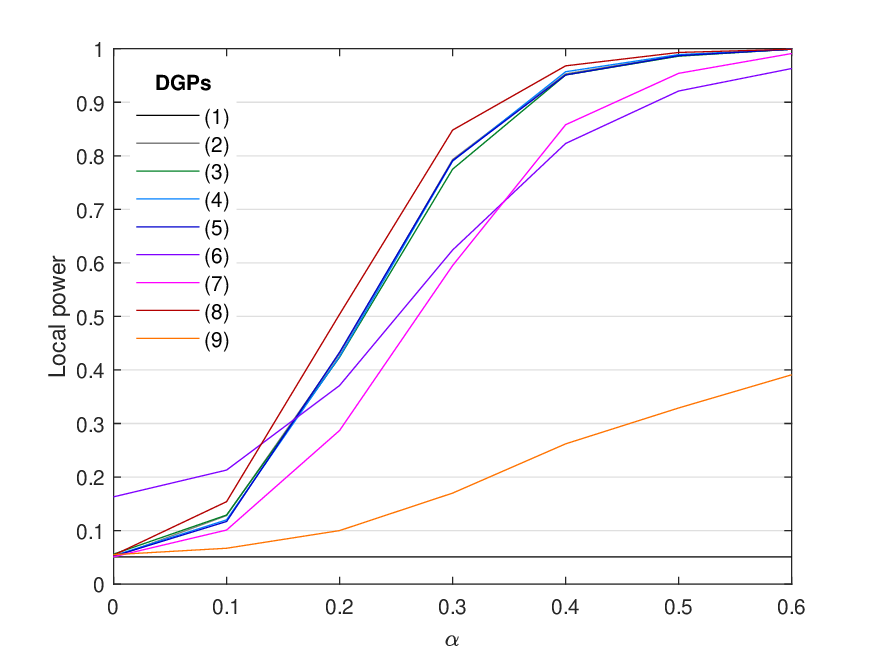}
\caption{$s_{\widehat{f},2}^{P}$ with $\nu_0 = 0.8$ $\lambda_1^0 = 1$, $\lambda_2^0 = 0.65$.}
\end{subfigure}
\medskip
\begin{subfigure}[b]{0.50\linewidth}
\includegraphics[width = 1.0\textwidth]{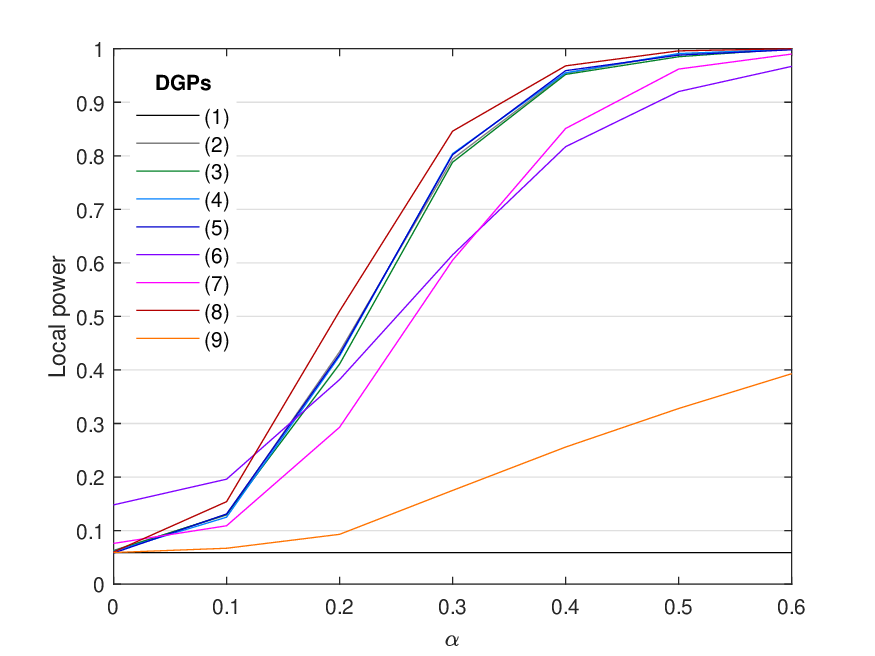}
\caption{$s_{\widehat{f},3}^{P}$ with $\nu_0 = 0.8$ $\lambda_1^0 = 1$, $\lambda_2^0 = 0.65$.}
\end{subfigure}
\hfill
\begin{subfigure}[b]{0.50\linewidth}
\includegraphics[width = 1.0\textwidth]{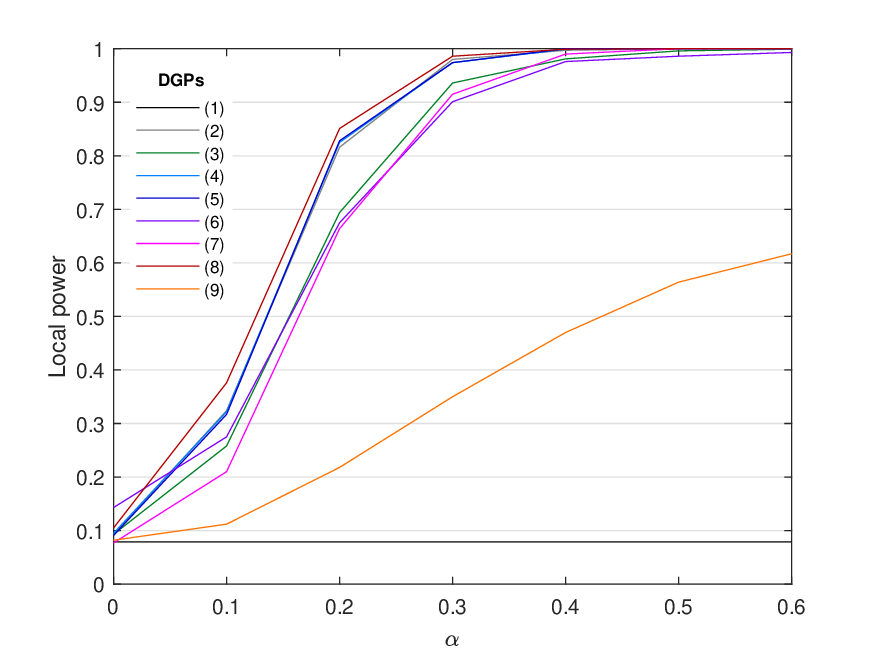}
\caption{$s_{\widehat{f},4}^{P}$ with $\nu_0 = 0.8$ $\lambda_1^0 = 0.65$, $\lambda_2^0 = 1$.}
\end{subfigure}
\caption{Evolution of local power (\%) of the four statistics for different DGPs and values of $\alpha$. Setting: $(N,T)$ = (200, 600), $h$ = 1, $\tau = 0.5$.}
\label{fig: LocalPowerG1G2G3G4}
\end{figure}

\noindent In general, it is unsurprising that the nondegenerate DGPs, namely (2) to (8), all exhibit improving local power as $\alpha$ moves away from zero. Note that the performance of DGP (8) with a break in loadings in the out-of-sample portion especially stands out.  We should remark that the results are not excellent in the case of DGPs (6) and (7). The former performs significantly worse across the group in terms of size and power, whereas the latter exhibits excellent test size but its local power seems to match that of DGP (6) for higher levels of $\alpha$. In fact, these statistics are constructed based on averages of the (squared) predictive regression residuals so it is natural to think that their estimation is affected by ARCH effects and fat tails. Nonetheless, our findings seem to indicate that these issues are only of finite sample nature and vanish asymptotically with increasing time dimension. With respect to the first DGP, whereas, the values are flat around the significant level and should be interpreted as test size. The reason is that the panel data has no factor structure nor any kind of dependence, so that the CAs-estimated factors should not improve upon the baseline specification. Finally, DGP (9) models the rank violation so that the CAs are inconsistent for the factor space. The statistical inference is not valid in this case and, indeed, the local power is visibly impaired.\\
\indent Finally, we apply another battery of simulation scenarios and report them in the Supplement in the interest of space. The interested reader should be informed in advance that, in line with previous findings, a similar analysis almost invariably applies to all four statistics across these additional settings, with excellent local power across the spectrum. The experimental evidence includes: 1) size experiments with a known number of factors ($m=r$), 2) the effect of different redundant blocks, 3) varying levels of persistence (also in the presence of serially correlated idiosyncratics), 4) violations of the MDS assumption of $u_{t}$. Overall, local power deteriorates in all of these settings in small samples, with the notable exception of 1) where the impact almost does not exist (but the size of $s_{\widehat{f},1}$ and $s_{\widehat{f},4}^{P}$ is improves significantly, as discussed above). To summarise, our CAs-based tests perform quite well under a plethora of empirically relevant economic scenarios. More importantly, our MC simulations seem to support two important theoretical contributions of this work to the factor model literature. First, our tests seem to be invariant to uncertainty in the number of factors and breaks in the loadings. Second, they can accommodate for moderately integrated systems of variables with acceptable loss of local power. 
\subsection{Extension to Multi-Step Ahead Forecasts}
Following Remark 1, we explore the evolution of the power for $h=1,4,12,24$. We consider HAC standard errors for all statistics under the empirically appealing DGP(2) and report the results in Figure 2. To our surprise, the statistic $s_{\widehat{f},1,HAC}$ - proposed by \cite{pitarakis2023direct} to handle multi-step forecasts - suffers from a loss in local power, while the remaining statistics seem to be unaffected. Despite this , $s_{\widehat{f},1,HAC}$ still outperforms $s_{\widehat{f},2,HAC}^P$ and $s_{\widehat{f},3,HAC}^P$ yet underperforms relative to $s_{\widehat{f},4,HAC}^P$. This is consistent to the previous results in our simulations. Indeed, the first and last statistics exhibit the best local power while the second and third statistics offer better statistical size, thus again corroborating the discussion on the trade-off between size and power.
\begin{figure}[H]
\begin{subfigure}[b]{0.5\linewidth} 
  
\includegraphics[width = 1.0\textwidth]{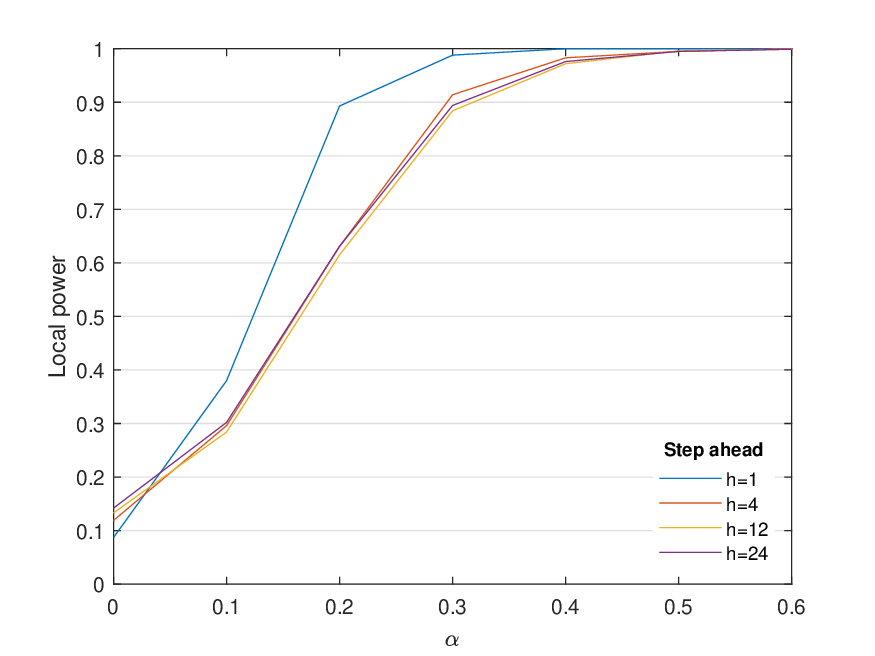}
\footnotesize \caption{$s_{\widehat{f},1,HAC}$ with $\mu_0 = 0.45$.}
\label{fig: hstepaheadForecastingEq1}
\end{subfigure}
\begin{subfigure}[b]{0.5\linewidth}

\includegraphics[width = 1.0\textwidth]{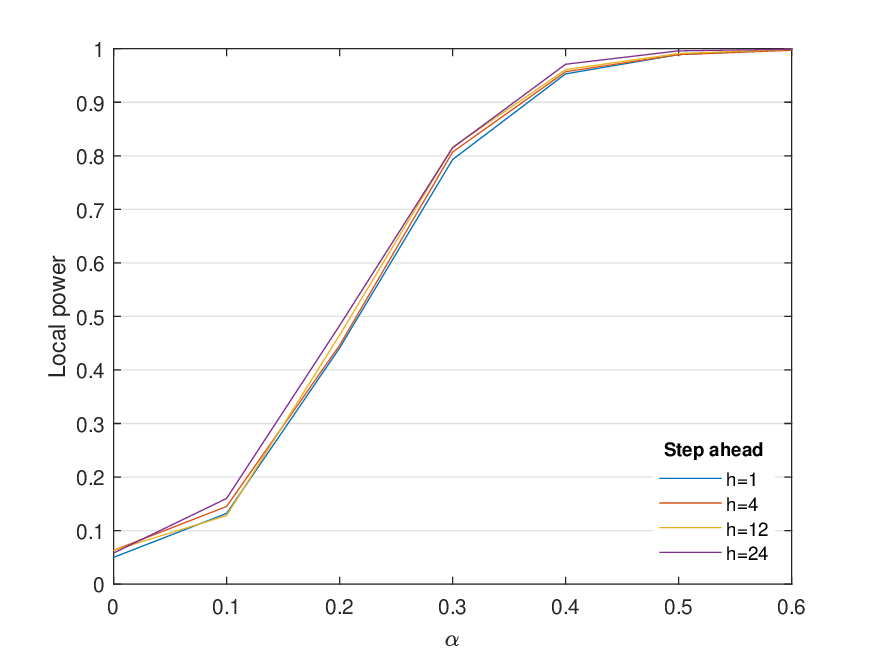}
\caption{$s_{\widehat{f},2,HAC}^{P}$ with $\nu_0 = 0.8$ $\lambda_1^0 = 1$, $\lambda_2^0 = 0.65$.}
\label{fig: hstepaheadForecastingEq2}
\end{subfigure}
\medskip
\begin{subfigure}[b]{0.5\linewidth}

\includegraphics[width = 1.0\textwidth]{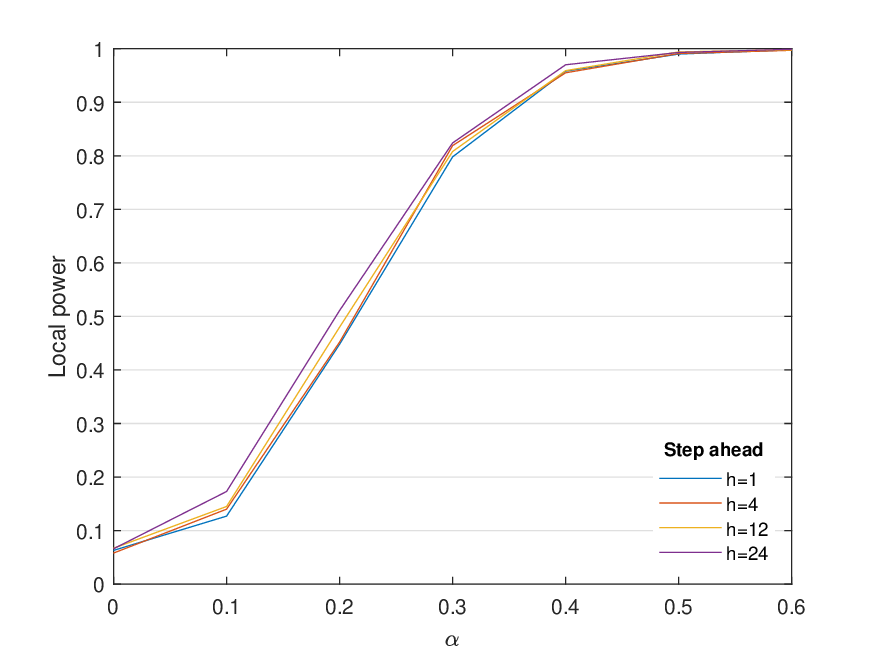}
\caption{$s_{\widehat{f},3,HAC}^{P}$ with $\nu_0 = 0.8$ $\lambda_1^0 = 1$, $\lambda_2^0 = 0.65$.}
\label{fig: hstepaheadForecastingEq3}
\end{subfigure}
\hfill
\begin{subfigure}[b]{0.5\linewidth}

\includegraphics[width = 1.0\textwidth]{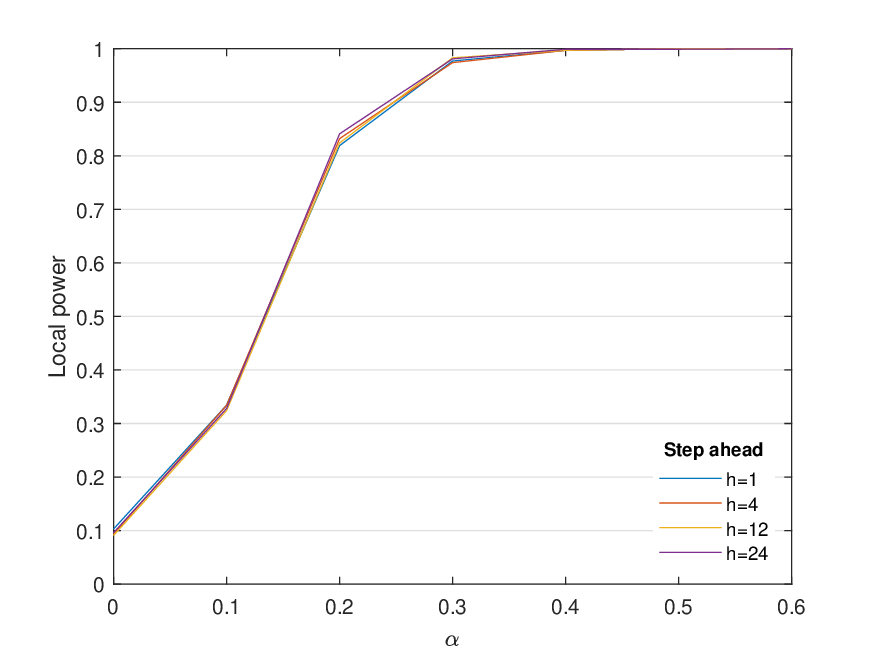}
\caption{$s_{\widehat{f},4,HAC}^{P}$ with $\nu_0 = 0.8$ $\lambda_1^0 = 0.65$, $\lambda_2^0 = 1$.}
\label{fig: hstepaheadForecastingEq4}
\end{subfigure}
\caption{
Same parameters as in Figure 1, with the focus on DGP (2) and   $\tau = 0.5$.}
\end{figure}

\section{Empirical Results} \label{Empirical_results}

In this section, we provide empirical evidence on the predictive power of CAs-estimated factors in macroeconomic forecasting. For this purpose, we use the novel, publicly available\footnote{Freely available at the page: \url{https://zenodo.org/records/14938961}} EA-MD-QD dataset of \cite{EAdataset}. The dataset is comprised of quarterly and monthly macroeconomic time series for both the Euro Area (EA) as a whole and its ten primary member countries, and it is updated on a monthly basis and constantly revised. To the best of our knowledge, this is the first article to use this dataset in the context of forecasting and out-of-sample (factor-based) model evaluation. In detail, the EA dataset consists of 118 time series - which include 71 and 47 variables collected at quarterly and monthly frequencies, respectively - that have been recorded starting from 2000:1 to 2024:4. Similarly to the FRED-MD dataset \citep{mccracken2016fred}, the variables are classified into 11 blocks: (1) National Accounts, (2) Labor Market Indicators, (3) Credit aggregates, (4) Labor Costs, (5) Exchange Rates, (6) Interest Rates (7) Industrial Production and Turnover, (8) Prices, (9) Confidence Indicators, (10) Monetary Aggregates and (11) Others. The country-level datasets have a comparable number of time series and the same classification groups, with the exception of Monetary Aggregates since it only recorded at EA level. We will apply our tests to the EA as a whole as well as a collection of individual countries, specifically Germany, France, Italy and Spain. 

We treat the data according to the thorough suggestions of \cite{EAdataset}. First, we aggregate the monthly data at the quarterly level to address the mixed-frequency nature of the variables. Next, while our testing framework is able to accomodate for different degrees of persistence in the time series, it is necessary to address the issue of missing values and outliers beforehand. To do so, we induce stationarity in the data according to the transformations recommended for each variable in Appendix A of \cite{EAdataset}. Subsequently, we impute the missing values using the EM algorithm of \cite{StockWatson2002} and follow standard screening for outliers as detailed by \cite{mccracken2016fred}. Completion of these steps results in a first dataset with fully treated stationary data. By reversing all first-differencing transformations, we also recover a second dataset with fully treated persistent data. 

In the forecasting exercises, we choose a collection of target variables in our regressions so to display the predictive power and usefulness of the estimated factors. The selected target variables include: real GDP (\textit{GDP}), industrial production index (\textit{IPMN}), harmonised index of consumer prices excluding energy and food (\textit{HICPNEF}), total unemployment (\textit{UNETOT}), 3-month interest rates (\textit{IRT3M}), 6-month interest rates (\textit{IRT6M}), long-term interest rates (\textit{LTIRT}), wages and salaries (\textit{WS}), general government total financial liabilities (\textit{GGFB}), non-financial corporations financial liabilities (\textit{NFCLB}), households total financial liabilities (\textit{HHLB}), real exchange rate against 42 industrial countries (\textit{REER42}), residential property prices (\textit{HPRC}), and share prices (\textit{SHIX}). We then extract the factors from stationary and persistent datasets (after removing the target variable to be forecasted) and use them to augment an AR(1) benchmark. Note that richer specifications can be attained by including, among other possibilities, other exogenous regressors or extra lags in the predictors. While doing so can improve the predictive accuracy, we do not explore it in what follows. In addition, in real life, a forecaster may decide to choose subsets of the blocks so to achieve a more parsimonious forecasting model. For instance, \cite{gonccalves2017tests} investigates the predictive content of factors for the equity premium associated with the S\&P 500 Composite Index by comparing all permutations of models that include a single lag of at least one of the eight estimated macroeconomic factors. However, we are not excessively concerned with this since our tests are invariant under the use of redundant blocks after all.

To guarantee the validity of some of the assumptions in Section (\ref{Assumptions}), we run some diagnostics on the estimated factor model. In the interest of space, we report the results when the EA real GDP is the target variable in the forecasting equation, but similar conclusions for all other variables can be drawn based on additional results in the last section of the Supplement. The first assumption that we examine is that the number of blocks is at least as large as the number of true factors, i.e. $m \geq r$. To estimate the number of factors, we use the eigenvalue-ratio (ER) criterion by \cite{ahn2013eigenvalue} with $r_{max} = 8$ and find one factor in the stationary data, which is in line with the findings of \cite{EAdataset}. Following \cite{OutputGap}, we also borrow this result in the context of the persistent data\footnote{To our knowledge, \cite{PilarPoncela2006} is the only paper in the literature that proposes a test for the number of non-stationary factors, although in a dynamic factor model framework.} since, as argued by \cite{OnatskiWang2021}, the number of factors in levels should coincide with the number of factors in first differences provided that no spurious effects are estimated. \cite{EAdataset} also implement the criterion of \cite{bai2002determining}, which select seven factors. Given that \cite{ahn2013eigenvalue} is often parsimonious in small sample while the information criterion of \cite{bai2002determining} tends to overselect the number of factors, a researcher should probably accept that the true number of factors in this dataset lies between 1 and 7. As discussed next, this number may also change as the result of structural breaks.

Another assumption that we make is that the factor loadings are constant over time. Since the inception of the euro currency and the start of the dataset, the economies of the European Union (EU) members have experienced systemic shocks such as the Great Financial Crisis, a sovereign debt crisis and the Covid-19 pandemic. Thus, we raise the question about the possibility of structural breaks in the factor model. In fact, the presence of $d$ breaks in the sample requires us to reformulate Equation (\ref{eq: factor model}) into a model with $(d+1) \times r$ factors, meaning that the number of CAs now needs to satisfy $m \geq (d+1) \times r$ unequivocably. In order to address our concerns about large shocks, we use the sup-LM statistic-based test of \cite{breakschendoladogonzalo} to detect big structural breaks in these loadings at unknown dates. In particular, there is evidence against the no break null hypothesis at the timestamps 2020:2 and 2021:4 at the 5\% significance level ($\widehat{d}=2$). Unsurprisingly, these dates exactly coincide with the beginning and the end of the national lockdowns imposed by EU countries, which had systemic and material repercussions on the entire Euro area. With this finding in mind, we then confront a consequential concern and inspect if $r$ changes due to these structural breaks in the factor loadings. Indeed, this concern is not new in the literature as it was previously raised, for instance, by \cite{breitung2011testing} and \cite{BaiNg2007} in the context of US macroeconomic data. Further, \cite{OutputGap} argues that the effect of the Covid shock is pervasive in the time series of most real macroeconomic variables and should therefore be treated as an additional latent common factor. In fact, the Covid shock induced a large shift not only in the levels \citep{MarozStockWatson, Ng2021} but also in the volatility \citep{CarrClarkMarcMert2022,LenzaPrim2021} of macroeconomic variables. Hence, we apply once more the ER criterion to the subsamples between the structural breaks: 1) 2000:2-2020:1, 2) 2020:2-2021:4, and 3) 2021:4 to the end of the sample. This check follows Section 4.4 of \cite{breakschendoladogonzalo} to address the possibility of factor heteroskedasticity. We find that $r = 1$ across all three subsamples. It should be noted that even if $r = 3$ over the entire sample, the number of CAs ($m=11$) is still not smaller than $(\widehat{d}+1) \times \widehat{r} = (2+1)\times 3=9$ so we conclude that we have sufficient estimated factors. Having said this, should any of the timestamps 2020:2 and 2021:4 correspond to a break in the factor covariance matrix, the sup-LM statistic will at worst provide an overestimate of the number of true breaks in the factor loadings, meaning that our tests are still robust.

Before turning to our equal predictive ability and encompassing tests, one more diagnostic step remains. Indeed, another assumption in our setup is that the idiosyncratic components in the Equation (\ref{eq: factor model}) are at most weakly dependent cross-sectionally and in time with stationary and persistent factors but serially uncorrelated with non-stationary predictors. To illustrate that the assumption of weak cross-sectional correlation in the idiosyncratics holds, we apply the sparse estimation methodology via adaptive thresholding of \cite{CaiLiu2011} to the covariance matrix of the panel residuals in the stationary and persistent cases. For the former case, we find that the estimated matrix is almost exactly diagonal with average (resp. average absolute) value of the cross-correlation of 0.002 (resp. 0.007), which suggests that the factors estimated by CAs are able to explain most of the variation in the co-movements of the cross-sectional units. For the latter case, whereas, the estimated matrix is non-diagonal but sparse with average (resp. average absolute) value of the cross-correlation of 0.005 (resp. 0.036). We regard this level of cross-sectional dependence as very weak overall. On the other hand, serial correlation in the idiosyncratics seems to be more assiduous according to standard Ljung-Box Q-testing. In the stationary dataset, we find average (resp. average absolute) serial correlations of order 1 equal to -0.002 (resp. 0.184). Conversely, the corresponding value is 0.558 (resp. 0.559) in the persistent case. While our findings are not of concern in the context of stationary data, they raise apprehension when it comes to persistent data since we do not allow for serial correlation in association with non-stationary regressors in our theory. To alleviate some distress, we report additional Monte Carlo simulations as a benchmark in the Supplement where we explore the effect of strong serial correlation in DGP (3) with $\tau \in \{0, 0.2, 0.4, 0.6, 0.8 \}$ on size and local power. Our results seem to indicate for all four statistics that, in spite of our rates of convergence, test size is fundamentally unaffected while local power suffers a small but acceptable loss. As a result, we suspect that our convergence rates may possibly be conservative in this scenario. Since time and cross-sectional dependence in DGP (3) is stronger than the one observed in the persistent data, we trust that our equal forecasting ability and encompassing tests are still able to convey practical insight into the predictive power of the estimated non-stationary factors. With all these pre-testing analysis in mind, we finally dedicate our attention to testing predictive accuracy. 

Upon estimation, we apply our four novel tests of equal predictive accuracy and encompassing to assess whether the predictive content of CAs improves upon the forecasting ability of the autoregressive baseline with the selected target variables. The results are reported in Table \ref{tab: empirical_results}.

\begin{table}[H] 
  \centering
    \begin{tabular}{ccccccccccc}
    \toprule
          &       & \multicolumn{4}{c}{Stationary} &       & \multicolumn{4}{c}{Persistent} \\
\cmidrule{3-6}\cmidrule{8-11}    EA &       & $s_{\widehat{f},1}$    & $s_{\widehat{f},2}^{P}$    & $s_{\widehat{f},3}^{P}$   & $s_{\widehat{f},4}^{P}$    &       & $s_{\widehat{f},1}$    & $s_{\widehat{f},2}^{P}$    & $s_{\widehat{f},3}^{P}$    & $s_{\widehat{f},4}^{P}$ \\
    \midrule
    \midrule
          &       &       &       &       &       &       &       &       &       &  \\
    GDP   &       & \textbf{0.000} & \textbf{0.005} & \textbf{0.001} & \textbf{0.025} &       & 0.201 & \textbf{0.089} & \textbf{0.062} & 0.999 \\
    IPMN  &       & \textbf{0.006} & 0.221 & 0.159 & 0.962 &       & \textbf{0.000} & 0.606 & 0.556 & 0.556 \\
    HICPNEF &       & \textbf{0.000} & \textbf{0.000} & \textbf{0.002} & \textbf{0.000} &       & \textbf{0.000} & \textbf{0.000} & \textbf{0.037} & \textbf{0.000} \\
    UNETOT &       & \textbf{0.000} & \textbf{0.001} & \textbf{0.000} & \textbf{0.000} &       & \textbf{0.000} & 0.131 & \textbf{0.096} & 0.994 \\
    IRT3M &       & 0.951 & 0.857 & 0.905 & 0.983 &       & \textbf{0.000} & 0.991 & 0.999 & 1.000 \\
    IRT6M &       & 0.665 & 0.890 & 0.935 & 0.995 &       & \textbf{0.005} & 0.995 & 0.999 & 1.000 \\
    LTIRT &       & 0.668 & 0.806 & 0.951 & 0.997 &       & 0.906 & 1.000 & 1.000 & 1.000 \\
    WS    &       & \textbf{0.000} & \textbf{0.010} & \textbf{0.004} & 0.122 &       & 0.746 & 0.114 & \textbf{0.093} & 1.000 \\
    GGLB  &       & \textbf{0.000} & 0.409 & 0.643 & 0.775 &       & \textbf{0.084} & 0.966 & 0.966 & 0.966 \\
    NFCLB &       & 0.105 & 0.979 & 0.989 & 1.000 &       & 0.551 & 0.984 & 0.992 & 1.000 \\
    HHLB  &       & 0.374 & 0.988 & 0.989 & 1.000 &       & 0.403 & 0.981 & 0.990 & 1.000 \\
    REER42 &       & 0.941 & 1.000 & 1.000 & 1.000 &       & 0.751 & 0.991 & 0.996 & 1.000 \\
    HPRC  &       & \textbf{0.000} & 0.781 & 0.978 & 0.949 &       & 0.131 & 0.977 & 0.988 & 1.000 \\
    SHIX  &       & 0.929 & 0.995 & 0.997 & 1.000 &       & 0.120 & 0.999 & 1.000 & 1.000 \\
          &       &       &       &       &       &       &       &       &       &  \\
    \bottomrule
    \end{tabular}%
    \caption{p-values of CA-based equal predictive accuracy and encompassing test statistics on EA quarterly data. Setting: $h=1$. The significance level is chosen at 10\%. }
  \label{tab: empirical_results}%
\end{table}

First and foremost, it is paramount to highlight that the predictive content inherently embedded in our CAs-estimated factors is particularly pronounced with target variables that play a key role in the design of political, fiscal or monetary policies. These variables are economic growth, inflation (excluding energy and food), unemployment and wages. Since monitoring inflation is at the core of the European Central Bank's mandate and policymaking, it is very valuable to observe that the estimated factors have formidable forecasting power for consumer prices under all tests and types of persistence. Focusing on individual tests, we see that $s_{\widehat{f},1}$ exhibits statistical significance across the largest collection of target variables. While this statistic exhibits power, we know from our simulations that it tends to overreject the null. Therefore, the support from other statistics, especially $s_{\widehat{f},2}^{P}$ and $s_{\widehat{f},3}^{P}$, strengthens the evidence for factor predictive power. Unsurprisingly to our theory, our statistics detect better overall forecast accuracy using stationary data, which is still against the popular belief that stationary data incurs a material loss of information content compared to non-stationary data. In fact, this finding was already anticipated by the fact that the convergence rates of our statistics slow down with increasing persistence in the data. Further, while the ultimate objective of our tests is to uncover the predictive specifications where the estimated factors improve upon the (autoregressive) baseline, it is also important to identify those situations where the factors are not informative about the future and should therefore be excluded from our forecasting considerations, as is rightfully and conventionally the case for the changes in, and the level of, the long-term interest rate, the exchange rate and the stock market returns. In fact, these three variables are notoriously difficult, if not impossible, to predict because - assuming the efficient market hypothesis holds - the expectations of rational, informed traders must already be priced into financial markets.

Nevertheless, we acknowledge that our results may inadvertedly be affected by structural breaks in the coefficients of the factor-augmented regression. Although we have dedicated considerable attention to breaks in the factor loadings, there exist no statistical procedure to detect structural instability à la \cite{CorradiSwanson1014} under the CAs framework. In fact, these authors propose a test for the joint hypothesis of structural stability of both factor loadings and regression coefficients in the factor-augmented forecasting model in a PC setting. As previously argued, the period associated with the Covid-19 pandemic should be treated by including an additional factor in the specification, which will by construction capture a portion of the shift in the level of the target variables. Hence, overestimating the number of CAs makes us robust to: 1) the possibility of breaks in the factor loadings, and consequently 2) an increase in the number of factors resulting from a large shock, but possibly also 3) a large break in the factor-augmented regression coefficients. However, it is not unreasonable to think that some concerns about the last point have not been dispelled once and for all. At the same time, our empirical results would not be interpretable fairly under the tests of \cite{clark2001tests} and \cite{mccracken2007asymptotics} due to their strong foundational assumptions about factor properties. \\
\indent We repeat the same forecasting exercise with data on Germany, France, Italy and Spain. The target variables are the same, with the exception of the 3-month and 6-month interest rates since they are not available at the country level. The results of our tests are reported in Table \ref{tab: empirical_results_ALL}. 
\begin{table}[H]
\centering

\begin{minipage}[t]{0.495\textwidth}
\centering
\resizebox{\textwidth}{!}{%
\begin{tabular}{ccccccccccc}
\toprule
      &       & \multicolumn{4}{c}{Stationary} &       & \multicolumn{4}{c}{Persistent} \\
\cmidrule{3-6}\cmidrule{8-11}
Germany & & $s_{\widehat{f},1}$ & $s_{\widehat{f},2}^{P}$ & $s_{\widehat{f},3}^{P}$ & $s_{\widehat{f},4}^{P}$ & & $s_{\widehat{f},1}$ & $s_{\widehat{f},2}^{P}$ & $s_{\widehat{f},3}^{P}$ & $s_{\widehat{f},4}^{P}$ \\
\midrule \midrule
GDP   & & \textbf{0.000} & \textbf{0.000} & \textbf{0.000} & \textbf{0.000} & & \textbf{0.072} & 0.144 & 0.107 & 0.991 \\
IPMN  & & \textbf{0.000} & \textbf{0.077} & 0.107 & 0.697 & & \textbf{0.022} & 0.885 & 0.927 & 0.999 \\
HICPNEF & & 0.434 & \textbf{0.099} & 0.255 & 0.965 & & \textbf{0.000} & \textbf{0.009} & 0.133 & 0.245 \\
UNETOT & & 0.635 & 1.000 & 1.000 & 1.000 & & \textbf{0.000} & 0.777 & 0.856 & 0.327 \\
LTIRT & & 0.999 & 0.980 & 0.996 & 1.000 & & 0.999 & 0.998 & 0.999 & 1.000 \\
WS    & & \textbf{0.000} & \textbf{0.000} & \textbf{0.000} & \textbf{0.009} & & \textbf{0.002} & \textbf{0.011} & \textbf{0.008} & 0.588 \\
GGLB  & & 0.351 & 0.435 & 0.522 & 0.971 & & 0.985 & 0.852 & 0.914 & 1.000 \\
NFCLB & & 0.650 & 0.999 & 1.000 & 1.000 & & \textbf{0.000} & 0.988 & 0.995 & 1.000 \\
HHLB  & & 0.928 & 0.822 & 0.640 & 1.000 & & \textbf{0.000} & \textbf{0.000} & \textbf{0.000} & \textbf{0.000} \\
REER42 & & 0.970 & 0.999 & 0.999 & 1.000 & & 0.677 & 1.000 & 1.000 & 1.000 \\
HPRC  & & 0.970 & 0.999 & 0.999 & 1.000 & & \textbf{0.000} & \textbf{0.000} & \textbf{0.000} & \textbf{0.000} \\
SHIX  & & 0.991 & 0.994 & 0.997 & 1.000 & & \textbf{0.000} & 0.866 & 0.882 & 1.000 \\
\bottomrule
\end{tabular}}
\end{minipage}%
\hfill
\begin{minipage}[t]{0.495\textwidth}
\centering
\resizebox{\textwidth}{!}{%
\begin{tabular}{ccccccccccc}
\toprule
      &       & \multicolumn{4}{c}{Stationary} &       & \multicolumn{4}{c}{Persistent} \\
\cmidrule{3-6}\cmidrule{8-11}
France & & $s_{\widehat{f},1}$ & $s_{\widehat{f},2}^{P}$ & $s_{\widehat{f},3}^{P}$ & $s_{\widehat{f},4}^{P}$ & & $s_{\widehat{f},1}$ & $s_{\widehat{f},2}^{P}$ & $s_{\widehat{f},3}^{P}$ & $s_{\widehat{f},4}^{P}$ \\
\midrule \midrule
GDP   & & \textbf{0.000} & \textbf{0.000} & \textbf{0.000} & \textbf{0.000} & & 0.948 & 0.222 & 0.193 & 0.999 \\
IPMN  & & \textbf{0.001} & \textbf{0.097} & \textbf{0.058} & 0.988 & & \textbf{0.020} & 0.248 & 0.190 & 0.931 \\
HICPNEF & & 0.621 & 0.230 & 0.552 & 0.999 & & \textbf{0.000} & \textbf{0.000} & \textbf{0.022} & \textbf{0.000} \\
UNETOT & & \textbf{0.000} & \textbf{0.000} & \textbf{0.000} & \textbf{0.000} & & \textbf{0.000} & \textbf{0.000} & \textbf{0.000} & \textbf{0.000} \\
LTIRT & & 0.382 & 0.999 & 1.000 & 1.000 & & \textbf{0.000} & 0.999 & 1.000 & 1.000 \\
WS    & & \textbf{0.000} & \textbf{0.014} & \textbf{0.006} & \textbf{0.095} & & 0.700 & \textbf{0.087} & \textbf{0.065} & 0.987 \\
GGLB  & & \textbf{0.000} & 0.987 & 0.998 & 1.000 & & \textbf{0.000} & 1.000 & 1.000 & 1.000 \\
NFCLB & & 0.999 & 0.994 & 0.998 & 1.000 & & \textbf{0.011} & 0.989 & 0.997 & 1.000 \\
HHLB  & & 0.992 & 1.000 & 1.000 & 1.000 & & 0.618 & 1.000 & 1.000 & 1.000 \\
REER42 & & 0.999 & 0.996 & 0.998 & 1.000 & & 0.971 & 0.999 & 1.000 & 1.000 \\
HPRC  & & \textbf{0.059} & 0.662 & 0.640 & 1.000 & & \textbf{0.000} & 0.779 & 0.882 & 0.999 \\
SHIX  & & 0.992 & 0.996 & 0.998 & 1.000 & & \textbf{0.024} & 0.996 & 0.998 & 1.000 \\
\bottomrule
\end{tabular}}
\end{minipage}

\vspace{1em}

\begin{minipage}[t]{0.495\textwidth}
\centering
\resizebox{\textwidth}{!}{%
\begin{tabular}{ccccccccccc}
\toprule
      &       & \multicolumn{4}{c}{Stationary} &       & \multicolumn{4}{c}{Persistent} \\
\cmidrule{3-6}\cmidrule{8-11}
Italy & & $s_{\widehat{f},1}$ & $s_{\widehat{f},2}^{P}$ & $s_{\widehat{f},3}^{P}$ & $s_{\widehat{f},4}^{P}$ & & $s_{\widehat{f},1}$ & $s_{\widehat{f},2}^{P}$ & $s_{\widehat{f},3}^{P}$ & $s_{\widehat{f},4}^{P}$ \\
\midrule \midrule
GDP   & & \textbf{0.000} & \textbf{0.000} & \textbf{0.000} & \textbf{0.000} & & \textbf{0.000} & \textbf{0.000} & \textbf{0.000} & \textbf{0.009} \\
IPMN  & & 0.793 & \textbf{0.096} & \textbf{0.062} & 1.000 & & \textbf{0.053} & 0.105 & \textbf{0.071} & 0.977 \\
HICPNEF & & \textbf{0.000} & 0.186 & 0.420 & 0.998 & & \textbf{0.000} & \textbf{0.000} & 0.101 & \textbf{0.000} \\
UNETOT & & \textbf{0.000} & \textbf{0.002} & \textbf{0.001} & \textbf{0.004} & & \textbf{0.000} & \textbf{0.031} & \textbf{0.026} & 0.817 \\
LTIRT & & 0.299 & 0.696 & 0.829 & 0.999 & & \textbf{0.002} & 0.963 & 0.993 & 1.000 \\
WS    & & \textbf{0.000} & \textbf{0.000} & \textbf{0.000} & \textbf{0.000} & & 0.239 & \textbf{0.025} & \textbf{0.015} & 0.951 \\
GGLB  & & \textbf{0.069} & \textbf{0.055} & 0.383 & \textbf{0.009} & & 0.155 & \textbf{0.081} & 0.461 & \textbf{0.000} \\
NFCLB & & 0.674 & 0.798 & 0.841 & 1.000 & & \textbf{0.002} & 0.999 & 1.000 & 1.000 \\
HHLB  & & \textbf{0.090} & 0.150 & 0.119 & 1.000 & & \textbf{0.021} & \textbf{0.086} & \textbf{0.038} & 0.981 \\
REER42 & & 0.871 & 0.919 & 0.963 & 1.000 & & \textbf{0.090} & 0.969 & 0.997 & 1.000 \\
HPRC  & & 0.136 & \textbf{0.018} & 0.161 & 0.126 & & \textbf{0.000} & \textbf{0.000} & \textbf{0.000} & \textbf{0.000} \\
SHIX  & & 0.911 & 0.979 & 0.982 & 1.000 & & 0.233 & 1.000 & 1.000 & 1.000 \\
\bottomrule
\end{tabular}}
\end{minipage}%
\hfill
\begin{minipage}[t]{0.495\textwidth}
\centering
\resizebox{\textwidth}{!}{%
\begin{tabular}{ccccccccccc}
\toprule
      &       & \multicolumn{4}{c}{Stationary} &       & \multicolumn{4}{c}{Persistent} \\
\cmidrule{3-6}\cmidrule{8-11}
Spain & & $s_{\widehat{f},1}$ & $s_{\widehat{f},2}^{P}$ & $s_{\widehat{f},3}^{P}$ & $s_{\widehat{f},4}^{P}$ & & $s_{\widehat{f},1}$ & $s_{\widehat{f},2}^{P}$ & $s_{\widehat{f},3}^{P}$ & $s_{\widehat{f},4}^{P}$ \\
\midrule \midrule
GDP   & & \textbf{0.074} & \textbf{0.052} & \textbf{0.032} & 0.606 & & 0.870 & 0.293 & 0.272 & 1.000 \\
IPMN  & & 0.895 & 0.352 & 0.328 & 1.000 & & \textbf{0.005} & 0.250 & 0.225 & 1.000 \\
HICPNEF & & 0.999 & 0.998 & 1.000 & 1.000 & & \textbf{0.000} & 0.830 & 0.996 & 0.851 \\
UNETOT & & 0.902 & 0.902 & 0.902 & 0.902 & & \textbf{0.000} & 0.705 & 0.703 & 0.535 \\
LTIRT & & 1.000 & 1.000 & 1.000 & 1.000 & & 0.243 & 0.993 & 0.999 & 1.000 \\
WS    & & \textbf{0.000} & \textbf{0.001} & \textbf{0.000} & \textbf{0.013} & & \textbf{0.002} & \textbf{0.040} & \textbf{0.028} & 0.850 \\
GGLB  & & \textbf{0.000} & 0.939 & 0.955 & 0.994 & & \textbf{0.000} & 0.726 & 0.763 & 0.452 \\
NFCLB & & \textbf{0.000} & 0.993 & 0.994 & 1.000 & & \textbf{0.000} & 1.000 & 1.000 & 1.000 \\
HHLB  & & 0.672 & 1.000 & 1.000 & 1.000 & & \textbf{0.000} & \textbf{0.000} & \textbf{0.000} & \textbf{0.000} \\
REER42 & & 0.358 & 1.000 & 1.000 & 1.000 & & 0.961 & 1.000 & 1.000 & 1.000 \\
HPRC  & & \textbf{0.000} & 1.000 & 1.000 & 1.000 & & \textbf{0.000} & 0.195 & 0.133 & \textbf{0.000} \\
SHIX  & & 0.997 & 0.999 & 0.999 & 1.000 & & \textbf{0.009} & 1.000 & 1.000 & 1.000 \\
\bottomrule
\end{tabular}}
\end{minipage}

\vspace{0.5em}
\caption{Results for Germany, France, Italy, and Spain. Same characteristics as in Table \ref{tab: empirical_results}.}
\label{tab: empirical_results_ALL}
\end{table}
Although the institutions of these countries are structurally distinct, the results will naturally be heterogeneous across countries. That is, the factors extracted from each country dataset are found to be informative about the future evolution of different variables. In sum, it seems from our results that the macroeconomic factors are still able to offer consistent forecasting insight into economic growth, wages and (possibly) consumer prices across the four countries, with the notable exceptions of Spain and Italy. With respect to Spain, the factors appear to show quite weak predictive ability while, for Italy, they anticipate the bulk of economic activity across most variables, excluding the stock market. As before, $s_{\widehat{f},1}$ detects the greatest number of specifications where our macroeconomic factors yield forecasting power. All things considered, we conclude that - although the CAs method may appear deceivingly simplistic at first glance - predictive regressions augmented with the estimated factors offer powerful predictive insight in the context of macroeconomic forecasting in both levels and differences, especially in the presence of possibly big structural breaks.

\section{Conclusion}
We present four novel tests of equal predictive accuracy and encompassing for factor-augmented regressions under a CAs setup by revisiting \cite{stauskas2022tests}. While their results in the context of tests by \cite{clark2001tests} bring important practical implications, merging CAs with a novel class of tests by \cite{pitarakis2023direct} and \cite{pitarakis2025novel} opens new possibilities. The overspecification of the number of factors has an asymptotically negligible influence, thereby eliminating the need to know the number of factors (as long as it is bounded from above by the number of averages). This goes against the usual trend in this literature according to which the excess CAs result in bias even asymptotically. In practice, the suggested upper bound is often substantial, as can be seen from our empirical application on the novel EA-MD-QD dataset. Further, observed predictors and factors can be persistent, and the location of structural breaks in the factor loadings need not be known. Ultimately, Monte Carlo simulations reveal an excellent small sample performance across many empirically relevant scenarios.
\newpage

\newpage

\section{Appendix}
\subsection{Auxiliary Results}
In the auxiliary lemmas and Lemma 1, we will use the stacked notation, such that $\*A\in \mathbb{R}^{(t-1)\times p}$ stacks $p\times 1$ vectors over time in a recursive fashion.
\renewcommand{\theequation}{A.\arabic{equation}}
\subsubsection{General Dependence vs. Uncorrelatedness in the Idiosyncratic Component}
\noindent \textbf{Lemma A.1} \textit{Under Assumptions 1-4, we have that 
\begin{align}
    \left\|\frac{1}{T^{1+\tau}}\*Z'\overline{\*E} \right\|=O_p(N^{-1/2}T^{-\tau/2}) \label{eq: lemma.a.1}
\end{align}
 as $(N,T)\to \infty$}.\\

\noindent \textbf{Proof.} To begin with, we can write $\frac{1}{T^{1+\tau}}\*Z'\overline{\*E}=\frac{1}{T^{1+\tau}}\sum_{t=1}^T\*z_t\overline{\*e}_t'$. For this and the upcoming auxiliary results, we use the total sample for $t=1,\ldots, T$, however, by Assumption 2 (c), the same rates hold in a recursive setup for each $t$. This is so, because by stacking over $s=1,\ldots,t$, we can write $\frac{1}{T^{1+\tau}}\sum_{s=1}^t\*z_s\overline{\*e}_{s}'=\left(\frac{t}{T}\right)^{1+\tau}\frac{1}{t^{1+\tau}}\sum_{s=1}^t\*z_s\overline{\*e}_{s}'$, where $t=\lfloor T\eta\rfloor$ for $\eta\in (0,1)$. Then for some positive $\epsilon$,
\begin{align}
     \mathbb{P}\left(\left\|\frac{1}{T^{1+\tau}}\sum_{t=1}^T\*z_t\overline{\*e}_t' \right\|>\epsilon \right)&\leq \epsilon^{-1}\mathbb{E}\left(\left\| \frac{1}{T^{1+\tau}}\sum_{t=1}^T\*z_t\overline{\*e}_t'\right\| \right)\notag\\
     &=\frac{1}{\epsilon \sqrt{N}}\mathbb{E}\left(\left\| \frac{1}{T^{1+\tau}}\sum_{t=1}^T\*z_t(\sqrt{N}\overline{\*e}_t')\right\| \right)\notag\\
     &\leq \frac{1}{\epsilon \sqrt{N}}\mathbb{E}\left( \frac{1}{T^{1+\tau}}\sum_{t=1}^T\left\|\*z_t\right\|\left\|(\sqrt{N}\overline{\*e}_t)\right\| \right)\notag\\
     &= \frac{1}{\epsilon \sqrt{N}} \frac{1}{T^{1+\tau}}\sum_{t=1}^T\mathbb{E}(\left\|\*z_t\right\|)\mathbb{E}\left(\left\|(\sqrt{N}\overline{\*e}_t)\right\|\right) \notag\\
     &\leq \frac{1}{\epsilon \sqrt{N}T^{\tau/2}} \left(\frac{1}{T^{1+\tau}}\sum_{t=1}^T\mathbb{E}(\left\|\*z_t\right\|)^2\right)^{1/2}\underbrace{\left(\frac{1}{T}\sum_{t=1}^T\mathbb{E}\left(\left\|(\sqrt{N}\overline{\*e}_t)\right\|\right)^2\right)^{1/2}}_{O(1)}\notag\\
     &=O(N^{-1/2}T^{-\tau/2}),
\end{align}
because 
\begin{align}
    \frac{1}{T^{1+\tau}}\sum_{t=1}^T\mathbb{E}(\left\|\*z_t\right\|)^2&\leq \frac{1}{T^{1+\tau}}\sum_{t=1}^T\mathbb{E}\left[\left(\sqrt{(\mathrm{tr}(\*z_t\*z_t'))} \right)^2\right]\notag\\
    &=\frac{1}{T^{1+\tau}}\sum_{t=1}^T\mathbb{E}(\mathrm{tr}[\*z_t\*z_t'])=O(1)
\end{align}
by the concavity of function appearing in Jensen's inequality. Note that this result gives a different rate than for similar terms in \cite{pitarakis2023direct}, which is brought down exactly by time dependence in $\{\*e_{i,t}\}$.  Also note that independence between the predictors and idiosyncratics is not strictly necessary, and we use it for simplicity. Indeed, 
\begin{align}
     \mathbb{P}\left(\left\|\frac{1}{T^{1+\tau}}\sum_{t=1}^T\*z_t\overline{\*e}_t' \right\|>\epsilon \right)&\leq \frac{1}{\epsilon \sqrt{N}}\mathbb{E}\left( \frac{1}{T^{1+\tau}}\sum_{t=1}^T\left\|\*z_t\right\|\left\|(\sqrt{N}\overline{\*e}_t)\right\| \right)\notag\\
     &\leq \frac{1}{\epsilon \sqrt{N}} \frac{1}{T^{1+\tau}}\sum_{t=1}^T\mathbb{E}\left( \left\|\*z_t \right\|^2\right)^{1/2}\mathbb{E}\left(\left\| \sqrt{N}\overline{\*e}_t\right\|^2\right)^{1/2}\notag\\
     &\leq \frac{\sup_t\mathbb{E}\left(\left\| \sqrt{N} \overline{\*e}_t\right\|^2\right)^{1/2}}{\epsilon \sqrt{N}T^{\tau/2}}\sup_t \mathbb{E}\left( \left\|T^{-\tau/2}\*z_t \right\|^2\right)^{1/2}\notag\\
     &=O(N^{-1/2}T^{-\tau/2}),
\end{align}
because $\sup_t \mathbb{E}\left( \left\|T^{-\tau/2}\*z_t \right\|^2\right)^{1/2}=O(1)$ by Lemma 3.1 in \cite{magdalinos2009limit}. 
\\

\noindent \textbf{Corollary A.1} \textit{Under Assumptions 1-4, but $\*e_{i,t}$ is uncorrelated over time, we have that 
\begin{align}
    \left\|\frac{1}{T^{1+\tau}}\*Z'\overline{\*E} \right\|=O_p(N^{-1/2}T^{-(1+\tau)/2}) \label{eq: corollary.a.1}
\end{align}
 as $(N,T)\to \infty$}.\\

\noindent \textbf{Proof.} Assume that $\*e_{i,t}$ is uncorrelated over time. Then by Markov's inequality we obtain 
\begin{align}\label{ZE_bound}
    \mathbb{P}\left(\left\|\frac{1}{T^{1+\tau}}\sum_{t=1}^T\*z_t\overline{\*e}_t' \right\|>\epsilon \right)&\leq \epsilon^{-2}\mathbb{E}\left(\left\|\frac{1}{T^{1+\tau}}\sum_{t=1}^T\*z_t\overline{\*e}_t' \right\|^2 \right)\notag\\
    &=\epsilon^{-2} \frac{1}{T^{1+\tau}}\mathbb{E}\left(\mathrm{tr}\left[\frac{1}{T^{1+\tau}}\sum_{t=1}^T\sum_{s=1}^T\*z_t\overline{\*e}_t'\overline{\*e}_s\*z_s' \right] \right)\notag\\
    &= \epsilon^{-2} \frac{1}{T^{1+\tau}} \left(\mathrm{tr}\left[\frac{1}{T^{1+\tau}}\sum_{t=1}^T\sum_{s=1}^T\mathbb{E}\left(\*z_s'\*z_t \overline{\*e}_t'\overline{\*e}_s\right) \right] \right)\notag\\
    &= \epsilon^{-2} \frac{1}{T^{1+\tau}} \left(\mathrm{tr}\left[\frac{1}{T^{1+\tau}}\sum_{t=1}^T\sum_{s=1}^T\mathbb{E}\left(\*z_t'\*z_s \right)\mathbb{E}\left(\overline{\*e}_t'\overline{\*e}_s\right) \right] \right)\notag\\
    &= \epsilon^{-2} \frac{1}{T^{1+\tau}} \left(\mathrm{tr}\left[\frac{1}{T^{1+\tau}}\sum_{t=1}^T\mathbb{E}\left(\*z_t'\*z_t \right)\mathbb{E}\left(\overline{\*e}_t'\overline{\*e}_t\right) \right] \right)\notag\\
    &=\epsilon^{-2} \frac{1}{NT^{1+\tau}} \left(\mathrm{tr}\left[\frac{1}{T^{1+\tau}}\sum_{t=1}^T\mathbb{E}\left(\*z_t'\*z_t \right)\mathbb{E}\left(N\overline{\*e}_t'\overline{\*e}_t\right) \right] \right)\notag\\
    &\leq \epsilon^{-2} \frac{\sup_{k_0\leq t\leq T}\mathbb{E}\left(N\overline{\*e}_t'\overline{\*e}_t\right)}{NT^{1+\tau}} \left(\mathrm{tr}\left[\frac{1}{T^{1+\tau}}\sum_{t=1}^T\mathbb{E}\left(\*z_t'\*z_t \right) \right] \right)\notag\\
    &\leq O(N^{-1}T^{-1-\tau})\times \underbrace{\mathrm{tr}\left[ \frac{1}{T^{1+\tau}}\sum_{t=1}^T\mathbb{E}\left(\*z_t\*z_t' \right)\right]}_{O(1)}\notag\\
    &=  O(N^{-1}T^{-1-\tau}),
\end{align}
which implies that $\left\|\frac{1}{T^{1+\tau}}\*Z'\overline{\*E} \right\|=O_p(N^{-1/2}T^{-(1+\tau)/2})$ if $\*e_{i,t}$ is uncorrelated over time. \\

\noindent \textbf{Lemma A.2} \textit{Under Assumptions 1-4, we have that 
\begin{align*}
    \left\| \frac{1}{T^{1+\tau}}\*Z'\*u\right\|=O_p(T^{-(1+\tau)/2})
\end{align*}
as $T\to \infty$.
}
\\

\noindent  \textbf{Proof.} By using similar steps as in Lemma A.1, we get 
\begin{align}
      \mathbb{P}\left(\left\|\frac{1}{T^{1+\tau}}\sum_{t=1}^T\*z_tu_{t+1} \right\|>\epsilon \right)&\leq \epsilon^{-2}\mathbb{E}\left(\left\|\frac{1}{T^{1+\tau}}\sum_{t=1}^{T}\*z_tu_{t+1} \right\|^2\right)\notag\\
      &= \epsilon^{-2} \frac{1}{T^{1+\tau}}\mathbb{E}\left(\frac{1}{T^{1+\tau}}\sum_{t=1}^T\sum_{s=1}^T\*z_t'\*z_su_{t+1}u_{s+1} \right)\notag\\
      &=  \epsilon^{-2} \frac{1}{T^{1+\tau}}\mathbb{E}\left(\frac{1}{T^{1+\tau}}\sum_{t=1}^T\*z_t'\*z_t\mathbb{E}(u_{t+1}^2|\mathcal{F}_{t}) \right)\notag\\
      &=  \frac{\sigma^2}{\epsilon^{2}T^{1+\tau}}\frac{1}{T^{1+\tau}}\sum_{t=1}^T\mathrm{tr}\left[\mathbb{E}(\*z_t\*z_t')\right]\notag\\
      &=O(T^{-(1+\tau)}),
\end{align}
which implies that $\left\|\frac{1}{T^{1+\tau}}\sum_{t=1}^T\*z_tu_{t+1} \right\|=O_p(T^{-(1+\tau)/2})$. Note that this rate is effectively assumed in \cite{pitarakis2023direct} and \cite{pitarakis2025novel}. However, it naturally follows from our more primitive assumptions.
\subsubsection{Extension to $MA(p)$ Idiosyncratics}
\noindent We can extend our analysis and proceed to develop a moving average analog of (\ref{ZE_bound}) by assuming that $\*e_{i,t}$ follows an $MA(p)$ process. \\

\noindent \textbf{Lemma A.3} \textit{Under Assumptions 1-4 and idiosyncratics following and $MA(p)$ process, we have that 
\begin{align}
    \left\|\frac{1}{T^{1+\tau}}\*Z'\overline{\*E} \right\|=O_p(N^{-1/2}T^{-(1+\tau)/2}) \label{eq: lemma.a.4}
\end{align}
 as $(N,T)\to \infty$}.\\

\noindent \textbf{Proof.} By applying the same chain of inequalities, we obtain 
\begin{align}
    \mathbb{P}&\left(\left\|\frac{1}{T^{1+\tau}}\sum_{t=1}^{T}\*z_t\overline{\*e}_t' \right\|>\epsilon \right)\leq \epsilon^{-2} \frac{1}{T^{1+\tau}} \left(\mathrm{tr}\left[\frac{1}{T^{1+\tau}}\sum_{t=1}^T\sum_{s=1}^T\mathbb{E}\left(\*z_t'\*z_s \right)\mathbb{E}\left(\overline{\*e}_t'\overline{\*e}_s\right) \right] \right)\notag\\
    &=\epsilon ^{-2}\frac{1}{T^{1+\tau}}\left(\frac{1}{T^{1+\tau}}\sum_{t=1}^T\sum_{s=\max\{1,t-p\}}^{\min \{T,t+p \}}\mathbb{E}\left(\*z_t'\*z_s \right)\mathbb{E}\left(\overline{\*e}_t'\overline{\*e}_s\right) \right)\notag\\
    &\leq \epsilon ^{-2}\frac{1}{T^{1+\tau}}\left(\frac{1}{T^{1+\tau}}\sum_{t=1}^T\sum_{s=\max\{1,t-p\}}^{\min \{T,t+p \}}\mathbb{E}\left(|\*z_t'\*z_s| \right)\left|\mathbb{E}\left(\overline{\*e}_t'\overline{\*e}_s\right)\right| \right)\notag\\
    &\leq \epsilon ^{-2}\frac{1}{NT^{1+\tau}}\sup_{t,s}\left|\mathbb{E}\left(N\overline{\*e}_t'\overline{\*e}_s\right)\right| \left(\frac{1}{T}\sum_{t=1}^T\sum_{s=\max\{1,t-p\}}^{\min \{T,t+p \}}\mathbb{E}\left(\left\|T^{-\tau/2}\*z_t\right\| \left\|T^{-\tau/2}\*z_s\right\| \right)\right)\notag\\
    &\leq \epsilon ^{-2}\frac{1}{NT^{1+\tau}}\sup_{t,s}\left|\mathbb{E}\left(N\overline{\*e}_t'\overline{\*e}_s\right)\right| \left(\frac{1}{T}\sum_{t=1}^T\sum_{s=\max\{1,t-p\}}^{\min \{T,t+p \}}\left[\mathbb{E}\left(\left\|T^{-\tau/2}\*z_t\right\|^2\right)\right] ^{1/2}\left[\mathbb{E}\left(\left\|T^{-\tau/2}\*z_s\right\|^2 \right)\right]^{1/2}\right)\notag\\
    & \leq \epsilon ^{-2}\frac{1}{NT^{1+\tau}}\sup_{t,s}\left|\mathbb{E}\left(N\overline{\*e}_t'\overline{\*e}_s\right)\right| \sup_t \mathbb{E}\left(\left\|T^{-\tau/2}\*z_t\right\|^2\right) \times B\notag\\
    &=O(N^{-1}T^{-1-\tau})
\end{align}
for some $B>0$. Here we used the fact that $\sup_{t}\left[\mathbb{E}\left(\left\|T^{-\tau/2}\*z_t \right\|^2\right)\right]^{1/2}\times \sup_{s}\left[\mathbb{E}\left(\left\|T^{-\tau/2}\*z_s \right\|^2\right)\right]^{1/2}=\sup_{t}\mathbb{E}\left(\left\|T^{-\tau/2}\*z_t \right\|^2\right)=O(1)$, and the fact that the sum over an index $s$ is necessarily bounded in case of $MA(h-1)$. This implies that $\left\|\frac{1}{T^{1+\tau}}\sum_{t=1}^{T}\*z_t\overline{\*e}_t' \right\|=O_p(N^{-1/2}T^{-(1+\tau)/2})$.
\subsection{Using the Rotation Matrix}
The case when $m > r$ is problematic. One reason for this is that $\overline{\boldsymbol{\Lambda}}$ is no longer invertible. However, we still need an equivalent of 
\begin{align}\label{m=r_object}
\overline{\+\Lambda}^{-1\prime}\widehat{\*f}_t=\*f_t+O_p(N^{-1/2}),
\end{align}
because it determines the object that is being estimated. The way we approach this issue is by introducing the following rotation matrix, which is chosen such that $\overline{\boldsymbol{\Lambda}}\overline{\mathbf{H}}= [\mathbf{I}_r, \mathbf{0}_{r\times (m-r)}]$ and that is going to play the same role as $\overline{\boldsymbol{\Lambda}}^{-1}$ under $m=r$:
\begin{equation}
    \overline{\mathbf{H}} =\left[\begin{array}{cc}\overline{\boldsymbol{\Lambda}}_r^{-1} & -\overline{\boldsymbol{\Lambda}}_r^{-1}\overline{\boldsymbol{\Lambda}}_{-r}\\
    \mathbf{0}_{(m-r)\times r} & \mathbf{I}_{m-r}\end{array}\right]=[\overline{\mathbf{H}}_r,\, \overline{\mathbf{H}}_{-r}]\in \mathbb{R}^{m\times m},
\end{equation}
where $\overline{\mathbf{H}}_r=[\overline{\boldsymbol{\Lambda}}_r^{-1\prime}, \mathbf{0}_{r\times (m-r)} ]'\in \mathbb{R}^{m\times r}$ and $\overline{\mathbf{H}}_{-r}=[ -\overline{\boldsymbol{\Lambda}}_{-r}'\overline{\boldsymbol{\Lambda}}_r^{-1\prime}, \mathbf{I}_{m-r}]'\in \mathbb{R}^{m\times (m-r)}$. If $m=r$, we define $\overline{\mathbf{H}}=\overline{\mathbf{H}}_r=\overline{\boldsymbol{\Lambda}}_r^{-1}=\overline{\boldsymbol{\Lambda}}^{-1}$. We further introduce $\mathbf{D}_N= \mathrm{diag}(\mathbf{I}_r , \sqrt{N}\mathbf{I}_{m-r}) \in \mathbb{R}^{m\times m}$
with $\mathbf{D}_N=\mathbf{I}_m$ if $m=r$. By pre-multiplying $\widehat{\mathbf{f}}_t$ by $\mathbf{D}_N\overline{\mathbf{H}}'$, we obtain
\begin{equation}\label{fhat0}
    \mathbf{D}_N\overline{\mathbf{H}}'\widehat{\mathbf{f}}_t=\widehat{\mathbf{f}}^0_t= \mathbf{D}_N\overline{\mathbf{H}}'\overline{\boldsymbol{\Lambda}}'\mathbf{f}_t+ \mathbf{D}_N\overline{\mathbf{H}}'\overline{\mathbf{e}}_t=\mathbf{f}^0_t+\overline{\mathbf{e}}^0_t,
\end{equation}
where $\mathbf{f}^0_t=[\mathbf{f}'_t, \mathbf{0}_{(m-r)\times 1}']'\in \mathbb{R}^{m}$ and $\overline{\mathbf{e}}^0_t=[\overline{\mathbf{e}}_{r,t}'\overline{\boldsymbol{\Lambda}}_r^{-1}, \sqrt{N}(\overline{\mathbf{e}}_{-r,t}- \overline{\boldsymbol{\Lambda}}_{-r}'\overline{\boldsymbol{\Lambda}}_r^{-1\prime}\overline{\mathbf{e}}_{r,t})']'=
[\overline{\mathbf{e}}^{0\prime}_{r,t}, \overline{\mathbf{e}}^{0\prime}_{-r,t}]'$ $\in \mathbb{R}^{m}$ with $\overline{\mathbf{e}}_{r,t}\in \mathbb{R}^{r}$ and $\overline{\mathbf{e}}_{-r,t}\in \mathbb{R}^{m-r}$ being the partitions of $\overline{\mathbf{e}}_t=[\overline{\mathbf{e}}_{r,t}', \overline{\mathbf{e}}_{-r,t}']'$. If $m=r$, then $\widehat{\mathbf{f}}^0_t=\overline{\boldsymbol{\Lambda}}^{-1\prime}\widehat{\mathbf{f}}_t$, $\mathbf{f}^0_t = \mathbf{f}_t$ and $\overline{\mathbf{e}}^0_t = \overline{\boldsymbol{\Lambda}}^{-1\prime}\overline{\mathbf{e}}_t$, and so we are back in (\ref{m=r_object}). Hence, since $\|\overline{\mathbf{e}}_{r,t}^0\| =O_p(N^{-1/2})$ and $\|\overline{\mathbf{e}}_{-r,t}^0\| =O_p(1)$, when $m > r$ we are no longer estimating $\mathbf{f}_t$ but rather $[\mathbf{f}_t', \overline{\mathbf{e}}_{-r,t}^{0\prime}]'$.\footnote{The scaling by $\sqrt{N}$ in $\mathbf{D}_N$ and hence in $\overline{\mathbf{e}}_{-r,t}^0$ is necessary, for without it $\overline{\mathbf{H}}'\widehat{\mathbf{f}}_t$ is estimating $[\mathbf{f}_t', \mathbf{0}_{(m-r)\times 1}']'$, whose second moment matrix is singular.} The fact that $\mathbf{f}_t$ is included in this object suggests that asymptotically, CAs should be able to account for the unknown factors even if $m > r$. By ensuring the existence of $\overline{\mathbf{H}}$, Assumption 5 makes this possible. However, we also note that because of the presence of $\overline{\mathbf{e}}_{-r,t}^0$, the asymptotic distribution theory will depend on whether $m=r$ or $m > r$.
\subsection{Proofs of the Main Results} 
\subsubsection{Proof of Lemma 1}\label{sec: Sec_2}
To prove Lemma 1, let us begin from the decomposition in \cite{stauskas2022tests} (proof of Lemma 1, Appendix A):
\\
\begin{eqnarray} \label{eq: decomp}
    &&\* D_T (\* Q_N^{-1}\boldsymbol{\widehat{\delta}}_t - \boldsymbol{\widetilde{\delta}_t^0}) \nonumber \\
    &=& \begin{bmatrix} \* D_{\* F} \left((\widehat{\* F}^{0\prime}\* M_{\* W} {\widehat{\* F}^0})^{+}\widehat{\* F}^{0\prime}\* M_{\* W} - \begin{bmatrix} ({\* F^\prime}\* M_{\* W} {\* F})^{+}{\* F^\prime}\* M_{\* W} \\ \* 0_{(m-r)\times(t-1)} \end{bmatrix} \right)\widehat{\* Z}^0 \boldsymbol{\delta}^0 \\ T^{\frac{1}{4}+\frac{\tau}{2}}\* I_q \bigl[({\* W^\prime}\* M_{\widehat{\* F}^0} {\* W})^{+}{\* W^\prime}\* M_{\widehat{\* F}^0} - ({\* W^\prime}\* M_{\* F} {\* W})^{+}{\* W^\prime}\* M_{\* F}
 \bigr]\widehat{\* Z}^0 \boldsymbol{\delta}^0 \end{bmatrix} \nonumber\\
    &-& \begin{bmatrix} \* D_{\* F} \left((\widehat{\* F}^{0\prime}\* M_{\* W} {\widehat{\* F}^0})^{+}\widehat{\* F}^{0\prime}\* M_{\* W} - \begin{bmatrix} ({\* F^\prime}\* M_{\* W} {\* F})^{+}{\* F^\prime}\* M_{\* W} \\ \* 0_{(m-r)\times(t-1)} \end{bmatrix} \right) \overline{\* E}_r^0 \boldsymbol{\alpha}  \\ T^{\frac{1}{4}+\frac{\tau}{2}}\* I_q \bigl[({\* W^\prime}\* M_{\widehat{\* F}^0} {\* W})^{+}{\* W^\prime}\* M_{\widehat{\* F}^0} - ({\* W^\prime}\* M_{\* F} {\* W})^{+}{\* W^\prime}\* M_{\* F}
 \bigr] \overline{\* E}_r^0 \boldsymbol{\alpha} \end{bmatrix} \nonumber\\ 
    &+& \begin{bmatrix} \* D_{\* F} \left((\widehat{\* F}^{0\prime}\* M_{\* W} {\widehat{\* F}^0})^{+}\widehat{\* F}^{0\prime}\* M_{\* W} - \begin{bmatrix} ({\* F^\prime}\* M_{\* W} {\* F})^{+}{\* F^\prime}\* M_{\* W} \\ \* 0_{(m-r)\times(t-1)} \end{bmatrix} \right) \*u \\ T^{\frac{1}{4}+\frac{\tau}{2}}\* I_q \bigl[({\* W^\prime}\* M_{\widehat{\* F}^0} {\* W})^{+}{\* W^\prime}\* M_{\widehat{\* F}^0} - ({\* W^\prime}\* M_{\* F} {\* W})^{+}{\* W^\prime}\* M_{\* F}
 \bigr] \*u \end{bmatrix} \nonumber\\
    &=& \begin{bmatrix} \mathrm{\* I} \\ \mathrm{\* I\*I} \end{bmatrix} - \begin{bmatrix} \mathrm{\* I\*I\*I} \\ \mathrm{\* IV} \end{bmatrix} + \begin{bmatrix} \mathrm{\* V} \\ \mathrm{\* V\*I} \end{bmatrix},
\end{eqnarray}
where $\* D_{T} = \mathrm{diag}(\* D_{\* F}, T^{\frac{1}{4}+\frac{\tau}{2}}\* I_q)$ with $\* D_{\* F}=\mathrm{diag}(T^{\frac{1}{4}+\frac{\tau}{2}}\* I_r, T^{1/4}\* I_{m-r})$ and the definitions of $\mathrm{\* I}-\mathrm{\* VI}$ are implicit. WE assume that $\*e_{i,t}$ is uncorrelated over time. Note that the matrices above stack observations over $s=1,\ldots, t-1$, therefore, we conduct the proof for a fixed $t$. However, under our assumptions, the rates will hold uniformly. 

Now we can work out the order of each of the terms by applying simple adjustments to the proofs of and by using the results in \cite{stauskas2022tests}. Indeed, starting with $\mathrm{\* I}$, and recalling that $\widehat{\* Z}^0 \boldsymbol{\delta}^0 = \widehat{\* F}^0{(\* D_N \overline{\* H}' \overline{\boldsymbol{\Lambda}}')^{+ \prime}}\boldsymbol{\alpha} + \* W\boldsymbol{\theta}$, $\boldsymbol{\alpha} = T^{-\frac{1}{4}-\frac{\tau}{2}}\boldsymbol{\alpha}^0$, $\* M_{\* W}\* W = \* 0_{(t-1)\times n}$, and the fact that $\widehat{\* F}^{0\prime}\* M_{\* W} {\widehat{\* F}^0}$ is a full rank matrix wp1, then
\begin{eqnarray} \label{eq: decomp11}
&&\* D_{\* F} (\widehat{\* F}^{0\prime}\* M_{\* W} {\widehat{\* F}^0})^{+}\widehat{\* F}^{0\prime}\* M_{\* W} \widehat{\* Z}^0 \boldsymbol{\delta}^0 \nonumber \\ 
&=& \* D_{\* F} (\widehat{\* F}^{0\prime}\* M_{\* W} {\widehat{\* F}^0})^{+}\widehat{\* F}^{0\prime}\* M_{\* W} \widehat{\* F}^0{(\* D_N \overline{\* H}' \overline{\boldsymbol{\Lambda}}')^{+ \prime}} \boldsymbol{\alpha} + \* D_{\* F} (\widehat{\* F}^{0\prime}\* M_{\* W} {\widehat{\* F}^0})^{+}\widehat{\* F}^{0\prime}\* M_{\* W} \* W\boldsymbol{\theta} \nonumber \\
&=& {(\* D_N \overline{\* H}' \overline{\boldsymbol{\Lambda}}')^{+ \prime}}\boldsymbol{\alpha}^0 = [\boldsymbol{\alpha}^{0 \prime},\* 0_{1\times (m-r))}]^\prime,
\end{eqnarray}
which holds wp1. Further recall that $({\* F^\prime}\* M_{\* W} {\* F})^{+}{\* F^\prime}\* M_{\* W} = (\* D_N \overline{\* H}' \overline{\boldsymbol{\Lambda}}')^\prime ({{{\* F}^0}^\prime}\* M_{\* W} {{\* F}^0})^{+}{{{\* F}^0}^\prime}\* M_{\* W}$, $\overline{\* E}^0(\* D_N \overline{\* H}' \overline{\boldsymbol{\Lambda}}')^\prime \boldsymbol{\alpha} =T^{\frac{1}{4}+\frac{\tau}{2}}\overline{\* E}_r^0 \boldsymbol{\alpha}^0$ and the fact that ${\* F^\prime}\* M_{\* W} {\* F}$ is full rank and positive definite wp1. Then
\begin{eqnarray} \label{eq: Sigmafw}
t^{-1-\tau}{\* F^\prime}\* M_{\* W} {\* F} &=& t^{-1-\tau}{\* F^\prime}{\* F}-t^{-1-\tau}{\* F^\prime}\* W(t^{-1-\tau}{\* W^\prime}{\* W})^{+}t^{-1-\tau}{\* W^\prime}{\* F} \nonumber \\
&\to_p& \boldsymbol{\Sigma}_{ff} - \boldsymbol{\Sigma}_{wf}^\prime\boldsymbol{\Sigma}_{ww}^{-1}\boldsymbol{\Sigma}_{fw} = \boldsymbol{\Sigma}_{f.w},
\end{eqnarray}
which is positive definite by assumption because $\boldsymbol{\Sigma}_{ZZ}$ is (Assumption 2). In addition, the rank of $T^{-1-\tau}{\* F^\prime}\* M_{\* W} {\* F}$ is invariant to changes in $T \to \infty$, so 
\begin{eqnarray*}
\big\|(t^{-1-\tau}{\* F^\prime}\* M_{\* W} {\* F})^{+}-\boldsymbol{\Sigma}_{f.w}^{+}\big\| = o_p(1).
\end{eqnarray*}
Thus, it follows that
\begin{eqnarray} \label{eq: FMF}
\big\|(t^{-1-\tau}{\* F^\prime}\* M_{\* W} {\* F})^{+}\big\| = O_p(1).
\end{eqnarray}
In a similar manner, $\big\|(t^{-1-\tau}{\* W^\prime} {\* W})^{+}\big\| = O_p(1)$. Turning now to $T^{-1-\tau}{\* F^\prime}\* M_{\* W}  \overline{\* E}_r^0$,
\begin{eqnarray} \label{eq: FMwE}
\big\|T^{-1-\tau}{\* F^\prime}\* M_{\* W}  \overline{\* E}_r^0\boldsymbol{\alpha}^0 \big\| &\leq & \big\|T^{-1-\tau}{\* F^\prime}\overline{\* E}_r^0 \big\| \big\|\boldsymbol{\alpha}^0\big\| - \big\|T^{-1-\tau}{\* F^\prime}\* W\big\|\big\|(T^{-1-\tau}{\* W^\prime} {\* W})^{+}\big\|\big\|T^{-1-\tau}{\* W^\prime}\overline{\* E}_r^0\big\| \big\|\boldsymbol{\alpha}^0\big\| \nonumber \\
&=&  O_p(N^{-\frac{1}{2}}T^{-\frac{1}{2} - \frac{\tau}{2}}).
\end{eqnarray}
By using these results,
\begin{eqnarray} \label{eq: decomp12}
&&\* T^{\frac{1}{4}+\frac{\tau}{2}}\*I_r ({\* F^\prime}\* M_{\* W} {\* F})^{+}{\* F^\prime}\* M_{\* W} \widehat{\* Z}^0 \boldsymbol{\delta}^0 \nonumber \\
&=& T^{\frac{1}{4}+\frac{\tau}{2}}\boldsymbol{\alpha} + T^{\frac{1}{4}+\frac{\tau}{2}}({\* F^\prime}\* M_{\* W} {\* F})^{+}{\* F^\prime}\* M_{\* W}  \overline{\* E}_r^0 \boldsymbol{\alpha} \nonumber \\
&=& \boldsymbol{\alpha}^0 + (T^{-1-\tau}{\* F^\prime}\* M_{\* W} {\* F})^{+}T^{-1-\tau}{\* F^\prime}\* M_{\* W}  \overline{\* E}_r^0 \boldsymbol{\alpha}^0 \nonumber \\
&=& \boldsymbol{\alpha}^0 + O_p(N^{-\frac{1}{2}}T^{-\frac{1}{2} - \frac{\tau}{2}}).   
\end{eqnarray}
By combining the results in (\ref{eq: decomp11}) and (\ref{eq: decomp12}), we obtain 
\begin{eqnarray}
\mathrm{\* I} &=& \* D_{\* F} \left((\widehat{\* F}^{0\prime}\* M_{\* W} {\widehat{\* F}^0})^{+}\widehat{\* F}^{0\prime}\* M_{\* W} - \begin{bmatrix} ({\* F^\prime}\* M_{\* W} {\* F})^{+}{\* F^\prime}\* M_{\* W} \\ \* 0_{(m-r)\times(t-1)} \end{bmatrix} \right)\widehat{\* Z}^0 \boldsymbol{\delta}^0 \nonumber \\
&=& \begin{bmatrix} \boldsymbol{\alpha}^0 \\ \* 0_{(m-r)\times 1} \end{bmatrix} - \left(\begin{bmatrix} \boldsymbol{\alpha}^0 \\ \* 0_{(m-r)\times 1} \end{bmatrix} + O_p(N^{-\frac{1}{2}}T^{-\frac{1}{2} - \frac{\tau}{2}})) \right) = O_p(N^{-\frac{1}{2}}T^{-\frac{1}{2} - \frac{\tau}{2}}).
 \end{eqnarray}\\
 
 Consider now $\mathrm{\*II}$. Since $\* M_{\widehat{\* F}^0} \widehat{\* F}^0 = \* 0_{(t-1)\times m}$ then it can be shown that 
\begin{eqnarray} \label{eq: decomp21}
&&T^{\frac{1}{4}+\frac{\tau}{2}} ({\* W^\prime}\* M_{\widehat{\* F}^0} {\* W})^{+}{\* W^\prime}\* M_{\widehat{\* F}^0}\widehat{\* Z}^0 \boldsymbol{\delta}^0 \nonumber \\
&=&T^{\frac{1}{4}+\frac{\tau}{2}} ({\* W^\prime}\* M_{\widehat{\* F}^0} {\* W})^{+}{\* W^\prime}\* M_{\widehat{\* F}^0}(\widehat{\* F}^0{(\* D_N \overline{\* H}' \overline{\boldsymbol{\Lambda}}')^{+ \prime}}\boldsymbol{\alpha} + \* W\boldsymbol{\theta}) \nonumber \\
&=&T^{\frac{1}{4}+\frac{\tau}{2}} ({\* W^\prime}\* M_{\widehat{\* F}^0} {\* W})^{+}{\* W^\prime}\* M_{\widehat{\* F}^0}\widehat{\* F}^0{(\* D_N \overline{\* H}' \overline{\boldsymbol{\Lambda}}')^{+ \prime}}\boldsymbol{\alpha} + T^{\frac{1}{4}+\frac{\tau}{2}} ({\* W^\prime}\* M_{\widehat{\* F}^0} {\* W})^{+}{\* W^\prime}\* M_{\widehat{\* F}^0}\* W\boldsymbol{\theta} \nonumber \\
&=& T^{\frac{1}{4}+\frac{\tau}{2}}\boldsymbol{\theta},
\end{eqnarray}
whereas using the definition and properties of $\* F^{0}$ together with previous results gives
\begin{eqnarray}
&&T^{\frac{1}{4}+\frac{\tau}{2}}({\* W^\prime}\* M_{\* F} {\* W})^{+}{\* W^\prime}\* M_{\* F}\widehat{\* Z}^0 \boldsymbol{\delta}^0 \nonumber \\
&=& T^{\frac{1}{4}+\frac{\tau}{2}}({\* W^\prime}\* M_{\* F} {\* W})^{+}{\* W^\prime}\* M_{\* F}\overline{\* E}_r^0 \boldsymbol{\alpha} + T^{\frac{1}{4}+\frac{\tau}{2}}\boldsymbol{\theta} \nonumber \\
&=& (T^{-1-\tau}{\* W^\prime}\* M_{\* F} {\* W})^{+}T^{-1-\tau}{\* W^\prime}\* M_{\* F}\overline{\* E}_r^0 \boldsymbol{\alpha}^0 + T^{\frac{1}{4}+\frac{\tau}{2}}\boldsymbol{\theta}.
\end{eqnarray}
Observe that ${\* W^\prime}\* M_{\* F} {\* W}$ is positive definite wp1 and
\begin{eqnarray}
t^{1+\tau}T^{-1-\tau}t^{-1-\tau}{\* W^\prime}\* M_{\* F} {\* W} &=&(t^{1+\tau}T^{-1-\tau}) [t^{-1-\tau} \* W^\prime \* W - t^{-1-\tau}{\* W^\prime}\* F(t^{-1-\tau}{\* F^\prime}{\* F})^{+}t^{-1-\tau}{\* F^\prime}{\* W}] \nonumber \\
&\to_p& s^{1+\tau}[\boldsymbol{\Sigma}_{ww} - \boldsymbol{\Sigma}_{fw}^\prime\boldsymbol{\Sigma}_{ff}^{-1}\boldsymbol{\Sigma}_{wf}] = s^{1+\tau}\boldsymbol{\Sigma}_{w.f}
\end{eqnarray}
in light of the fact that 
\begin{eqnarray} \label{eq: tTs}
\sup_{1 \leq t \leq T} \quad \sup_{(t-1)T^{-1} \leq s \leq tT^{-1}} |(tT^{-1})^k - s^k| = O(T^{-1}), \quad \forall k >0.
\end{eqnarray}
Since the rank of $T^{-1-\tau}{\* W^\prime}\* M_{\* F} {\* W}$ does not change as $T \to \infty$, then 
\begin{eqnarray}
\big\|(T^{-1-\tau}{\* W^\prime}\* M_{\* F} {\* W})^{+}-s^{-1-\tau}\boldsymbol{\Sigma}_{w.f}^{+}\big\| = o_p(1).
\end{eqnarray}
Thus, it follows that
\begin{eqnarray} \label{eq: WMfW}
\big\|(T^{-1-\tau}{\* W^\prime}\* M_{\* F} {\* W})^{+}\big\| = O_p(1).
\end{eqnarray}
Further, since 
\begin{eqnarray} \label{eq: decomp22}
&& \big\|T^{-1-\tau}{\* W^\prime}\* M_{\* F}  \overline{\* E}_r^0\boldsymbol{\alpha}^0 \big\| \nonumber \\
&& \leq \big\|T^{-1-\tau}{\* W^\prime}\overline{\* E}_r^0 \big\| \big\|\boldsymbol{\alpha}^0\big\| + \big\|T^{-1-\tau}{\* W^\prime}\* F\big\|\big\|(T^{-1-\tau}{\* F^\prime} {\* F})^{+}\big\|\big\|T^{-1-\tau}{\* F^\prime}\overline{\* E}_r^0\big\| \big\|\boldsymbol{\alpha}^0\big\| \nonumber\\
&& =  O_p(N^{-\frac{1}{2}}T^{-\frac{1}{2} - \frac{\tau}{2}}),
\end{eqnarray}
implying that
\begin{eqnarray*}
&&\big\|(T^{-1-\tau}{\* W^\prime}\* M_{\* F} {\* W})^{+}T^{-1-\tau}{\* W^\prime}\* M_{\* F}\overline{\* E}_r^0 \boldsymbol{\alpha}^0\big\| \\
&\leq& \big\|(T^{-1-\tau}{\* W^\prime}\* M_{\* F} {\* W})^{+}\big\| \big\|T^{-1-\tau}{\* W^\prime}\* M_{\* F}  \overline{\* E}_r^0\big\|\big\|\boldsymbol{\alpha}^0 \big\| =O_p(N^{-\frac{1}{2}}T^{-\frac{1}{2} - \frac{\tau}{2}})
\end{eqnarray*}
so that
\begin{eqnarray} \label{eq: decomp23}
&&T^{\frac{1}{4}+\frac{\tau}{2}}({\* W^\prime}\* M_{\* F} {\* W})^{+}{\* W^\prime}\* M_{\* F}\widehat{\* Z}^0 \boldsymbol{\delta}^0 \nonumber \\
&=& ({\* W^\prime}\* M_{\* F} {\* W})^{+}{\* W^\prime}\* M_{\* F}\overline{\* E}_r^0 \boldsymbol{\alpha}^0 + T^{\frac{1}{4}+\frac{\tau}{2}}\boldsymbol{\theta} = T^{\frac{1}{4}+\frac{\tau}{2}}\boldsymbol{\theta} + O_p(N^{-\frac{1}{2}}T^{-\frac{1}{2} - \frac{\tau}{2}})
\end{eqnarray}
 Combining the results in (\ref{eq: decomp21}) and (\ref{eq: decomp23}) yields
\begin{eqnarray}
 \mathrm{\*{II}}&=&T^{\frac{1}{4}+\frac{\tau}{2}}\bigl[({\* W^\prime}\* M_{\widehat{\* F}^0} {\* W})^{+}{\* W^\prime}\* M_{\widehat{\* F}^0} - ({\* W^\prime}\* M_{\* F} {\* W})^{+}{\* W^\prime}\* M_{\* F}
 \bigr]\widehat{\* Z}^0 \boldsymbol{\delta}^0 \nonumber \\
 &=& T^{\frac{1}{4}+\frac{\tau}{2}}\boldsymbol{\theta} - \bigl[T^{\frac{1}{4}+\frac{\tau}{2}}\boldsymbol{\theta} + O_p(N^{-\frac{1}{2}}T^{-\frac{1}{2} - \frac{\tau}{2}})\bigr] \nonumber \\
 &=& O_p(N^{-\frac{1}{2}}T^{-\frac{1}{2} - \frac{\tau}{2}}).
\end{eqnarray}
 
We now continue onto $\mathrm{\*III}$ and $\mathrm{\*IV}$. It is useful to recall that $\widehat{\* F}^0 = \* F^{0}+\overline{\* E}^0$ is correlated with both $\overline{\* E}_r^0$ and $\* W$ so that $\big\|T^{-1-\tau}{\* W^\prime}\widehat{\* F}^0 \big\| = \big\|T^{-1-\tau}{{\widehat{\*F}^{0 \prime}}}{\* W}\big\| = O_p(1)$. Similarly, define $\* S$ to be 
\begin{eqnarray}
\* S = \begin{bmatrix}
\* F^\prime {\* F} & \* 0_{r \times (m-r)} \\
\* 0_{(m-r)\times r} & \overline{\* E}_{-r}^{0 \prime} \overline{\* E}_{-r}^{0}.
\end{bmatrix}
\end{eqnarray}
Note that $T^{-1-\tau}\* F^\prime {\* F}$ is positive definite. Further, $T^{-1}\overline{\* E}_{-r}^{0 \prime} \overline{\* E}_{-r}^{0}$ is positive definite wp1, as well \citep{stauskas2022tests}. Similarly to \cite{karabiyik2017role}, define $\* G=\mathrm{diag}(T^{\frac{1}{2} + \frac{\tau}{2}}\*I_r, T^{\frac{1}{2}}\*I_{m-r})$ and consider the decomposition
\begin{eqnarray}
\* G^{-1}\widehat{\* F}^{0 \prime} \widehat{\* F}^{0}\* G^{-1} = \* G^{-1}\* F^{0 \prime} \* F^{0}\* G^{-1} + \* G^{-1}\* F^{0 \prime}\overline{\* E}^0\* G^{-1} + \* G^{-1}\overline{\* E}^{0 \prime}\* F^{0} \* G^{-1} + \* G^{-1} \overline{\* E}^{0 \prime}\overline{\* E}^0\* G^{-1}.
\end{eqnarray}
In detail,  
\begin{eqnarray}
\* G^{-1}\* F^{0 \prime} \* F^{0}\* G^{-1} &=& \begin{bmatrix}
T^{-1-\tau}\* F^\prime {\* F} & \* 0_{r \times (m-r)} \\
\* 0_{(m-r)\times r} & \* 0_{(m-r)\times (m-r)}
\end{bmatrix}, 
\end{eqnarray}
\begin{eqnarray}
\* G^{-1}\* F^{0 \prime} \overline{\* E}^0\* G^{-1} &=&\begin{bmatrix}
T^{-1-\tau}\* F^\prime \overline{\* E}_r\boldsymbol{\overline{\Lambda}}_r^{-1} & N^{\frac{1}{2}}T^{-1-\frac{\tau}{2}}(\* F^\prime \overline{\* E}_{-r} - \* F^\prime \overline{\* E}_r\boldsymbol{\overline{\Lambda}}_r^{-1}\boldsymbol{\overline{\Lambda}}_{-r} ) \\
\* 0_{(m-r)\times r} & \* 0_{(m-r)\times (m-r)}
\end{bmatrix},
\end{eqnarray}
\begin{eqnarray}
\* G^{-1}\overline{\* E}^{0 \prime}\* F^{0}\* G^{-1} &=&\begin{bmatrix}
T^{-1-\tau}(\boldsymbol{\overline{\Lambda}}_r^{-1})^\prime \overline{\* E}_r^\prime\* F & \* 0_{(m-r)\times r} \\
N^{\frac{1}{2}}T^{-1-\frac{\tau}{2}}(\overline{\* E}_{-r}^\prime\* F - \boldsymbol{\overline{\Lambda}_{-r}^\prime}(\boldsymbol{\overline{\Lambda}}_r^{-1})^\prime \overline{\* E}_r^\prime\* F) & \* 0_{(m-r)\times (m-r)}
\end{bmatrix},
\end{eqnarray}
and
\begin{eqnarray}
\* G^{-1}\overline{\* E}^{0 \prime}\overline{\* E}^0\* G^{-1} &=& \begin{bmatrix}
T^{-1-\tau} \overline{\* E}_{r}^{0 \prime}\overline{\* E}_{r}^{0} & T^{-1-\frac{\tau}{2}}\overline{\* E}_{r}^{0 \prime}\overline{\* E}_{-r}^{0}  \\
T^{-1-\frac{\tau}{2}}\overline{\* E}_{-r}^{0 \prime}\overline{\* E}_{r}^{0} & T^{-1}\overline{\* E}_{-r}^{0 \prime}\overline{\* E}_{-r}^{0}
\end{bmatrix} 
\end{eqnarray}
where 
\begin{eqnarray}
\overline{\* E}_{r}^{0 \prime}\overline{\* E}_{r}^{0} &=& (\boldsymbol{\overline{\Lambda}}_r^{-1})^\prime \overline{\* E}_r^\prime\overline{\* E}_r(\boldsymbol{\overline{\Lambda}}_r^{-1}) \nonumber \\
\overline{\* E}_{r}^{0 \prime}\overline{\* E}_{-r}^{0} &=& N^{\frac{1}{2}}((\boldsymbol{\overline{\Lambda}}_r^{-1})^\prime\overline{\* E}_r^\prime \overline{\* E}_{-r} - (\boldsymbol{\overline{\Lambda}}_r^{-1})^\prime\overline{\* E}_r^\prime \overline{\* E}_r\boldsymbol{\overline{\Lambda}}_r^{-1}\boldsymbol{\overline{\Lambda}}_{-r}) \nonumber \\
\overline{\* E}_{-r}^{0 \prime}\overline{\* E}_{r}^{0} &=& N^{\frac{1}{2}}(\overline{\* E}_{-r}^\prime\overline{\* E}_r\boldsymbol{\overline{\Lambda}}_r^{-1} - \boldsymbol{\overline{\Lambda}_{-r}^\prime}(\boldsymbol{\overline{\Lambda}}_r^{-1})^\prime \overline{\* E}_r^\prime\overline{\* E}_r\boldsymbol{\overline{\Lambda}}_r^{-1}) \nonumber \\
\overline{\* E}_{-r}^{0 \prime}\overline{\* E}_{-r}^{0} &=& N(\overline{\* E}_{-r}^{\prime}\overline{\* E}_{-r} - \overline{\* E}_{-r}^{\prime}\overline{\* E}_r\boldsymbol{\overline{\Lambda}}_r^{-1}\boldsymbol{\overline{\Lambda}}_{-r} -\boldsymbol{\overline{\Lambda}_{-r}^\prime}(\boldsymbol{\overline{\Lambda}}_r^{-1})^\prime \overline{\* E}_r^\prime \overline{\* E}_{-r} + \boldsymbol{\overline{\Lambda}_{-r}^\prime}(\boldsymbol{\overline{\Lambda}}_r^{-1})^\prime \overline{\* E}_r^\prime\overline{\* E}_r\boldsymbol{\overline{\Lambda}}_r^{-1}\boldsymbol{\overline{\Lambda}}_{-r} )\nonumber
\end{eqnarray}
Now notice that $\big\|T^{-1 - \frac{\tau}{2}}\* F^\prime \overline{\* E}\big\|$ is of order $O_p(N^{-\frac{1}{2}}T^{-\frac{1}{2}})$, which implies that $\big\|\* G^{-1}\* F^{0 \prime} \overline{\* E}^0\* G^{-1}\big\|$ and $\big\|\* G^{-1}\overline{\* E}^{0 \prime}\* F^{0}\* G^{-1}\big\|$ are both of order $O_p(T^{-\frac{1}{2}})$ by symmetry arguments by Corollary A.1. However, by Lemma A.1 $\big\|\* G^{-1}\overline{\* E}^{0 \prime}\* F^{0}\* G^{-1}\big\|=O_p(1)$ if the idiosyncratics are serially correlated. Finally, it remains to address the terms in $T^{-1}\overline{\* E}^{0 \prime} \overline{\* E}^0$: 
\begin{eqnarray} \label{eq: decomp 211}
\big\|T^{-1-\tau}(\boldsymbol{\overline{\Lambda}}_r^{-1})^\prime \overline{\* E}_r^\prime\overline{\* E}_r(\boldsymbol{\overline{\Lambda}}_r^{-1}) \big\| &\leq& \big\|\boldsymbol{\overline{\Lambda}}_r^{-1} \big\|^2 \big\|T^{-1-\tau}\overline{\* E}_r^\prime \overline{\* E}_r\big\| \nonumber \\
&=& O_p(N^{-1}T^{-\tau}),
\end{eqnarray}
by virtue of the fact that $\big\|T^{-1} \overline{\*E}^\prime \overline{\*E}\big\|$ is $O_p(N^{-1})$  while  $\big\|\boldsymbol{\overline{\Lambda}}_r \big\|$ and $\big\|\boldsymbol{\overline{\Lambda}}_r^{-1} \big\|$ are $O_p(1)$. In turn, (\ref{eq: decomp 211}) implies
\begin{eqnarray}
\big\|N^{\frac{1}{2}}T^{-1-\frac{\tau}{2}}((\boldsymbol{\overline{\Lambda}}_r^{-1})^\prime\overline{\* E}_r^\prime \overline{\* E}_{-r} - (\boldsymbol{\overline{\Lambda}}_r^{-1})^\prime\overline{\* E}_r^\prime \overline{\* E}_r\boldsymbol{\overline{\Lambda}}_r^{-1}\boldsymbol{\overline{\Lambda}}_{-r}) \big\| &\leq& N^{\frac{1}{2}} \big\|\boldsymbol{\overline{\Lambda}}_r^{-1} \big\| \big\|T^{-1-\frac{\tau}{2}}\overline{\* E}_r^\prime \overline{\* E}_{-r}\big\| \nonumber\\
&+& N^{\frac{1}{2}} \big\|\boldsymbol{\overline{\Lambda}}_r^{-1} \big\|^2 \big\|T^{-1-\frac{\tau}{2}}\overline{\* E}_r^\prime \overline{\* E}_{r}\big\|\big\|\boldsymbol{\overline{\Lambda}}_{-r} \big\| \nonumber\\
&=& O_p(N^{-\frac{1}{2}}T^{-\frac{\tau}{2}})
\end{eqnarray}
By symmetry, it also follows that $\big\|N^{\frac{1}{2}}T^{-1-\frac{\tau}{2}}(\overline{\* E}_{-r}^\prime\overline{\* E}_r\boldsymbol{\overline{\Lambda}}_r^{-1} - \boldsymbol{\overline{\Lambda}_{-r}^\prime}(\boldsymbol{\overline{\Lambda}}_r^{-1})^\prime \overline{\* E}_r^\prime\overline{\* E}_r\boldsymbol{\overline{\Lambda}}_r^{-1}) \big\|= O_p(N^{-\frac{1}{2}}T^{-\frac{\tau}{2}})$. Moreover,      
\begin{eqnarray}
&&\big\|NT^{-1}(\overline{\* E}_{-r}^{\prime}\overline{\* E}_{-r} - \overline{\* E}_{-r}^{\prime}\overline{\* E}_r\boldsymbol{\overline{\Lambda}}_r^{-1}\boldsymbol{\overline{\Lambda}}_{-r} -\boldsymbol{\overline{\Lambda}_{-r}^\prime}(\boldsymbol{\overline{\Lambda}}_r^{-1})^\prime \overline{\* E}_r^\prime \overline{\* E}_{-r} + \boldsymbol{\overline{\Lambda}_{-r}^\prime}(\boldsymbol{\overline{\Lambda}}_r^{-1})^\prime \overline{\* E}_r^\prime\overline{\* E}_r\boldsymbol{\overline{\Lambda}}_r^{-1}\boldsymbol{\overline{\Lambda}}_{-r} )\nonumber \big\| \\
&&\leq N\big\|T^{-1}\overline{\* E}_{-r}^{\prime}\overline{\* E}_{-r}\big\| + N\big\|T^{-1}\overline{\* E}_{-r}^{\prime}\overline{\* E}_r\big\| \big\|\boldsymbol{\overline{\Lambda}}_r^{-1}\big\|\big\|\boldsymbol{\overline{\Lambda}}_{-r}\big\| +N\big\|\boldsymbol{\overline{\Lambda}}_r^{-1}\big\|\big\|\boldsymbol{\overline{\Lambda}}_{-r}\big\|\big\|T^{-1}\overline{\* E}_r^{\prime}\overline{\* E}_{-r}\big\| \nonumber \\
&&+N\big\|\boldsymbol{\overline{\Lambda}}_r^{-1}\big\|^2\big\|\boldsymbol{\overline{\Lambda}}_{-r}\big\|^2\big\|T^{-1}\overline{\* E}_{-r}^{\prime}\overline{\* E}_{-r}\big\| \nonumber \\
&&= O_p(1).
\end{eqnarray}
Thus, the probability limit of $\* G^{-1}\widehat{\* F}^{0 \prime} \widehat{\* F}^{0}\* G^{-1}$ is then derived by combining the above results:
\begin{eqnarray} \label{eq: F0F0order}
\big\|\* G^{-1}\widehat{\* F}^{0 \prime} \widehat{\* F}^{0}\* G^{-1} - \* G^{-1}\* S\* G^{-1} \big\| = O_p(N^{-\frac{1}{2}}T^{-\frac{\tau}{2}}) + O_p(T^{-\frac{1}{2}})
\end{eqnarray}
That is, we obtain
\begin{eqnarray} \label{eq: decompF0F0}
\* G^{-1}\widehat{\* F}^{0 \prime} \widehat{\* F}^{0}\* G^{-1} = \* G^{-1}\* S\* G^{-1} + O_p(N^{-\frac{1}{2}}T^{-\frac{\tau}{2}}) + O_p(T^{-\frac{1}{2}}),
\end{eqnarray}
Since $T^{-1-\tau}\* F^\prime {\* F}$ and $T^{-1}\overline{\* E}_{-r}^{0 \prime} \overline{\* E}_{-r}^{0}$ are both positive definite wp1, then so is $\boldsymbol{\Sigma}_{\* F^{0}}$. Further, the rank of $\* G^{-1}\* S\* G^{-1}$ is $m$, which is also the rank of $\* G^{-1}\widehat{\* F}^{0 \prime} \widehat{\* F}^{0}\* G^{-1}$. It then follows that as $N, T \to \infty$,
\begin{eqnarray} \label{eq: rankF0F0}
rk [\* G^{-1}\widehat{\* F}^{0 \prime} \widehat{\* F}^{0}\* G^{-1}] \to rk \* G^{-1}\* S\* G^{-1}
\end{eqnarray}
almost surely. Following \cite{karabiyik2017role}, (\ref{eq: decompF0F0}) and (\ref{eq: rankF0F0}) lead to the key result 
\begin{eqnarray} \label{eq: orderF0F0prime}
\big\|(\* G^{-1}\widehat{\* F}^{0 \prime} \widehat{\* F}^{0}\* G^{-1})^{+} - (\* G^{-1}\* S\* G^{-1})^{+} \big\| = O_p(N^{-\frac{1}{2}}T^{-\frac{\tau}{2}}) + O_p(T^{-\frac{1}{2}}).
\end{eqnarray}
An important implication of this result is that 
\begin{eqnarray}
\big\|(\* G^{-1}\widehat{\* F}^{0 \prime} \widehat{\* F}^{0}\* G^{-1})^{+} \big\| = O_p(1).
\end{eqnarray}
Further, note that $\left\| \* G^{-1}\right\|\leq M T^{-1/2}$ for some positive $M$ and 
\begin{eqnarray} \label{eq: WF0hatGminus1}
\left\|T^{-\frac{1}{2}-\frac{\tau}{2}}\* W^\prime \widehat{\* F}^{0} \* G^{-1}\right\| &\leq& \left\|T^{-\frac{1}{2}-\frac{\tau}{2}}\* W^\prime \* F^{0} \* G^{-1}\right\| + \left\|T^{-\frac{1}{2}-\frac{\tau}{2}}\* W^\prime \overline{\* E}^{0} \* G^{-1}\right\| \nonumber \\
&\leq& \left\|T^{-\frac{1}{2}-\frac{\tau}{2}}\* W^\prime \* F^{0}\* G^{-1}\right\|  + \left\|T^{-\frac{1}{2}-\frac{\tau}{2}}\* W^\prime \overline{\* E}^{0}\right\| \left\|\* G^{-1}\right\| \nonumber \\
&\leq& \left\|T^{-1-\tau}\* W^\prime \* F\right\| + M T^{-\frac{1}{2}} \left\|T^{-\frac{1}{2}-\frac{\tau}{2}}\* W^\prime \overline{\* E}^{0}\right\| \nonumber \\
&=& O_p(1) + O_p(T^{-\frac{1}{2}}) = O_p(1).
\end{eqnarray}
In turn, it follows by using (A.77) in the Supplement of \cite{stauskas2022tests} and (\ref{eq: WF0hatGminus1}) that 
\begin{eqnarray}
&&\big\|T^{-1-\tau}{\* W^\prime}\* M_{\widehat{\* F}^0} {\* W} - t^{-1-\tau}{\* W^\prime}\* M_{\* F^{0}} {\* W} \big\| \nonumber \\
&& = \big\|T^{-1-\tau}{\* W^\prime}(\* M_{\widehat{\* F}^0}-\* M_{\* F^{0}}) {\* W} \big\| \nonumber \\
&& \leq \big\|T^{-1-\frac{\tau}{2}}\* W^\prime \overline{\* E}_{-r}^{0}(T^{-1}\overline{\* E}_{-r}^{0 \prime}\overline{\* E}_{-r}^{0})^{+}T^{-1-\frac{\tau}{2}}\overline{\* E}_{-r}^{0 \prime} \* W\big\| + \big\|T^{-1-\tau}\* W^\prime \overline{\* E}_{r}^{0}(T^{-1-\tau}\* F^{\prime}\* F)^{+}T^{-1-\tau}\overline{\* E}_{r}^{0 \prime} \* W\big\| \nonumber \\
&& + \big\|T^{-1-\tau}\* W^\prime \overline{\* E}_{r}^{0}(T^{-1-\tau}\* F^{\prime}\* F)^{+}T^{-1-\tau}\* F^{\prime} \* W\big\| + \big\|T^{-1-\tau}\* W^\prime \* F(T^{-1-\tau}\* F^{\prime}\* F)^{+}T^{-1-\tau}\overline{\* E}_{r}^{0 \prime} \* W\big\| \nonumber \\
&& + \big\|T^{-\frac{1}{2}-\frac{\tau}{2}}\* W^\prime \widehat{\* F}^{0} \* G^{-1} [(\* G^{-1}\widehat{\* F}^{0 \prime} \widehat{\* F}^{0}\* G^{-1})^{+} - (\* G^{-1}\* S\* G^{-1})^{+}] \* G^{-1}T^{-\frac{1}{2}-\frac{\tau}{2}}\widehat{\* F}^{0 \prime}\* W \big\| \nonumber \\
&& = \big\|T^{-1-\frac{\tau}{2}}\* W^\prime \overline{\* E}_{-r}^{0}\big\|\big\|(T^{-1}\overline{\* E}_{-r}^{0 \prime}\overline{\* E}_{-r}^{0})^{+}\big\|\big\|T^{-1-\frac{\tau}{2}}\overline{\* E}_{-r}^{0 \prime} \* W\big\| + \big\|T^{-1-\tau}\* W^\prime \overline{\* E}_{r}^{0}\big\|\big\|(T^{-1-\tau}\* F^{\prime}\* F)^{+}\big\|\big\|T^{-1-\tau}\overline{\* E}_{r}^{0 \prime} \* W\big\| \nonumber \\
&& + \big\|T^{-1-\tau}\* W^\prime \overline{\* E}_{r}^{0}\big\|\big\|(T^{-1-\tau}\* F^{\prime}\* F)^{+}\big\|\big\|T^{-1-\tau}\* F^{\prime} \* W\big\| + \big\|T^{-1-\tau}\* W^\prime \* F\big\|\big\|(T^{-1-\tau}\* F^{\prime}\* F)^{+}\big\|\big\|T^{-1-\tau}\overline{\* E}_{r}^{0 \prime} \* W\big\| \nonumber \\
&& + \big\|T^{-\frac{1}{2}-\frac{\tau}{2}}\* W^\prime \widehat{\* F}^0\* G^{-1}\big\|\big\|(\* G^{-1}\widehat{\* F}^{0 \prime} \widehat{\* F}^{0}\* G^{-1})^{+} - (\* G^{-1}\* S\* G^{-1})^{+}\big\|\big\|\* G^{-1} T^{-\frac{1}{2}-\frac{\tau}{2}}\widehat{\* F}^{0 \prime}\* W \big\| \nonumber \\
&& = O_p(N^{-\frac{1}{2}}T^{-\frac{\tau}{2}}) + O_p(T^{-\frac{1}{2}}),
\end{eqnarray}
due to the independence of $\* W$ and $\overline{\* E}^{0}$ together with (\ref{eq: orderF0F0prime}). Since $T^{-1-\tau}{\* W^\prime}\* M_{\widehat{\* F}^0} {\* W}$ and $T^{-1-\tau}{\* W^\prime}\* M_{\* F^{0}} {\* W}$ are both positive definite wp1, we also have that 
\begin{eqnarray}
\big\|(T^{-1-\tau}{\* W^\prime}\* M_{\widehat{\* F}^0} {\* W})^{+} - (T^{-1-\tau}{\* W^\prime}\* M_{\* F^{0}} {\* W})^{+} \big\| = O_p(N^{-\frac{1}{2}}T^{-\frac{\tau}{2}}) + O_p(T^{-\frac{1}{2}})
\end{eqnarray}
so that 
\begin{eqnarray} \label{eq: WMfhatW}
\big\|(T^{-1-\tau}{\* W^\prime}\* M_{\widehat{\* F}^0} {\* W})^{+}\big\| = O_p(1).
\end{eqnarray}
A second important implication of (\ref{eq: orderF0F0prime}) is that
\begin{eqnarray} \label{eq: WMfhatEr}
\big\|T^{-1-\tau}\* W^\prime\* M_{\widehat{\* F}^0} \overline{\* E}_{r}^{0} \big\| &\leq& \big\|T^{-1-\tau}\* W^\prime\overline{\* E}_{r}^{0}\big\| + \big\|T^{-\frac{1}{2}-\frac{\tau}{2}}\* W^\prime \widehat{\* F}^0\* G^{-1}\big\|\big\|(\* G^{-1}\widehat{\* F}^{0 \prime} \widehat{\* F}^{0}\* G^{-1})^{+} \big\|\big\|\* G^{-1}T^{-\frac{1}{2}-\frac{\tau}{2}} \widehat{\* F}^{0\prime}\overline{\* E}_{r}^{0}\big\| \nonumber \\
&=& \big\|T^{-1-\tau}\* W^\prime\overline{\* E}_{r}^{0}\big\| + \big\|T^{-\frac{1}{2}-\frac{\tau}{2}}\* W^\prime \widehat{\* F}^0\* G^{-1}\big\|\big\|(\* G^{-1}\widehat{\* F}^{0 \prime} \widehat{\* F}^{0}\* G^{-1})^{+} \big\|\big\|\* G^{-1}T^{-\frac{1}{2}-\frac{\tau}{2}}\* F^{0\prime}\overline{\* E}_{r}^{0}\big\| \nonumber \\
&+& \big\|T^{-\frac{1}{2}-\frac{\tau}{2}}\* W^\prime \widehat{\* F}^0\* G^{-1}\big\|\big\|(\* G^{-1}\widehat{\* F}^{0 \prime} \widehat{\* F}^{0}\* G^{-1})^{+} \big\|\big\|\* G^{-1}\big\|\big\|T^{-\frac{1}{2}-\frac{\tau}{2}}\overline{\* E}^{0\prime}\overline{\* E}_{r}^{0}\big\| \nonumber \\
&\leq& \big\|T^{-1-\tau}\* W^\prime\overline{\* E}_{r}^{0}\big\| + \big\|T^{-\frac{1}{2}-\frac{\tau}{2}}\* W^\prime \widehat{\* F}^0\* G^{-1}\big\|\big\|(\* G^{-1}\widehat{\* F}^{0 \prime} \widehat{\* F}^{0}\* G^{-1})^{+} \big\|\big\|{\*G^{-1}}T^{-\frac{1}{2}-\frac{\tau}{2}}\* F^{0\prime}\overline{\* E}_{r}^{0}\big\| \nonumber \\
&+& M{T^{-\frac{\tau}{2}}}\big\|T^{-\frac{1}{2}-\frac{\tau}{2}}\* W^\prime \widehat{\* F}^0\* G^{-1}\big\|\big\|(\* G^{-1}\widehat{\* F}^{0 \prime} \widehat{\* F}^{0}\* G^{-1})^{+} \big\|\big\|{T^{-1}}\overline{\* E}^{0\prime}\overline{\* E}_{r}^{0}\big\| \nonumber \\
&=& O_p(N^{-\frac{1}{2}}T^{-\frac{1}{2}-\frac{\tau}{2}}) + O_p(N^{-\frac{1}{2}}T^{-\frac{1}{2}-\frac{\tau}{2}}) + O_p(N^{-\frac{1}{2}}T^{-\frac{\tau}{2}}) = O_p(N^{-\frac{1}{2}}T^{-\frac{\tau}{2}})
\end{eqnarray} similarly to (\ref{eq: decomp22}). As a result of (\ref{eq: WMfW}), (\ref{eq: decomp22}), (\ref{eq: WMfhatW}) and (\ref{eq: WMfhatEr}), we have
\begin{eqnarray} \label{eq: decomp41}
\big\|({\* W^\prime}\* M_{\widehat{\* F}^0} {\* W})^{+}{\* W^\prime}\* M_{\widehat{\* F}^0}\overline{\* E}_r^0\big\| &\leq& \big\|(T^{-1-\tau}{\* W^\prime}\* M_{\widehat{\* F}^0} {\* W})^{+}\big\|\big\|T^{-1-\tau}{\* W^\prime}\* M_{\widehat{\* F}^0}\overline{\* E}_r^0\big\| \nonumber \\
&=& O_p(N^{-\frac{1}{2}}T^{-\frac{\tau}{2}})
\end{eqnarray}
and 
\begin{eqnarray} \label{eq: decomp42}
\big\|({\* W^\prime}\* M_{\* F^{0}} {\* W})^{+}{\* W^\prime}\* M_{\* F^{0}}\overline{\* E}_r^0\big\| &\leq& \big\|(T^{-1-\tau}{\* W^\prime}\* M_{\* F^{0}} {\* W})^{+}\big\|\big\|T^{-1-\tau}{\* W^\prime}\* M_{\* F^{0}}\overline{\* E}_r^0\big\| \nonumber \\
&=& O_p(N^{-\frac{1}{2}}T^{-\frac{1}{2} - \frac{\tau}{2}}).
\end{eqnarray}
Using (\ref{eq: decomp41}) and (\ref{eq: decomp42}), we obtain 
\begin{eqnarray}
\big\|\mathrm{\*{IV}}\big\|&=&T^{\frac{1}{4}+\frac{\tau}{2}}\big\|\bigl[({\* W^\prime}\* M_{\widehat{\* F}^0} {\* W})^{+}{\* W^\prime}\* M_{\widehat{\* F}^0} - ({\* W^\prime}\* M_{\* F} {\* W})^{+}{\* W^\prime}\* M_{\* F}
 \bigr] \overline{\* E}_r^0 \boldsymbol{\alpha}\big\| \nonumber \\
 &\leq& \big\|({\* W^\prime}\* M_{\widehat{\* F}^0} {\* W})^{+}{\* W^\prime}\* M_{\widehat{\* F}^0}\overline{\* E}_r^0\big\| \big\|\boldsymbol{\alpha}^0\big\| + \big\|({\* W^\prime}\* M_{\* F} {\* W})^{+}{\* W^\prime}\* M_{\* F}\overline{\* E}_r^0\big\| \big\|\boldsymbol{\alpha}^0\big\| \nonumber \\
 &=& O_p(N^{-\frac{1}{2}}T^{-\frac{\tau}{2}}).
 \end{eqnarray}
For $\mathrm{\*{III}}$, recall from (\ref{eq: decomp12}) that $\big\|(T^{-1-\tau}{\* F^\prime}\* M_{\* W} {\* F})^{+}T^{-1-\tau}{\* F^\prime}\* M_{\* W}\overline{\* E}_r^0\boldsymbol{\alpha} \big\| = O_p(N^{-\frac{1}{2}}T^{-\frac{1}{2} - \frac{\tau}{2}})$. Further, we use the fact that $\big\| \* D_{\* F}\* G^{-1}(\* G^{-1}\widehat{\* F}^{0\prime}\* M_{\* W} {\widehat{\* F}^0}\* G^{-1})^{+}\* G^{-1}\widehat{\* F}^{0\prime}\* M_{\* W}\overline{\* E}_r^0\boldsymbol{\alpha}\big\| = O_p(N^{-\frac{1}{2}}T^{-\tau/2})$. Indeed, given that $\* D_{\* F}\* G^{-1} =T^{-\frac{1}{4}}\* I_m$, we have
\begin{eqnarray}
\big\|T^{-\frac{1}{4}}\* G^{-1} \widehat{\* F}^{0\prime}\* M_{\* W}\overline{\* E}_r^0\boldsymbol{\alpha}\big\| &=& \big\|T^{-\frac{1}{2}-\frac{\tau}{2}}\* G^{-1} \widehat{\* F}^{0\prime}\* M_{\* W}\overline{\* E}_r^0\boldsymbol{\alpha}^0\big\| \nonumber \\
&\leq& \big\|T^{-\frac{1}{2}-\frac{\tau}{2}}\* G^{-1} \widehat{\* F}^{0\prime}\overline{\* E}_r^0\boldsymbol{\alpha}^0\big\| + \big\|T^{-1-\tau} \widehat{\* F}^{0\prime}\* W\big\|\big\|(T^{-1-\tau}\* W^{\prime}\* W)^{+}\big\|\big\|T^{-\frac{1}{2}-\frac{\tau}{2}}\* G^{-1}\* W^{\prime}\overline{\* E}_r^0\boldsymbol{\alpha}^0\big\| \nonumber \\
&\leq& \big\|T^{-(1+\tau)} \* F^{\prime}\overline{\* E}_r^0\boldsymbol{\alpha}^0\big\| + \big\|T^{-\frac{1}{2}-\frac{\tau}{2}}\* G^{-1} \overline{\* E}^{\prime}\overline{\* E}_r^0\boldsymbol{\alpha}^0\big\| \nonumber \\
&+& \big\|T^{-1-\tau} \widehat{\* F}^{0\prime}\* W\big\|\big\|(T^{-1-\tau}\* W^{\prime}\* W)^{+}\big\|\big\|T^{-\frac{1}{2}-\frac{\tau}{2}}\* G^{-1}\* W^{\prime}\overline{\* E}_r^0\boldsymbol{\alpha}^0\big\| \nonumber \\
&\leq& \big\|T^{-(1+\tau)} \* F^{\prime}\overline{\* E}_r^0\big\|\big\|\boldsymbol{\alpha}^0\big\| + MT^{-\frac{\tau}{2}}\big\|T^{-1} \overline{\* E}^{\prime}\overline{\* E}_r^0\big\|\big\|\boldsymbol{\alpha}^0\big\| \nonumber \\
&+& \big\|T^{-1-\tau} \widehat{\* F}^{0\prime}\* W\big\|\big\|(T^{-1-\tau}\* W^{\prime}\* W)^{+}\big\|\big\|T^{-\frac{1}{2}-\frac{\tau}{2}}\* G^{-1}\* W^{\prime}\overline{\* E}_r^0\big\|\big\|\boldsymbol{\alpha}^0\big\| \nonumber \\
&=& O_p(N^{-\frac{1}{2}}T^{-\frac{\tau}{2}}) + O_p(N^{-1}T^{-\frac{\tau}{2}}) + O_p(N^{-\frac{1}{2}}T^{-\frac{1}{2}-\frac{\tau}{2}}) \nonumber \\
&=& O_p(N^{-\frac{1}{2}}T^{-\frac{\tau}{2}}).
\end{eqnarray}

With respect to $(\* G^{-1}\widehat{\* F}^{0\prime}\* M_{\* W} {\widehat{\* F}^0}\* G^{-1})^{+}$, we proceed analogously to (\ref{eq: F0F0order}). Let $\* S_{\* W}$ be defined as
\begin{eqnarray} \label{eq: S_MW}
\* S_{\* M_{\* W}} &=& \begin{bmatrix}
\* F^\prime \* M_{\* W} {\* F} & \* 0_{r \times (m-r)} \\
\* 0_{(m-r)\times r} & \overline{\* E}_{-r}^{0 \prime}\* M_{\* W} \overline{\* E}_{-r}^{0}
\end{bmatrix} \nonumber \\
&=&\begin{bmatrix}
\* F^\prime{\* F} & \* 0_{r \times (m-r)} \\
\* 0_{(m-r)\times r} & \overline{\* E}_{-r}^{0 \prime}\overline{\* E}_{-r}^{0}
\end{bmatrix} - \begin{bmatrix}
\* F^\prime \* W(\* W^{\prime}\* W)^{+}\* W^{\prime} {\* F} & \* 0_{r \times (m-r)} \\
\* 0_{(m-r)\times r} & \overline{\* E}_{-r}^{0 \prime}\* W(\* W^{\prime}\* W)^{+}\* W^{\prime} \overline{\* E}_{-r}^{0}
\end{bmatrix} \nonumber \\
&=&\* S - \* S_{\* W}.
\end{eqnarray} 
and exploiting the definition of $\* M_{\* W}$ yields the decomposition
\begin{eqnarray}
\* G^{-1}\widehat{\* F}^{0\prime}\* M_{\* W} {\widehat{\* F}^0}\* G^{-1} &=& \* G^{-1}\widehat{\* F}^{0\prime}{\widehat{\* F}^0}\* G^{-1} - \* G^{-1}\widehat{\* F}^{0\prime}\* W(\* W^{\prime}\* W)^{+}\* W^{\prime}{\widehat{\* F}^0}\* G^{-1}.
\end{eqnarray}
Thus,
\begin{eqnarray}\label{FMF-SM}
&& \big\|\* G^{-1}\widehat{\* F}^{0\prime}\* M_{\* W} {\widehat{\* F}^0}\* G^{-1} - \* G^{-1}\* S_{\* M_{\* W}}\* G^{-1} \big\| \nonumber \\
&& \leq \big\| \* G^{-1}\widehat{\* F}^{0\prime}{\widehat{\* F}^0}\* G^{-1} - \* G^{-1}\* S\* G^{-1}\big\| + \big\|\* G^{-1}\widehat{\* F}^{0\prime}\* W(\* W^{\prime}\* W)^{+}\* W^{\prime}{\widehat{\* F}^0}\* G^{-1} - \* G^{-1}\* S_{\* W}\* G^{-1}\big\| \nonumber \\
&=& O_p(N^{-\frac{1}{2}}T^{-\frac{\tau}{2}}) + O_p(T^{-\frac{1}{2}}),
\end{eqnarray}
using the fact that $\big\|T^{-1-\tau}\* F^{0 \prime}\overline{\* E}^0\big\|$ and $\big\|T^{-1-\tau}\*W^{\prime}\overline{\* E}^0\big\|$ are both of order $ O_p(T^{-\frac{1}{2}-\frac{\tau}{2}})$, while $\big\|T^{-1}\overline{\* E}^{0 \prime}\overline{\* E}^0\big\|$ is of order $ O_p(1)$, which ensures that the first term on the right-hand side of (\ref{FMF-SM}) drives the order (i.e. the second one does not dominate). Indeed the slowest decaying component of the second term is 
\begin{align}
   &\left\|T^{-1-\tau}\*F^{\prime}\*W(T^{-(1+\tau)}\*W'\*W)^+T^{-1-\tau/2}\*W'\overline{\*E}_{-r}^0 \right\|\\
   &\leq T^{-1/2}\left\|T^{-(1+\tau)}\*F^{\prime}\*W \right\|\left\|(T^{-(1+\tau)}\*W'\*W)^+ \right\|\left\| T^{-(1+\tau)/2}\*W'\overline{\*E}_{-r}^0\right\|\notag\\
   &=O_p(T^{-1/2}).
\end{align}
Therefore, we obtain
\begin{eqnarray} \label{eq: decompF0hatMWF0hat}
\* G^{-1}\widehat{\* F}^{0\prime}\* M_{\* W} {\widehat{\* F}^0}\* G^{-1} = \* G^{-1}\* S_{\* M_{\* W}}\* G^{-1} + O_p(N^{-\frac{1}{2}}T^{-\frac{\tau}{2}}) + O_p(T^{-\frac{1}{2}}).
\end{eqnarray}
Since $\* G^{-1}\* S\* G^{-1}$ and $\* G^{-1}\* S_{\* W}\* G^{-1}$ are positive definite wp1, then so are $\* G^{-1}\* S_{\* M_{\* W}}\* G^{-1}$ and $\* G^{-1}\widehat{\* F}^{0\prime}\* M_{\* W} {\widehat{\* F}^0}\* G^{-1}$. Further, the rank of $\* G^{-1}\* S_{\* M_{\* W}}\* G^{-1}$is $m$, which is also the rank of $\* G^{-1}\widehat{\* F}^{0\prime}\* M_{\* W} {\widehat{\* F}^0}\* G^{-1}$. It then follows that as $N, T \to \infty$,
\begin{eqnarray} \label{eq: rankF0hatMWF0hat}
rk [\* G^{-1}\widehat{\* F}^{0\prime}\* M_{\* W} {\widehat{\* F}^0}\* G^{-1}] \to rk \* G^{-1}\* S_{\* M_{\* W}}\* G^{-1}
\end{eqnarray}
almost surely. Following \cite{karabiyik2017role}, (\ref{eq: decompF0hatMWF0hat}) and (\ref{eq: rankF0hatMWF0hat}) lead to the key result 
\begin{eqnarray} \label{eq: orderF0hatMWF0hat}
\big\|(\* G^{-1}\widehat{\* F}^{0\prime}\* M_{\* W} {\widehat{\* F}^0}\* G^{-1})^{+} - (\* G^{-1}\* S_{\* M_{\* W}}\* G^{-1})^{+} \big\| = O_p(N^{-\frac{1}{2}}T^{-\frac{\tau}{2}}) + O_p(T^{-\frac{1}{2}}).
\end{eqnarray}
An important implication of this result is that 
\begin{eqnarray} \label{eq: rankF0hatMWF0Hat}
\big\|(\* G^{-1}\widehat{\* F}^{0\prime}\* M_{\* W} {\widehat{\* F}^0}\* G^{-1})^{+} \big\| = O_p(1).
\end{eqnarray}
Combining (\ref{eq: rankF0hatMWF0Hat}) and (\ref{eq: decompF0hatMWF0hat}) directly implies that 
\begin{eqnarray}
\big\| \* D_{\* F}(\widehat{\* F}^{0\prime}\* M_{\* W} {\widehat{\* F}^0})^{+}\widehat{\* F}^{0\prime}\* M_{\* W}\overline{\* E}_r^0\boldsymbol{\alpha}\big\| &\leq& \big\|(\* G^{-1}\widehat{\* F}^{0\prime}\* M_{\* W} {\widehat{\* F}^0}\* G^{-1})^{+} \big\| \big\|T^{-\frac{1}{4}}\* G^{-1} \widehat{\* F}^{0\prime}\* M_{\* W}\overline{\* E}_r^0\boldsymbol{\alpha}\big\| \nonumber \\
&=& O_p(N^{-\frac{1}{2}}T^{-\tau/2}),
\end{eqnarray}
as required. 
Finally, we are able to obtain the order $\mathrm{\*III}$ by using the definition of $\+\alpha$ and $\*D_\*F\*G^{-1}=\*I_m T^{-1/4}$:
\begin{eqnarray}
\big\|\mathrm{\*{III}}\big\|&=&\bigg\|\* D_{\* F}\left((\widehat{\* F}^{0\prime}\* M_{\* W} {\widehat{\* F}^0})^{+}\widehat{\* F}^{0\prime}\* M_{\* W} - \begin{bmatrix} ({\* F^\prime}\* M_{\* W} {\* F})^{+}{\* F^\prime}\* M_{\* W} \\ \* 0_{(m-r)\times(t-1)} \end{bmatrix} \right) \overline{\* E}_r^0 \boldsymbol{\alpha}\bigg\| \nonumber \\
&\leq& \bigg\|(\*G^{-1}\widehat{\* F}^{0\prime}\* M_{\* W} {\widehat{\* F}^0}\*G^{-1})^{+}\*G^{-1}T^{-\frac{1}{2}-\frac{\tau}{2}}\widehat{\* F}^{0\prime}\* M_{\* W}\overline{\* E}_r^0\bigg\|\big\|\boldsymbol{\alpha}^0\big\| + \bigg\|\begin{bmatrix} (T^{-(1+\tau)}{\* F^\prime}\* M_{\* W} {\* F})^{+}T^{-(1+\tau)}{\* F^\prime}\* M_{\* W} \\ \* 0_{(m-r)\times(t-1)} \end{bmatrix} \overline{\* E}_r^0\bigg\| \big\|\boldsymbol{\alpha}^0\big\| \nonumber \\
 &=& O_p((NT)^{-\frac{1}{2}}) + O_p(N^{-\frac{1}{2}}T^{-\tau}).
\end{eqnarray}

Moving on to $\mathrm{\*V}$, consider the expansion 
\begin{eqnarray} \label{eq: decomp5}
\mathrm{\*V} &=& \* D_{\* F} \bigl[(\widehat{\* F}^{0\prime}\* M_{\* W} {\widehat{\* F}^0})^{+} - \* S_{\* M_{\* W}}^{+}\bigr]( \widehat{\* F}^{0\prime} - {\* F}^{0\prime})\* M_{\* W}\*u + \* D_{\* F} \bigl[(\widehat{\* F}^{0\prime}\* M_{\* W} {\widehat{\* F}^0})^{+} - \* S_{\* M_{\* W}}^{+}\bigr] {\* F}^{0\prime}\* M_{\* W}\*u \nonumber \\
&+& \* D_{\* F}\* S_{\* M_{\* W}}^{+}( \widehat{\* F}^{0\prime} - {\* F}^{0\prime})\* M_{\* W}\*u \nonumber \\
&=& \* D_{\* F} \bigl[(\widehat{\* F}^{0\prime}\* M_{\* W} {\widehat{\* F}^0})^{+} - \* S_{\* M_{\* W}}^{+}\bigr]( \widehat{\* F}^{0\prime} - {\* F}^{0\prime})\* M_{\* W}\*u + \* D_{\* F} \bigl[(\widehat{\* F}^{0\prime}\* M_{\* W} {\widehat{\* F}^0})^{+} - \* S_{\* M_{\* W}}^{+}\bigr] {\* F}^{0\prime}\* M_{\* W}\*u \nonumber \\
&+& \* D_{\* F}\* S_{\* M_{\* W}}^{+}\overline{\* E}^{0 \prime}\* M_{\* W}\*u,
\end{eqnarray}
where $\* S_{\* M_{\* W}}$ is defined as in (\ref{eq: S_MW}). Beginning from the third term on the right-hand side of (\ref{eq: decomp5}),
\begin{eqnarray} \label{eq: DFSMWE0barMWu}
\* D_{\* F}\* S_{\* M_{\* W}}^{+}\overline{\* E}^{0 \prime}\* M_{\* W}\*u &=& \* D_{\* F} \* G^{-1} \begin{bmatrix}
(T^{-1-\tau}\* F^\prime \* M_{\* W} {\* F})^{+} & \* 0_{r \times (m-r)} \\
\* 0_{(m-r)\times r} & (T^{-1}\overline{\* E}_{-r}^{0 \prime}\* M_{\* W} \overline{\* E}_{-r}^{0})^{+}
\end{bmatrix} \* G^{-1} \begin{bmatrix} \overline{\* E}_r^{0 \prime}\* M_{\* W}\*u \\ \overline{\* E}_{-r}^{0 \prime}\* M_{\* W}\*u \end{bmatrix} \nonumber \\
&=&  \begin{bmatrix} (T^{-1-\tau}\* F^\prime \* M_{\* W} {\* F})^{+} T^{-\frac{3}{4}-\frac{\tau}{2}} \overline{\* E}_r^{0 \prime}\* M_{\* W}\*u \\ (T^{-1}\overline{\* E}_{-r}^{0 \prime}\* M_{\* W} \overline{\* E}_{-r}^{0})^{+}T^{-\frac{3}{4}} \overline{\* E}_{-r}^{0 \prime}\* M_{\* W}\*u \end{bmatrix} \nonumber \\
&=& O_p(T^{-\frac{1}{4}}),
\end{eqnarray}
where the order of the remainder is due to the fact that 
\begin{eqnarray}
\big\|T^{-\frac{3}{4}-\frac{\tau}{2}} \overline{\* E}_r^{0 \prime}\* M_{\* W}\*u \big\| &\leq& \big\| T^{-\frac{3}{4}-\frac{\tau}{2}} \overline{\* E}_r^{0 \prime}\*u \big\| + \big\| T^{-1-\tau}\overline{\* E}_r^{0 \prime}\* W\big\|\big\|(T^{-1-\tau}\* W^{\prime}\* W)^{+}\big\|\big\|T^{-\frac{3}{4}-\frac{\tau}{2}}\* W^{\prime}\*u \big\| \nonumber \\
&=& O_p(N^{-\frac{1}{2}}T^{-\frac{1}{4}-\frac{\tau}{2}}),
\end{eqnarray}
and $\big\|(T^{-1}\overline{\* E}_{-r}^{0 \prime}\* M_{\* W} \overline{\* E}_{-r}^{0})^{+}T^{-\frac{1}{2}} \overline{\* E}_{-r}^{0 \prime}\* M_{\* W}\*u\big\| = O_p(1)$ from \cite{stauskas2022tests}. Note that it can be simplified slightly. Firstly, 
\begin{align}
    \left\|T^{-1/2} \overline{\* E}_{-r}^{0 \prime}\* M_{\* W}\*u\right\|\leq\left\| T^{-1/2}\overline{\* E}_{-r}^{0 \prime}\*u\right\|+&T^{-1/2}\left\|T^{-(1+\tau)/2} \overline{\* E}_{-r}^{0 \prime}\*W\right\|\left\|\left(T^{-(1+\tau)}\*W'\*W \right)^+ \right\|\notag\\
    &\times \left\|T^{-(1+\tau)/2}\*W'\*u \right\|\notag\\
    &=\left\| T^{-1/2}\overline{\* E}_{-r}^{0 \prime}\*u\right\|+O_p(T^{-1/2}),
\end{align}
and 
\begin{align}
    \left\|T^{-1}\overline{\* E}_{-r}^{0 \prime}\* M_{\* W} \overline{\* E}_{-r}^{0} \right\|&\leq \left\|T^{-1}\overline{\* E}_{-r}^{0 \prime} \overline{\* E}_{-r}^{0} \right\|+T^{-1}\left\|T^{-(1+\tau)/2} \overline{\* E}_{-r}^{0 \prime}\*W\right\|^2\left\|\left(T^{-(1+\tau)}\*W'\*W \right)^+ \right\|\notag\\
    &=\left\|T^{-1}\overline{\* E}_{-r}^{0 \prime} \overline{\* E}_{-r}^{0} \right\|+O_p(T^{-1}).
\end{align}
This leads to 
\begin{align}
    \* D_{\* F}\* S_{\* M_{\* W}}^{+}\overline{\* E}^{0 \prime}\* M_{\* W}\*u 
&=  \begin{bmatrix} (T^{-1-\tau}\* F^\prime \* M_{\* W} {\* F})^{+} T^{-\frac{3}{4}-\frac{\tau}{2}} \overline{\* E}_r^{0 \prime}\* M_{\* W}\*u \\ (T^{-1}\overline{\* E}_{-r}^{0 \prime}\* M_{\* W} \overline{\* E}_{-r}^{0})^{+}T^{-\frac{3}{4}} \overline{\* E}_{-r}^{0 \prime}\* M_{\* W}\*u \end{bmatrix}\notag\\
&=\begin{bmatrix}
    \*0_r\\
    (T^{-1}\overline{\* E}_{-r}^{0 \prime}\* M_{\* W} \overline{\* E}_{-r}^{0})^+T^{-\frac{3}{4}}\overline{\* E}_{-r}^{0 \prime}\*u
\end{bmatrix}+O_p(T^{-3/4})\notag\\
&=\begin{bmatrix}
    \*0_r\\
    T^{-1/4}(T^{-1}\overline{\*E}^{0\prime}_{-r}\overline{\*E}_{-r}^{0})^{+}T^{-1/2}\overline{\*E}^{0\prime}_{-r}\*u
\end{bmatrix}+O_p(T^{-1/2})\\
&=\begin{bmatrix}
    \*0_r\\
   T^{-1/4}\*v_t
\end{bmatrix}+O_p(T^{-1/2})=O_p(T^{-1/4}).
\end{align}
Here, we define 
\begin{align}
    \*v_t=(T^{-1}\overline{\*E}^{0\prime}_{-r}\overline{\*E}_{-r}^{0})^{+}T^{-1/2}\overline{\*E}^{0\prime}_{-r}\*u&=\left(\frac{1}{T}\sum_{s=1}^{t-1}\overline{\*e}_{-r,s}^{0}\overline{\*e}_{-r,s}^{0\prime}\right)^+\frac{1}{\sqrt{T}}\sum_{s=1}^{t-1}\overline{\*e}_{-r,s}^0u_{s+1}\notag\\
    &= \left(\frac{1}{T}\sum_{s=1}^{t-1}\overline{\*e}_{-r,s}^{0}\overline{\*e}_{-r,s}^{0\prime}\right)^{-1}\frac{1}{\sqrt{T}}\sum_{s=1}^{t-1}\overline{\*e}_{-r,s}^0u_{s+1},
\end{align}
which can be interpreted as a least squares regression coefficient of the forecast error on the error arising  from the excess $m-r$ CAs. With respect to the first term on the right-hand side of (\ref{eq: decomp5}), we proceed as follows. Because of the results in (\ref{eq: orderF0hatMWF0hat}) and (\ref{eq: DFSMWE0barMWu}), we then have
\begin{eqnarray} \label{eq: decomp51}
&& \big\|\* D_{\* F} \bigl[(\widehat{\* F}^{0\prime}\* M_{\* W} {\widehat{\* F}^0})^{+} - \* S_{\* M_{\* W}}^{+}\bigr]( \widehat{\* F}^{0\prime} - {\* F}^{0\prime})\* M_{\* W}\*u \big\| \nonumber \\
&& \leq \underbrace{\big\|(\* G^{-1}\widehat{\* F}^{0\prime}\* M_{\* W} {\widehat{\* F}^0}\* G^{-1})^{+} - (\* G^{-1}\* S_{\* M_{\* W}}\* G^{-1})^{+}\big\|}_{ O_p(N^{-\frac{1}{2}}T^{-\frac{\tau}{2}}) + O_p(T^{-\frac{1}{2}})}\big\|T^{-\frac{1}{4}}\* G^{-1}\overline{\* E}^{0 \prime}\* M_{\* W}\*u \big\| \nonumber \\
&& = O_p(N^{-\frac{1}{2}}T^{-\frac{1}{4}-\frac{\tau}{2}}) + O_p(T^{-\frac{3}{4}}).
\end{eqnarray}
Finally, the order of the remaining term in (\ref{eq: decomp5}) can be shown to be  
\begin{eqnarray} \label{eq: decomp52}
&& \big\|\* D_{\* F} \bigl[\* G^{-1}(\* G^{-1}\widehat{\* F}^{0\prime}\* M_{\* W} {\widehat{\* F}^0}\* G^{-1})^{+} - (\* G^{-1}\* S_{\* M_{\* W}}^{+}\* G^{-1})\bigr]\* G^{-1} {\* F}^{0\prime}\* M_{\* W}\*u\big\| \nonumber \\
&& \leq \big\|(\* G^{-1}\widehat{\* F}^{0\prime}\* M_{\* W} {\widehat{\* F}^0}\* G^{-1})^{+} - (\* G^{-1}\* S_{\* M_{\* W}}^{+}\* G^{-1})\big\| \big\|T^{-\frac{1}{4}}\* G^{-1}{\* F}^{0\prime}\* M_{\* W}\*u\big\| \nonumber \\
&& \leq \big\|(\* G^{-1}\widehat{\* F}^{0\prime}\* M_{\* W} {\widehat{\* F}^0}\* G^{-1})^{+} - (\* G^{-1}\* S_{\* M_{\* W}}^{+}\* G^{-1})\big\| \big\|T^{-\frac{3}{4}-\frac{\tau}{2}}{\* F}^{0\prime}\*u\big\| \nonumber \\
&& + \big\|(\* G^{-1}\widehat{\* F}^{0\prime}\* M_{\* W} {\widehat{\* F}^0}\* G^{-1})^{+} - (\* G^{-1}\* S_{\* M_{\* W}}^{+}\* G^{-1})\big\| \big\|T^{-1-\tau}{\* F}^{0\prime}\* W\big\|\big\|(T^{-1-\tau}\* W^{\prime}\* W)^{+}\big\|\big\|T^{-\frac{3}{4}-\frac{\tau}{2}}\* W^{\prime}\*u\big\| \nonumber \\
&& = O_p(N^{-\frac{1}{2}}T^{-\frac{1}{4}-\frac{\tau}{2}}) + O_p(T^{-\frac{3}{4}}). 
\end{eqnarray}
Hence, putting together the results in (\ref{eq: DFSMWE0barMWu}), (\ref{eq: decomp51}) and (\ref{eq: decomp52}) gives the order of $\mathrm{\*V}$:
\begin{eqnarray}
\big\|\mathrm{\*V}\big\|&=& O_p(T^{-\frac{1}{4}}).
\end{eqnarray}

It remains to address the term $\mathrm{\*V\*I}$. Recall the decomposition
\begin{eqnarray} \label{eq: decomp6}
\mathrm{\*V\*I} &=& T^{\frac{1}{4}+\frac{\tau}{2}} \bigl[({\* W^\prime}\* M_{\widehat{\* F}^0} {\* W})^{+} -({\* W^\prime}\* M_{\* F^{0}} {\* W})^{+}\bigr] {\* W^\prime}(\* M_{\widehat{\* F}^0}  - \* M_{\* F^{0}})\*u \nonumber \\
&+& T^{\frac{1}{4}+\frac{\tau}{2}} \bigl[({\* W^\prime}\* M_{\widehat{\* F}^0} {\* W})^{+} -({\* W^\prime}\* M_{\* F^{0}} {\* W})^{+}\bigr] {\* W^\prime}\* M_{\* F^{0}}\*u \nonumber \\
&+& T^{\frac{1}{4}+\frac{\tau}{2}}({\* W^\prime}\* M_{\widehat{\* F}^0} {\* W})^{+}{\* W^\prime}(\* M_{\widehat{\* F}^0}  - \* M_{\* F^{0}})\*u.
\end{eqnarray}
Using (A.77) in the Supplement of \cite{stauskas2022tests}, note that
\begin{eqnarray}
&& \big\|T^{-\frac{1}{2}-\frac{\tau}{2}} {\* W^\prime}(\* M_{\widehat{\* F}^0}  - \* M_{\* F^{0}})\*u\big\| \nonumber \\
&& \leq \big\|T^{-\frac{1}{2}-\frac{\tau}{2}} {\* W^\prime}(\overline{\* E}_{-r}^{0}(\overline{\* E}_{-r}^{0 \prime}\overline{\* E}_{-r}^{0})^{+} \overline{\* E}_{-r}^{0 \prime} + \overline{\* E}_{r}^{0}(\* F^{\prime}\* F)^{+} \overline{\* E}_{r}^{0 \prime} + \overline{\* E}_{r}^{0}(\* F^{\prime}\* F)^{+} \* F^{\prime} + \* F(\* F^{\prime}\* F)^{+} \overline{\* E}_{r}^{0 \prime}+ \widehat{\* F}^0[(\widehat{\* F}^{0 \prime}\widehat{\* F}^0)^{+}+\* S^{+}]\widehat{\* F}^{0 \prime})\*u\big\|  \nonumber \\
&& \leq \big\|T^{-1-\frac{\tau}{2}} {\* W^\prime}\overline{\* E}_{-r}^{0}\big\|\big\|(T^{-1}\overline{\* E}_{-r}^{0 \prime}\overline{\* E}_{-r}^{0})^{+}\big\|\big\| T^{-\frac{1}{2}}\overline{\* E}_{-r}^{0 \prime}\*u\big\| + T^{-\tau}\big\|T^{-1-\frac{\tau}{2}} {\* W^\prime}\overline{\* E}_{r}^{0}\big\|\big\|(T^{-1-\tau}\* F^{\prime}\* F)^{+} \big\|\big\| T^{-\frac{1}{2}}\overline{\* E}_{r}^{0 \prime}\*u\big\| \nonumber \\
&& + T^{-\frac{\tau}{2}}\big\|T^{-1-\frac{\tau}{2}} {\* W^\prime}\overline{\* E}_{r}^{0}\big\|\big\|(T^{-1-\tau} \* F^{\prime}\* F)^{+} \big\|\big\| T^{-\frac{1}{2}-\frac{\tau}{2}} \* F^{\prime}\*u\big\| + T^{-\frac{\tau}{2}}\big\|T^{-1-\tau} {\* W^\prime}\* F\big\|\big\|(T^{-1-\tau}\* F^{\prime}\* F)^{+}\big\|\big\|T^{-\frac{1}{2}}\overline{\* E}_{r}^{0 \prime}\*u\big\| \nonumber \\
&& + \big\|T^{-\frac{1}{2}-\frac{\tau}{2}} {\* W^\prime}\widehat{\* F}^0\* G^{-1}\big\|\big\|(\* G^{-1}\widehat{\* F}^{0 \prime}\widehat{\* F}^0\* G^{-1})^{+} +(\* G^{-1}\* S\* G^{-1})^{+}\big\|\big\|\* G^{-1}\widehat{\* F}^{0 \prime}\*u\big\| \nonumber \\
&& = O_p(T^{-\frac{1}{2}}) + O_p(N^{-1}T^{-\frac{\tau}{2}}) + O_p(N^{-\frac{1}{2}}T^{-\frac{1}{2}-\frac{\tau}{2}}) + O_p(N^{-\frac{1}{2}}T^{-\frac{\tau}{2}}) +  O_p(N^{-\frac{1}{2}}T^{-\frac{\tau}{2}}) + O_p(T^{-\frac{1}{2}}) \nonumber \\
&& = O_p(T^{-\frac{1}{2}}) + O_p(N^{-\frac{1}{2}}T^{-\frac{\tau}{2}}),
\end{eqnarray}
by (\ref{eq: orderF0F0prime}) and the independence of $\* W$ and $\overline{\* E}$. By putting this latter result together with $\big\|(T^{-1-\tau}{\* W^\prime}\* M_{\widehat{\* F}^0} {\* W})^{+} - (T^{-1-\tau}{\* W^\prime}\* M_{\* F^{0}} {\* W})^{+} \big\| = O_p(N^{-\frac{1}{2}}T^{-\frac{\tau}{2}}) + O_p(T^{-\frac{1}{2}})$, we can show that (\ref{eq: decomp6}) reduces to

\begin{eqnarray} 
\big\|\mathrm{\*V\*I}\big\| &\leq& T^{-\frac{1}{4}}\big\|(T^{-1-\tau}{\* W^\prime}\* M_{\widehat{\* F}^0} {\* W})^{+} -(T^{-1-\tau}{\* W^\prime}\* M_{\* F^{0}} {\* W})^{+}\big\| \big\|T^{-\frac{1}{2}-\frac{\tau}{2}}{\* W^\prime}(\* M_{\widehat{\* F}^0}  - \* M_{\* F^{0}})\*u\big\| \nonumber \\
&+& T^{-\frac{1}{4}} \big\|(T^{-1-\tau}{\* W^\prime}\* M_{\widehat{\* F}^0} {\* W})^{+} -(T^{-1-\tau}{\* W^\prime}\* M_{\* F^{0}} {\* W})^{+}\big\| \big\|T^{-\frac{1}{2}-\frac{\tau}{2}}{\* W^\prime}\* M_{\* F^{0}}\*u\big\| \nonumber \\
&+& T^{-\frac{1}{4}} \big\|(T^{-1-\tau}{\* W^\prime}\* M_{\widehat{\* F}^0} {\* W})^{+}\big\|\big\|T^{-\frac{1}{2}-\frac{\tau}{2}}{\* W^\prime}(\* M_{\widehat{\* F}^0}  - \* M_{\* F^{0}})\*u\big\| \nonumber\\
&=& O_p(T^{-\frac{3}{4}}) + O_p(N^{-\frac{1}{2}}T^{-\frac{1}{4}-\frac{\tau}{2}}).
\end{eqnarray}
With the results of $\mathrm{\*I}-\mathrm{\*V\*I}$ in mind, we finally obtain 
\begin{eqnarray} 
\* D_{T}(\* Q_N^{-1}\boldsymbol{\widehat{\delta}}_t - \boldsymbol{\widetilde{\delta}_t^0}) &=& \begin{bmatrix} \mathrm{\* I} \\ \mathrm{\* I\*I} \end{bmatrix} - \begin{bmatrix} \mathrm{\* I\*I\*I} \\ \mathrm{\* I\*V} \end{bmatrix} + \begin{bmatrix} \mathrm{\* V} \\ \mathrm{\* V\*I} \end{bmatrix} \nonumber \\
&=& O_p(N^{-\frac{1}{2}}T^{-\frac{\tau}{2}}) + O_p(T^{-\frac{1}{4}}),
\end{eqnarray}
thus concluding the lemma. $\blacksquare$

\subsubsection{Proof of Lemma 2}

We now prove Lemma 2. For the reference, we will demonstrate the following statements that constitute $|\xi_{N,T,j}|=o_p(1)$ for $j=1,\ldots,4$: 
\begin{enumerate}[label=(\alph*)]
    \item $\frac{1}{\sqrt{n}}\sum_{t=k_0}^{k_0+m_0-1}\widetilde{u}_{1,t+1}(\widetilde{u}_{2,t+1}- \widehat{u}_{2,t+1}) = o_p(1)$
    \item $\frac{1}{\sqrt{n}}\sum_{t=k_0}^{k_0+l_2^0-1}(\widetilde{u}_{2,t+1}-\widehat{u}_{2,t+1})^2 = o_p(1)$  
    \item $\frac{1}{\sqrt{n}}\sum_{t=k_0}^{k_0+l_2^0-1}\widetilde{u}_{2,t+1}(\widetilde{u}_{2,t+1}-\widehat{u}_{2,t+1}) = o_p(1)$
    \item $\frac{1}{n}\sum_{l_2}\frac{n}{l_2}\frac{1}{\sqrt{n}}\sum_{t=k_0}^{k_0+l_2-1}(\widetilde{u}_{2,t+1}-\widehat{u}_{2,t+1})^2 = o_p(1)$  
    \item $\frac{1}{n}\sum_{l_2}\frac{n}{l_2}\frac{1}{\sqrt{n}}\sum_{t=k_0}^{k_0+l_2-1}\widetilde{u}_{2,t+1}(\widetilde{u}_{2,t+1}-\widehat{u}_{2,t+1}) = o_p(1)$
    \item $|\widetilde{\phi}^2-\widehat{\phi}^2| = o_p(1)$.
\end{enumerate}
Consider (a). Beginning from the definitions of $\widetilde{u}_{1,t+1}$ and $\widehat{u}_{2,t+1}$ and the fact that $\boldsymbol{\widetilde{\delta}}_t^\prime\*z_t = \boldsymbol{\widetilde{\delta}}_t^{0 \prime}\* Q_N^\prime\widehat{\*z}_t^0 - \boldsymbol{\widetilde{\alpha}_t}^\prime \overline{\* e}_{r,t}^0$:
\begin{eqnarray} \label{eq: decomputildeuhat}
\widetilde{u}_{2,t+1}- \widehat{u}_{2,t+1} &=& \boldsymbol{\widehat{\delta}}_t^\prime\widehat{\*z}_t - \boldsymbol{\widetilde{\delta}}_t^\prime\*z_t \nonumber \\
&=& \boldsymbol{\widetilde{\delta}}_t^\prime\widehat{\*z}_t - \boldsymbol{\widetilde{\delta}}_t^{0 \prime´}\* Q_N^\prime\widehat{\*z}_t + \boldsymbol{\widetilde{\alpha}}_t^\prime \overline{\* e}_{r,t}^0 \nonumber \\
&=& \boldsymbol{\widehat{\delta}}_t^\prime\* Q_N^{-1 \prime}\widehat{\*z}_t^0 - \boldsymbol{\widetilde{\delta}_t^{0 \prime´}}\widehat{\*z}_t^0 + \boldsymbol{\widetilde{\alpha}}_t^\prime \overline{\* e}_{r,t}^0 \nonumber \\
&=& (\* Q_N^{-1}\boldsymbol{\widehat{\delta}}_t - \boldsymbol{\widetilde{\delta}}_t^{0})^{\prime}\widehat{\*z}_t^0 + \boldsymbol{\widetilde{\alpha}}_t^\prime \overline{\* e}_{r,t}^0, \nonumber \\
&=& (\* Q_N^{-1}\boldsymbol{\widehat{\delta}}_t - \boldsymbol{\widetilde{\delta}}_t^{0})^{\prime}\widehat{\*z}_t^0 + (\boldsymbol{\widetilde{\alpha}}_t-\boldsymbol{\alpha})^\prime  \overline{\* e}_{r,t}^0 + \boldsymbol{\alpha}^\prime \overline{\* e}_{r,t}^0
\end{eqnarray} 
the first term of which Lemma 1 is concerned with. With respect to the last term on the right-hand side of (\ref{eq: decomputildeuhat}), whereas, recall that $\big\|\overline{\* e}_{r,t}^0 \big\| = O_p(N^{-\frac{1}{2}})$ and observe that
\begin{eqnarray} \label{eq: ratealphahat}
\big\| \boldsymbol{\widetilde{\alpha}_t} - \boldsymbol{\alpha} \big\| &\leq& \big\|(t^{-1-\tau}\* F^\prime\* M_{\* W}\* F)^{-1} \big\| \big\|t^{-1-\tau}\* F^\prime\* M_{\* W}\*u \big\| \nonumber \\
&=& O_p(T^{-\frac{1}{2}-\frac{\tau}{2}}).
\end{eqnarray}
Then it follows that
\begin{eqnarray}
\big\|\boldsymbol{\widetilde{\alpha}_t}^\prime \overline{\* e}_{r,t}^0\big\| &=& \big\|(\boldsymbol{\widetilde{\alpha}_t}-\boldsymbol{\alpha})^\prime  \overline{\* e}_{r,t}^0 + \boldsymbol{\alpha}^\prime \overline{\* e}_{r,t}^0\big\| \nonumber \\
&\leq& \big\|(\boldsymbol{\widetilde{\alpha}_t}-\boldsymbol{\alpha})^\prime \big\|\big\| \overline{\* e}_{r,t}^0\big\| + \big\|\boldsymbol{\alpha}^\prime \big\|\big\| \overline{\* e}_{r,t}^0\big\| \nonumber \\
&=& O_p(N^{-\frac{1}{2}} T^{-\frac{1}{2}-\frac{\tau}{2}})
\end{eqnarray}
which holds uniformly in $t$, thus meaning that the last term on the right-hand side of (\ref{eq: decomputildeuhat}) is negligible. 
Let us consider the last two summations in (3.3). First, using the definition $\widetilde{u}_{1,t+1} = u_{t+1}-(\boldsymbol{\widetilde{\theta}}_t-\boldsymbol{\theta})^\prime \* w_{t}+\boldsymbol{\alpha}^\prime \mathrm{\* f}_t$ and the result in (\ref{eq: decomputildeuhat}) gives
\begin{eqnarray} \label{eq: decompfirstsum}
&& \frac{1}{\sqrt{n}}\sum_{t=k_0}^{k_0+m_0-1}\widetilde{u}_{1,t+1}(\widetilde{u}_{2,t+1}- \widehat{u}_{2,t+1}) \nonumber \\
&=& \frac{1}{\sqrt{n}}\sum_{t=k_0}^{k_0+m_0-1} u_{t+1} (\widetilde{u}_{2,t+1}- \widehat{u}_{2,t+1}) - \frac{1}{\sqrt{n}}\sum_{t=k_0}^{k_0+m_0-1} (\boldsymbol{\widetilde{\theta}}_t-\boldsymbol{\theta})^\prime \* w_{t} (\widetilde{u}_{2,t+1}- \widehat{u}_{2,t+1}) \nonumber \\ 
&+& \frac{1}{\sqrt{n}}\sum_{t=k_0}^{k_0+m_0-1} \boldsymbol{\alpha}^\prime \mathrm{\* f}_t (\widetilde{u}_{2,t+1}- \widehat{u}_{2,t+1}) \nonumber \\
&=& \frac{1}{\sqrt{n}}\sum_{t=k_0}^{k_0+m_0-1} u_{t+1} ((\* Q_N^{-1}\boldsymbol{\widehat{\delta}}_t - \boldsymbol{\widetilde{\delta}}_t^{0})^{\prime}\widehat{\*z}_t^0 + (\boldsymbol{\widetilde{\alpha}}_t-\boldsymbol{\alpha})^\prime  \overline{\* e}_{r,t}^0 + \boldsymbol{\alpha}^\prime \overline{\* e}_{r,t}^0) - \frac{1}{\sqrt{n}}\sum_{t=k_0}^{k_0+m_0-1} (\boldsymbol{\widetilde{\theta}}_t-\boldsymbol{\theta})^\prime \* w_{t} (\widetilde{u}_{2,t+1}- \widehat{u}_{2,t+1}) \nonumber \\
&+& \frac{1}{\sqrt{n}}\sum_{t=k_0}^{k_0+m_0-1} \boldsymbol{\alpha}^\prime \mathrm{\* f}_t (\widetilde{u}_{2,t+1}- \widehat{u}_{2,t+1}) \nonumber \\
&=& \frac{1}{\sqrt{n}}\sum_{t=k_0}^{k_0+m_0-1} u_{t+1} (\* Q_N^{-1}\boldsymbol{\widehat{\delta}}_t - \boldsymbol{\widetilde{\delta}}_t^{0})^{\prime} \widehat{\*z}_t^0 + \frac{1}{\sqrt{n}}\sum_{t=k_0}^{k_0+m_0-1} (\boldsymbol{\widetilde{\alpha}_t}-\boldsymbol{\alpha})^\prime \overline{\* e}_{r,t}^0u_{t+1} + \frac{1}{\sqrt{n}}\sum_{t=k_0}^{k_0+m_0-1} \boldsymbol{\alpha}^\prime \overline{\* e}_{r,t}^0u_{t+1} \nonumber \\
&-& \frac{1}{\sqrt{n}}\sum_{t=k_0}^{k_0+m_0-1} (\boldsymbol{\widetilde{\theta}}_t-\boldsymbol{\theta})^\prime \* w_{t} (\widetilde{u}_{2,t+1}- \widehat{u}_{2,t+1}) + \frac{1}{\sqrt{n}}\sum_{t=k_0}^{k_0+m_0-1} \boldsymbol{\alpha}^\prime \mathrm{\* f}_t (\widetilde{u}_{2,t+1}- \widehat{u}_{2,t+1}) \nonumber \\
&=& \mathrm{\* I} + \mathrm{\*II} + \mathrm{\*III} - \mathrm{\*IV} + \mathrm{\*V},
\end{eqnarray}
where the definitions of $\mathrm{\* I}$-$\mathrm{\*V}$ are implicit. Begining from $\mathrm{\*II}$, and using the fact that $\sqrt{\frac{T}{n}}=\frac{1}{\sqrt{1-\pi_0}}+O(T^{-1/2})$, we have
\begin{align}
   & \left|\frac{1}{\sqrt{n}}\sum_{t=k_0}^{k_0+m_0-1} 
 (\boldsymbol{\widetilde{\alpha}_t}-\boldsymbol{\alpha})^\prime \overline{\* e}_{r,t}^0u_{t+1} \right|\notag\\
 &=\frac{1}{\sqrt{N}}\left| \frac{1}{\sqrt{n}}\frac{1}{T^{1/2+\tau/2}}\sum_{t=k_0}^{k_0+m_0-1} 
 T^{1/2+\tau/2}(\boldsymbol{\widetilde{\alpha}_t}-\boldsymbol{\alpha})^\prime (\sqrt{N}\overline{\* e}_{r,t}^0)u_{t+1}\right|\notag\\
 &=T^{-\tau/2}\frac{1}{\sqrt{N}}\left| \frac{T}{\sqrt{nT}}\frac{1}{T}\sum_{t=k_0}^{k_0+m_0-1} 
 T^{1/2+\tau/2}(\boldsymbol{\widetilde{\alpha}_t}-\boldsymbol{\alpha})^\prime (\sqrt{N}\overline{\* e}_{r,t}^0)u_{t+1}\right|\notag\\
 &\leq T^{-\tau/2}\frac{1}{\sqrt{N}}\sqrt{\frac{T}{n}}\sup_{t}\left\| T^{1/2+\tau/2}(\boldsymbol{\widetilde{\alpha}_t}-\boldsymbol{\alpha}) \right\| \frac{1}{T}\sum_{t=k_0}^{k_0+m_0-1} 
  |(\sqrt{N}\overline{\* e}_{r,t}^0)u_{t+1}| \notag\\
  &= O_p(N^{-\frac{1}{2}}T^{-\frac{\tau}{2}}),
\end{align}
because $\sup_{t} \| T^{\frac{1}{2}+\frac{\tau}{2}}(\boldsymbol{\widetilde{\alpha}_t}-\boldsymbol{\alpha}) \| = O_p(1)$ as in (38) and (46) of \citealp{pitarakis2023direct}, and the term with the summation is of the same order. In a similar manner, the order of $\mathrm{\*III}$ is obtained as follows:
\begin{eqnarray}
&&\left|\frac{1}{\sqrt{n}}\sum_{t=k_0}^{k_0+m_0-1} \boldsymbol{\alpha}^\prime \overline{\* e}_{r,t}^0u_{t+1}\right| \nonumber \\
&\leq& \frac{1}{\sqrt{N}T^{\frac{1}{4}+\frac{\tau}{2}}} \sqrt{\frac{T}{n}}\big\|\boldsymbol{\alpha}^{0}\big\| \bigg\|\frac{1}{\sqrt{T}}\sum_{t=k_0}^{k_0+m_0-1} (\sqrt{N}\overline{\* e}_{r,t}^0)u_{t+1}\bigg\| \nonumber \\
&=& O_p(N^{-\frac{1}{2}}T^{-\frac{1}{4}-\frac{\tau}{2}})
\end{eqnarray}
under the same logic. Let us now turn our attention to $\mathrm{\* I}$:
\begin{eqnarray}
&&\left|\frac{1}{\sqrt{n}}\sum_{t=k_0}^{k_0+m_0-1} u_{t+1} (\* Q_N^{-1}\boldsymbol{\widehat{\delta}}_t - \boldsymbol{\widetilde{\delta}}_t^{0})^{\prime}\* D_T \* D_T^{-1} \widehat{\*z}_t^0\right| \nonumber \\
&& = T^{-\frac{1}{4}}\left|\frac{1}{\sqrt{n}}\sum_{t=k_0}^{k_0+m_0-1} (\* Q_N^{-1}\boldsymbol{\widehat{\delta}}_t - \boldsymbol{\widetilde{\delta}}_t^{0})^{\prime}\* D_T T^{\frac{1}{4}} \* D_T^{-1} \widehat{\*z}_t^0 u_{t+1}\right| \nonumber \\
&&= T^{\frac{\tau}{2}-\frac{1}{2}}\left|\frac{1}{\sqrt{n}}\sum_{t=k_0}^{k_0+m_0-1} \left[(\* Q_N^{-1}\boldsymbol{\widehat{\delta}}_t - \boldsymbol{\widetilde{\delta}}_t^{0})^{\prime}\* D_TT^{1/4}\right] (T^{1/4}\* D_T^{-1}) T^{-\tau/2}\widehat{\*z}_t^0 u_{t+1}\right| \nonumber \\
&& = O_p(T^{(\tau-1)/2}),
\end{eqnarray}
 which holds for all $\tau \in (0,1)$ by noting that the summand, $\left[(\* Q_N^{-1}\boldsymbol{\widehat{\delta}}_t - \boldsymbol{\widetilde{\delta}}_t^{0})^{\prime}\* D_TT^{1/4}\right] (T^{1/4}\* D_T^{-1}) T^{-\tau/2}\widehat{\*z}_t^0 u_{t+1}$, is a heterogeneous martingale difference process by Assumption 1. In particular, $\left\|(\* Q_N^{-1}\boldsymbol{\widehat{\delta}}_t - \boldsymbol{\widetilde{\delta}}_t^{0})^{\prime}\* D_TT^{1/4} \right\|=O_p(1)$ by Lemma 1, while 
\begin{align}
    \sup_t\left\|T^{-\tau/2}\widehat{\*z}_t^0 \right\|\leq  \sup_t\left\| T^{-\tau/2} \*z_t\right\|+ \sup_t\left\|T^{-\tau/2} \overline{\*e}_t^0 \right\|=O_p(1),
\end{align}
as $sup_t\left\| T^{-\tau/2} \*z_t\right\|=O_p(1)$ based on \cite{magdalinos2009limit}. For $\mathrm{\*IV}$, whereas, we can start by showing that, as $T \to \infty$, we have
\begin{eqnarray}
T^{\frac{1}{4}+\frac{\tau}{2}}(\boldsymbol{\widetilde{\theta}}_t-\boldsymbol{\theta}) &=& T^{\frac{1}{4}+\frac{\tau}{2}} (T^{-1-\tau}\* W^\prime\* W)^{-1} T^{-1-\tau}\* W^\prime (\* F\boldsymbol{\alpha} + \*u) \nonumber \\
&=& (t^{-1-\tau}\* W^\prime\* W)^{-1} t^{-1-\tau}\* W^\prime \* F\boldsymbol{\alpha}^0 + (T^{1+\tau} t^{-1-\tau} )T^{\frac{1}{4}+\frac{\tau}{2}} (t^{-1-\tau}\* W^\prime\* W)^{-1} T^{-1-\tau}\* W^\prime \*u \nonumber \\
&=& (t^{-1-\tau}\* W^\prime\* W)^{-1} t^{-1-\tau}\* W^\prime \* F\boldsymbol{\alpha}^0 + T^{-\frac{1}{4}}(T^{1+\tau} t^{-1-\tau}) (t^{-1-\tau}\* W^\prime\* W)^{-1} T^{-\frac{1}{2}-\frac{\tau}{2}}\* W^\prime \*u \nonumber \\
&=& (t^{-1-\tau}\* W^\prime\* W)^{-1} t^{-1-\tau}\* W^\prime \* F\boldsymbol{\alpha}^0 +  O_p(T^{-\frac{1}{4}})\nonumber \\
&\to_p& \boldsymbol{\Sigma}_{ww}^{-1}\boldsymbol{\Sigma}_{wf}\boldsymbol{\alpha}^0,
\end{eqnarray}
because of the result in (\ref{eq: tTs}) and the fact that $\big\|(t^{-1-\tau}\* W^\prime\* W)^{-1} T^{-\frac{1}{2}-\frac{\tau}{2}}\* W^\prime \*u \big\|= O_p(1)$. Hence, we have that $\big\|T^{\frac{1}{4}+\frac{\tau}{2}}(\boldsymbol{\widetilde{\theta}}_t-\boldsymbol{\theta})\big\| = O_p(1)$. To be more precise, we have that $\sup_{t} \big\|T^{\frac{1}{4}+\frac{\tau}{2}}(\boldsymbol{\widetilde{\theta}}_t-\boldsymbol{\theta})\big\| = O_p(1)$. Further, substituting (\ref{eq: decomputildeuhat}) in $\mathrm{\*IV}$ yields
\begin{eqnarray}
&& \frac{1}{\sqrt{n}}\sum_{t=k_0}^{k_0+m_0-1} (\boldsymbol{\widetilde{\theta}}_t-\boldsymbol{\theta})^\prime \* w_{t} (\widetilde{u}_{2,t+1}- \widehat{u}_{2,t+1}) \nonumber \\
&& = \frac{1}{\sqrt{n}}\sum_{t=k_0}^{k_0+m_0-1} (\boldsymbol{\widetilde{\theta}}_t-\boldsymbol{\theta})^\prime \* w_{t} ((\* Q_N^{-1}\boldsymbol{\widehat{\delta}}_t - \boldsymbol{\widetilde{\delta}}_t^{0})^{\prime}\widehat{\*z}_t^0 + (\boldsymbol{\widetilde{\alpha}}_t-\boldsymbol{\alpha})^\prime  \overline{\* e}_{r,t}^0 + \boldsymbol{\alpha}^\prime \overline{\* e}_{r,t}^0) \nonumber \\
&& = \frac{1}{\sqrt{n}}\sum_{t=k_0}^{k_0+m_0-1} (\boldsymbol{\widetilde{\theta}}_t-\boldsymbol{\theta})^\prime \* w_{t} (\* Q_N^{-1}\boldsymbol{\widehat{\delta}}_t - \boldsymbol{\widetilde{\delta}}_t^{0})^{\prime}\widehat{\*z}_t^0 + \frac{1}{\sqrt{n}}\sum_{t=k_0}^{k_0+m_0-1} (\boldsymbol{\widetilde{\theta}}_t-\boldsymbol{\theta})^\prime \* w_{t}(\boldsymbol{\widetilde{\alpha}}_t-\boldsymbol{\alpha})^\prime  \overline{\* e}_{r,t}^0 \nonumber \\
&& + \frac{1}{\sqrt{n}}\sum_{t=k_0}^{k_0+m_0-1} (\boldsymbol{\widetilde{\theta}}_t-\boldsymbol{\theta})^\prime \* w_{t}\boldsymbol{\alpha}^\prime \overline{\* e}_{r,t}^0.
\end{eqnarray}
Here, the third term in this decomposition becomes
\begin{eqnarray}
&&\left| \frac{1}{\sqrt{n}}\sum_{t=k_0}^{k_0+m_0-1} (\boldsymbol{\widetilde{\theta}}_t-\boldsymbol{\theta})^\prime \* w_{t} \boldsymbol{\alpha}^\prime \overline{\* e}_{r,t}^0 \right| \nonumber \\
&& = \left| T^{-\frac{1}{4}-\frac{\tau}{2}}\frac{1}{\sqrt{n}}\sum_{t=k_0}^{k_0+m_0-1} T^{\frac{1}{4}+\frac{\tau}{2}} (\boldsymbol{\widetilde{\theta}}_t-\boldsymbol{\theta})^\prime T^{-\frac{1}{4}-\frac{\tau}{2}}\* w_{t} T^{\frac{1}{4}+\frac{\tau}{2}} \boldsymbol{\alpha}^\prime \overline{\* e}_{r,t}^0 \right| \nonumber \\
&& \leq T^{-\frac{1}{4}-\frac{\tau}{2}} \big\|\boldsymbol{\alpha}^0\big\| \sup_{k_0 \leq t \leq k_0+m_0 -1} \big\|T^{\frac{1}{4}+\frac{\tau}{2}}(\boldsymbol{\widetilde{\theta}}_t-\boldsymbol{\theta})\big\| \left|\frac{1}{\sqrt{n}}\sum_{t=k_0}^{k_0+m_0-1}  T^{-\frac{1}{4}-\frac{\tau}{2}}\* w_{t} \overline{\*e}_{r,t}^{0 \prime} \right| \nonumber \\
&& \leq \big\|\boldsymbol{\alpha}^0\big\| \sup_{k_0 \leq t \leq k_0+m_0 -1} \big\|T^{\frac{1}{4}+\frac{\tau}{2}}(\boldsymbol{\widetilde{\theta}}_t-\boldsymbol{\theta})\big\| \sqrt{\frac{T}{n}} \frac{1}{T^{1+\tau}}\sum_{t=k_0}^{k_0+m_0-1}  \big\|\* w_{t}\big\| \big\|\overline{\* e}_{r,t}^0\big\| \nonumber \\
&& \leq T^{-\frac{\tau}{2}}\big\|\boldsymbol{\alpha}^0\big\| \sup_{k_0 \leq t \leq k_0+m_0 -1} \big\|T^{\frac{1}{4}+\frac{\tau}{2}}(\boldsymbol{\widetilde{\theta}}_t-\boldsymbol{\theta})\big\| \sqrt{\frac{T}{n}} \left(\frac{1}{T^{1+\tau}}\sum_{t=k_0}^{k_0+m_0-1}  \big\|\* w_{t}\big\|^2\right)^{\frac{1}{2}} \left(\frac{1}{T}\sum_{t=k_0}^{k_0+m_0-1}  \big\|\overline{\* e}_{r,t}^0\big\|^2\right)^{\frac{1}{2}} \nonumber \\
&& = O_p(N^{-\frac{1}{2}}T^{-\frac{\tau}{2}}).
\end{eqnarray}
Moving on to the second term, we follow a similar argument:
\begin{eqnarray}
&&\left| \frac{1}{\sqrt{n}}\sum_{t=k_0}^{k_0+m_0-1} (\boldsymbol{\widetilde{\theta}}_t-\boldsymbol{\theta})^\prime \* w_{t} (\boldsymbol{\widetilde{\alpha}}_t-\boldsymbol{\alpha})^\prime \overline{\* e}_{r,t}^0 \right| \nonumber \\ 
&& = \left| T^{-\frac{1}{4}-\frac{\tau}{2}}\frac{1}{\sqrt{n}}\sum_{t=k_0}^{k_0+m_0-1} T^{\frac{1}{4}+\frac{\tau}{2}} (\boldsymbol{\widetilde{\theta}}_t-\boldsymbol{\theta})^\prime T^{-\frac{1}{2}-\frac{\tau}{2}}\* w_{t} T^{\frac{1}{2}+\frac{\tau}{2}} (\boldsymbol{\widetilde{\alpha}}_t-\boldsymbol{\alpha})^\prime \overline{\* e}_{r,t}^0 \right| \nonumber \\
&& \leq T^{-\frac{1}{4}} \sup_{k_0 \leq t \leq k_0+m_0 -1} \big\|T^{\frac{1}{2}+\frac{\tau}{2}} (\boldsymbol{\widetilde{\alpha}}_t-\boldsymbol{\alpha})\big\| \sup_{k_0 \leq t \leq k_0+m_0 -1} \big\|T^{\frac{1}{4}+\frac{\tau}{2}}(\boldsymbol{\widetilde{\theta}}_t-\boldsymbol{\theta})\big\| \left|\frac{1}{\sqrt{n}}\sum_{t=k_0}^{k_0+m_0-1}  T^{-\frac{1}{2}-\tau}\* w_{t} \overline{\*e}_{r,t}^{0 \prime} \right| \nonumber \\
&& \leq T^{-\frac{1}{4}} \sup_{k_0 \leq t \leq k_0+m_0 -1} \big\|T^{\frac{1}{2}+\frac{\tau}{2}} (\boldsymbol{\widetilde{\alpha}}_t-\boldsymbol{\alpha})\big\| \sup_{k_0 \leq t \leq k_0+m_0 -1} \big\|T^{\frac{1}{4}+\frac{\tau}{2}}(\boldsymbol{\widetilde{\theta}}_t-\boldsymbol{\theta})\big\| \sqrt{\frac{T}{n}} \frac{1}{T^{1+\tau}}\sum_{t=k_0}^{k_0+m_0-1}  \big\|\* w_{t}\big\| \big\|\overline{\* e}_{r,t}^0\big\| \nonumber \\
&& \leq T^{-\frac{1}{4}-\frac{\tau}{2}}\sup_{k_0 \leq t \leq k_0+m_0 -1} \big\|T^{\frac{1}{2}+\frac{\tau}{2}} (\boldsymbol{\widetilde{\alpha}}_t-\boldsymbol{\alpha})\big\| \times  \sup_{k_0 \leq t \leq k_0+m_0 -1} \big\|T^{\frac{1}{4}+\frac{\tau}{2}}(\boldsymbol{\widetilde{\theta}}_t-\boldsymbol{\theta})\big\| \sqrt{\frac{T}{n}} \left(\frac{1}{T^{1+\tau}}\sum_{t=k_0}^{k_0+m_0-1}  \big\|\* w_{t}\big\|^2\right)^{\frac{1}{2}} \nonumber \\ 
&& \times \left(\frac{1}{T}\sum_{t=k_0}^{k_0+m_0-1}  \big\|\overline{\* e}_{r,t}^0\big\|^2\right)^{\frac{1}{2}}\nonumber \\
&& = O_p(N^{-\frac{1}{2}}T^{-\frac{1}{4}-\frac{\tau}{2}}).
\end{eqnarray}
It remains to look at the first term, and to obtain an order result sufficient for our analysis, we cannot use rough bounds straight away. By letting $\*R_T=T^{1/4}\*D_T^{-1}=\mathrm{diag}(\*0_{r\times r}, \*I_{m-r},\*0_{q\times 1})+O(T^{-\tau/2})=O(1)$, we have 
\begin{eqnarray} \label{eq: decompthetawzhatQdelta}
&& \frac{1}{\sqrt{n}}\sum_{t=k_0}^{k_0+m_0-1} (\boldsymbol{\widetilde{\theta}}_t-\boldsymbol{\theta})^\prime \* w_{t} \widehat{\*z}_t^{0 \prime} (\* Q_N^{-1}\boldsymbol{\widehat{\delta}}_t - \boldsymbol{\widetilde{\delta}}_t^{0}) \nonumber \\
&& =  \frac{1}{\sqrt{n}}\sum_{t=k_0}^{k_0+m_0-1} T^{\frac{1}{4}+\frac{\tau}{2}}(\boldsymbol{\widetilde{\theta}}_t-\boldsymbol{\theta})^\prime T^{-\frac{1}{4}-\frac{\tau}{2}} \* w_{t} \widehat{\*z}_t^{0 \prime} \* D_T^{-1}\* D_T(\* Q_N^{-1}\boldsymbol{\widehat{\delta}}_t - \boldsymbol{\widetilde{\delta}}_t^{0}) \nonumber \\
&& =\sqrt{\frac{T}{n}}\frac{1}{T}\sum_{t=k_0}^{k_0+m_0-1}T^{\frac{1}{4}+\frac{\tau}{2}}(\boldsymbol{\widetilde{\theta}}_t-\boldsymbol{\theta})^\prime T^{-\tau}\* w_{t} \widehat{\*z}_t^{0 \prime}\*R_TT^{\tau/2}\*D_T(\* Q_N^{-1}\boldsymbol{\widehat{\delta}}_t - \boldsymbol{\widetilde{\delta}}_t^{0})\notag\\
&&=\sqrt{\frac{T}{n}}\frac{1}{T}\sum_{t=k_0}^{k_0+m_0-1}\left(\*R_T T^{\tau/2}\*D_T(\* Q_N^{-1}\boldsymbol{\widehat{\delta}}_t - \boldsymbol{\widetilde{\delta}}_t^{0})\otimes T^{\frac{1}{4}+\frac{\tau}{2}}(\boldsymbol{\widetilde{\theta}}_t-\boldsymbol{\theta}) \right)'\mathrm{vec}\left( T^{-\tau}\* w_{t} \widehat{\*z}_t^{0 \prime}\right)\notag\\
&&= \sqrt{\frac{T}{n}}\frac{1}{T}\sum_{t=k_0}^{k_0+m_0-1}\left(T^{\tau/2-1/4}\*v_t^0\otimes T^{\frac{1}{4}+\frac{\tau}{2}}(\boldsymbol{\widetilde{\theta}}_t-\boldsymbol{\theta}) \right)'\mathrm{vec}\left( T^{-\tau}\* w_{t} \widehat{\*z}_t^{0 \prime}\right) \notag\\
&&+ O_p(N^{-1/2})+O_p(T^{(\tau-1)/2}).
\end{eqnarray}
This simplification requires an additional explanation. Here, we use our Lemma 1, where we find that $\sup_{k_0\leq t\leq k_0+m_0-1}\left\| \*D_T(\* Q_N^{-1}\boldsymbol{\widehat{\delta}}_t - \boldsymbol{\widetilde{\delta}}_t^{0})\right\|=O_p(N^{-1/2}T^{-\tau/2})+O_p(T^{-1/4})$. Alternatively, this can be represented as
\begin{eqnarray}
  \*D_T(\* Q_N^{-1}\boldsymbol{\widehat{\delta}}_t - \boldsymbol{\widetilde{\delta}}_t^{0})&&=[
    \*0_r',
   T^{-1/4}\*v_t',
    \*0_{q}'
]' +O_p(N^{-1/2}T^{-\tau/2})+O_p(T^{-1/2})\notag\\&&= T^{-1/4}\*v_t^0+O_p(N^{-1/2}T^{-\tau/2})++O_p(T^{-1/2}), 
\end{eqnarray}
where indeed the slowest decaying ($O_p(T^{-1/4})$) term is $T^{-1/4}\*v_t$, and the second slowest term is $O_p(T^{-1/2})$, as per thorough inspection of Lemma 1. Therefore,
\begin{eqnarray}
    T^{\tau/2}\*D_T(\* Q_N^{-1}\boldsymbol{\widehat{\delta}}_t - \boldsymbol{\widetilde{\delta}}_t^{0})=T^{\tau/2-1/4}\*v_t^0+O_p(N^{-1/2})+O_p(T^{(\tau-1)/2}).
\end{eqnarray}
Eventually, by using the structure of $\*R_T$, we have that 
\begin{eqnarray}
    T^{\tau/2}\*R_T\*D_T(\* Q_N^{-1}\boldsymbol{\widehat{\delta}}_t - \boldsymbol{\widetilde{\delta}}_t^{0})=T^{\tau/2-1/4}\*v_t^0+O_p(N^{-1/2})+O_p(T^{(\tau-1)/2}),
\end{eqnarray}
as well. We can implement this substitution, because $\left\| T^{-\tau}\* w_{t} \widehat{\*z}_t^{0 \prime}\right\|=O_p(1)$ uniformly in $t$, according to Lemma 3.1 of \cite{magdalinos2009limit}, and also $\sup_{k_0\leq t\leq k_0+m_0-1}\left\|T^{\frac{1}{4}+\frac{\tau}{2}}(\boldsymbol{\widetilde{\theta}}_t-\boldsymbol{\theta}) \right\|=O_p(1)$. Therefore, we obtain 
\begin{eqnarray}
   && \frac{1}{\sqrt{n}}\sum_{t=k_0}^{k_0+m_0-1} (\boldsymbol{\widetilde{\theta}}_t-\boldsymbol{\theta})^\prime \* w_{t} \widehat{\*z}_t^{0 \prime} (\* Q_N^{-1}\boldsymbol{\widehat{\delta}}_t - \boldsymbol{\widetilde{\delta}}_t^{0})\notag\\
   &&=\sqrt{\frac{T}{n}}\frac{1}{T}\sum_{t=k_0}^{k_0+m_0-1}\left(T^{\tau/2-1/4}\*v_t^0\otimes T^{\frac{1}{4}+\frac{\tau}{2}}(\boldsymbol{\widetilde{\theta}}_t-\boldsymbol{\theta}) \right)'\mathrm{vec}\left( T^{-\tau}\* w_{t} \widehat{\*z}_t^{0 \prime}\right) + O_p(N^{-1/2})+O_p(T^{(\tau-1)/2})\notag\\
   &&= \sqrt{\frac{T}{n}}\frac{1}{T}\sum_{t=k_0}^{k_0+m_0-1}\left(T^{\tau/2-1/4}\*v_t^0\otimes T^{\frac{1}{4}+\frac{\tau}{2}}(\boldsymbol{\widetilde{\theta}}_t-\boldsymbol{\theta}) \right)'\mathrm{vec}\left( T^{-\tau}\* w_{t} \*z_t^{0 \prime}\right)\notag\\
   &&+\sqrt{\frac{T}{n}}\frac{1}{T}\sum_{t=k_0}^{k_0+m_0-1}\left(T^{\tau/2-1/4}\*v_t^0\otimes T^{\frac{1}{4}+\frac{\tau}{2}}(\boldsymbol{\widetilde{\theta}}_t-\boldsymbol{\theta}) \right)'\mathrm{vec}\left( T^{-\tau}\* w_{t} \widehat{\*z}_t^{0 \prime}-T^{-\tau}\* w_{t} \*z_t^{0 \prime}\right) + O_p(N^{-1/2})+O_p(T^{(\tau-1)/2})\notag\\
   &&= \sqrt{\frac{T}{n}}\frac{1}{T}\sum_{t=k_0}^{k_0+m_0-1}\left(T^{\tau/2-1/4}\*v_t^0\otimes T^{\frac{1}{4}+\frac{\tau}{2}}(\boldsymbol{\widetilde{\theta}}_t-\boldsymbol{\theta}) \right)'\mathrm{vec}\left( T^{-\tau}\* w_{t} \*z_t^{0 \prime}\right) + O_p(N^{-1/2}) +O_p(T^{(\tau-1)/2})+O_p(T^{-1/4})\notag\\
   &&=T^{\tau/2-1/4}\sqrt{\frac{T}{n}}\frac{1}{T}\sum_{t=k_0}^{k_0+m_0-1}T^{\frac{1}{4}+\frac{\tau}{2}}(\boldsymbol{\widetilde{\theta}}_t-\boldsymbol{\theta})'T^{-\tau}\*w_t\*z_t^{0\prime}\*v_t^0+ O_p(N^{-1/2}) +O_p(T^{(\tau-1)/2})+O_p(T^{-1/4})\notag\\
   &&=O_p(N^{-1/2}) + +O_p(T^{(\tau-1)/2})+O_p(T^{-1/4}),
\end{eqnarray}
because $\*z_t^{0\prime}\*v_t^0=0$ algebraically. Notice that 
\begin{eqnarray}
&&\left|\sqrt{\frac{T}{n}}\frac{1}{T}\sum_{t=k_0}^{k_0+m_0-1}\left(T^{\tau/2-1/4}\*v_t^0\otimes T^{\frac{1}{4}+\frac{\tau}{2}}(\boldsymbol{\widetilde{\theta}}_t-\boldsymbol{\theta}) \right)'\mathrm{vec}\left( T^{-\tau}\* w_{t} \widehat{\*z}_t^{0 \prime}-T^{-\tau}\* w_{t} \*z_t^{0 \prime}\right) \right|\notag\\
&&\leq T^{\tau/2-1/4}\underbrace{\sup_{k_0\leq t\leq k_0+m_0-1}\left\| T^{-\tau}\* w_{t} \widehat{\*z}_t^{0 \prime}-T^{-\tau}\* w_{t} \*z_t^{0 \prime}\right\|}_{O_p(T^{-\tau/2})}\sqrt{\frac{T}{n}}\frac{1}{T}\sum_{t=k_0}^{k_0+m_0-1}\left\|\left(\*v_t^0\otimes T^{\frac{1}{4}+\frac{\tau}{2}}(\boldsymbol{\widetilde{\theta}}_t-\boldsymbol{\theta}) \right)\right\|\notag\\
&&=O_p(T^{-1/4}),
\end{eqnarray}
since $T^{-\tau/2}\widehat{\*z}_t^0=T^{-\tau/2}\*z_t^0+O_p(T^{-\tau/2})$ uniformly in $t$ by Lemma 3.1 in \cite{magdalinos2009limit}, because the idiosyncratic component is stationary. Hence, $\mathrm{\*IV}$ reduces to
\begin{eqnarray}
\frac{1}{\sqrt{n}}\sum_{t=k_0}^{k_0+m_0-1} (\boldsymbol{\widetilde{\theta}}_t-\boldsymbol{\theta})^\prime \* w_{t} (\widetilde{u}_{2,t+1}- \widehat{u}_{2,t+1}) = O_p(N^{-\frac{1}{2}}) + O_p(T^{(\tau-1)/2})+ O_p(T^{-\frac{1}{4}}),
\end{eqnarray}
which is negligible for all $\tau\in (0,1)$. Finally, the order of $\mathrm{\*V}$ can be obtained by employing an argument similar $\mathrm{\*IV}$. That is,
\begin{eqnarray}
&& \left|\frac{1}{\sqrt{n}}\sum_{t=k_0}^{k_0+m_0-1} \boldsymbol{\alpha}^\prime \mathrm{\* f}_t (\widetilde{u}_{2,t+1}- \widehat{u}_{2,t+1})\right| \nonumber \\
&& = \left|\frac{1}{\sqrt{n}}\sum_{t=k_0}^{k_0+m_0-1} \boldsymbol{\alpha}^\prime \mathrm{\* f}_t (\* Q_N^{-1}\boldsymbol{\widehat{\delta}}_t - \boldsymbol{\widetilde{\delta}}_t^{0})^{\prime}\widehat{\*z}_t^0\right| + \left|\frac{1}{\sqrt{n}}\sum_{t=k_0}^{k_0+m_0-1} \boldsymbol{\alpha}^\prime \mathrm{\* f}_t (\boldsymbol{\widetilde{\alpha}}_t-\boldsymbol{\alpha})^\prime  \overline{\* e}_{r,t}^0\right| \nonumber \\
&& + \left|\frac{1}{\sqrt{n}}\sum_{t=k_0}^{k_0+m_0-1} \boldsymbol{\alpha}^\prime \mathrm{\* f}_t\boldsymbol{\alpha}^\prime \overline{\* e}_{r,t}^0\right| \nonumber \\
&& = O_p(N^{-\frac{1}{2}}) + O_p(T^{(\tau-1)/2})+ O_p(T^{-\frac{1}{4}}).
\end{eqnarray}
To derive this order, again, we need more work than relying on rough bounds. Notice that we can re-write the first term, because it is a scalar. By letting $\*R_T=T^{1/4}\*D_T^{-1}$, we obtain
\begin{eqnarray}
&& \frac{1}{\sqrt{n}}\sum_{t=k_0}^{k_0+m_0-1} \boldsymbol{\alpha}^\prime \mathrm{\* f}_t (\* Q_N^{-1}\boldsymbol{\widehat{\delta}}_t - \boldsymbol{\widetilde{\delta}}_t^{0})^{\prime}\widehat{\*z}_t^0 \nonumber \\
&& = \frac{1}{\sqrt{n}}\sum_{t=k_0}^{k_0+m_0-1}(\* Q_N^{-1}\boldsymbol{\widehat{\delta}}_t - \boldsymbol{\widetilde{\delta}}_t^{0})^{\prime}\widehat{\*z}_t^0\*f_t'\+\alpha\notag\\
&&= T^{-\frac{1}{4}-\frac{\tau}{2}}\frac{1}{\sqrt{n}}\sum_{t=k_0}^{k_0+m_0-1}(\* Q_N^{-1}\boldsymbol{\widehat{\delta}}_t - \boldsymbol{\widetilde{\delta}}_t^{0})^{\prime}\*D_T\*D_T^{-1}\widehat{\*z}_t^0\*f_t'\+\alpha^0\notag\\
&&= T^{-1/2}T^{-\tau/2}\frac{1}{\sqrt{n}}\sum_{t=k_0}^{k_0+m_0-1}(\* Q_N^{-1}\boldsymbol{\widehat{\delta}}_t - \boldsymbol{\widetilde{\delta}}_t^{0})^{\prime}\*D_T \*R_T\widehat{\*z}_t^0\*f_t'\+\alpha^0\notag\\
&&= \sqrt{\frac{T}{n}}\frac{1}{T}\sum_{t=k_0}^{k_0+m_0-1}\left(\+\alpha^0 \otimes T^{\tau/2}(\* Q_N^{-1}\boldsymbol{\widehat{\delta}}_t - \boldsymbol{\widetilde{\delta}}_t^{0})^{\prime}\*D_T\*R_T \right)'\mathrm{vec}\left(T^{-\tau} \widehat{\*z}_t^0\*f_t'\right)\notag\\
&&= \sqrt{\frac{T}{n}}\frac{1}{T}\sum_{t=k_0}^{k_0+m_0-1}\left(\+\alpha^0 \otimes T^{\tau/2-1/4}\*v_t^0\right)'\mathrm{vec}\left(T^{-\tau} \widehat{\*z}_t^0\*f_t'\right)+O_p(N^{-1/2})+O_p(T^{(\tau-1)/2}),
\end{eqnarray}
where $\left\|T^{-\tau}\widehat{\*z}_t^0\*f_t' \right\|=O_p(1)$ uniformly in $t$ by Lemma 3.1 in \cite{magdalinos2009limit}. Again, we used $\sup_{k_0\leq t\leq k_0+m_0-1}\left\| \*D_T(\* Q_N^{-1}\boldsymbol{\widehat{\delta}}_t - \boldsymbol{\widetilde{\delta}}_t^{0})\right\|=O_p(N^{-1/2}T^{-\tau/2})+O_p(T^{-1/4})$ to uncover the leading term, and  $\*R_T=\mathrm{diag}(\*0_{r\times r}, \*I_{m-r}, \*0_{q\times q})+O(T^{-\tau/2})$. Therefore, 
\begin{eqnarray}\label{long_argm2}
    &&\frac{1}{\sqrt{n}}\sum_{t=k_0}^{k_0+m_0-1} \boldsymbol{\alpha}^\prime \mathrm{\* f}_t (\* Q_N^{-1}\boldsymbol{\widehat{\delta}}_t - \boldsymbol{\widetilde{\delta}}_t^{0})^{\prime}\widehat{\*z}_t^0 \nonumber \\
    &&=\sqrt{\frac{T}{n}}\frac{1}{T}\sum_{t=k_0}^{k_0+m_0-1}\left(\+\alpha^0 \otimes T^{\tau/2-1/4}\*v_t^0\right)'\mathrm{vec}\left(T^{-\tau} \widehat{\*z}_t^0\*f_t'\right)+O_p(N^{-1/2})+O_p(T^{(\tau-1)/2})\notag\\
    &&=\sqrt{\frac{T}{n}}\frac{1}{T}\sum_{t=k_0}^{k_0+m_0-1}\left(\+\alpha^0 \otimes T^{\tau/2-1/4}\*v_t^0\right)'\mathrm{vec}\left(T^{-\tau} \*z_t^0\*f_t'\right)\notag\\
    &&+\sqrt{\frac{T}{n}}\frac{1}{T}\sum_{t=k_0}^{k_0+m_0-1}\left(\+\alpha^0 \otimes T^{\tau/2-1/4}\*v_t^0\right)'\mathrm{vec}\left(T^{-\tau} \widehat{\*z}_t^0\*f_t'-T^{-\tau}\*z_t^0\*f_t'\right)+O_p(N^{-1/2})+O_p(T^{(\tau-1)/2})\notag\\
    &&=\sqrt{\frac{T}{n}}\frac{1}{T}\sum_{t=k_0}^{k_0+m_0-1}\left(\+\alpha^0 \otimes T^{\tau/2-1/4}\*v_t^0\right)'\mathrm{vec}\left(T^{-\tau} \*z_t^0\*f_t'\right)+O_p(N^{-1/2})+O_p(T^{(\tau-1)/2})+O_p(T^{-1/4})\notag\\
    &&= T^{\tau/2-1/4}\sqrt{\frac{T}{n}}\frac{1}{T}\sum_{t=k_0}^{k_0+m_0-1}\*v_t^{0\prime}T^{-\tau}\*z_t^0\*f_t'\+\alpha +O_p(N^{-1/2})+O_p(T^{(\tau-1)/2})+O_p(T^{-1/4})\notag\\
    &&=O_p(N^{-1/2})+O_p(T^{(\tau-1)/2})+O_p(T^{-1/4}),
\end{eqnarray}
because, again $\*v_t^{0\prime}\*z_t^0=0$ algebraically, and which is negligible for all $\tau \in (0,1)$. The term $O_p(T^{-1/4})$ is obtained from 
\begin{eqnarray}
&&\left|\sqrt{\frac{T}{n}}\frac{1}{T}\sum_{t=k_0}^{k_0+m_0-1}\left(\+\alpha^0 \otimes T^{\tau/2-1/4}\*v_t^0\right)'\mathrm{vec}\left(T^{-\tau} \widehat{\*z}_t^0\*f_t'-T^{-\tau}\*z_t^0\*f_t'\right) \right|\notag\\
&&\leq T^{\tau/2-1/4}\underbrace{\sup_{k_0\leq t\leq k_0+m_0-1}\left\| T^{-\tau} \widehat{\*z}_t^0\*f_t'-T^{-\tau}\*z_t^0\*f_t'\right\|}_{O_p(T^{-\tau/2})}\sqrt{\frac{T}{n}}\frac{1}{T}\sum_{t=k_0}^{k_0+m_0-1}\left\|\left(\+\alpha^0 \otimes \*v_t^0\right)\right\|=O_p(T^{-1/4}),
\end{eqnarray}
as expected from the fact that $T^{-\tau/2}\widehat{\*z}^0_t=T^{-\tau/2}\*z_t^0+O_p(T^{-\tau/2})$ uniformly in $t$. The remaining components follow:
\begin{eqnarray}
&&\left| \frac{1}{\sqrt{n}}\sum_{t=k_0}^{k_0+m_0-1} \boldsymbol{\alpha}^\prime \mathrm{\* f}_t (\boldsymbol{\widetilde{\alpha}}_t-\boldsymbol{\alpha})^\prime \overline{\* e}_{r,t}^0 \right| \nonumber \\ 
&& = \left| \frac{1}{\sqrt{n}}\sum_{t=k_0}^{k_0+m_0-1} T^{\frac{1}{4}+\frac{\tau}{2}} \boldsymbol{\alpha}^{\prime} T^{-\frac{3}{4}-\tau}\mathrm{\* f}_t T^{\frac{1}{2}+\frac{\tau}{2}} (\boldsymbol{\widetilde{\alpha}}_t-\boldsymbol{\alpha})^\prime \overline{\*e}_{r,t}^{0 \prime} \right| \nonumber \\
&& \leq \big\|\boldsymbol{\alpha}^0 \big\| \sup_{k_0 \leq t \leq k_0+m_0 -1} \big\|T^{\frac{1}{2}+\frac{\tau}{2}} (\boldsymbol{\widetilde{\alpha}}_t-\boldsymbol{\alpha})\big\| \left|\frac{1}{\sqrt{n}}\sum_{t=k_0}^{k_0+m_0-1}  T^{-\frac{3}{4}-\tau} \mathrm{\* f}_t \overline{\* e}_{r,t}^0 \right| \nonumber \\
&& \leq T^{-\frac{1}{4}}\big\|\boldsymbol{\alpha}^0 \big\| \sup_{k_0 \leq t \leq k_0+m_0 -1}\big\|T^{\frac{1}{2}+\frac{\tau}{2}} (\boldsymbol{\widetilde{\alpha}}_t-\boldsymbol{\alpha})\big\| \sqrt{\frac{T}{n}} \frac{1}{T^{1+\tau}}\sum_{t=k_0}^{k_0+m_0-1}  \big\|\mathrm{\* f}_t\big\| \big\|\overline{\* e}_{r,t}^0\big\| \nonumber \\
&& \leq T^{-\frac{1}{4}-\frac{\tau}{2}} \big\|\boldsymbol{\alpha}^0 \big\| \sup_{k_0 \leq t \leq k_0+m_0 -1} \big\|T^{\frac{1}{2}+\frac{\tau}{2}} (\boldsymbol{\widetilde{\alpha}}_t-\boldsymbol{\alpha})\big\| \sqrt{\frac{T}{n}} \left(\frac{1}{T^{1+\tau}}\sum_{t=k_0}^{k_0+m_0-1}  \big\|\mathrm{\* f}_t\big\|^2\right)^{\frac{1}{2}} \left(\frac{1}{T}\sum_{t=k_0}^{k_0+m_0-1}  \big\|\overline{\* e}_{r,t}^0\big\|^2\right)^{\frac{1}{2}} \nonumber \\ 
&& = O_p(N^{-\frac{1}{2}} T^{-\frac{1}{4}-\frac{\tau}{2}}),
\end{eqnarray}
and
\begin{eqnarray}
&&\left| \frac{1}{\sqrt{n}}\sum_{t=k_0}^{k_0+m_0-1} \boldsymbol{\alpha}^\prime \mathrm{\* f}_t \boldsymbol{\alpha}^\prime \overline{\* e}_{r,t}^0 \right| \nonumber \\
&& = \left| \frac{1}{\sqrt{n}}\sum_{t=k_0}^{k_0+m_0-1} T^{\frac{1}{4}+\frac{\tau}{2}} \boldsymbol{\alpha}^\prime T^{-\frac{1}{2}-\tau}\mathrm{\* f}_t T^{\frac{1}{4}+\frac{\tau}{2}} \boldsymbol{\alpha}^\prime \overline{\* e}_{r,t}^0 \right| \nonumber \\
&& \leq \big\|\boldsymbol{\alpha}^0\big\|^2 \left|\frac{1}{\sqrt{n}}\sum_{t=k_0}^{k_0+m_0-1}  T^{-\frac{1}{2}-\tau}\mathrm{\* f}_t \overline{\* e}_{r,t}^0 \right| \nonumber \\
&& \leq \big\|\boldsymbol{\alpha}^0\big\|^2 \sqrt{\frac{T}{n}} \frac{1}{T^{1+\tau}}\sum_{t=k_0}^{k_0+m_0-1}  \big\|\mathrm{\* f}_t\big\| \big\|\overline{\* e}_{r,t}^0\big\| \nonumber \\
&& \leq T^{-\frac{\tau}{2}}\big\|\boldsymbol{\alpha}^0\big\|^2 \sqrt{\frac{T}{n}} \left(\frac{1}{T^{1+\tau}}\sum_{t=k_0}^{k_0+m_0-1}  \big\|\mathrm{\* f}_t\big\|^2\right)^{\frac{1}{2}} \left(\frac{1}{T}\sum_{t=k_0}^{k_0+m_0-1}  \big\|\overline{\* e}_{r,t}^0\big\|^2\right)^{\frac{1}{2}} \nonumber \\
&& = O_p(N^{-\frac{1}{2}}T^{-\frac{\tau}{2}}).
\end{eqnarray}

Therefore, combining the orders of $\mathrm{\* I}$-$\mathrm{\*V}$ gives
\\
\begin{eqnarray}
\frac{1}{\sqrt{n}}\sum_{t=k_0}^{k_0+m_0-1}\widetilde{u}_{1,t+1}(\widetilde{u}_{2,t+1}- \widehat{u}_{2,t+1}) &=& O_p(N^{-\frac{1}{2}}) + +O_p(T^{(\tau-1)/2})+O_p(T^{-\frac{1}{4}}).
\end{eqnarray}
which is negligible for all $\tau \in \left(0,1\right)$. \\

\noindent Now, let us turn our attention to Lemma 2 (b) by recalling that $\widetilde{u}_{2,t+1}- \widehat{u}_{2,t+1} = (\* Q_N^{-1}\boldsymbol{\widehat{\delta}}_t - \boldsymbol{\widetilde{\delta}}_t^{0})^{\prime}\widehat{\*z}_t^0 + (\boldsymbol{\widetilde{\alpha}}_t-\boldsymbol{\alpha})^\prime  \overline{\* e}_{r,t}^0 + \boldsymbol{\alpha}^\prime \overline{\* e}_{r,t}^0$ and proceeding as follows:
\begin{eqnarray}
\frac{1}{\sqrt{n}}\sum_{t=k_0}^{k_0+l_2^0-1}(\widetilde{u}_{2,t+1}- \widehat{u}_{2,t+1})^2 &\leq& \frac{3}{\sqrt{n}}\sum_{t=k_0}^{k_0+m_0-1}((\* Q_N^{-1}\boldsymbol{\widehat{\delta}}_t - \boldsymbol{\widetilde{\delta}}_t^{0})^{\prime}\widehat{\*z}_t^0)^2 +\frac{3}{\sqrt{n}}\sum_{t=k_0}^{k_0+m_0-1}((\boldsymbol{\widetilde{\alpha}}_t-\boldsymbol{\alpha})^\prime  \overline{\* e}_{r,t}^0)^2 \nonumber \\
&+& \frac{3}{\sqrt{n}}\sum_{t=k_0}^{k_0+m_0-1}(\boldsymbol{\alpha}^\prime \overline{\* e}_{r,t}^0)^2 \nonumber \\
&=& 3(U_1 + U_2 + U_3).
\end{eqnarray}
Starting from $U_1$, and letting $\*r_t=O_p(N^{-1/2})+O_p(T^{(\tau-1)/2})$ for short-hand notation, we obtain
\begin{align}\label{eq: U1}
    U_1 &= \frac{1}{\sqrt{n}}\sum_{t=k_0}^{k_0+l_2^0-1}((\* Q_N^{-1}\boldsymbol{\widehat{\delta}}_t - \boldsymbol{\widetilde{\delta}}_t^{0})^{\prime}\widehat{\*z}_t^0)^2\notag\\
&=\frac{1}{\sqrt{n}}\sum_{t=k_0}^{k_0+l_2^0-1}((\* Q_N^{-1}\boldsymbol{\widehat{\delta}}_t - \boldsymbol{\widetilde{\delta}}_t^{0})^{\prime}\* D_T\* D_T^{-1}\widehat{\*z}_t^0)^2\notag\\
    &= \frac{1}{\sqrt{n}}\sum_{t=k_0}^{k_0+l_2^0-1}\left( \*D_T(\* Q_N^{-1}\boldsymbol{\widehat{\delta}}_t - \boldsymbol{\widetilde{\delta}}_t^{0})\otimes\*D_T(\* Q_N^{-1}\boldsymbol{\widehat{\delta}}_t - \boldsymbol{\widetilde{\delta}}_t^{0}) \right)'\mathrm{vec}(\*D_T^{-1}\widehat{\*z}_t^0\widehat{\*z}_t^{0\prime}\*D_T^{-1})\notag\\
    &= \sqrt{\frac{T}{n}}\frac{1}{T}\sum_{t=k_0}^{k_0+l_2^0-1}\left( T^{\tau/2}\*D_T(\* Q_N^{-1}\boldsymbol{\widehat{\delta}}_t - \boldsymbol{\widetilde{\delta}}_t^{0})\otimes T^{\tau/2}\*D_T(\* Q_N^{-1}\boldsymbol{\widehat{\delta}}_t - \boldsymbol{\widetilde{\delta}}_t^{0}) \right)'\left (\*R_T \otimes \*R_T\right)\notag\\
    &\times \mathrm{vec}\left(T^{-\tau}\widehat{\*z}_t^0\widehat{\*z}_t^{0\prime} \right)\notag\\
    &=\sqrt{\frac{T}{n}}\frac{1}{T}\sum_{t=k_0}^{k_0+l_2^0-1}\left( T^{\tau/2-1/4}\*v_t^0\otimes T^{\tau/2}\*D_T(\* Q_N^{-1}\boldsymbol{\widehat{\delta}}_t - \boldsymbol{\widetilde{\delta}}_t^{0}) \right)'\left (\*R_T \otimes \*R_T\right) \mathrm{vec}\left(T^{-\tau}\widehat{\*z}_t^0\widehat{\*z}_t^{0\prime} \right)\notag\\
    &+\sqrt{\frac{T}{n}}\frac{1}{T}\sum_{t=k_0}^{k_0+l_2^0-1}\left( \*r_t\otimes T^{\tau/2}\*D_T(\* Q_N^{-1}\boldsymbol{\widehat{\delta}}_t - \boldsymbol{\widetilde{\delta}}_t^{0}) \right)'\left (\*R_T \otimes \*R_T\right)\times \mathrm{vec}\left(T^{-\tau}\widehat{\*z}_t^0\widehat{\*z}_t^{0\prime} \right)\notag\\
    &= \sqrt{\frac{T}{n}}\frac{1}{T}\sum_{t=k_0}^{k_0+l_2^0-1}\left( T^{\tau/2-1/4}\*v_t^0\otimes T^{\tau/2}\*D_T(\* Q_N^{-1}\boldsymbol{\widehat{\delta}}_t - \boldsymbol{\widetilde{\delta}}_t^{0}) \right)'\left (\*R_T \otimes \*R_T\right) \mathrm{vec}\left(T^{-\tau}\widehat{\*z}_t^0\widehat{\*z}_t^{0\prime} \right)\notag\\
    &+\sqrt{\frac{T}{n}}\frac{1}{T}\sum_{t=k_0}^{k_0+l_2^0-1}\left( \*r_t\otimes T^{\tau/2}\*D_T(\* Q_N^{-1}\boldsymbol{\widehat{\delta}}_t - \boldsymbol{\widetilde{\delta}}_t^{0}) \right)'\left (\*R_T \otimes \*R_T\right)\times \mathrm{vec}\left(T^{-\tau}\*z_t^0\*z_t^{0\prime}\right)\notag\\
    &+\sqrt{\frac{T}{n}}\frac{1}{T}\sum_{t=k_0}^{k_0+l_2^0-1}\left( \*r_t\otimes \*D_T(\* Q_N^{-1}\boldsymbol{\widehat{\delta}}_t - \boldsymbol{\widetilde{\delta}}_t^{0}) \right)'\left (\*R_T \otimes \*R_T\right)\times T^{\tau/2}\mathrm{vec}\left(T^{-\tau}\widehat{\*z}_t^0\widehat{\*z}_t^{0\prime}- T^{-\tau}\*z_t^0\*z_t^{0\prime}\right)\notag\\
    &= \sqrt{\frac{T}{n}}\frac{1}{T}\sum_{t=k_0}^{k_0+l_2^0-1}\left( T^{\tau/2-1/4}\*v_t^0\otimes T^{\tau/2}\*D_T(\* Q_N^{-1}\boldsymbol{\widehat{\delta}}_t - \boldsymbol{\widetilde{\delta}}_t^{0}) \right)'\left (\*R_T \otimes \*R_T\right) \mathrm{vec}\left(T^{-\tau}\widehat{\*z}_t^0\widehat{\*z}_t^{0\prime} \right)\notag\\
    &+\sqrt{\frac{T}{n}}\frac{1}{T}\sum_{t=k_0}^{k_0+l_2^0-1}\left( \*r_t\otimes T^{\tau/2}\*D_T(\* Q_N^{-1}\boldsymbol{\widehat{\delta}}_t - \boldsymbol{\widetilde{\delta}}_t^{0}) \right)'\left (\*R_T \otimes \*R_T\right)\times \mathrm{vec}\left(T^{-\tau}\*z_t^0\*z_t^{0\prime}\right)\notag\\
    &+O_p(N^{-1}T^{-\tau/2})+O_p((NT)^{-1/2})+O_p(N^{-1/2}T^{-1/4})+O_p(T^{\tau/2-3/4})\notag\\
    &=\sqrt{\frac{T}{n}}\frac{1}{T}\sum_{t=k_0}^{k_0+l_2^0-1}\left( T^{\tau/2-1/4}\*v_t^0\otimes T^{\tau/2}\*D_T(\* Q_N^{-1}\boldsymbol{\widehat{\delta}}_t - \boldsymbol{\widetilde{\delta}}_t^{0}) \right)'\left (\*R_T \otimes \*R_T\right) \mathrm{vec}\left(T^{-\tau}\widehat{\*z}_t^0\widehat{\*z}_t^{0\prime} \right)\notag\\
    &+\underbrace{\sqrt{\frac{T}{n}}\frac{1}{T}\sum_{t=k_0}^{k_0+l_2^0-1}\left( \*r_t\otimes T^{\tau/2-1/4}\*v_t^0 \right)'\left (\*R_T \otimes \*R_T\right)\times \mathrm{vec}\left(T^{-\tau}\*z_t^0\*z_t^{0\prime}\right)}_{=0}\notag\\
    &+\sqrt{\frac{T}{n}}\frac{1}{T}\sum_{t=k_0}^{k_0+l_2^0-1}\left( \*r_t\otimes \*r_t \right)'\left (\*R_T \otimes \*R_T\right)\times \mathrm{vec}\left(T^{-\tau}\*z_t^0\*z_t^{0\prime}\right)\notag\\
    &+O_p(N^{-1}T^{-\tau/2})+O_p((NT)^{-1/2})+O_p(N^{-1/2}T^{-1/4})+O_p(T^{\tau/2-3/4})\notag\\
    &=\sqrt{\frac{T}{n}}\frac{1}{T}\sum_{t=k_0}^{k_0+l_2^0-1}\left( T^{\tau/2-1/4}\*v_t^0\otimes T^{\tau/2}\*D_T(\* Q_N^{-1}\boldsymbol{\widehat{\delta}}_t - \boldsymbol{\widetilde{\delta}}_t^{0}) \right)'\left (\*R_T \otimes \*R_T\right) \mathrm{vec}\left(T^{-\tau}\widehat{\*z}_t^0\widehat{\*z}_t^{0\prime} \right)\notag\\
    &+O_p(N^{-1})+O_p(T^{\tau-1})\notag\\
    &= \sqrt{\frac{T}{n}}\frac{1}{T}\sum_{t=k_0}^{k_0+l_2^0-1}\left( T^{\tau/2-1/4}\*v_t^0\otimes T^{\tau/2-1/4}\*v_t^0 \right)'\left (\*R_T \otimes \*R_T\right) \mathrm{vec}\left(T^{-\tau}\widehat{\*z}_t^0\widehat{\*z}_t^{0\prime} \right)\notag\\
    &+ \sqrt{\frac{T}{n}}\frac{1}{T}\sum_{t=k_0}^{k_0+l_2^0-1}\left( T^{\tau/2-1/4}\*v_t^0\otimes \*r_t\right)'\left (\*R_T \otimes \*R_T\right) \mathrm{vec}\left(T^{-\tau}\widehat{\*z}_t^0\widehat{\*z}_t^{0\prime} \right)+O_p(N^{-1})+O_p(T^{\tau-1})\notag\\
    &=\sqrt{\frac{T}{n}}\frac{1}{T}\sum_{t=k_0}^{k_0+l_2^0-1}\left( T^{\tau/2-1/4}\*v_t^0\otimes T^{\tau/2-1/4}\*v_t^0 \right)'\left (\*R_T \otimes \*R_T\right) \mathrm{vec}\left(T^{-\tau}\widehat{\*z}_t^0\widehat{\*z}_t^{0\prime} \right)\notag\\
    &+ \underbrace{\sqrt{\frac{T}{n}}\frac{1}{T}\sum_{t=k_0}^{k_0+l_2^0-1}\left( T^{\tau/2-1/4}\*v_t^0\otimes \*r_t\right)'\left (\*R_T \otimes \*R_T\right) \mathrm{vec}\left(T^{-\tau}\*z_t^0\*z_t^{0\prime} \right)}_{=0}\notag\\
    &+\sqrt{\frac{T}{n}}\frac{1}{T}\sum_{t=k_0}^{k_0+l_2^0-1}\left( T^{-1/4}\*v_t^0\otimes \*r_t\right)'\left (\*R_T \otimes \*R_T\right) T^{\tau/2}\mathrm{vec}\left(T^{-\tau}\widehat{\*z}_t^0\widehat{\*z}_t^{0\prime}-T^{-\tau}\*z_t^0\*z_t^{0\prime} \right)\notag\\
    &+O_p(N^{-1})+O_p(T^{\tau-1})\notag\\
    &= \sqrt{\frac{T}{n}}\frac{1}{T}\sum_{t=k_0}^{k_0+l_2^0-1}\left( T^{\tau/2-1/4}\*v_t^0\otimes T^{\tau/2-1/4}\*v_t^0 \right)'\left (\*R_T \otimes \*R_T\right) \mathrm{vec}\left(T^{-\tau}\widehat{\*z}_t^0\widehat{\*z}_t^{0\prime} \right) \notag\\
&+O_p(N^{-1})+O_p(N^{-1/2}T^{-1/4})+O_p(T^{\tau-1})
\end{align}
where we used the same approximation arguments as in (\ref{eq: decompthetawzhatQdelta}) and (\ref{long_argm2}) iteratively in the substitutions along the fact that $\sup_{k_0\leq t\leq k_0+m_0-1}T^{\tau/2}\left\|T^{-\tau}\widehat{\*z}_t^0\widehat{\*z}_t^{0\prime}-T^{-\tau}\*z_t^0\*z_t^{0\prime}  \right\|=O_p(1)$, because now we use squares of the approximating terms. Subsequently, 
\begin{eqnarray}
    &&U_1 = \frac{1}{\sqrt{n}}\sum_{t=k_0}^{k_0+l_2^0-1}((\* Q_N^{-1}\boldsymbol{\widehat{\delta}}_t - \boldsymbol{\widetilde{\delta}}_t^{0})^{\prime}\widehat{\*z}_t^0)^2\notag\\
    &&=\sqrt{\frac{T}{n}}\frac{1}{T}\sum_{t=k_0}^{k_0+l_2^0-1}\left(T^{\tau/2-1/4}\*v_t^0\otimes T^{\tau/2-1/4}\*v_t^0 \right)'\mathrm{vec}\left(T^{-\tau}\*z_t^0\*z_t^{0\prime} \right)\notag\\
    &&+ \sqrt{\frac{T}{n}}\frac{1}{T}\sum_{t=k_0}^{k_0+l_2^0-1}\left(T^{\tau/2-1/4}\*v_t^0\otimes T^{\tau/2-1/4}\*v_t^0 \right)'\mathrm{vec}\left(T^{-\tau}\widehat{\*z}_t^0\widehat{\*z}_t^{0\prime}-\*z_t^0\*z_t^{0\prime} \right)\notag\\
    &&+O_p(N^{-1})+O_p(N^{-1/2}T^{-1/4})+O_p(T^{\tau-1})\notag\\
    &&=\sqrt{\frac{T}{n}}\frac{1}{T}\sum_{t=k_0}^{k_0+l_2^0-1}\left(T^{\tau/2-1/4}\*v_t^0\otimes T^{\tau/2-1/4}\*v_t^0 \right)'\mathrm{vec}\left(T^{-\tau}\*z_t^0\*z_t^{0\prime} \right)+O_p(N^{-1})+O_p(N^{-1/2}T^{-1/4})+O_p(T^{(\tau-1)/2})\notag\\
    &&= T^{\tau-1/2}\sqrt{\frac{T}{n}}\frac{1}{T}\sum_{t=k_0}^{k_0+l_2^0-1}\*v_t^{0\prime}T^{-\tau}\*z_t^0\*z_t^{0\prime}\*v_t^0+O_p(N^{-1})+O_p(N^{-1/2}T^{-1/4})+O_p(T^{(\tau-1)/2})\notag\\
    &&=O_p(N^{-1})+O_p(N^{-1/2}T^{-1/4})+O_p(T^{(\tau-1)/2}),
\end{eqnarray}
 since $\*v_t^{0\prime}\*z_t^0=0$ exactly. The slowest decaying term, which dominates $O_p(T^{\tau-1})$ is obtained by 
 \begin{eqnarray}
     &&\left| \sqrt{\frac{T}{n}}\frac{1}{T}\sum_{t=k_0}^{k_0+l_2^0-1}\left(T^{\tau/2-1/4}\*v_t^0\otimes T^{\tau/2-1/4}\*v_t^0 \right)'\mathrm{vec}\left(T^{-\tau}\widehat{\*z}_t^0\widehat{\*z}_t^{0\prime}-\*z_t^0\*z_t^{0\prime} \right) \right|\notag\\
     &&\leq T^{\tau-1/2}\underbrace{\sup_{k_0\leq t\leq k_0+l_2^0-1}\left\|T^{-\tau}\widehat{\*z}_t^0\widehat{\*z}_t^{0\prime}-\*z_t^0\*z_t^{0\prime} \right\|}_{O_p(T^{-\tau/2})}\sqrt{\frac{T}{n}}\frac{1}{T}\sum_{t=k_0}^{k_0+l_2^0-1}\left\|\left(\*v_t^0\otimes \*v_t^0 \right) \right\|\notag\\
     &&=O_p(T^{\tau/2-1/2})=o_p(1)
 \end{eqnarray}
as required for any $\tau \in (0,1)$. Note terms of this nature can decay very slowly for $\tau \approx 1$. Next,
\begin{eqnarray} \label{eq: U2}
U_2 &=& \frac{1}{\sqrt{n}}\sum_{t=k_0}^{k_0+l_2^0-1}(T^{\frac{1}{2}+\frac{\tau}{2}}(\boldsymbol{\widetilde{\alpha}}_t-\boldsymbol{\alpha})^\prime  T^{-\frac{1}{2}-\frac{\tau}{2}}\overline{\* e}_{r,t}^0)^2 \nonumber \\
& \leq & T^{-\frac{1}{2}-\tau}\sup_{k_0 \leq t \leq k_0+m_0 -1} \big\|T^{\frac{1}{2}+\frac{\tau}{2}}(\boldsymbol{\widetilde{\alpha}}_t-\boldsymbol{\alpha})\big\|^2 \sqrt{\frac{T}{n}}\frac{1}{T}\sum_{t=k_0}^{k_0+l_2^0-1}\big\|\overline{\* e}_{r,t}^0\big\|^2 \nonumber \\
&=& O_p(N^{-1}T^{-\frac{1}{2}-\tau}).
\end{eqnarray}
Moving on to $U_3$, 
\begin{eqnarray} \label{eq: U3}
U_3 &=& \frac{1}{\sqrt{n}}\sum_{t=k_0}^{k_0+l_2^0-1}(T^{\frac{1}{4}+\frac{\tau}{2}}\boldsymbol{\alpha}^\prime  T^{-\frac{1}{4}-\frac{\tau}{2}}\overline{\* e}_{r,t}^0)^2 \nonumber \\
& \leq & T^{-\tau}\sup_{k_0 \leq t \leq k_0+l_2^0 -1} \big\|\boldsymbol{\alpha}^{0}\big\|^2 \sqrt{\frac{T}{n}}\frac{1}{T}\sum_{t=k_0}^{k_0+m_0-1}\big\|\overline{\* e}_{r,t}^0\big\|^2 \nonumber \\
&=& O_p(N^{-1}T^{-\tau}).
\end{eqnarray}
Combining (\ref{eq: U1})-(\ref{eq: U3}) yields
\begin{eqnarray}
\frac{1}{\sqrt{n}}\sum_{t=k_0}^{k_0+l_2^0-1}(\widetilde{u}_{2,t+1}- \widehat{u}_{2,t+1})^2 = O_p(N^{-1}) +O_p(N^{-1/2}T^{-1/4})+ O_p(T^{(\tau-1)/2}).
\end{eqnarray}
Then we show that the term in (c) is negligible. Indeed, using the definition $\widetilde{u}_{2,t+1} = u_{t+1}-(\boldsymbol{\widetilde{\delta}}_t-\boldsymbol{\delta})^\prime \*z_{t}$ and the result in (\ref{eq: decomputildeuhat}) gives
\begin{eqnarray} \label{eq: decompu2(u2-u2)First}
&&\frac{1}{\sqrt{n}}\sum_{t=k_0}^{k_0+l_2^0-1}\widetilde{u}_{2,t+1}(\widetilde{u}_{2,t+1}-\widehat{u}_{2,t+1}) \nonumber \\
&& = \frac{1}{\sqrt{n}}\sum_{t=k_0}^{k_0+l_2^0-1}u_{t+1}((\* Q_N^{-1}\boldsymbol{\widehat{\delta}}_t - \boldsymbol{\widetilde{\delta}}_t^{0})^{\prime}\widehat{\*z}_t^0 + (\boldsymbol{\widetilde{\alpha}}_t-\boldsymbol{\alpha})^\prime  \overline{\* e}_{r,t}^0 + \boldsymbol{\alpha}^\prime \overline{\* e}_{r,t}^0) \nonumber \\
&&+ \frac{1}{\sqrt{n}}\sum_{t=k_0}^{k_0+l_2^0-1}(\boldsymbol{\widetilde{\delta}}_t-\boldsymbol{\delta})^\prime \*z_{t}((\* Q_N^{-1}\boldsymbol{\widehat{\delta}}_t - \boldsymbol{\widetilde{\delta}}_t^{0})^{\prime}\widehat{\*z}_t^0 + (\boldsymbol{\widetilde{\alpha}}_t-\boldsymbol{\alpha})^\prime  \overline{\* e}_{r,t}^0 + \boldsymbol{\alpha}^\prime \overline{\* e}_{r,t}^0) \nonumber \\
&& = \frac{1}{\sqrt{n}}\sum_{t=k_0}^{k_0+l_2^0-1}(\* Q_N^{-1}\boldsymbol{\widehat{\delta}}_t - \boldsymbol{\widetilde{\delta}}_t^{0})^{\prime}\widehat{\*z}_t^0u_{t+1} + \frac{1}{\sqrt{n}}\sum_{t=k_0}^{k_0+l_2^0-1} (\boldsymbol{\widetilde{\alpha}}_t-\boldsymbol{\alpha})^\prime \overline{\* e}_{r,t}^0u_{t+1} + \frac{1}{\sqrt{n}}\sum_{t=k_0}^{k_0+l_2^0-1} \boldsymbol{\alpha}^\prime \overline{\* e}_{r,t}^0u_{t+1} \nonumber \\
&& + \frac{1}{\sqrt{n}}\sum_{t=k_0}^{k_0+l_2^0-1}(\boldsymbol{\widetilde{\delta}}_t-\boldsymbol{\delta})^\prime \*z_{t}(\* Q_N^{-1}\boldsymbol{\widehat{\delta}}_t - \boldsymbol{\widetilde{\delta}}_t^{0})^{\prime}\widehat{\*z}_t^0 + \frac{1}{\sqrt{n}}\sum_{t=k_0}^{k_0+l_2^0-1}(\boldsymbol{\widetilde{\delta}}_t-\boldsymbol{\delta})^\prime \*z_{t}(\boldsymbol{\widetilde{\alpha}}_t-\boldsymbol{\alpha})^\prime  \overline{\* e}_{r,t}^0 \nonumber \\
&& + \frac{1}{\sqrt{n}}\sum_{t=k_0}^{k_0+l_2^0-1}(\boldsymbol{\widetilde{\delta}}_t-\boldsymbol{\delta})^\prime \*z_{t}\boldsymbol{\alpha}^\prime \overline{\* e}_{r,t}^0
\end{eqnarray}
Note that the first three terms in this expression are the same as in (\ref{eq: decompfirstsum}), so we only address the other three terms. Before looking at the fourth term, observe that
\begin{eqnarray} \label{eq: ratedeltahat}
\big\|\boldsymbol{\widetilde{\delta}}_t-\boldsymbol{\delta}\big\| &\leq& \big\|(t^{-1-\tau}\*Z^\prime\*Z)^{-1}\big\|\big\|t^{-1-\tau}\*Z^\prime\*u\big\| \nonumber \\
&=& O_p(T^{-\frac{1}{2}-\frac{\tau}{2}}).
\end{eqnarray}
due to fact that $t^{-1-\tau}\*Z^\prime\*Z \to_p \boldsymbol{\Sigma}_{ZZ}$ which is positive definite wp1 (Assumption 2). Hence,
\begin{eqnarray}
&&\left|\frac{1}{\sqrt{n}}\sum_{t=k_0}^{k_0+l_2^0-1}(\boldsymbol{\widetilde{\delta}}_t-\boldsymbol{\delta})^\prime \*z_{t}(\* Q_N^{-1}\boldsymbol{\widehat{\delta}}_t - \boldsymbol{\widetilde{\delta}}_t^{0})^{\prime}\widehat{\*z}_t^0\right|\nonumber \\
&& = \left|\frac{1}{\sqrt{n}}\sum_{t=k_0}^{k_0+l_2^0-1}T^{\frac{1}{2}+\frac{\tau}{2}}(\boldsymbol{\widetilde{\delta}}_t-\boldsymbol{\delta})^\prime T^{-\frac{1}{2}-\frac{\tau}{2}}\*z_{t}\* D_T^{-1}\* D_T(\* Q_N^{-1}\boldsymbol{\widehat{\delta}}_t - \boldsymbol{\widetilde{\delta}}_t^{0})^{\prime}\widehat{\*z}_t^0\right| \nonumber \\
&& \leq \big\|\* D_T^{-1}\big\| \sup_{k_0 \leq t \leq k_0+l_2^0 -1} \big\| T^{\frac{1}{2}+\frac{\tau}{2}}(\boldsymbol{\widetilde{\delta}}_t-\boldsymbol{\delta})\big\| \sup_{k_0 \leq t \leq k_0+l_2^0 -1} \big\| T^{\frac{\tau}{2}}\* D_T(\* Q_N^{-1}\boldsymbol{\widehat{\delta}}_t - \boldsymbol{\widetilde{\delta}}_t^{0})\big\| \sqrt{\frac{T}{n}} \frac{1}{T^{1+\tau}}\sum_{t=k_0}^{k_0+l_2^0-1} \big\|\*z_{t}\big\|\big\|\widehat{\*z}_t^0\big\|\nonumber \\
&& \leq MT^{-\frac{1}{4}}\sup_{k_0 \leq t \leq k_0+l_2^0 -1} \big\| T^{\frac{1}{2}+\frac{\tau}{2}}(\boldsymbol{\widetilde{\delta}}_t-\boldsymbol{\delta})\big\| \sup_{k_0 \leq t \leq k_0+l_2^0 -1} \big\| T^{\frac{\tau}{2}}\* D_T(\* Q_N^{-1}\boldsymbol{\widehat{\delta}}_t - \boldsymbol{\widetilde{\delta}}_t^{0})\big\| \sqrt{\frac{T}{n}} \left(\frac{1}{T^{1+\tau}}\sum_{t=k_0}^{k_0+l_2^0-1}  \big\|\*z_{t}\big\|^2\right)^{\frac{1}{2}} \nonumber \\
&& \times \left(\frac{1}{T^{1+\tau}}\sum_{t=k_0}^{k_0+l_2^0-1}  \big\|\widehat{\*z}_t^0\big\|^2\right)^{\frac{1}{2}} \nonumber \\
&& = O_p(N^{-\frac{1}{2}}T^{-\frac{1}{4}}) + O_p(T^{\frac{\tau}{2}-\frac{1}{2}}),
\end{eqnarray}
which is negligible for all values of $\tau \in \left(0,1\right)$. With respect to the fifth term in (\ref{eq: decompu2(u2-u2)First}), by using the results in (\ref{eq: ratealphahat}) and (\ref{eq: ratedeltahat}), we have that
\begin{eqnarray}
&& \left|\frac{1}{\sqrt{n}}\sum_{t=k_0}^{k_0+l_2^0-1}(\boldsymbol{\widetilde{\delta}}_t-\boldsymbol{\delta})^\prime \*z_{t}(\boldsymbol{\widetilde{\alpha}}_t-\boldsymbol{\alpha})^\prime  \overline{\* e}_{r,t}^0\right| \nonumber \\
&& = \left|\frac{1}{\sqrt{n}}\sum_{t=k_0}^{k_0+l_2^0-1}T^{\frac{1}{2}+\frac{\tau}{2}}(\boldsymbol{\widetilde{\delta}}_t-\boldsymbol{\delta})^\prime T^{-1-\tau}\*z_{t}T^{\frac{1}{2}+\frac{\tau}{2}}(\boldsymbol{\widetilde{\alpha}}_t-\boldsymbol{\alpha})^\prime  \overline{\* e}_{r,t}^0\right| \nonumber \\
&& \leq T^{-\frac{1}{2}} \sup_{k_0 \leq t \leq k_0+l_2^0 -1} \big\| T^{\frac{1}{2}+\frac{\tau}{2}}(\boldsymbol{\widetilde{\delta}}_t-\boldsymbol{\delta})\big\| \sup_{k_0 \leq t \leq k_0+l_2^0 -1} \big\| T^{\frac{1}{2}+\frac{\tau}{2}}(\boldsymbol{\widetilde{\alpha}}_t-\boldsymbol{\alpha})\big\| \sqrt{\frac{T}{n}} \frac{1}{T^{1+\tau}}\sum_{t=k_0}^{k_0+l_2^0-1} \big\|\*z_{t}\big\|\big\|\overline{\* e}_{r,t}^0\big\| \nonumber \\
&& \leq T^{-\frac{1}{2}-\frac{\tau}{2}} \sup_{k_0 \leq t \leq k_0+l_2^0 -1} \big\| T^{\frac{1}{2}+\frac{\tau}{2}}(\boldsymbol{\widetilde{\delta}}_t-\boldsymbol{\delta})\big\| \sup_{k_0 \leq t \leq k_0+l_2^0 -1} \big\| T^{\frac{1}{2}+\frac{\tau}{2}}(\boldsymbol{\widetilde{\alpha}}_t-\boldsymbol{\alpha})\big\| \sqrt{\frac{T}{n}} \left(\frac{1}{T^{1+\tau}}\sum_{t=k_0}^{k_0+l_2^0-1}  \big\|\*z_{t}\big\|^2\right)^{\frac{1}{2}} \nonumber \\
&& \times \left(\frac{1}{T}\sum_{t=k_0}^{k_0+l_2^0-1}  \big\|\overline{\* e}_{r,t}^0\big\|^2\right)^{\frac{1}{2}} \nonumber \\
&& = O_p(N^{-\frac{1}{2}} T^{-\frac{1}{2}-\frac{\tau}{2}}).
\end{eqnarray}
In a similar manner, the last term is 
\begin{eqnarray}
&& \left| \frac{1}{\sqrt{n}}\sum_{t=k_0}^{k_0+l_2^0-1}(\boldsymbol{\widetilde{\delta}}_t-\boldsymbol{\delta})^\prime \*z_{t}\boldsymbol{\alpha}^\prime \overline{\* e}_{r,t}^0 \right| \nonumber \\
&& =  T^{-\frac{3}{4}-\tau}\left| \frac{1}{\sqrt{n}}\sum_{t=k_0}^{k_0+l_2^0-1}T^{\frac{1}{2}+\frac{\tau}{2}}(\boldsymbol{\widetilde{\delta}}_t-\boldsymbol{\delta})^\prime \*z_{t}T^{\frac{1}{4}+\frac{\tau}{2}}\boldsymbol{\alpha}^\prime \overline{\* e}_{r,t}^0 \right| \nonumber \\
&& =  T^{-\frac{1}{4}} \big\|\boldsymbol{\alpha}^0 \big\| \sup_{k_0 \leq t \leq k_0+l_2^0 -1} \big\| T^{\frac{1}{2}+\frac{\tau}{2}}(\boldsymbol{\widetilde{\delta}}_t-\boldsymbol{\delta})\big\| \sqrt{\frac{T}{n}} \frac{1}{T^{1+\tau}}\sum_{t=k_0}^{k_0+l_2^0-1} \big\|\*z_{t}\big\|\big\|\overline{\* e}_{r,t}^0\big\| \nonumber \\
&& \leq  T^{-\frac{1}{4}-\frac{\tau}{2}} \big\|\boldsymbol{\alpha}^0 \big\| \sup_{k_0 \leq t \leq k_0+l_2^0 -1} \big\| T^{\frac{1}{2}+\frac{\tau}{2}}(\boldsymbol{\widetilde{\delta}}_t-\boldsymbol{\delta})\big\| \sqrt{\frac{T}{n}} \left(\frac{1}{T^{1+\tau}}\sum_{t=k_0}^{k_0+l_2^0-1}  \big\|\*z_{t}\big\|^2\right)^{\frac{1}{2}} \left(\frac{1}{T}\sum_{t=k_0}^{k_0+l_2^0-1}  \big\|\overline{\* e}_{r,t}^0\big\|^2\right)^{\frac{1}{2}} \nonumber \\
&& = O_p(N^{-\frac{1}{2}}T^{-\frac{1}{4}-\frac{\tau}{2}})
\end{eqnarray}
Putting all of these results together reduces (\ref{eq: decompu2(u2-u2)First}) to
\begin{eqnarray} \label{eq: decompu2(u2-u2)Second}
\frac{1}{\sqrt{n}}\sum_{t=k_0}^{k_0+l_2^0-1}\widetilde{u}_{2,t+1}(\widetilde{u}_{2,t+1}-\widehat{u}_{2,t+1}) =  O_p(N^{-\frac{1}{2}}T^{-\frac{1}{4}}) + O_p(T^{\frac{\tau}{2}-\frac{1}{2}}) + O_p(N^{-\frac{1}{2}} T^{-\frac{\tau}{2}}),
\end{eqnarray}
which is negligible for all values of $\tau \in \left(0,1\right)$. \\

\noindent Parts (d) and (e) follow directly based on the logic in \cite{pitarakis2025novel}, and in particular, the passage from (A.6) to (A.8), which demonstrates that the remainder still vanishes inside of the integral. That is 
\begin{align}
    \left|\frac{1}{n}\sum_{l_2=\lfloor n\nu_0 \rfloor}^n\frac{n}{l_2}\frac{1}{\sqrt{n}}\sum_{t=k_0}^{k_0+l_2-1}(\widetilde{u}_{2,t+1}-\widehat{u}_{2,t+1})^2\right|&\leq \sup_{\lfloor n\nu_0 \rfloor \leq t\leq n}\left|\frac{1}{\sqrt{n}}\sum_{t=k_0}^{k_0+l_2-1}(\widetilde{u}_{2,t+1}-\widehat{u}_{2,t+1})^2 \right|\frac{1}{n}\sum_{l_2=\lfloor n\nu_0 \rfloor}^n\frac{n}{l_2}\notag\\
    &\leq \sup_{\lfloor n\nu_0 \rfloor \leq t\leq n}\left|\frac{1}{\sqrt{n}}\sum_{t=k_0}^{k_0+l_2-1}(\widetilde{u}_{2,t+1}-\widehat{u}_{2,t+1})^2 \right| \int_{\lambda_2=\nu_0}^1\frac{1}{\lambda_2}d\lambda_2\notag\\
    &+ o_p(1)\notag\\
    &=o_p(1),
\end{align}
and for (e) the argument is identical. \\

\noindent Finally, consider (f). Let us define $\phi^2=\lim_{T\to \infty}Var\left[\frac{1}{\sqrt{n}}\sum_{t=k_0}^{T-1}(u_{2,t+1}^2-\mathbb{E}(u_{2,t+1}^2))\right]$. By Assumption 1, the error process $u_t$ is a martingale difference sequence  so that 
\begin{align} \label{eq: phisquared}
\phi^2=\lim_{T\to \infty}\frac{1}{n}\sum_{t=k_0}^{T-1}\mathbb{E}(u_{2,t+1}^2-\mathbb{E}(u_{2,t+1}^2))^2, 
\end{align}
Then the natural unfeasible and feasible estimators of (\ref{eq: phisquared}) are obtained as
\begin{align}
\widehat{\phi}^2=\frac{1}{n}\sum_{t=k_0}^{T-1}\left(\widehat{u}_{2,t+1}^2-\frac{1}{n}\sum_{t=k_0}^{T-1}\widehat{u}_{2,t+1}^2\right)^2, 
\end{align}
and
\begin{eqnarray}
\widetilde{\phi}^2= \frac{1}{n}\sum_{t=k_0}^{T-1}\left(\widetilde{u}^2_{2,t+1} -\frac{1}{n}\sum_{t=k_0}^{T-1}\widetilde{u}_{2,t+1}^2 \right)^2
\end{eqnarray}
respectively. To demonstrate (e), it is convenient to start from the decomposition of \cite{margaritella2024new}:
\begin{eqnarray} \label{eq: MSdecomp}
\widetilde{\phi}^2 - \widehat{\phi}^2 &=& \frac{2}{n}\sum_{t=k_0}^{T-1}\left(\widetilde{u}^2_{2,t+1} -\frac{1}{n}\sum_{t=k_0}^{T-1}\widetilde{u}_{2,t+1}^2 \right)\left( (\widetilde{u}^2_{2,t+1}-\widehat{u}^2_{2,t+1}) + \left[\frac{1}{n}\sum_{t=k_0}^{T-1}(\widetilde{u}_{2,t+1}^2-\widehat{u}_{2,t+1}^2) \right] \right)\notag\\
&-&\frac{1}{n}\sum_{t=k_0}^{T-1}\left( (\widetilde{u}^2_{2,t+1}-\widehat{u}^2_{2,t+1}) + \left[\frac{1}{n}\sum_{t=k_0}^{T-1}(\widetilde{u}_{2,t+1}^2-\widehat{u}_{2,t+1}^2) \right] \right)^2 \nonumber \\
&=& \frac{2}{n}\sum_{t=k_0}^{T-1} \widetilde{u}^2_{2,t+1}(\widetilde{u}^2_{2,t+1}-\widehat{u}^2_{2,t+1}) - 2\left(\frac{1}{n}\sum_{t=k_0}^{T-1}\widetilde{u}_{2,t+1}^2 \right)\frac{1}{n}\sum_{t=k_0}^{T-1}(\widetilde{u}^2_{2,t+1}-\widehat{u}^2_{2,t+1}) \nonumber \\
&+& 2\left(\frac{1}{n}\sum_{t=k_0}^{T-1}(\widetilde{u}_{2,t+1}^2-\widehat{u}_{2,t+1}^2) \right)\frac{1}{n}\sum_{t=k_0}^{T-1} \widetilde{u}^2_{2,t+1} - 2\left(\frac{1}{n}\sum_{t=k_0}^{T-1}\widetilde{u}_{2,t+1}^2 \right)\frac{1}{n}\sum_{t=k_0}^{T-1}(\widetilde{u}_{2,t+1}^2-\widehat{u}_{2,t+1}^2) \nonumber \\
&-& \frac{1}{n}\sum_{t=k_0}^{T-1}\left(\widetilde{u}^2_{2,t+1}-\widehat{u}^2_{2,t+1}\right)^2 - \left(\frac{1}{n}\sum_{t=k_0}^{T-1}(\widetilde{u}_{2,t+1}^2-\widehat{u}_{2,t+1}^2) \right)^2 \nonumber \\
&-&2\left(\frac{1}{n}\sum_{t=k_0}^{T-1}(\widetilde{u}_{2,t+1}^2-\widehat{u}_{2,t+1}^2) \right)\frac{1}{n}\sum_{t=k_0}^{T-1}(\widetilde{u}^2_{2,t+1}-\widehat{u}^2_{2,t+1}) \nonumber \\
&=& \frac{2}{n}\sum_{t=k_0}^{T-1} \widetilde{u}^2_{2,t+1}(\widetilde{u}^2_{2,t+1}-\widehat{u}^2_{2,t+1})- 2\left(\frac{1}{n}\sum_{t=k_0}^{T-1}\widetilde{u}_{2,t+1}^2 \right)\frac{1}{n}\sum_{t=k_0}^{T-1}(\widetilde{u}_{2,t+1}^2-\widehat{u}_{2,t+1}^2) \nonumber \\
&-& \frac{1}{n}\sum_{t=k_0}^{T-1}\left(\widetilde{u}^2_{2,t+1}-\widehat{u}^2_{2,t+1}\right)^2 - 3\left(\frac{1}{n}\sum_{t=k_0}^{T-1}(\widetilde{u}_{2,t+1}^2-\widehat{u}_{2,t+1}^2) \right)^2
\end{eqnarray}
which is consistent if the remainder is negligible. For this purpose, observe that the difference of squares can be rewritten as $\widetilde{u}^2_{2,t+1}-\widehat{u}^2_{2,t+1}= 2\widetilde{u}_{2,t+1}(\widetilde{u}_{2,t+1}-\widehat{u}_{2,t+1}) - (\widetilde{u}_{2,t+1}-\widehat{u}_{2,t+1})^2$ so that the fourth term in (\ref{eq: MSdecomp}) becomes
\begin{eqnarray} \label{eq: diffsquaresu2}
\left|\frac{1}{n}\sum_{t=k_0}^{T-1}(\widetilde{u}_{2,t+1}^2-\widehat{u}_{2,t+1}^2)\right| &=& 2\left|\frac{1}{n}\sum_{t=k_0}^{T-1}\widetilde{u}_{2,t+1}(\widetilde{u}_{2,t+1}-\widehat{u}_{2,t+1})\right| + \left|\frac{1}{n}\sum_{t=k_0}^{T-1} (\widetilde{u}_{2,t+1}-\widehat{u}_{2,t+1})^2\right| \notag\\
&=& o_p(n^{-\frac{1}{2}})
\end{eqnarray}
by part (a) and (b) of this lemma. Next, we can show that $\widetilde{u}_{2,t+1}^2$ is a consistent estimator of $u_{t+1}^2$ uniformly in $t$:
\begin{eqnarray}
\sup_{k_0 \leq t \leq T -1} \widetilde{u}_{2,t+1}^2 &=& \sup_{k_0 \leq t \leq T -1} (u_{t+1}-(\boldsymbol{\widetilde{\delta}}_t-\boldsymbol{\delta})^\prime \*z_{t})^2 \nonumber \\
&\leq& 2\sup_{k_0 \leq t \leq T -1}u_{t+1}^2 + 2\sup_{k_0 \leq t \leq T -1}((\boldsymbol{\widetilde{\delta}}_t-\boldsymbol{\delta})^\prime \*z_{t})^2 \nonumber \\
&\leq& 2\sup_{k_0 \leq t \leq T -1}u_{t+1}^2 +2T^{-1}\sup_{k_0 \leq t \leq T -1}\big\|T^{\frac{1}{2}+\frac{\tau}{2}}(\boldsymbol{\widetilde{\delta}}_t-\boldsymbol{\delta})\big\|^2\sup_{k_0 \leq t \leq T -1}\big\|T^{-\frac{\tau}{2}}\*z_{t}\big\|^2 \nonumber \\
&=& 2\sup_{k_0 \leq t \leq T -1}u_{t+1}^2 + O_p(T^{-1}) = O_p(1),
\end{eqnarray}
where $\big\|T^{-\frac{\tau}{2}}\*z_{t}\big\|^2=O_p(1)$ follows from the results of Lemma 3.1 of \cite{magdalinos2009limit}. Now, this expression implies that the first and second term in (\ref{eq: MSdecomp}) are
\begin{eqnarray}
\left| \frac{1}{n}\sum_{t=k_0}^{T-1} \widetilde{u}^2_{2,t+1}(\widetilde{u}^2_{2,t+1}-\widehat{u}^2_{2,t+1}) \right| &\leq& 2\sup_{k_0 \leq t \leq T -1} \widetilde{u}_{2,t+1}^2 \frac{1}{n}\sum_{t=k_0}^{T-1} (\widetilde{u}^2_{2,t+1}-\widehat{u}^2_{2,t+1}) \nonumber \\
&=& o_p(n^{-\frac{1}{2}})
\end{eqnarray}
and 
\begin{eqnarray}
\left| \left(\frac{1}{n}\sum_{t=k_0}^{T-1}\widetilde{u}_{2,t+1}^2 \right)\frac{1}{n}\sum_{t=k_0}^{T-1}(\widetilde{u}_{2,t+1}^2-\widehat{u}_{2,t+1}^2) \right| &\leq& \sup_{k_0 \leq t \leq T -1} \widetilde{u}_{2,t+1}^2 \frac{1}{n}\sum_{t=k_0}^{T-1} (\widetilde{u}^2_{2,t+1}-\widehat{u}^2_{2,t+1}) \nonumber \\
&=& o_p(n^{-\frac{1}{2}}).
\end{eqnarray}
Hence, it remains to look at the third term. From previous results, we see that 
\begin{eqnarray} \label{eq: squareofsquareddiff}
\frac{1}{n}\sum_{t=k_0}^{T-1}\left(\widetilde{u}^2_{2,t+1}-\widehat{u}^2_{2,t+1}\right)^2 &=& \frac{1}{n}\sum_{t=k_0}^{T-1}\left(2\widetilde{u}_{2,t+1}(\widetilde{u}_{2,t+1}-\widehat{u}_{2,t+1}) - (\widetilde{u}_{2,t+1}-\widehat{u}_{2,t+1})^2\right)^2 \nonumber \\ 
&=& \frac{4}{n}\sum_{t=k_0}^{T-1} \widetilde{u}^2_{2,t+1}(\widetilde{u}^2_{2,t+1}-\widehat{u}^2_{2,t+1}) - \frac{4}{n}\sum_{t=k_0}^{T-1}\widetilde{u}_{2,t+1}(\widetilde{u}_{2,t+1}-\widehat{u}_{2,t+1})^3 \nonumber \\
&+&\frac{1}{n}\sum_{t=k_0}^{T-1}(\widetilde{u}_{2,t+1}-\widehat{u}_{2,t+1})^4
\end{eqnarray}
Here, 
\begin{eqnarray}
\left|\frac{1}{n}\sum_{t=k_0}^{T-1} \widetilde{u}^2_{2,t+1}(\widetilde{u}^2_{2,t+1}-\widehat{u}^2_{2,t+1})\right| &\leq& \sup_{k_0 \leq t \leq T -1} \widetilde{u}_{2,t+1}^2\left|\frac{1}{n}\sum_{t=k_0}^{T-1} (\widetilde{u}^2_{2,t+1}-\widehat{u}^2_{2,t+1})\right| \nonumber \\
&=& o_p(n^{-\frac{1}{2}}),
\end{eqnarray}
while using the result in (\ref{eq: decomputildeuhat}) gives
\begin{eqnarray}
&&\left|\frac{1}{n}\sum_{t=k_0}^{T-1}\widetilde{u}_{2,t+1}(\widetilde{u}_{2,t+1}-\widehat{u}_{2,t+1})^3\right| \nonumber \\
&& = \left|\frac{1}{n}\sum_{t=k_0}^{T-1}\widetilde{u}_{2,t+1}(\widetilde{u}_{2,t+1}-\widehat{u}_{2,t+1})^2((\* Q_N^{-1}\boldsymbol{\widehat{\delta}}_t - \boldsymbol{\widetilde{\delta}}_t^{0})^{\prime}\widehat{\*z}_t^0 + (\boldsymbol{\widetilde{\alpha}}_t-\boldsymbol{\alpha})^\prime  \overline{\* e}_{r,t}^0 + \boldsymbol{\alpha}^\prime \overline{\* e}_{r,t}^0)\right|\nonumber \\
&& \leq \big\|\* D_T^{-1} \big\| \sup_{k_0 \leq t \leq T -1} \left|\widetilde{u}_{2,t+1}\right| \sup_{k_0 \leq t \leq T -1} \big\| \* D_T(\* Q_N^{-1}\boldsymbol{\widehat{\delta}}_t - \boldsymbol{\widetilde{\delta}}_t^{0})\big\|  \frac{T}{n}\frac{1}{T}\sum_{t=k_0}^{T-1}(\widetilde{u}_{2,t+1}-\widehat{u}_{2,t+1})^2\big\|\widehat{\*z}_t^0\big\| \nonumber \\
&& + T^{-\frac{1}{2}-\frac{\tau}{2}}\sup_{k_0 \leq t \leq T -1} \left|\widetilde{u}_{2,t+1}\right| \sup_{k_0 \leq t \leq T -1} \big\| T^{\frac{1}{2}+\frac{\tau}{2}}(\boldsymbol{\widetilde{\alpha}}_t-\boldsymbol{\alpha})\big\| \frac{T}{n}\frac{1}{T}\sum_{t=k_0}^{T-1}(\widetilde{u}_{2,t+1}-\widehat{u}_{2,t+1})^2 \big\|\overline{\* e}_{r,t}^0\big\| \nonumber \\
&& + T^{-\frac{1}{4}-\frac{\tau}{2}}\big\| \boldsymbol{\alpha}^0\big\|\sup_{k_0 \leq t \leq T -1} \left|\widetilde{u}_{2,t+1}\right| \frac{T}{n}\frac{1}{T}\sum_{t=k_0}^{T-1}(\widetilde{u}_{2,t+1}-\widehat{u}_{2,t+1})^2 \big\|\overline{\* e}_{r,t}^0\big\| \nonumber \\
&& \leq MT^{-\frac{1}{4}} \frac{T}{n} \sup_{k_0 \leq t \leq T -1} \left|\widetilde{u}_{2,t+1}\right| \sup_{k_0 \leq t \leq T -1} \big\| T^{\frac{\tau}{2}}\* D_T(\* Q_N^{-1}\boldsymbol{\widehat{\delta}}_t - \boldsymbol{\widetilde{\delta}}_t^{0})\big\| \left(\frac{1}{T}\sum_{t=k_0}^{T-1}(\widetilde{u}_{2,t+1}-\widehat{u}_{2,t+1})^4 \right)^{\frac{1}{2}} \left(\frac{1}{T^{1+\tau}}\sum_{t=k_0}^{T-1}\big\|\widehat{\*z}_t^0\big\|^2 \right)^{\frac{1}{2}} \nonumber \\
&& + T^{-\frac{1}{2}-\frac{\tau}{2}}\frac{T}{n}\sup_{k_0 \leq t \leq T -1} \left|\widetilde{u}_{2,t+1}\right| \sup_{k_0 \leq t \leq T -1} \big\| T^{\frac{1}{2}+\frac{\tau}{2}}(\boldsymbol{\widetilde{\alpha}}_t-\boldsymbol{\alpha})\big\| \left(\frac{1}{T}\sum_{t=k_0}^{T-1}(\widetilde{u}_{2,t+1}-\widehat{u}_{2,t+1})^4 \right)^{\frac{1}{2}} \left(\frac{1}{T}\sum_{t=k_0}^{T-1}\big\|\overline{\* e}_{r,t}^0\big\|^2 \right)^{\frac{1}{2}} \nonumber \\
&& + T^{-\frac{1}{4}-\frac{\tau}{2}}\frac{T}{n}\big\| \boldsymbol{\alpha}^0\big\|\sup_{k_0 \leq t \leq T -1} \left|\widetilde{u}_{2,t+1}\right| \left(\frac{1}{T}\sum_{t=k_0}^{T-1}(\widetilde{u}_{2,t+1}-\widehat{u}_{2,t+1})^4 \right)^{\frac{1}{2}} \left(\frac{1}{T}\sum_{t=k_0}^{T-1}\big\|\overline{\* e}_{r,t}^0\big\|^2 \right)^{\frac{1}{2}}  \nonumber \\
&& = o_p(N^{-\frac{1}{2}}T^{-\frac{1}{4}}) +o_p(T^{\frac{\tau}{2}-\frac{1}{2}}) + o_p(N^{-\frac{1}{2}}T^{-\frac{1}{2}-\frac{\tau}{2}}) + o_p(N^{-\frac{1}{2}}T^{-\frac{1}{4}-\frac{\tau}{2}}) \nonumber \\
&& = o_p(N^{-\frac{1}{2}}T^{-\frac{1}{4}}) +o_p(T^{\frac{\tau}{2}-\frac{1}{2}}).
\end{eqnarray}
since one can show that $\frac{1}{T}\sum_{t=k_0}^{T-1}(\widetilde{u}_{2,t+1}-\widehat{u}_{2,t+1})^4 = o_p(1)$. Indeed, start by noticing that
\begin{eqnarray}
\frac{1}{T^{2+2\tau}}\sum_{t=k_0}^{T-1}\big\|\widehat{\*z}_t^0\big\|^4 \leq \left(\frac{1}{T^{1+\tau}}\sum_{t=k_0}^{T-1}\big\|\widehat{\*z}_t^0\big\|^2\right)^2 = O_p(1).
\end{eqnarray}
Substituting the expression (\ref{eq: decomputildeuhat}) into the quartic term of (\ref{eq: squareofsquareddiff}) then gives
\begin{eqnarray} \label{eq: quarticdifference}
&& \left|\frac{1}{n}\sum_{t=k_0}^{T-1}(\widetilde{u}_{2,t+1}-\widehat{u}_{2,t+1})^4\right| \nonumber \\
&& = \left|\frac{1}{n}\sum_{t=k_0}^{T-1}((\* Q_N^{-1}\boldsymbol{\widehat{\delta}}_t - \boldsymbol{\widetilde{\delta}}_t^{0})^{\prime}\widehat{\*z}_t^0 + (\boldsymbol{\widetilde{\alpha}}_t-\boldsymbol{\alpha})^\prime  \overline{\* e}_{r,t}^0 + \boldsymbol{\alpha}^\prime \overline{\* e}_{r,t}^0)^4\right| \nonumber \\
&& \leq \frac{4}{n}\sum_{t=k_0}^{T-1}\left|(\* Q_N^{-1}\boldsymbol{\widehat{\delta}}_t - \boldsymbol{\widetilde{\delta}}_t^{0})^{\prime}\widehat{\*z}_t^0\right|^4 + \frac{4}{n}\sum_{t=k_0}^{T-1}\left|(\boldsymbol{\widetilde{\alpha}}_t-\boldsymbol{\alpha})^\prime  \overline{\* e}_{r,t}^0\right|^4 + \frac{4}{n}\sum_{t=k_0}^{T-1}\left|\boldsymbol{\alpha}^\prime \overline{\* e}_{r,t}^0\right|^4 \nonumber \\
&& \leq \frac{4}{n}\sum_{t=k_0}^{T-1}\big\|\* D_T(\* Q_N^{-1}\boldsymbol{\widehat{\delta}}_t - \boldsymbol{\widetilde{\delta}}_t^{0})\big\|^4\big\|\* D_T^{-1}\widehat{\*z}_t^0\big\|^4 \nonumber \\
&& + T^{-2 - 2\tau}\frac{4}{n}\sum_{t=k_0}^{T-1}\big\|T^{\frac{1}{2}+\frac{\tau}{2}}(\boldsymbol{\widetilde{\alpha}}_t-\boldsymbol{\alpha})\big\|^4\big\| \overline{\* e}_{r,t}^0\big\|^4 \nonumber \\
&& + T^{-1-2\tau}\frac{4}{n}\sum_{t=k_0}^{T-1}\big\|T^{\frac{1}{4}+\frac{\tau}{2}}\boldsymbol{\alpha}\big\|^4 \big\|\overline{\* e}_{r,t}^0\big\|^4 \nonumber \\
&\leq& 4 \frac{T}{n}\big\|\* D_T^{-1}\big\|^4 \sup_{k_0 \leq t \leq T -1} \big\|T^{\frac{\tau}{2}}\* D_T(\* Q_N^{-1}\boldsymbol{\widehat{\delta}}_t - \boldsymbol{\widetilde{\delta}}_t^{0})\big\|^4 \frac{1}{T^{1+2\tau}}\sum_{t=k_0}^{T-1}\big\|\widehat{\*z}_t^0\big\|^4 \nonumber \\
&& + 4T^{-2 - 2\tau} \frac{T}{n} \sup_{k_0 \leq t \leq T -1} \big\|T^{\frac{1}{2}+\frac{\tau}{2}}(\boldsymbol{\widetilde{\alpha}}_t-\boldsymbol{\alpha})\big\|^4 \frac{1}{T}\sum_{t=k_0}^{T-1}\big\| \overline{\* e}_{r,t}^0\big\|^4 \nonumber \\
&& +T^{-1-2\tau} \frac{T}{n} \big\|\boldsymbol{\alpha}^0\big\|^4\frac{1}{T}\sum_{t=k_0}^{T-1} \big\|\overline{\* e}_{r,t}^0\big\|^4 \nonumber \\
&& \leq MT^{-1} \frac{T}{n} \sup_{k_0 \leq t \leq T -1} \big\|T^{\frac{\tau}{2}}\* D_T(\* Q_N^{-1}\boldsymbol{\widehat{\delta}}_t - \boldsymbol{\widetilde{\delta}}_t^{0})\big\|^4 \frac{1}{T^{2+2\tau}}\sum_{t=k_0}^{T-1}\big\|\widehat{\*z}_t^0\big\|^4 + O_p(N^{-2}T^{-1-2\tau}) \nonumber \\
&& = O_p(N^{-2}) + O_p(T^{2\tau-1}) \nonumber \\
&& = o_p(1),
\end{eqnarray}
which is negligible if $\tau \in \left(0,\frac{1}{2}\right)$. Hence, 
\begin{eqnarray}
\left|\frac{1}{n}\sum_{t=k_0}^{T-1}\left(\widetilde{u}^2_{2,t+1}-\widehat{u}^2_{2,t+1}\right)^2\right| &=& o_p(1).
\end{eqnarray}
Therefore, combining all of the results in (\ref{eq: diffsquaresu2})-(\ref{eq: quarticdifference}) leads to
\begin{eqnarray} \label{eq: diffsquarephi}
\left|\widetilde{\phi}^2 - \widehat{\phi}^2\right| = o_p(1),
\end{eqnarray}
as desired. $\blacksquare$ 
\subsection{Additional Results: Theory}
\subsubsection{A Comment on Multi-Step Forecast Errors}
In $h$-step-ahead forecast context, $\mathbb{E}(u_{t+h}|\mathcal{F}_t)=0$ implies $MA(h-1)$ dependence structure in $\{u_{t+h}\}$. In this section, we explore the rates of convergence of the key components in our proofs under multi-step ahead forecasts. The results provided below directly imply that the asymptotic equivalence result holds under $h$-step-ahead forecasts for $h>1$.
\subsubsection{Convergence Rates}
\textbf{Lemma A.4} \textit{Under Assumptions 1-4 and forecast error following an $MA(h-1)$ process, we have that 
\begin{align*}
   \left\| \frac{1}{T^{1+\tau}} \*Z'\*u\right\|=O_p(T^{-(1+\tau)/2})
\end{align*}
as $T\to \infty$}. \\

\noindent \textbf{Proof}.  By Markov's inequality, 
\begin{align}\label{Zu_MA}
    \mathbb{P}&\left(\left\|\frac{1}{T^{1+\tau}}\sum_{t=1}^{T-h}\*z_tu_{t+h}\right\| >\epsilon \right)\leq \epsilon^{-2}\mathbb{E}\left(\left\|\frac{1}{T^{1+\tau}}\sum_{t=1}^{T-h}\*z_tu_{t+h} \right\|^2\right)\notag\\
    &=\epsilon^{-2}\mathbb{E}\left(\frac{1}{T^{2+2\tau}}\sum_{t=1}^{T-h}\sum_{s=1}^{T-h}\*z_t'\*z_su_{t+h}u_{s+h} \right)\notag\\
    &=\frac{1}{T^{2+2\tau}}\sum_{t=1}^{T-h}\sum_{s=1}^{T-h}\mathbb{E}\left(\*z_t'\*z_s \right)\mathbb{E}(u_{t+h}u_{s+h})\notag\\
    &=\frac{1}{T^{2+2\tau}}\sum_{t=1}^{T-h}\sum_{s=\max\{ 1,t-h+1\}}^{\min \{ T-h,t+h-1\}}\mathbb{E}\left(\*z_{t}'\*z_{s}\right)\mathbb{C}ov(u_{t+h},u_{s+h})\notag\\
    &\leq \frac{1}{T^{2+2\tau}}\sum_{t=1}^{T-h}\sum_{s=\max\{ 1,t-h+1\}}^{\min \{ T-h,t+h-1\}}\mathbb{E}\left(|\*z_{t}'\*z_{s}|\right)\left|\mathbb{C}ov(u_{t+h},u_{s+h})\right|\notag\\
    &\leq \sup_{t,s}\left|\mathbb{C}ov(u_{t+h},u_{s+h})\right| \frac{1}{T^{2+2\tau}}\sum_{t=1}^{T-h}\sum_{s=\max\{ 1,t-h+1\}}^{\min \{ T-h,t+h-1\}}\mathbb{E}\left(|\*z_{t}'\*z_{s}|\right)\notag\\
    &\leq \sup_{t,s}\left|\mathbb{C}ov(u_{t+h},u_{s+h})\right| \frac{1}{T^{2+2\tau}}\sum_{t=1}^{T-h}\sum_{s=\max\{ 1,t-h+1\}}^{\min \{ T-h,t+h-1\}}\mathbb{E}\left(\left\|\*z_{t}\right\|\left\|\*z_{s}\right\|\right)\notag\\
    &\leq \sup_{t,s}\left|\mathbb{C}ov(u_{t+h},u_{s+h})\right| \frac{1}{T^{2+2\tau}}\sum_{t=1}^{T-h}\left[\mathbb{E}\left(\left\|\*z_t \right\|^2\right) \right]^{1/2}\sum_{s=\max\{ 1,t-h+1\}}^{\min \{ T-h,t+h-1\}} \left[\mathbb{E}\left(\left\|\*z_{s} \right\|^2\right) \right]^{1/2}\notag\\
&=\sup_{t,s}\left|\mathbb{C}ov(u_{t+h},u_{s+h})\right| \frac{1}{T^{2+\tau}}\sum_{t=1}^{T-h}\left[\mathbb{E}\left(\left\|T^{-\tau/2}\*z_t \right\|^2\right) \right]^{1/2}\sum_{s=\max\{ 1,t-h+1\}}^{\min \{ T-h,t+h-1\}} \left[\mathbb{E}\left(\left\|T^{-\tau/2}\*z_{s} \right\|^2\right) \right]^{1/2}\notag\\
&=T^{-1-\tau} \sup_{t,s}\left|\mathbb{C}ov(u_{t+h},u_{s+h})\right|\sup_{t}\mathbb{E}\left(\left\|T^{-\tau/2}\*z_t \right\|^2\right) C \frac{T-h-1}{T}\notag\\
&=O(T^{-1-\tau}),
\end{align}
for some positive $C$. Here we used the fact that $\sup_{t}\left[\mathbb{E}\left(\left\|T^{-\tau/2}\*z_t \right\|^2\right)\right]^{1/2}\times \sup_{s}\left[\mathbb{E}\left(\left\|T^{-\tau/2}\*z_s \right\|^2\right)\right]^{1/2}=\sup_{t}\mathbb{E}\left(\left\|T^{-\tau/2}\*z_t \right\|^2\right)=O(1)$, and the fact that the sum over an index $s$ is necessarily bounded in case of $MA(h-1)$ process. This implies that $\left\|\frac{1}{T^{1+\tau}}\sum_{t=1}^{T-h}\*z_tu_{t+h} \right\|=O_p(T^{-(1+\tau)/2})$.
\subsubsection{Verification of Uniform Rates}
In several steps of our proofs, especially that of Lemma 2, we need to have $\sup_{t}\left\|\widetilde{\+\delta}_t-\+\delta \right\|=O_p(T^{-(1+\tau)/2})$ (and several similar least squares deviations) to demonstrate that the factor estimation error is negligible as $(N,T)\to \infty$. Clearly, $\left\|\widetilde{\+\delta}_{\lfloor T\kappa \rfloor}-\+\delta\right\|\leq \left\|\left(\frac{1}{T^{1+\tau}}\sum_{s=1}^{{\lfloor T\kappa \rfloor}-h}\*z_s\*z_s'\right)^{-1} \right\| \left\| \frac{1}{T^{1+\tau}}\sum_{s=1}^{{\lfloor T\kappa \rfloor}-h}\*z_su_{s+h}\right\|$, where the denominator is bounded uniformly by Assumption 2 (c). For the numerator, we can use the same argument as in A.11 of \cite{magdalinos2009limit}, where for all $t=1,\ldots, T$ (equivalently, $\kappa\in (0,1)$), we create a bound: 
\begin{align}\label{unif_bound_check}
    \mathbb{E}\left(\left\| \frac{1}{T^{1+\tau}}\sum_{s=1}^{{\lfloor T\kappa \rfloor}-h}\*z_su_{s+h}\right\|^2 \right)&=\frac{1}{T^{2+2\tau}}\sum_{s=1}^{{\lfloor T\kappa \rfloor}-h}\sum_{j=1}^{\lfloor T\kappa \rfloor-h}\mathbb{E}(\*z_s'\*z_j)\mathbb{C}ov(u_{s+h},u_{j+h})\notag\\
    &=\frac{1}{T^{2+2\tau}}\sum_{s=1}^{\lfloor T\kappa \rfloor-h}\sum_{j=\max\{ 1,s-h+1\}}^{\min \{ \lfloor T\kappa \rfloor-h,s+h-1\}}\mathbb{E}(\*z_s'\*z_j)\mathbb{C}ov(u_{s+h},u_{j+h})\notag\\
    &\leq \frac{1}{T^{2+2\tau}}\sum_{s=1}^{\lfloor T\kappa \rfloor-h}\sum_{j=\max\{ 1,s-h+1\}}^{\min \{ \lfloor T\kappa \rfloor-h,s+h-1\}}\mathbb{E}(|\*z_s'\*z_j|)|\mathbb{C}ov(u_{s+h},u_{j+h})|\notag\\
    &\leq \frac{1}{T^{2+2\tau}}\sum_{s=1}^{T-h}\sum_{j=\max\{ 1,s-h+1\}}^{\min \{ T-h,s+h-1\}}\mathbb{E}(|\*z_s'\*z_j|)|\mathbb{C}ov(u_{s+h},u_{j+h})|\notag\\
    &=O(T^{-1-\tau}),
\end{align}
where we created the second inequality by simply summing more positive terms. The rest follows from (\ref{Zu_MA}). Similar least squares deviations can be checked in an analogous fashion. For example $\left\|\widetilde{\+\alpha}_{\lfloor T \kappa \rfloor} - \+\alpha \right\|\leq \left\|\left(\frac{1}{T^{1+\tau}}\*F'\*M_{\*W}\*F \right)^{-1} \right\|\left\| \frac{1}{T^{1+\tau}}\*F'\*M_{\*W}\*u\right\|$, where the denominator is bounded by our assumptions, while the numerator is negligible uniformly in $t$, as well, because 
\begin{align}
   \left\|\frac{1}{T^{1+\tau}}\*F'\*M_{\*W}\*u \right\|\leq \underbrace{\left\|\frac{1}{T^{1+\tau}}\sum_{s=1}^{\lfloor T \kappa \rfloor-h}\*f_s u_{s+h} \right\|}_{O_p(T^{-(1+\tau)/2})}&+\left\|\frac{1}{T^{1+\tau}}\sum_{s=1}^{\lfloor T \kappa \rfloor-h} \*f_s\*w_s'\right\|\left\|\left(\frac{1}{T^{1+\tau}}\sum_{s=1}^{\lfloor T \kappa \rfloor-h} \*w_s\*w_s'\right)^{-1} \right\| \notag\\
   &\times \underbrace{\left\|\frac{1}{T^{1+\tau}}\sum_{s=1}^{\lfloor T \kappa \rfloor-h}\*w_s u_{s+h} \right\|}_{O_p(T^{-(1+\tau)/2})}=O_p(T^{-(1+\tau)/2}),
\end{align}
where the rates come immediately form \ref{unif_bound_check}. 
\subsubsection{Power Enhancements}
\noindent The statistics $s_{f,j}$ for $j=2,3,4$ can be power-enhanced, as argued in \cite{pitarakis2025novel}. The power adjustment is derived by replacing $\widetilde{u}_{2,t+1}^2$ with $\breve{u}_{2,t+1}=\widetilde{u}_{2,t+1}^2-(\widetilde{u}_{1,t+1}-\widetilde{u}_{2,t+1})^2$. Clearly, we only have $\widehat{u}_{2,t+1}^2$, therefore, we work with the feasible power-adjustment term. For $s_{\widehat{f},2}$, for instance, this means that we need to add 
\begin{align*}
    \widehat{\triangle}_2=\frac{1}{\widehat{\omega}_2}\frac{1}{\lambda_2^0}\frac{1}{\sqrt{n}}\sum_{t=k_0}^{k_0+l_2^0-1}(\widetilde{u}_{1,t+1}-\widehat{u}_{2,t+1})^2
\end{align*}
to the initial statistic to obtain 
\begin{align*}
    s_{\widehat{f},2}^{P}=s_{\widehat{f},2}+  \widehat{\triangle}_2.
\end{align*}
Because $s_{\widehat{f},2}=s_{f,2}+o_p(1)$, we need to demonstrate that $  \widehat{\triangle}_2=\triangle_2+o_p(1)$ to obtain $ s_{\widehat{f},2}^{Power}= s_{f,2}^{Power}+o_p(1)$. It is enough to see that 
\begin{align}
    \frac{1}{\sqrt{n}}\sum_{t=k_0}^{k_0+l_2^0-1}(\widetilde{u}_{1,t+1}-\widehat{u}_{2,t+1})^2&=\frac{1}{\sqrt{n}}\sum_{t=k_0}^{k_0+l_2^0-1}(\widetilde{u}_{1,t+1}-\widetilde{u}_{2,t+1}+\widetilde{u}_{2,t+1}-\widehat{u}_{2,t+1})^2\notag\\
    &=\frac{1}{\sqrt{n}}\sum_{t=k_0}^{k_0+l_2^0-1}(\widetilde{u}_{1,t+1}-\widetilde{u}_{2,t+1})^2+\frac{2}{\sqrt{n}}\sum_{t=k_0}^{k_0+l_2^0-1}(\widetilde{u}_{1,t+1}-\widetilde{u}_{2,t+1})(\widetilde{u}_{2,t+1}-\widehat{u}_{2,t+1})\notag\\
    &+\frac{1}{\sqrt{n}}\sum_{t=k_0}^{k_0+l_2^0-1}(\widetilde{u}_{1,t+1}-\widehat{u}_{2,t+1})^2\notag\\
    &=\frac{1}{\sqrt{n}}\sum_{t=k_0}^{k_0+l_2^0-1}(\widetilde{u}_{1,t+1}-\widetilde{u}_{2,t+1})^2+o_p(1),
\end{align}
because the last term is negligible based on Lemma 2. Moreover, by Cauchy-Schwarz inequality,
\begin{align}
    \Bigg|\frac{1}{\sqrt{n}}\sum_{t=k_0}^{k_0+l_2^0-1}(\widetilde{u}_{1,t+1}-\widetilde{u}_{2,t+1})(\widetilde{u}_{1,t+1}&-\widehat{u}_{2,t+1}) \Bigg| \leq \frac{1}{\sqrt{n}}\sum_{t=k_0}^{k_0+l_2^0-1}|\widetilde{u}_{1,t+1}-\widetilde{u}_{2,t+1}||\widetilde{u}_{2,t+1}-\widehat{u}_{2,t+1}|\notag\\
    &\leq \left(\frac{1}{\sqrt{n}}\sum_{t=k_0}^{k_0+l_2^0-1}(\widetilde{u}_{1,t+1}-\widetilde{u}_{2,t+1})^2\right)^{1/2}\left(\frac{1}{\sqrt{n}}\sum_{t=k_0}^{k_0+l_2^0-1}(\widetilde{u}_{2,t+1}-\widehat{u}_{2,t+1})^2 \right)^{1/2}\notag\\
    &=o_p(1),
\end{align}
due to Lemma 2 again, and the first component coming from the infeasible adjustment term, which is clearly bounded. The power adjustment terms for $s_{\widehat{f},3}$ and $s_{\widehat{f},4}$ are analogous. 
\subsubsection{Breaks in Loadings} 
We provide a discussion on how we can accommodate structural breaks in factor loadings. Let the loadings break at some time points $D_1=\lfloor \phi_1 T \rfloor,\ldots, D_d=\lfloor \phi_d T \rfloor$ with $\phi_1<\phi_2<\ldots<\phi_d \in (0,1)$ for a finite $d$, such that
\begin{align*}
\boldsymbol{\Lambda}_{i,t}'\mathbf{f}_t=\mathbb{I}(t<D_1)\boldsymbol{\Lambda}_{1,i}'\mathbf{f}_t +\left(\sum_{j=1}^d\mathbb{I}(t\geq D_j)\+\Lambda_{j+1,i}'\right)\*f_t.
\end{align*}
For simplicity of the further analysis, let us assume a single break without a loss of generality. According to \cite{breitung2011testing}, we can represent this as 
\begin{align}
    \boldsymbol{\Lambda}_{i,t}'\mathbf{f}_t = \*Q_{i}'\mathbf{g}_t,
\end{align}
where $\*Q_{i} = [\boldsymbol{\Lambda}_{1,i}', \boldsymbol{\Lambda}_{2,i}']'\in \mathbb{R}^{2r\times m}$ and $\mathbf{g}_t = [\mathbb{I}(t<D)\mathbf{f}_t', \mathbb{I}(t\geq D)\mathbf{f}_t']'\in \mathbb{R}^{2r}$. This means that the model with breaking loadings and $r$ factors can be written as a model without break but with $2r$ factors. We can therefore reparameterize the model for $\mathbf{x}_{i,t}$ in terms of $\mathbf{g}_t$:
\begin{align}
    \*x_{i,t}=\*Q_i'\*g_t+\*e_{i,t}.
\end{align}
We reparameterize the model for $y_{t+1}$ in the same way:
\begin{align}\label{rep_x}
y_{t+1}=\boldsymbol{\theta}'\mathbf{w}_t+\boldsymbol{\alpha}'\mathbf{f}_t+u_{t+1}=\boldsymbol{\theta}'\mathbf{w}_t+\mathbf{a}'\mathbf{g}_t+u_{t+1},
\end{align}
where $\mathbf{a}= \+\iota_2 \otimes T^{-1/4-\tau/2}\boldsymbol{\alpha}^0=T^{-1/4-\tau/2}\mathbf{a}^0$, where $\+\iota$ is a vector of ones. Based on (\ref{rep_x}), we have 
\begin{align}
\widehat{\*g}_t=\overline{\*Q}'\*g_t+\overline{\*e}_t,
\end{align}
where $\mathrm{rk}(\overline{\*Q})=2r$. This means that our estimated object in the asymptotic analysis is 
\begin{align}
\widehat{\mathbf{g}}^0_t=\mathbf{D}_N\overline{\mathbf{M}}'\widehat{\mathbf{g}}_t= \mathbf{D}_N\overline{\mathbf{M}}'\overline{\boldsymbol{\Lambda}}_t'\mathbf{f}_t+ \mathbf{D}_N\overline{\mathbf{M}}'\overline{\mathbf{e}}_t = \mathbf{D}_N\overline{\mathbf{M}}'\overline{\*Q}'\mathbf{g}_t+ \mathbf{D}_N\overline{\mathbf{M}}'\overline{\mathbf{e}}_t  =\mathbf{g}^0_t+\overline{\mathbf{e}}^0_t,
\end{align}
where $\mathbf{g}^0_t=[\mathbf{g}'_t, \mathbf{0}_{(m-2r)\times 1}']'\in \mathbb{R}^{m}$, $\overline{\mathbf{e}}^0_t=[\overline{\mathbf{e}}_{2r,t}'\overline{\*Q}_{2r}^{-1}, \sqrt{N}(\overline{\mathbf{e}}_{-2r,t}- \overline{\*Q}_{-2r}'\overline{\*Q}_{2r}^{-1\prime}\overline{\mathbf{e}}_{2r,t})']'=
[\overline{\mathbf{e}}^{0\prime}_{2r,t}, \overline{\mathbf{e}}^{0\prime}_{-2r,t}]'$ $\in \mathbb{R}^{m}$, $\mathbf{D}_N=\mathrm{diag}(\sqrt{N}\mathbf{I}_{2r}, \mathbf{I}_{m-2r})$ and 
\begin{equation}
    \overline{\*M} =\left[\begin{array}{cc}\overline{\*Q}_{2r}^{-1} & -\overline{\*Q}_{2r}^{-1}\overline{\*Q}_{-2r}\\
    \mathbf{0}_{(m-2r)\times 2r} & \mathbf{I}_{m-2r}\end{array}\right]=[\overline{\*M}_{2r},\, \overline{\*M}_{-2r}]\in \mathbb{R}^{m\times m}.
\end{equation}
Note that every loss differential in the asymptotic expansion of the feasible statistic in \cite{pitarakis2023direct} and \cite{pitarakis2025novel} is an appropriately normalized version of 
\begin{align}
\sum_{t=k_0}^{k_0+b_0}\widehat{d}_t,\hspace{2mm} \text{where} \hspace{3mm} \widehat{d}_t\in \{\widetilde{u}_{1,t+1}(\widetilde{u}_{2,t+1}- \widehat{u}_{2,t+1}),\widetilde{u}_{2,t+1}(\widetilde{u}_{2,t+1}- \widehat{u}_{2,t+1}), (\widetilde{u}_{2,t+1}- \widehat{u}_{2,t+1})^2\}
\end{align}
and $b^0$ is a tuning parameter which splits the out-of-sample part of observations, such that $b^0T^{-1}=O(1)$. Clearly, $b_0\in \{m_0, l_1^0, l_2^0 \}$. Consider that $D\in (1, k_0]$. Then we use  $\widehat{\*g}_t$ instead of $\widehat{\*f}_t$ in the analysis from the beginning and thus the results of Lemma 1 carry over, which makes Lemma 2 hold, as well. Therefore, the asymptotic results do not change. Next, let $D\in (k_0, k_0+b_0]$. This means that 
\begin{align}\label{break_split}
    \sum_{t=k_0}^{k_0+b_0}\widehat{d}_t= \sum_{t=k_0}^{D-1}\widehat{d}_t + \sum_{t=D}^{k_0+b_0}\widehat{d}_t,
\end{align}
where $\sum_{t=k_0}^{D-1}\widehat{d}_t$ is based on the sample portion without the break, and therefore it behaves as typical loss differential analyzed under Lemma 2 (function of $\widehat{\*f}_t$), because $DT^{-1}=\phi+ O(T^{-1})$. Importantly, $\sum_{t=D}^{k_0+b_0}\widehat{d}_t$ behaves in the same way, because its summands are functions of $\widehat{\*g}_t$. As long as $\mathrm{rk}(\overline{\*Q})=2r$, the rates will stay the same, and the final asymptotic result will remain unchanged. \\

\noindent Every variation of the location of $b_0$ will lead to the same conclusions. In the extreme case, let $b_0\in \{l_1^0, l_2^0\}$, where each is strictly smaller than 1 (equivalent to trimming some last out-of-sample observations). Then if $D\in [l_j, T-1]$ for $j=1,2$, the break never appears in our analysis, which implies that Lemma 1 and Lemma 2 carry over again. \\

\noindent It is important to comment on the difference from \cite{stauskas2022tests}, where it must be that $D\in (1, k_0]$, i.e. only in the in-sample period. Take the situation in (\ref{break_split}). Because of a very different structure of the statistics that appear in \cite{clark2001tests}, both components are $O_p(1)$ instead of $o_p(1)$. In that case, the presence of the break affects the variance of an already highly non-standard asymptotic distribution. Because the precise location of the break is unknown, simulating the distribution becomes a notorious challenge. This is not an issue in the current case, because the loss differentials in the expansions of the feasible statistics will be negligible irrespective of the break location. 

\subsection{Additional Results: Monte Carlo Simulations}
\subsubsection{Test Size with Known Number of Factors}

In Table \ref{tab: Size with known factors}, we report the results of test size when the true number of factors, which is $r=1$, is known so that we can set $m=r$ and avoid redundant CAs. As in the main manuscript, we consider a setup under DGP (2) and two levels of persistence, i.e. $\tau = 0.2$ (mild) and $\tau = 0.6$ (moderate). We observe that, relative to the case with $m=3$, all test sizes inch closer to the significance level of 5\% for all combinations of $N$ and $T$, with pronounced improvements for $s_{\widehat{f},1}$ and $s_{\widehat{f},4}^{P}$. This outcome suggests that while redundant blocks do not yield an effect on the local power of our statistics, they inflate their test size, especially under small $T$ and for $s_{\widehat{f},1}$ and $s_{\widehat{f},4}^{P}$. In order to address this issue and alleviate any concerns about the inflated test size, it is sufficient to apply the Information Criterion (IC) by \cite{de2024cross}. The study of \cite{ditzen2025selection} explored this IC in the context of non-stationary factors and concluded that it remains consistent, but $N$ and $T$ have to be large, unless the persistence is moderate. Once the redundant blocks are omitted from the factor-augmented forecasting equation, our four tests of forecast accuracy and encompassing enjoy acceptable size and excellent local power. 

\begin{table}[H]
  \centering
    \begin{tabular}{rrrrrrrrrrrrrrr}
    \multicolumn{7}{c}{$\tau$ = 0.2}                         &       & \multicolumn{7}{c}{$\tau$ = 0.6} \\
\cmidrule{1-7}\cmidrule{9-15}    \multicolumn{1}{c}{$N$} & \multicolumn{1}{c}{$T$} &       & \multicolumn{1}{c}{$s_{\widehat{f},1}$} & \multicolumn{1}{c}{$s_{\widehat{f},2}^{P}$} & \multicolumn{1}{c}{$s_{\widehat{f},3}^{P}$} & \multicolumn{1}{c}{$s_{\widehat{f},4}^{P}$} &       & \multicolumn{1}{c}{$N$} & \multicolumn{1}{c}{$T$} &       & \multicolumn{1}{c}{$s_{\widehat{f},1}$} & \multicolumn{1}{c}{$s_{\widehat{f},2}^{P}$} & \multicolumn{1}{c}{$s_{\widehat{f},3}^{P}$} & \multicolumn{1}{c}{$s_{\widehat{f},4}^{P}$} \\
\cmidrule{1-7}\cmidrule{9-15}          &       &       &       &       &       &       &       &       &       &       &       &       &       &  \\
    \multicolumn{1}{c}{10} & \multicolumn{1}{c}{50} &       & \multicolumn{1}{c}{14.5} & \multicolumn{1}{c}{8.2} & \multicolumn{1}{c}{9.4} & \multicolumn{1}{c}{12.9} &       & \multicolumn{1}{c}{10} & \multicolumn{1}{c}{50} &       & \multicolumn{1}{c}{15.2} & \multicolumn{1}{c}{8.4} & \multicolumn{1}{c}{9.3} & \multicolumn{1}{c}{12.9} \\
    \multicolumn{1}{c}{20} & \multicolumn{1}{c}{50} &       & \multicolumn{1}{c}{15.9} & \multicolumn{1}{c}{8.6} & \multicolumn{1}{c}{9.6} & \multicolumn{1}{c}{13.5} &       & \multicolumn{1}{c}{20} & \multicolumn{1}{c}{50} &       & \multicolumn{1}{c}{15.8} & \multicolumn{1}{c}{8.5} & \multicolumn{1}{c}{9.4} & \multicolumn{1}{c}{13.6} \\
    \multicolumn{1}{c}{50} & \multicolumn{1}{c}{50} &       & \multicolumn{1}{c}{14.6} & \multicolumn{1}{c}{8.8} & \multicolumn{1}{c}{9.8} & \multicolumn{1}{c}{13.8} &       & \multicolumn{1}{c}{50} & \multicolumn{1}{c}{50} &       & \multicolumn{1}{c}{16.0} & \multicolumn{1}{c}{8.9} & \multicolumn{1}{c}{9.8} & \multicolumn{1}{c}{14.2} \\
    \multicolumn{1}{c}{100} & \multicolumn{1}{c}{50} &       & \multicolumn{1}{c}{15.8} & \multicolumn{1}{c}{9.1} & \multicolumn{1}{c}{10.1} & \multicolumn{1}{c}{14.6} &       & \multicolumn{1}{c}{100} & \multicolumn{1}{c}{50} &       & \multicolumn{1}{c}{16.8} & \multicolumn{1}{c}{9.1} & \multicolumn{1}{c}{10.2} & \multicolumn{1}{c}{14.3} \\
    \multicolumn{1}{c}{200} & \multicolumn{1}{c}{50} &       & \multicolumn{1}{c}{15.9} & \multicolumn{1}{c}{9.0} & \multicolumn{1}{c}{10.0} & \multicolumn{1}{c}{13.9} &       & \multicolumn{1}{c}{200} & \multicolumn{1}{c}{50} &       & \multicolumn{1}{c}{16.9} & \multicolumn{1}{c}{9.3} & \multicolumn{1}{c}{10.2} & \multicolumn{1}{c}{14.2} \\
    \multicolumn{1}{c}{10} & \multicolumn{1}{c}{100} &       & \multicolumn{1}{c}{9.7} & \multicolumn{1}{c}{6.2} & \multicolumn{1}{c}{6.9} & \multicolumn{1}{c}{9.6} &       & \multicolumn{1}{c}{10} & \multicolumn{1}{c}{100} &       & \multicolumn{1}{c}{11.4} & \multicolumn{1}{c}{6.6} & \multicolumn{1}{c}{7.1} & \multicolumn{1}{c}{9.8} \\
    \multicolumn{1}{c}{20} & \multicolumn{1}{c}{100} &       & \multicolumn{1}{c}{9.7} & \multicolumn{1}{c}{6.1} & \multicolumn{1}{c}{6.8} & \multicolumn{1}{c}{10.1} &       & \multicolumn{1}{c}{20} & \multicolumn{1}{c}{100} &       & \multicolumn{1}{c}{11.5} & \multicolumn{1}{c}{6.5} & \multicolumn{1}{c}{7.0} & \multicolumn{1}{c}{10.4} \\
    \multicolumn{1}{c}{50} & \multicolumn{1}{c}{100} &       & \multicolumn{1}{c}{10.7} & \multicolumn{1}{c}{6.4} & \multicolumn{1}{c}{6.8} & \multicolumn{1}{c}{10.2} &       & \multicolumn{1}{c}{50} & \multicolumn{1}{c}{100} &       & \multicolumn{1}{c}{11.2} & \multicolumn{1}{c}{6.4} & \multicolumn{1}{c}{6.9} & \multicolumn{1}{c}{10.0} \\
    \multicolumn{1}{c}{100} & \multicolumn{1}{c}{100} &       & \multicolumn{1}{c}{10.7} & \multicolumn{1}{c}{6.5} & \multicolumn{1}{c}{6.8} & \multicolumn{1}{c}{10.2} &       & \multicolumn{1}{c}{100} & \multicolumn{1}{c}{100} &       & \multicolumn{1}{c}{12.0} & \multicolumn{1}{c}{6.5} & \multicolumn{1}{c}{6.8} & \multicolumn{1}{c}{10.7} \\
    \multicolumn{1}{c}{200} & \multicolumn{1}{c}{100} &       & \multicolumn{1}{c}{9.2} & \multicolumn{1}{c}{6.0} & \multicolumn{1}{c}{6.7} & \multicolumn{1}{c}{10.2} &       & \multicolumn{1}{c}{200} & \multicolumn{1}{c}{100} &       & \multicolumn{1}{c}{11.9} & \multicolumn{1}{c}{6.5} & \multicolumn{1}{c}{7.0} & \multicolumn{1}{c}{10.6} \\
    \multicolumn{1}{c}{10} & \multicolumn{1}{c}{200} &       & \multicolumn{1}{c}{8.1} & \multicolumn{1}{c}{5.1} & \multicolumn{1}{c}{5.8} & \multicolumn{1}{c}{8.2} &       & \multicolumn{1}{c}{10} & \multicolumn{1}{c}{200} &       & \multicolumn{1}{c}{8.4} & \multicolumn{1}{c}{5.1} & \multicolumn{1}{c}{5.7} & \multicolumn{1}{c}{8.1} \\
    \multicolumn{1}{c}{20} & \multicolumn{1}{c}{200} &       & \multicolumn{1}{c}{9.8} & \multicolumn{1}{c}{5.3} & \multicolumn{1}{c}{5.7} & \multicolumn{1}{c}{8.2} &       & \multicolumn{1}{c}{20} & \multicolumn{1}{c}{200} &       & \multicolumn{1}{c}{8.8} & \multicolumn{1}{c}{5.2} & \multicolumn{1}{c}{5.7} & \multicolumn{1}{c}{8.5} \\
    \multicolumn{1}{c}{50} & \multicolumn{1}{c}{200} &       & \multicolumn{1}{c}{8.5} & \multicolumn{1}{c}{5.2} & \multicolumn{1}{c}{5.8} & \multicolumn{1}{c}{8.6} &       & \multicolumn{1}{c}{50} & \multicolumn{1}{c}{200} &       & \multicolumn{1}{c}{8.8} & \multicolumn{1}{c}{5.3} & \multicolumn{1}{c}{5.8} & \multicolumn{1}{c}{9.0} \\
    \multicolumn{1}{c}{100} & \multicolumn{1}{c}{200} &       & \multicolumn{1}{c}{8.4} & \multicolumn{1}{c}{5.1} & \multicolumn{1}{c}{5.7} & \multicolumn{1}{c}{8.5} &       & \multicolumn{1}{c}{100} & \multicolumn{1}{c}{200} &       & \multicolumn{1}{c}{9.0} & \multicolumn{1}{c}{5.3} & \multicolumn{1}{c}{5.8} & \multicolumn{1}{c}{8.9} \\
    \multicolumn{1}{c}{200} & \multicolumn{1}{c}{200} &       & \multicolumn{1}{c}{8.8} & \multicolumn{1}{c}{5.3} & \multicolumn{1}{c}{5.9} & \multicolumn{1}{c}{8.4} &       & \multicolumn{1}{c}{200} & \multicolumn{1}{c}{200} &       & \multicolumn{1}{c}{8.9} & \multicolumn{1}{c}{5.1} & \multicolumn{1}{c}{5.9} & \multicolumn{1}{c}{8.9} \\
    \multicolumn{1}{c}{500} & \multicolumn{1}{c}{500} &       & \multicolumn{1}{c}{5.4} & \multicolumn{1}{c}{5.3} & \multicolumn{1}{c}{5.4} & \multicolumn{1}{c}{6.4} &       & \multicolumn{1}{c}{500} & \multicolumn{1}{c}{500} &       & \multicolumn{1}{c}{7.3} & \multicolumn{1}{c}{5.5} & \multicolumn{1}{c}{5.5} & \multicolumn{1}{c}{7.3} \\
          &       &       &       &       &       &       &       &       &       &       &       &       &       &  \\
\cmidrule{1-7}\cmidrule{9-15}    \end{tabular}%
\caption{Size (\%) under DGP(2) and persistence.}
  \label{tab: Size with known factors}%
\end{table}%

\subsubsection{Power Curves under Redundant Blocks}
As anticipated by our results in Section 3 of the manuscript, the asymptotic distributions of the tests are invariant under a misspecification of the number of factors. Our MC simulations highlight this theoretical finding very clearly in panel (b). In this experiment run on DGP (2) with $N = 200$ $T=600$ and $\mu_0 = 0.45$, the true number of factors is $r=1$ but we consider three possible choices for the number of blocks, namely $m \in \{1,2,3\}$, so to account for a correct specification as well as overspecification. It is clear from the figure that the power curves corresponding to these three choices are almost undistinguishable numerically, especially for the higher values of $\alpha$. Peculiarly, more blocks also yield a slightly better local power in the vicinity of 0. However, we notice that each additional block costs as a marginal size distortion when $\alpha =0$, but we believe that it is a price that we are willing to accept for the insensitivity of the local power to an overspecification in the number of blocks. That is, we prefer the redundancy of including an irrelevant regressor in our predictive equation than the risk of omitting a relevant one.

\begin{figure}[H]
\begin{subfigure}[b]{0.55\linewidth} 
  \hspace*{-1cm}
\includegraphics[width = 1.0\textwidth]{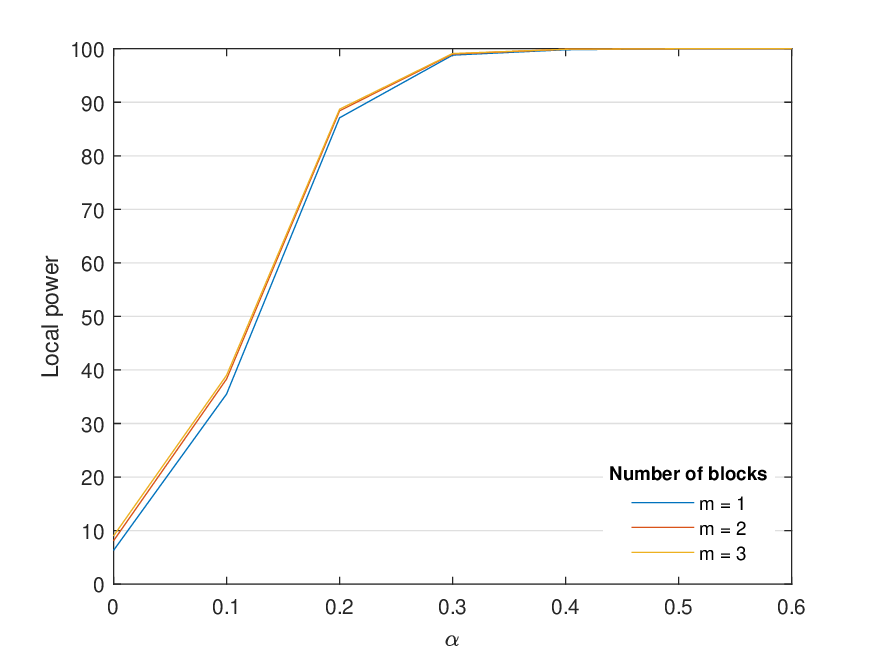}
\footnotesize \caption{$s_{\widehat{f},1}$ with $\mu_0 = 0.45$.}
\end{subfigure}
\begin{subfigure}[b]{0.55\linewidth}
  \hspace*{-1cm}
\includegraphics[width = 1.0\textwidth]{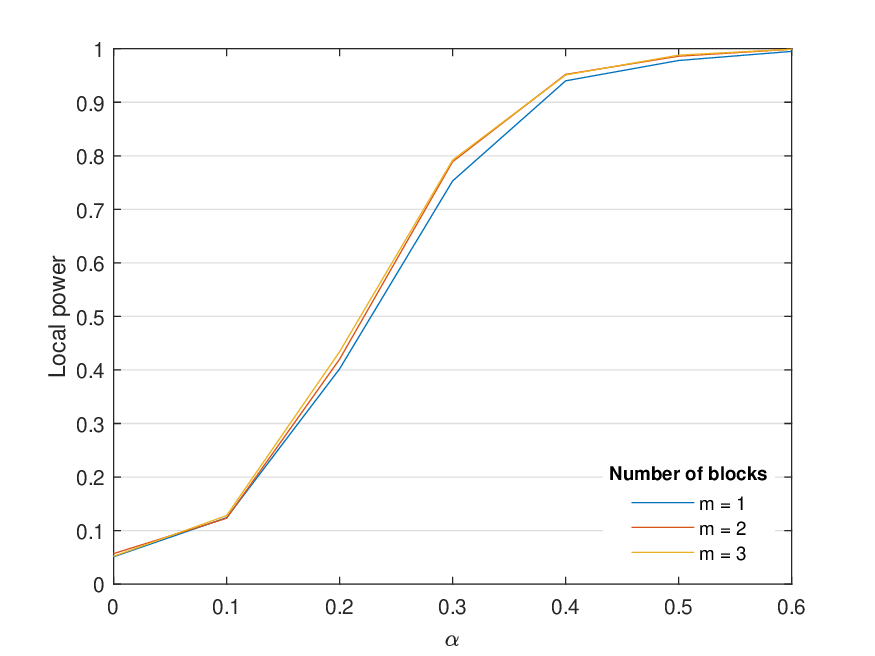}
\caption{$s_{\widehat{f},2}^{P}$ with $\nu_0 = 0.8$ $\lambda_1^0 = 1$, $\lambda_2^0 = 0.65$.}
\end{subfigure}
\medskip
\begin{subfigure}[b]{0.55\linewidth}
  \hspace*{-1cm}
\includegraphics[width = 1.0\textwidth]{Power_alpha_num_m_plot_G2_tau05.eps}
\caption{$s_{\widehat{f},3}^{P}$ with $\nu_0 = 0.8$ $\lambda_1^0 = 1$, $\lambda_2^0 = 0.65$.}
\end{subfigure}
\hfill
\begin{subfigure}[b]{0.55\linewidth}
  \hspace*{-1cm}
\includegraphics[width = 1.0\textwidth]{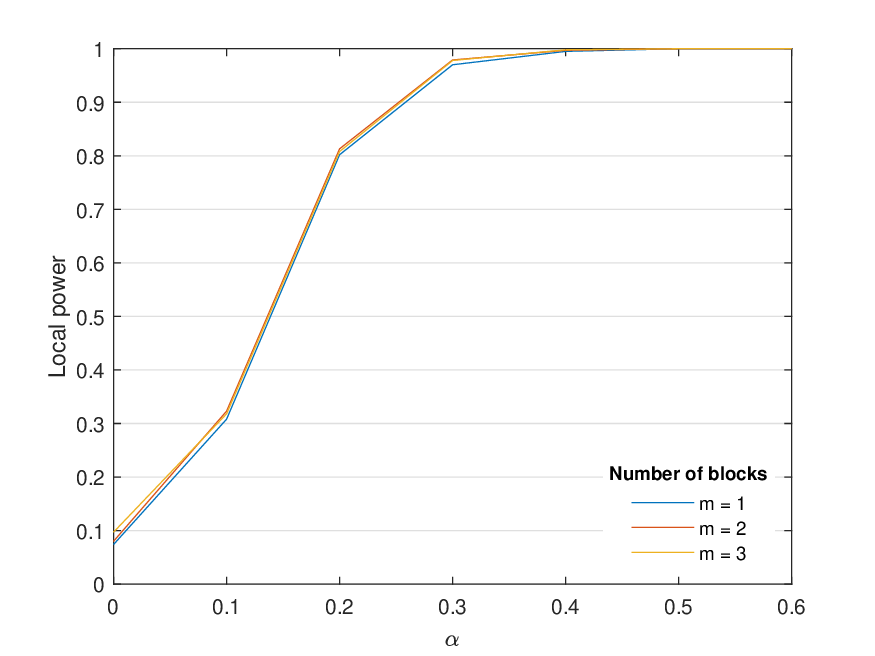}
\caption{$s_{\widehat{f},4}^{P}$ with $\nu_0 = 0.8$ $\lambda_1^0 = 0.65$, $\lambda_2^0 = 1$.}
\end{subfigure}
\caption{Evolution of local power (\%) of the four statistics for different DGPs and values of $\alpha$. Setting: $(N,T)$ = (200, 600), $h$ = 1,$\tau = 0.5$.}
\label{fig: LocalPowerG1G2G3G4}
\end{figure}

\subsubsection{Power Curves under Different Persistence Levels}
\begin{figure}[H]
\begin{subfigure}[b]{0.55\linewidth} 
  \hspace*{-1cm}
\includegraphics[width = 1.0\textwidth]{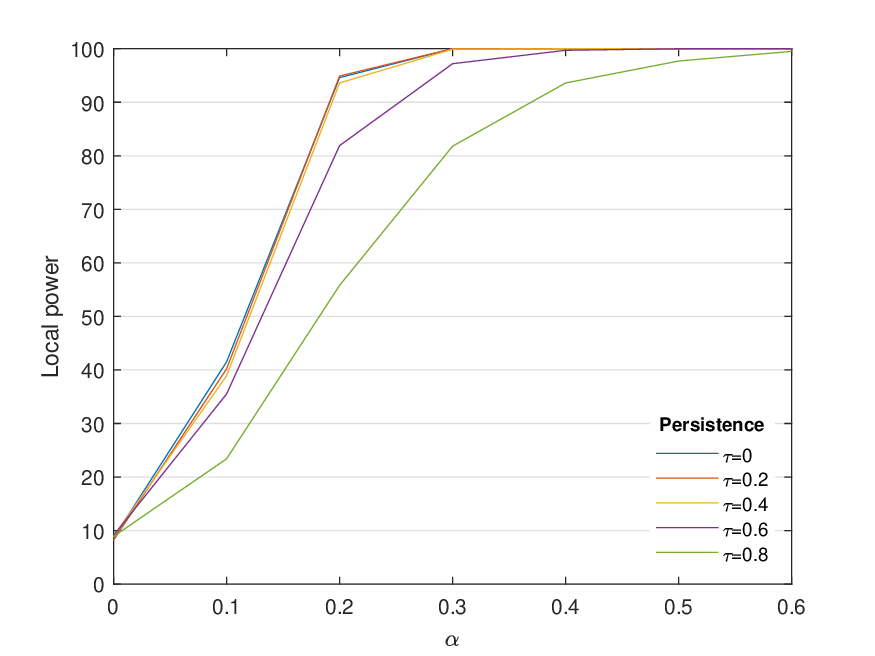}
\footnotesize \caption{$s_{\widehat{f},1}$ with $\mu_0 = 0.45$.}
\end{subfigure}
\begin{subfigure}[b]{0.55\linewidth}
  \hspace*{-1cm}
\includegraphics[width = 1.0\textwidth]{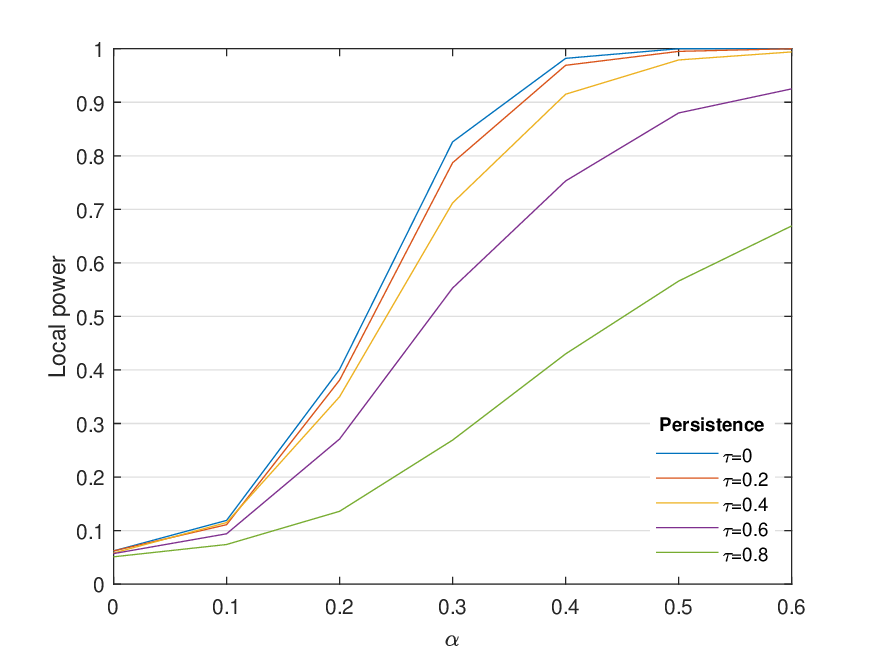}
\caption{$s_{\widehat{f},2}^{P}$ with $\nu_0 = 0.8$ $\lambda_1^0 = 1$, $\lambda_2^0 = 0.65$.}
\end{subfigure}
\medskip
\begin{subfigure}[b]{0.55\linewidth}
  \hspace*{-1cm}
\includegraphics[width = 1.0\textwidth]{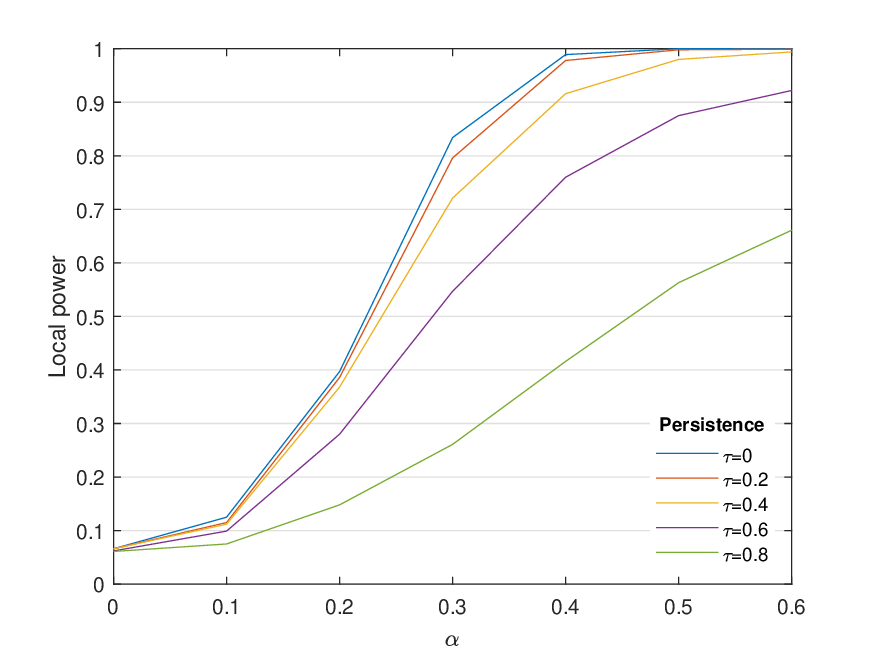}
\caption{$s_{\widehat{f},3}^{P}$ with $\nu_0 = 0.8$ $\lambda_1^0 = 1$, $\lambda_2^0 = 0.65$.}
\end{subfigure}
\hfill
\begin{subfigure}[b]{0.55\linewidth}
  \hspace*{-1cm}
\includegraphics[width = 1.0\textwidth]{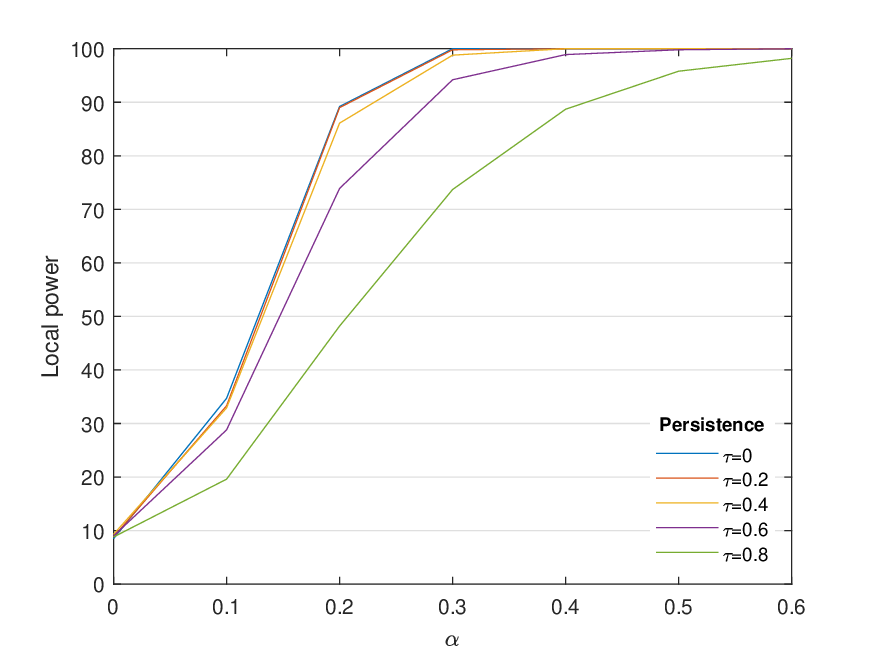}
\caption{$s_{\widehat{f},4}^{P}$ with $\nu_0 = 0.8$ $\lambda_1^0 = 0.65$, $\lambda_2^0 = 1$.}
\end{subfigure}
\caption{Evolution of local power (\%) of the four statistics for different values $\alpha$ and $\tau$. Setting: DGP (2), $(N,T)$ = (200, 600), $h$ = 1.}
\label{fig: LocalPowerPersistence}
\end{figure}

We then turn our attention to the effects of higher degrees of persistency on test size and local power. As in the main manuscript, we consider varying levels of persistence with $\tau \in \{0, 0.2, 0.4, 0.6, 0.8\}$. The results for the four statistics are presented in Figure \ref{fig: LocalPowerPersistence}. It is clear from these graphical representations that regressor persistency has no implications for test size, but it causes a deterioration in local power for greater values of $\tau$ which is anticipated by our theory. As in the above, it seems that the first statistics outperforms its peers as well as suffer from less decay in local power due to nonstationarity. 

\subsubsection{Power Curves under Serial Correlation in Idiosyncratics and Persistence}
Furthermore, we investigate how our tests perform in the case of persistent data and serial correlation in the idiosyncratic components of the factor model under DGP (3) and different values of $\tau$. Recall in fact that, under DGP (3), all panel idiosyncratic components $\* e_{i,t}$ exhibit autocorrelation of order 1 in the range $(0.6, 1)$, meaning that they have strong persistence in time. Further, as outlined in the proof of Lemma 1 in Section \ref{sec: Sec_2}, the presence of serial correlation in the idiosyncratics reduces the rate of $\frac{1}{T^{1+\tau}}\sum_{t=1}^T\*z_t\overline{\*e}_t'$ from $O_p(N^{-1/2}T^{-(1+\tau)/2})$ in Equation (\ref{eq: corollary.a.1}) to $O_p(N^{-1/2}T^{-\tau/2})$ in Equation (\ref{eq: lemma.a.1}). If $\tau \in (0, 1)$ in addition, we have to rule out serial correlation in order to have asymptotically valid tests. However, we recognise that it is unrealistic to do so in empirical applications, particularly in the context of macroeconomic data. Hence, a Monte Carlo experiment under DGP (3) and different values of $\tau$ will serve as a robust benchmark should any potential concerns arise. The results on the local power of the four statistics are presented in Figures 7 to 10. All things considered, it is very clear that, while the local power worsens with the persistence as before, the results are still very satisfactory and seem to suggest that our rates in the case with serial correlation and persistence might possibly be conservative, if anything. Please note that the average level of the serial correlation in DGP (3) is materially higher than that which is estimated in all combinations of target variable and geography in the empirical application, as shown below in Section \ref{sec: Section8}. Hence, we argue that these simulations serves as trustworthy benchmarks against which we can compare the findings in our empirical analysis.

\begin{figure}[H]
\begin{subfigure}[b]{0.55\linewidth} 
  \hspace*{-1cm}
\includegraphics[width = 1.0\textwidth]{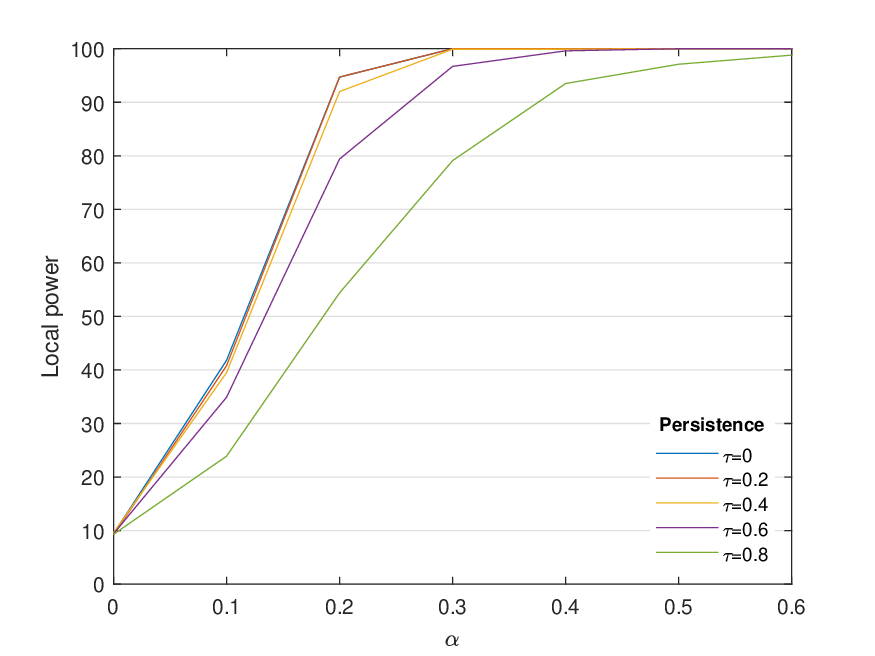}
\footnotesize \caption{$s_{\widehat{f},1,HAC}$ with $\mu_0 = 0.45$.}
\label{fig: persistserialcorrEq1}
\end{subfigure}
\begin{subfigure}[b]{0.55\linewidth}
  \hspace*{-1cm}
\includegraphics[width = 1.0\textwidth]{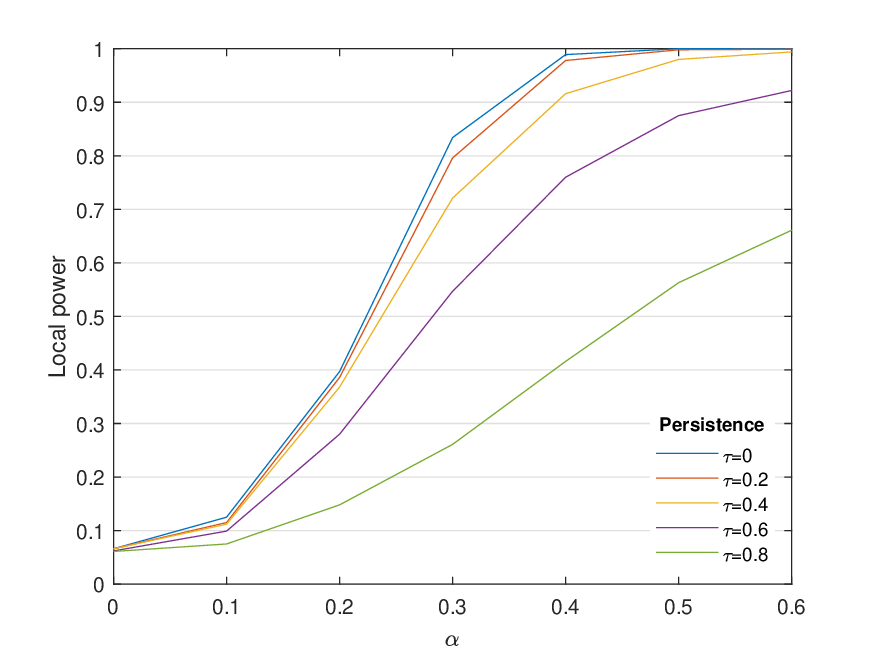}
\caption{$s_{\widehat{f},2,HAC}^{P}$ with $\nu_0 = 0.8$ $\lambda_1^0 = 1$, $\lambda_2^0 = 0.65$.}
\label{fig: persistserialcorrEq2}
\end{subfigure}
\medskip
\begin{subfigure}[b]{0.55\linewidth}
  \hspace*{-1cm}
\includegraphics[width = 1.0\textwidth]{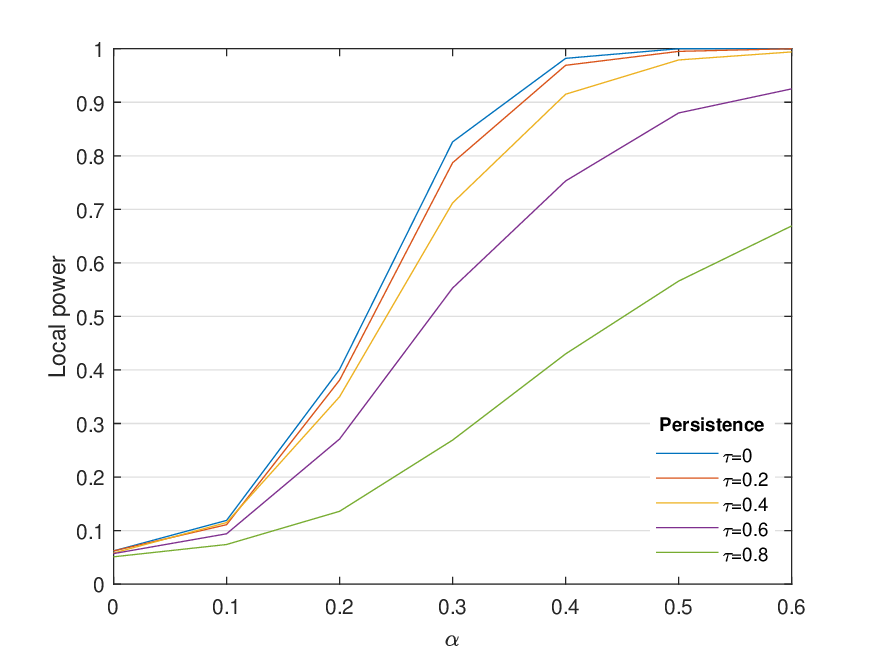}
\caption{$s_{\widehat{f},3,HAC}^{P}$ with $\nu_0 = 0.8$ $\lambda_1^0 = 1$, $\lambda_2^0 = 0.65$.}
\label{fig: persistserialcorrEq3}
\end{subfigure}
\hfill
\begin{subfigure}[b]{0.55\linewidth}
  \hspace*{-1cm}
\includegraphics[width = 1.0\textwidth]{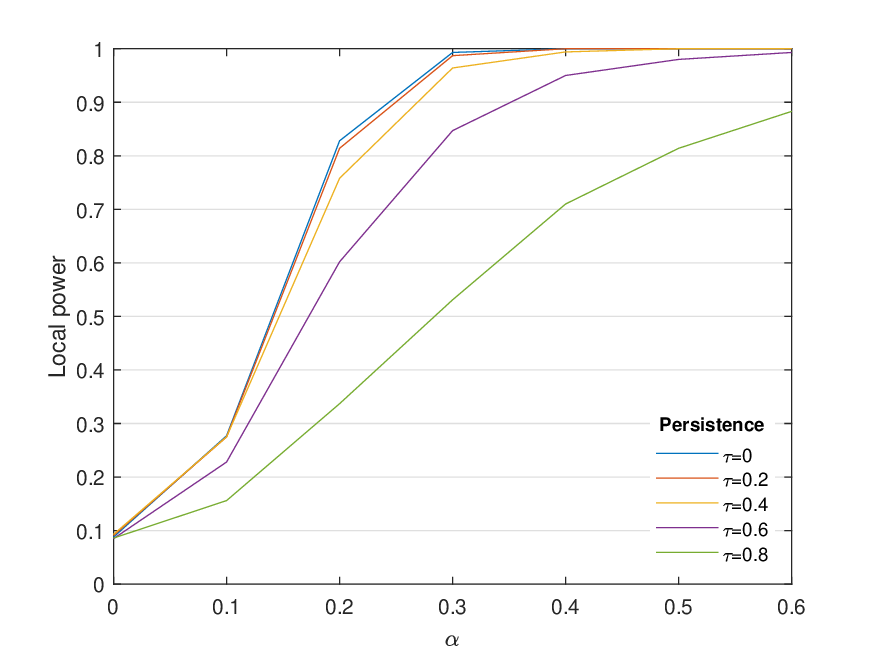}
\caption{$s_{\widehat{f},4,HAC}^{P}$ with $\nu_0 = 0.8$ $\lambda_1^0 = 0.65$, $\lambda_2^0 = 1$.}
\label{fig: persistserialcorrEq4}
\end{subfigure}
\caption{Evolution of local power (\%) of the four HAC-corrected statistics for different values $\alpha$ and $\tau$. Setting: DGP (3), $(N,T)$ = (200, 600), $h$ = 1.}
\end{figure}

\subsubsection{Power Curves: Serial Correlation in Forecast Errors}

\begin{figure}[H]
\begin{subfigure}[b]{0.55\linewidth} 
  \hspace*{-1cm}
\includegraphics[width = 1\textwidth]{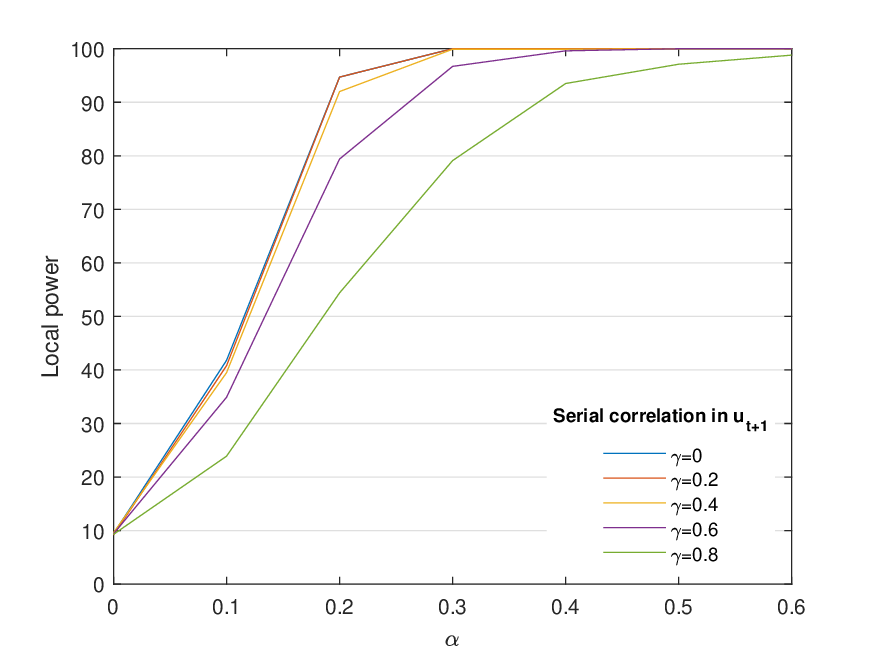}
\footnotesize \caption{$s_{\widehat{f},1,HAC}$ with $\mu_0 = 0.45$.}
\label{fig: SerCorrForecastingEq1}
\end{subfigure}
\begin{subfigure}[b]{0.55\linewidth} 
  \hspace*{-1cm}
\includegraphics[width = 1\textwidth]{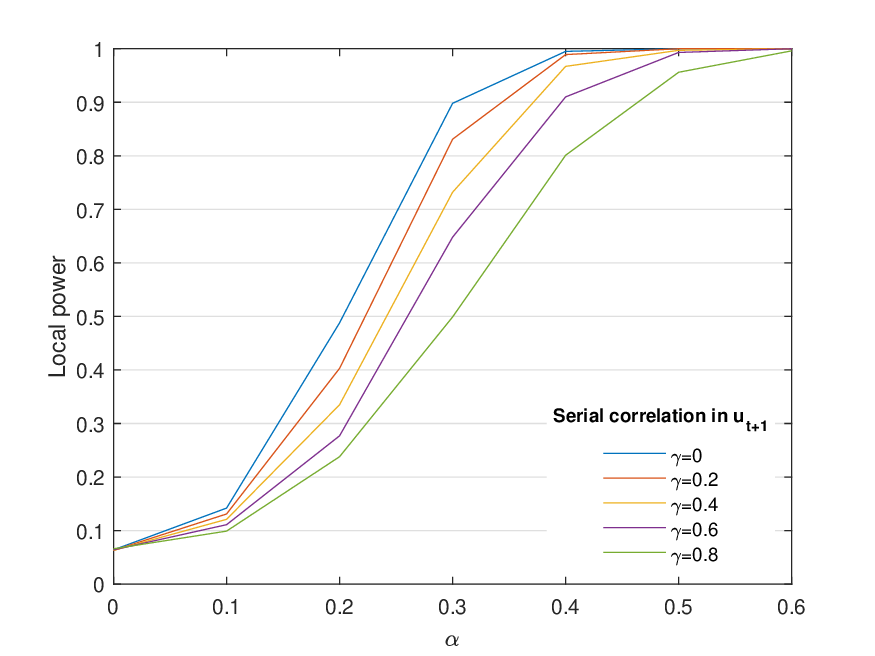}
\caption{$s_{\widehat{f},2,HAC}^{P}$ with $\nu_0 = 0.8$ $\lambda_1^0 = 1$, $\lambda_2^0 = 0.65$.}
\label{fig: SerCorrForecastingEq2}
\end{subfigure}
\medskip
\begin{subfigure}[b]{0.55\linewidth}
  \hspace*{-1cm}
\includegraphics[width = 1\textwidth]{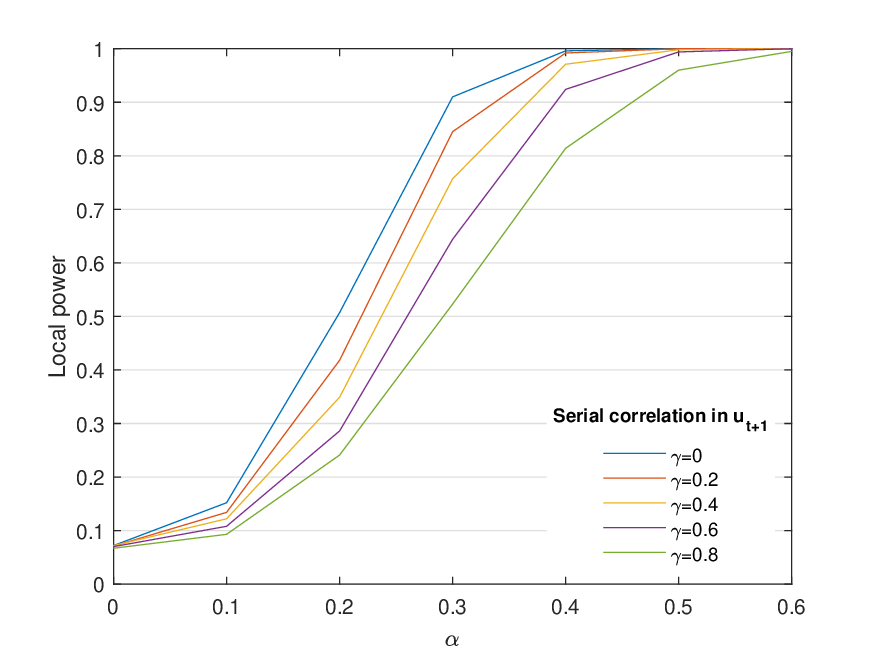}
\caption{$s_{\widehat{f},3,HAC}^{P}$ with $\nu_0 = 0.8$ $\lambda_1^0 = 1$, $\lambda_2^0 = 0.65$.}
\label{fig: SerCorrForecastingEq3}
\end{subfigure}
\hfill
\begin{subfigure}[b]{0.55\linewidth}
  \hspace*{-1cm}
\includegraphics[width = 1\textwidth]{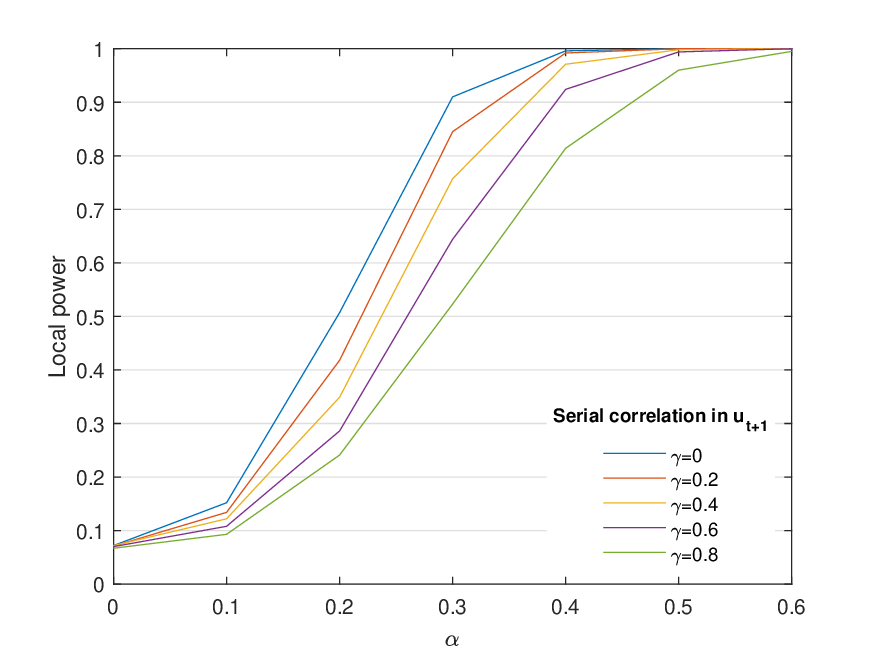}
\caption{$s_{\widehat{f},4,HAC}^{P}$ with $\nu_0 = 0.8$ $\lambda_1^0 = 0.65$, $\lambda_2^0 = 1$.}
\label{fig: SerCorrForecastingEq4}
\end{subfigure}
\caption{Evolution of local power (\%) of the four HAC-corrected statistics for different values $\alpha$ and $\gamma$. Setting: DGP (2), $(N,T)$ = (200, 600), $h$ = 1, $\tau = 0$.}
\label{fig: SerCorrForecasting}
\end{figure}

We consider the impact of a violation of the MDS assumption on the testing framework of \cite{pitarakis2025novel} and \cite{pitarakis2023direct}. We allow the error term in the forecasting equation in the main text to follow an $AR(1)$ process, and we estimate our four statistics with HAC-corrected standard errors (subscript $HAC$) based on the celebrated methodology of \cite{NWHAC}. The bandwidth value is chosen to be $\lfloor1.2 \times T^{1/3}\rfloor$. We report the power evolution of the HAC-corrected statistics in Figure \ref{fig: SerCorrForecasting}, where the autoregressive coefficient $\gamma$ takes values in $\{0, 0.1, 0.2, 0.3, 0.4, 0.5, 0.8\}$. Unsurprisingly, our results indicate that, despite the denominator of our statistics being suitably corrected, the power deteriorates materially with departures from the MDS assumption in the forecasting equation. This is particularly true for $s_{\widehat{f},2,HAC}^{P}$ and $s_{\widehat{f},3,HAC}^{P}$ where the effect on the local power is visible, while $s_{\widehat{f},1,HAC}$ and $s_{\widehat{f},4,HAC}^{P}$ seem to be more robust to the issue. On contrast, the effect on the test size of the four statistics is nearly inexistent. Unreported results, however, point at no difference in the local power evolution between our original statistics and their HAC-corrected counterparts. In fact, serial correlation in the process $u_{t}$ directly influences the numerator of our statistics but does not enter into the calculations of the long-run variance normaliser. Nonetheless, it is paramount to remember that this problem can be readily ironed out by adding a sufficient number of lags of the target variable in the forecasting regression so to capture all residual serial correlation. \\


\subsubsection{Power Curves: Multi-Step Ahead Forecasts with $\tau = 0$}

\begin{figure}[H]
\begin{subfigure}[b]{0.5\linewidth} 
  
\includegraphics[width = 1.0\textwidth]{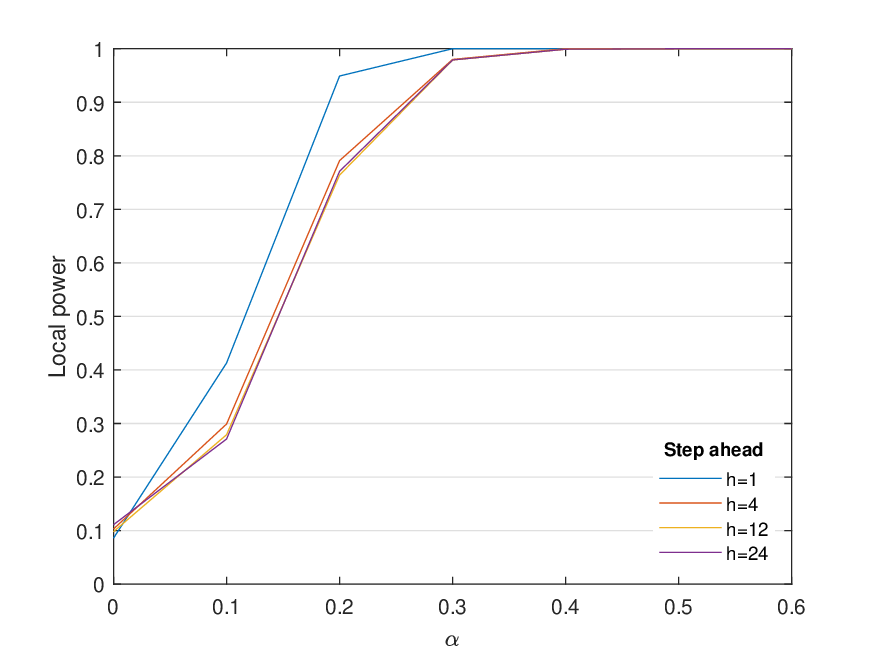}
\footnotesize \caption{$s_{\widehat{f},1,HAC}$ with $\mu_0 = 0.45$.}
\label{fig: hstepaheadForecastingEq1}
\end{subfigure}
\begin{subfigure}[b]{0.5\linewidth}

\includegraphics[width = 1.0\textwidth]{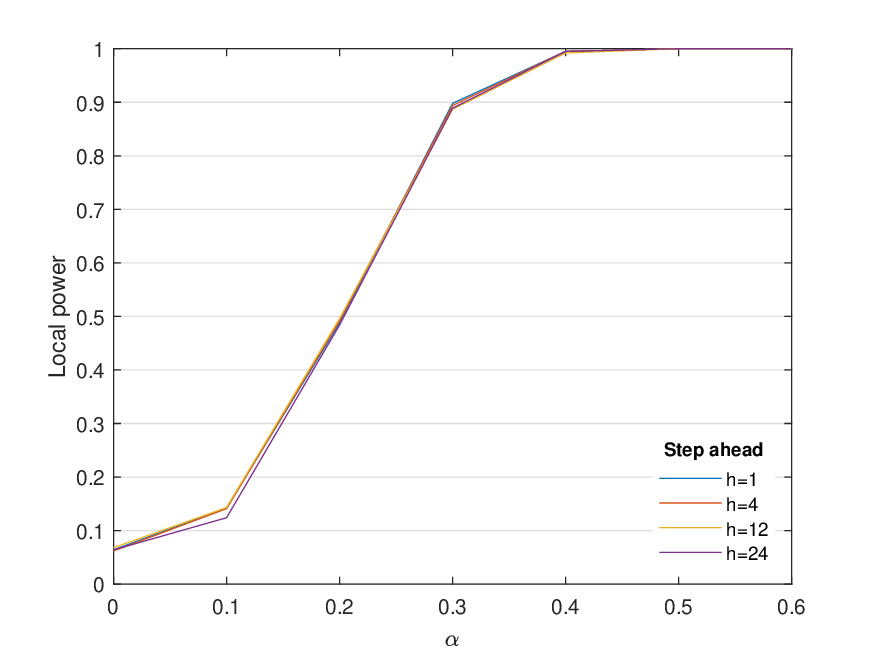}
\caption{$s_{\widehat{f},2,HAC}^{P}$ with $\nu_0 = 0.8$ $\lambda_1^0 = 1$, $\lambda_2^0 = 0.65$.}
\label{fig: hstepaheadForecastingEq2}
\end{subfigure}
\medskip
\begin{subfigure}[b]{0.5\linewidth}

\includegraphics[width = 1.0\textwidth]{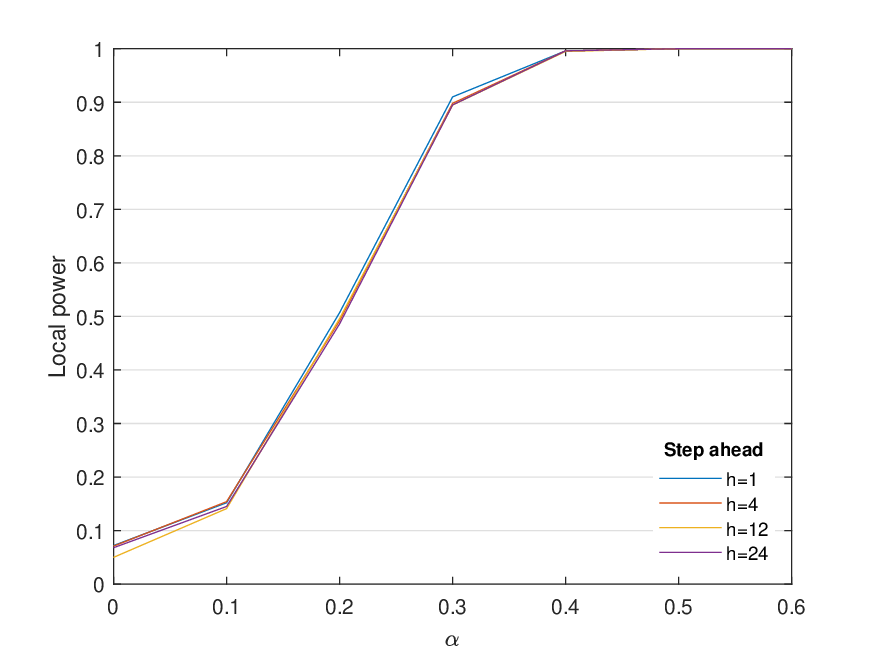}
\caption{$s_{\widehat{f},3,HAC}^{P}$ with $\nu_0 = 0.8$ $\lambda_1^0 = 1$, $\lambda_2^0 = 0.65$.}
\label{fig: hstepaheadForecastingEq3}
\end{subfigure}
\hfill
\begin{subfigure}[b]{0.5\linewidth}

\includegraphics[width = 1.0\textwidth]{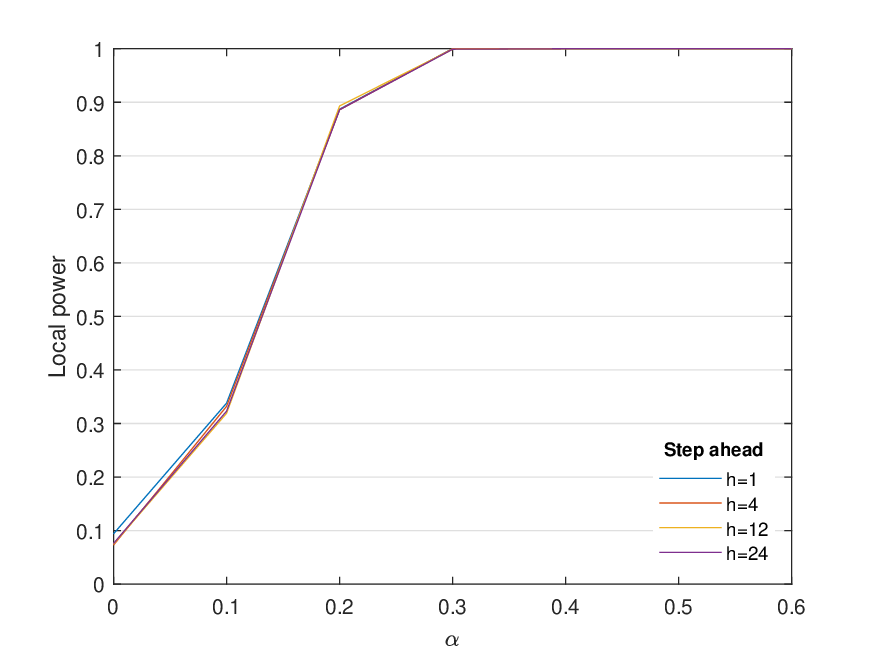}
\caption{$s_{\widehat{f},4,HAC}^{P}$ with $\nu_0 = 0.8$ $\lambda_1^0 = 0.65$, $\lambda_2^0 = 1$.}
\label{fig: hstepaheadForecastingEq4}
\end{subfigure}
\caption{
Same parameters as in Figure 1, with the focus on DGP (2) and   $\tau = 0$.}
\label{fig: hstepaheadForecasting}
\end{figure}

Similarly to the main manuscript, we report the power evolution of our tests in a multi-step ahead environment for $h=1,4,12,24$ with stationary predictors, i.e. $\tau = 0$. We consider HAC standard errors for all statistics under DGP(2) and report the results in Figure \ref{fig: hstepaheadForecasting}. Since there is no persistence in the predictors, the local power improves upon the simulations in the main manuscript where $\tau$ is set at 0.5.

\subsection{Additional Results and Diagnostic Tests for the Empirical Application} \label{sec: Section8}

In this section, we report additional pre-estimation results and diagnostics from the empirical application for illustration completeness. In particular, Tables \ref{tab: FirstHalf} and \ref{tab: SecondHalf} report the following statistics for each combination of target variable and geography: estimated number of factors ("ER"), number of breaks ("\# breaks") and break dates ("break dates") in the loadings of the factor panel, as well as average serial correlation ("SC"), average absolute serial correlation("Abs SC"), pair-wise correlations ("XC") and absolute correlations ("Abs XC") in the idiosyncratic panel data residuals.

\begin{table}
  \centering
    \resizebox{15.8cm}{!}{\begin{tabular}{ccccccccccccccc}
    \multicolumn{7}{c}{Stationary}                        &       & \multicolumn{7}{c}{Persistent} \\
          &       &       &       &       &       &       &       &       &       &       &       &       &       &  \\
    Statistic &       & EA    & Germany & France & Italy & Spain &       & Statistic &       & EA    & Germany & France & Italy & Spain \\
\cmidrule{1-7}\cmidrule{9-15}          &       &       &       &       &       &       &       &       &       &       &       &       &       &  \\
          &       & \multicolumn{5}{c}{GDP }              &       &       &       & \multicolumn{5}{c}{GDP } \\
    ER    &       & 1     & 2     & 1     & 1     & 1     &       & ER    &       & 1     & 2     & 1     & 1     & 1 \\
    \# breaks &       & 2     & 1     & 2     & 1     & 0     &       & \# breaks &       & 2     & 1     & 2     & 1     & 0 \\
    break dates &       & \multicolumn{1}{p{3.59em}}{2020:2\newline{}2021:4} & 2001:3 & \multicolumn{1}{p{3.59em}}{2009:3\newline{}2010:4} & 2021:4 & -     &       & break dates &       & \multicolumn{1}{p{3.455em}}{2020:2\newline{}2021:4} & \multicolumn{1}{p{3.455em}}{2020:2\newline{}2021:4} & 2001:3 & \multicolumn{1}{p{3.455em}}{2009:3\newline{}2010:4} & 2021:4 \\
    SC &       & -0.002 & 0.139 & 0.103 & 0.095 & 0.117 &       & SC &       & 0.558 & 0.580 & 0.558 & 0.534 & 0.511 \\
    Abs SC &       & 0.184 & 0.288 & 0.309 & 0.262 & 0.260 &       & Abs SC &       & 0.559 & 0.580 & 0.558 & 0.535 & 0.511 \\
    XC &       & 0.002 & 0.002 & 0.002 & 0.000 & 0.000 &       & XC &       & 0.005 & 0     & 0.002 & 0.001 & 0.007 \\
    Abs XC &       & 0.007 & 0.006 & 0.002 & 0.000 & 0.000 &       & Abs XC &       & 0.036 & 0     & 0.014 & 0.001 & 0.024 \\
          &       & \multicolumn{5}{c}{IPMN}              &       &       &       & \multicolumn{5}{c}{IPMN} \\
    ER    &       & 1     & 2     & 1     & 1     & 1     &       & ER    &       & 1     & 2     & 1     & 1     & 1 \\
    \# breaks &       & 2     & 1     & 2     & 1     & 0     &       & \# breaks &       & 2     & 1     & 2     & 1     & 0 \\
    break dates &       & \multicolumn{1}{p{3.59em}}{2020:2\newline{}2021:4} & 2001:3 & \multicolumn{1}{p{3.59em}}{2009:3\newline{}2010:4} & 2021:4 & -     &       & break dates &       & \multicolumn{1}{p{3.455em}}{2020:2\newline{}2021:4} & 2001:3 & \multicolumn{1}{p{3.455em}}{2009:3\newline{}2010:4} & 2021:4 & - \\
    SC &       & -0.003 & 0.141 & 0.102 & 0.097 & 0.117 &       & SC &       & 0.555 & 0.589 & 0.564 & 0.532 & 0.513 \\
    Abs SC &       & 0.183 & 0.287 & 0.305 & 0.265 & 0.261 &       & Abs SC &       & 0.556 & 0.589 & 0.564 & 0.532 & 0.513 \\
    XC &       & 0.002 & 0.003 & 0.002 & 0.000 & 0.000 &       & XC &       & 0.005 & 0.000 & 0.002 & 0.006 & 0.007 \\
    Abs XC &       & 0.007 & 0.006 & 0.002 & 0.000 & 0.000 &       & Abs XC &       & 0.040 & 0.000 & 0.019 & 0.031 & 0.022 \\
          &       & \multicolumn{5}{c}{HICPNEF}           &       &       &       & \multicolumn{5}{c}{HICPNEF} \\
    ER    &       & 1     & 2     & 1     & 1     & 1     &       & ER    &       & 1     & 2     & 1     & 1     & 1 \\
    \# breaks &       & 2     & 1     & 2     & 1     & 0     &       & \# breaks &       & 2     & 1     & 2     & 1     & 0 \\
    break dates &       & \multicolumn{1}{p{3.59em}}{2020:2\newline{}2021:4} & 2001:3 & \multicolumn{1}{p{3.59em}}{2009:3\newline{}2010:4} & 2021:4 & -     &       & break dates &       & \multicolumn{1}{p{3.455em}}{2020:2\newline{}2021:4} & 2001:3 & \multicolumn{1}{p{3.455em}}{2009:3\newline{}2010:4} & 2021:4 & - \\
    SC &       & 0.001 & 0.142 & 0.111 & 0.097 & 0.122 &       & SC &       & 0.555 & 0.585 & 0.560 & 0.534 & 0.515 \\
    Abs SC &       & 0.181 & 0.286 & 0.306 & 0.265 & 0.256 &       & Abs SC &       & 0.555 & 0.585 & 0.560 & 0.535 & 0.515 \\
    XC &       & 0.002 & 0.003 & 0.002 & 0.000 & 0.000 &       & XC &       & 0.004 & 0     & 0.002 & 0.005 & 0.005 \\
    Abs XC &       & 0.007 & 0.006 & 0.002 & 0.000 & 0.000 &       & Abs XC &       & 0.033 & 0     & 0.013 & 0.030 & 0.021 \\
          &       & \multicolumn{5}{c}{UNETOT}            &       &       &       & \multicolumn{5}{c}{UNETOT} \\
    ER    &       & 1     & 2     & 1     & 1     & 1     &       & ER    &       & 1     & 2     & 1     & 1     & 1 \\
    \# breaks &       & 2     & 1     & 2     & 1     & 0     &       & \# breaks &       & 2     & 1     & 2     & 1     & 0 \\
    break dates &       & \multicolumn{1}{p{3.59em}}{2020:2\newline{}2021:4} & 2001:3 & \multicolumn{1}{p{3.59em}}{2009:3\newline{}2010:4} & 2021:4 & -     &       & break dates &       & \multicolumn{1}{p{3.455em}}{2020:2\newline{}2021:4} & 2001:3 & \multicolumn{1}{p{3.455em}}{2009:3\newline{}2010:4} & 2021:4 & - \\
    SC &       & -0.005 & 0.133 & 0.107 & 0.100 & 0.115 &       & SC &       & 0.555 & 0.587 & 0.564 & 0.535 & 0.518 \\
    Abs SC &       & 0.181 & 0.286 & 0.313 & 0.262 & 0.258 &       & Abs SC &       & 0.555 & 0.587 & 0.564 & 0.536 & 0.518 \\
    XC &       & 0.002 & 0.003 & 0.002 & 0.000 & 0.000 &       & XC &       & 0.005 & 0.000 & 0.002 & 0.001 & 0.006 \\
    Abs XC &       & 0.006 & 0.006 & 0.002 & 0.000 & 0.000 &       & Abs XC &       & 0.035 & 0.001 & 0.013 & 0.002 & 0.023 \\
          &       & \multicolumn{5}{c}{IRT3M}             &       &       &       & \multicolumn{5}{c}{IRT3M} \\
    ER    &       & 1     & 2     & 1     & 1     & 1     &       & ER    &       & 1     & 2     & 1     & 1     & 1 \\
    \# breaks &       & 2     & 1     & 2     & 1     & 0     &       & \# breaks &       & 2     & 1     & 2     & 1     & 0 \\
    break dates &       & \multicolumn{1}{p{3.59em}}{2020:2\newline{}2021:4} & 2001:3 & \multicolumn{1}{p{3.59em}}{2009:3\newline{}2010:4} & 2021:4 & -     &       & break dates &       & \multicolumn{1}{p{3.455em}}{2020:2\newline{}2021:4} & 2001:3 & \multicolumn{1}{p{3.455em}}{2009:3\newline{}2010:4} & 2021:4 & - \\
    SC &       & -0.004 & -     & -     & -     & -     &       & SC &       & 0.551 & -     & -     & -     & - \\
    Abs SC &       & 0.187 & -     & -     & -     & -     &       & Abs SC &       & 0.552 & -     & -     & -     & - \\
    XC &       & 0.002 & -     & -     & -     & -     &       & XC &       & 0.004 & -     & -     & -     & - \\
    Abs XC &       & 0.007 & -     & -     & -     & -     &       & Abs XC &       & 0.032 & -     & -     & -     & - \\
          &       & \multicolumn{5}{c}{IRT6M}             &       &       &       & \multicolumn{5}{c}{IRT6M} \\
    ER    &       & 1     & 2     & 1     & 1     & 1     &       & ER    &       & 1     & 2     & 1     & 1     & 1 \\
    \# breaks &       & 2     & 1     & 2     & 1     & 0     &       & \# breaks &       & 2     & 1     & 2     & 1     & 0 \\
    break dates &       & \multicolumn{1}{p{3.59em}}{2020:2\newline{}2021:4} & 2001:3 & \multicolumn{1}{p{3.59em}}{2009:3\newline{}2010:4} & 2021:4 & -     &       & break dates &       & \multicolumn{1}{p{3.455em}}{2020:2\newline{}2021:4} & 2001:3 & \multicolumn{1}{p{3.455em}}{2009:3\newline{}2010:4} & 2021:4 & - \\
    SC &       & -0.006 & -     & -     & -     & -     &       & SC &       & 0.552 & -     & -     & -     & - \\
    Abs SC &       & 0.185 & -     & -     & -     & -     &       & Abs SC &       & 0.552 & -     & -     & -     & - \\
    XC &       & 0.002 & -     & -     & -     & -     &       & XC &       & 0.003 & -     & -     & -     & - \\
    Abs XC &       & 0.006 & -     & -     & -     & -     &       & Abs XC &       & 0.030 & -     & -     & -     & - \\
          &       & \multicolumn{5}{c}{LTIRT}             &       &       &       & \multicolumn{5}{c}{LTIRT} \\
    ER    &       & 1     & 2     & 1     & 1     & 1     &       & ER    &       & 1     & 2     & 1     & 1     & 1 \\
    \# breaks &       & 2     & 1     & 2     & 1     & 0     &       & \# breaks &       & 2     & 1     & 2     & 1     & 0 \\
    break dates &       & \multicolumn{1}{p{3.59em}}{2020:2\newline{}2021:4} & 2001:3 & \multicolumn{1}{p{3.59em}}{2009:3\newline{}2010:4} & 2021:4 & -     &       & break dates &       & \multicolumn{1}{p{3.455em}}{2020:2\newline{}2021:4} & 2001:3 & \multicolumn{1}{p{3.455em}}{2009:3\newline{}2010:4} & 2021:4 & - \\
    SC &       & 0.002 & 0.134 & 0.107 & 0.096 & 0.112 &       & SC &       & 0.558 & 0.621 & 0.593 & 0.559 & 0.539 \\
    Abs SC &       & 0.179 & 0.288 & 0.307 & 0.258 & 0.259 &       & Abs SC &       & 0.559 & 0.621 & 0.593 & 0.559 & 0.539 \\
    XC &       & 0.002 & 0.002 & 0.002 & 0.001 & 0.000 &       & XC &       & 0.006 & 0     & 0.004 & 0.001 & 0.000 \\
    Abs XC &       & 0.007 & 0.024 & 0.002 & 0.001 & 0.000 &       & Abs XC &       & 0.038 & 0     & 0.024 & 0.001 & 0.000 \\
          &       &       &       &       &       &       &       &       &       &       &       &       &       &  \\
    \bottomrule
    \end{tabular}}%
    \caption{\footnotesize "ER" refers to the estimated number of factors based on the eigenvalue ratio of \cite{ahn2013eigenvalue}. "\# breaks" and "break dates" are obtained using the sup-LM test statistics of \cite{breakschendoladogonzalo}. "SC" and "Abs SC" refer to the average and average absolute serial correlations of order 1 in the idiosyncratic panel data residuals that are significant at a 5\% level under the Ljung-Box tests. Similarly, "XC" and "Abs XC" refer to the average and average absolute pair-wise correlations in the idiosyncratic panel data residuals using the thresholding procedure of \cite{CaiLiu2011}.}
  \label{tab: FirstHalf}%
\end{table}%

\begin{table}
  \centering
    \resizebox{15.6cm}{!}{
    \begin{tabular}{ccccccccccccccc}
    \multicolumn{7}{c}{Stationary}                        &       & \multicolumn{7}{c}{Persistent} \\
          &       &       &       &       &       &       &       &       &       &       &       &       &       &  \\
    Statistic &       & EA    & Germany & France & Italy & Spain &       & Statistic &       & EA    & Germany & France & Italy & Spain \\
\cmidrule{1-7}\cmidrule{9-15}          &       &       &       &       &       &       &       &       &       &       &       &       &       &  \\
          &       & \multicolumn{5}{c}{WS}                &       &       &       & \multicolumn{5}{c}{WS} \\
    ER    &       & 1     & 2     & 1     & 1     & 1     &       & ER    &       & 1     & 2     & 1     & 1     & 1 \\
    \# breaks &       & 2     & 1     & 2     & 1     & 0     &       & \# breaks &       & 2     & 1     & 2     & 1     & 0 \\
    break dates &       & \multicolumn{1}{p{3.59em}}{2020:2\newline{}2021:4} & 2001:3 & \multicolumn{1}{p{3.59em}}{2009:3\newline{}2010:4} & 2021:4 & -     &       & break dates &       & \multicolumn{1}{p{3.455em}}{2020:2\newline{}2021:4} & 2001:3 & \multicolumn{1}{p{3.455em}}{2009:3\newline{}2010:4} & 2021:4 & - \\
    SC &       & -0.002 & 0.134 & 0.101 & 0.099 & 0.118 &       & SC &       & 0.557 & 0.579 & 0.566 & 0.533 & 0.519 \\
    Abs SC &       & 0.183 & 0.294 & 0.306 & 0.266 & 0.262 &       & Abs SC &       & 0.557 & 0.579 & 0.566 & 0.534 & 0.519 \\
    XC &       & 0.002 & 0.002 & 0.002 & 0.000 & 0.000 &       & XC &       & 0.005 & 0.000 & 0.001 & 0.007 & 0.008 \\
    Abs XC &       & 0.007 & 0.006 & 0.002 & 0.000 & 0.000 &       & Abs XC &       & 0.036 & 0.000 & 0.015 & 0.030 & 0.029 \\
          &       & \multicolumn{5}{c}{GGLB}              &       &       &       & \multicolumn{5}{c}{GGLB} \\
    ER    &       & 1     & 2     & 1     & 1     & 1     &       & ER    &       & 1     & 2     & 1     & 1     & 1 \\
    \# breaks &       & 2     & 1     & 2     & 1     & 0     &       & \# breaks &       & 2     & 1     & 2     & 1     & 0 \\
    break dates &       & \multicolumn{1}{p{3.59em}}{2020:2\newline{}2021:4} & 2001:3 & \multicolumn{1}{p{3.59em}}{2009:3\newline{}2010:4} & 2021:4 & -     &       & break dates &       & \multicolumn{1}{p{3.455em}}{2020:2\newline{}2021:4} & 2001:3 & \multicolumn{1}{p{3.455em}}{2009:3\newline{}2010:4} & 2021:4 & - \\
    SC &       & -0.002 & 0.137 & 0.109 & 0.094 & 0.118 &       & SC &       & 0.555 & 0.588 & 0.564 & 0.504 & 0.516 \\
    Abs SC &       & 0.183 & 0.291 & 0.309 & 0.259 & 0.262 &       & Abs SC &       & 0.555 & 0.588 & 0.564 & 0.505 & 0.516 \\
    XC &       & 0.002 & 0.003 & 0.002 & 0.000 & 0.000 &       & XC &       & 0.004 & 0.000 & 0.002 & 0     & 0.004 \\
    Abs XC &       & 0.007 & 0.006 & 0.002 & 0.000 & 0.000 &       & Abs XC &       & 0.028 & 0.001 & 0.014 & 0     & 0.017 \\
          &       & \multicolumn{5}{c}{NFCLB}             &       &       &       & \multicolumn{5}{c}{NFCLB} \\
    ER    &       & 1     & 2     & 1     & 1     & 1     &       & ER    &       & 1     & 2     & 1     & 1     & 1 \\
    \# breaks &       & 2     & 1     & 2     & 1     & 0     &       & \# breaks &       & 2     & 1     & 2     & 1     & 0 \\
    break dates &       & \multicolumn{1}{p{3.59em}}{2020:2\newline{}2021:4} & 2001:3 & \multicolumn{1}{p{3.59em}}{2009:3\newline{}2010:4} & 2021:4 & -     &       & break dates &       & \multicolumn{1}{p{3.455em}}{2020:2\newline{}2021:4} & 2001:3 & \multicolumn{1}{p{3.455em}}{2009:3\newline{}2010:4} & 2021:4 & - \\
    SC &       & -0.002 & 0.140 & 0.108 & 0.098 & 0.119 &       & SC &       & 0.551 & 0.583 & 0.569 & 0.532 & 0.516 \\
    Abs SC &       & 0.183 & 0.289 & 0.313 & 0.266 & 0.262 &       & Abs SC &       & 0.552 & 0.583 & 0.569 & 0.533 & 0.516 \\
    XC &       & 0.002 & 0.003 & 0.002 & 0.000 & 0.000 &       & XC &       & 0.003 & 0     & 0.002 & 0.001 & 0.006 \\
    Abs XC &       & 0.006 & 0.006 & 0.002 & 0.000 & 0.000 &       & Abs XC &       & 0.038 & 0     & 0.015 & 0.001 & 0.022 \\
          &       & \multicolumn{5}{c}{HHLB}              &       &       &       & \multicolumn{5}{c}{HHLB} \\
    ER    &       & 1     & 2     & 1     & 1     & 1     &       & ER    &       & 1     & 2     & 1     & 1     & 1 \\
    \# breaks &       & 2     & 1     & 2     & 1     & 0     &       & \# breaks &       & 2     & 1     & 2     & 1     & 0 \\
    break dates &       & \multicolumn{1}{p{3.59em}}{2020:2\newline{}2021:4} & 2001:3 & \multicolumn{1}{p{3.59em}}{2009:3\newline{}2010:4} & 2021:4 & -     &       & break dates &       & \multicolumn{1}{p{3.455em}}{2020:2\newline{}2021:4} & 2001:3 & \multicolumn{1}{p{3.455em}}{2009:3\newline{}2010:4} & 2021:4 & - \\
    SC &       & -0.002 & 0.138 & 0.107 & 0.099 & 0.122 &       & SC &       & 0.555 & 0.583 & 0.572 & 0.538 & 0.519 \\
    Abs SC &       & 0.184 & 0.291 & 0.311 & 0.266 & 0.259 &       & Abs SC &       & 0.555 & 0.583 & 0.572 & 0.539 & 0.519 \\
    XC &       & 0.002 & 0.003 & 0.001 & 0.000 & 0.000 &       & XC &       & 0.005 & 0.000 & 0.003 & 0.001 & 0.006 \\
    Abs XC &       & 0.007 & 0.006 & 0.001 & 0.000 & 0.000 &       & Abs XC &       & 0.038 & 0.001 & 0.029 & 0.002 & 0.026 \\
          &       & \multicolumn{5}{c}{REER42}            &       &       &       & \multicolumn{5}{c}{REER42} \\
    ER    &       & 1     & 2     & 1     & 1     & 1     &       & ER    &       & 1     & 2     & 1     & 1     & 1 \\
    \# breaks &       & 2     & 1     & 2     & 1     & 0     &       & \# breaks &       & 2     & 1     & 2     & 1     & 0 \\
    break dates &       & \multicolumn{1}{p{3.59em}}{2020:2\newline{}2021:4} & 2001:3 & \multicolumn{1}{p{3.59em}}{2009:3\newline{}2010:4} & 2021:4 & -     &       & break dates &       & \multicolumn{1}{p{3.455em}}{2020:2\newline{}2021:4} & 2001:3 & \multicolumn{1}{p{3.455em}}{2009:3\newline{}2010:4} & 2021:4 & - \\
    SC &       & -0.004 & 0.134 & 0.104 & 0.093 & 0.111 &       & SC &       & 0.556 & 0.601 & 0.622 & 0.529 & 0.524 \\
    Abs SC &       & 0.178 & 0.283 & 0.309 & 0.267 & 0.259 &       & Abs SC &       & 0.556 & 0.601 & 0.622 & 0.529 & 0.524 \\
    XC &       & 0.002 & 0.002 & 0.002 & 0.001 & 0.000 &       & XC &       & 0.005 & 0     & 0.001 & 0.002 & 0.001 \\
    Abs XC &       & 0.008 & 0.020 & 0.002 & 0.001 & 0.000 &       & Abs XC &       & 0.038 & 0     & 0.006 & 0.024 & 0.001 \\
          &       & \multicolumn{5}{c}{HPRC}              &       &       &       & \multicolumn{5}{c}{HPRC} \\
    ER    &       & 1     & 2     & 1     & 1     & 1     &       & ER    &       & 1     & 2     & 1     & 1     & 1 \\
    \# breaks &       & 2     & 1     & 2     & 1     & 0     &       & \# breaks &       & 2     & 1     & 2     & 1     & 0 \\
    break dates &       & \multicolumn{1}{p{3.59em}}{2020:2\newline{}2021:4} & 2001:3 & \multicolumn{1}{p{3.59em}}{2009:3\newline{}2010:4} & 2021:4 & -     &       & break dates &       & \multicolumn{1}{p{3.455em}}{2020:2\newline{}2021:4} & 2001:3 & \multicolumn{1}{p{3.455em}}{2009:3\newline{}2010:4} & 2021:4 & - \\
    SC &       & 0.002 & 0.139 & 0.109 & 0.101 & 0.122 &       & SC &       & 0.549 & 0.558 & 0.568 & 0.523 & 0.512 \\
    Abs SC &       & 0.180 & 0.289 & 0.308 & 0.263 & 0.257 &       & Abs SC &       & 0.550 & 0.558 & 0.568 & 0.523 & 0.512 \\
    XC &       & 0.002 & 0.003 & 0.002 & 0.000 & 0.000 &       & XC &       & 0.001 & 0     & 0.003 & 0.002 & 0.005 \\
    Abs XC &       & 0.007 & 0.007 & 0.002 & 0.000 & 0.000 &       & Abs XC &       & 0.004 & 0     & 0.021 & 0.002 & 0.019 \\
          &       & \multicolumn{5}{c}{SHIX}              &       &       &       & \multicolumn{5}{c}{SHIX} \\
    ER    &       & 1     & 2     & 1     & 1     & 1     &       & ER    &       & 1     & 2     & 1     & 1     & 1 \\
    \# breaks &       & 2     & 1     & 2     & 1     & 0     &       & \# breaks &       & 2     & 1     & 2     & 1     & 0 \\
    break dates &       & \multicolumn{1}{p{3.59em}}{2020:2\newline{}2021:4} & 2001:3 & \multicolumn{1}{p{3.59em}}{2009:3\newline{}2010:4} & 2021:4 & -     &       & break dates &       & \multicolumn{1}{p{3.455em}}{2020:2\newline{}2021:4} & 2001:3 & \multicolumn{1}{p{3.455em}}{2009:3\newline{}2010:4} & 2021:4 & - \\
    SC &       & -0.004 & 0.134 & 0.107 & 0.094 & 0.109 &       & SC &       & 0.538 & 0.609 & 0.602 & 0.559 & 0.522 \\
    Abs SC &       & 0.178 & 0.290 & 0.309 & 0.267 & 0.260 &       & Abs SC &       & 0.539 & 0.609 & 0.602 & 0.559 & 0.522 \\
    XC &       & 0.002 & 0.002 & 0.002 & 0.000 & 0.000 &       & XC &       & 0.006 & 0.000 & 0.004 & 0.005 & 0.008 \\
    Abs XC &       & 0.008 & 0.006 & 0.003 & 0.000 & 0.000 &       & Abs XC &       & 0.044 & 0.000 & 0.027 & 0.029 & 0.032 \\
          &       &       &       &       &       &       &       &       &       &       &       &       &       &  \\
    \bottomrule
    \end{tabular}}%
    \caption{\footnotesize "ER" refers to the estimated number of factors based on the eigenvalue ratio of \cite{ahn2013eigenvalue}. "\# breaks" and "break dates" are obtained using the sup-LM test statistics of \cite{breakschendoladogonzalo}. "SC" and "Abs SC" refer to the average and average absolute serial correlations of order 1 in the idiosyncratic panel data residuals that are significant at a 5\% level under the Ljung-Box tests. Similarly, "XC" and "Abs XC" refer to the average and average absolute pair-wise correlations in the idiosyncratic panel data residuals using the thresholding procedure of \cite{CaiLiu2011}.}
  \label{tab: SecondHalf}%
\end{table}
\newpage
\bibliographystyle{apalike}
\bibliography{refs}
\end{document}